
\input harvmac
\input amssym.def
\input amssym.tex
\input epsf.tex

\font\teneurm=eurm10 \font\seveneurm=eurm7 \font\fiveeurm=eurm5
\newfam\eurmfam
\textfont\eurmfam=\teneurm \scriptfont\eurmfam=\seveneurm
\scriptscriptfont\eurmfam=\fiveeurm
\def\eurm#1{{\fam\eurmfam\relax#1}}

\font\teneusm=eusm10 \font\seveneusm=eusm7 \font\fiveeusm=eusm5
\newfam\eusmfam
\textfont\eusmfam=\teneusm
\scriptfont\eusmfam=\seveneusm
\scriptscriptfont\eusmfam=\fiveeusm
\def\eusm#1{{\fam\eusmfam\relax#1}}

\font\tencmmib=cmmib10 \skewchar\tencmmib='177
\font\sevencmmib=cmmib7 \skewchar\sevencmmib='177
\font\fivecmmib=cmmib5 \skewchar\fivecmmib='177
\newfam\cmmibfam
\textfont\cmmibfam=\tencmmib \scriptfont\cmmibfam=\sevencmmib
\scriptscriptfont\cmmibfam=\fivecmmib
\def\cmmib#1{{\fam\cmmibfam\relax#1}}

\font\fivebf=cmbx10 scaled 500 
\font\sevenbf=cmbx10 scaled 700 
\font\tenbf=cmbx10             
\font\fivemb=cmmib10 scaled 500 
\font\sevenmb=cmmib10 scaled 700 
\font\tenmb=cmmib10      
\def\boldmath{\textfont0=\tenbf\scriptfont0=\sevenbf\scriptscriptfont0=\fivebf
\textfont1=\tenmb\scriptfont1=\sevenmb\scriptscriptfont1=\fivemb}

\def\^{{\wedge}}
\def\*{{\star}}
\def\bar{\overline}

\def\ad#1{{\rm ad}#1}
\def\e#1{{\rm e}^{\, #1}}
\def\wt#1{\widetilde#1}

\def\A{{\rm A}}
\def\B{{\rm B}}

\def\Cl{\mathop{\rm Cl}}

\def\Hom{\mathop{\rm Hom}}
\def\Td{\mathop{\rm Td}}
\def\Lie{\mathop{\rm Lie}}
\def\Vol{\mathop{\rm Vol}}
\def\Map{\mathop{\rm Map}}
\def\Sym{\mathop{\rm Sym}}

\def\diag{\mathop{\rm diag}}
\def\mod{\mathop{\rm mod}}
\def\Res{\mathop{\rm Res}}
\def\sgn{\mathop{\rm sign}}
\def\rk{\mathop{\rm rk}}
\def\ch{{\mathop{\rm ch}}}
\def\lk{\mathop{\rm lk}}

\def\BC{{\Bbb C}}

\def\BP{{\Bbb P}}

\def\BR{{\Bbb R}}

\def\BW{{\Bbb W}}
\def\BZ{{\Bbb Z}}

\def\CA{{\cal A}}
\def\CC{{\cal C}}
\def\CD{{\cal D}}
\def\CE{{\cal E}}
\def\CF{{\cal F}}
\def\CG{{\cal G}}
\def\CH{{\cal H}}

\def\CL{{\cal L}}

\def\CN{{\cal N}}
\def\CO{{\cal O}}

\def\CS{{\cal S}}

\def\CZ{{\cal Z}}

\def\Fe{{\frak e}}
\def\Fg{{\frak g}}
\def\Fh{{\frak h}}

\def\Fs{{\frak s}}
\def\Ft{{\frak t}}
\def\Fu{{\frak u}}

\def\FC{{\frak C}}
\def\FL{{\frak L}}

\def\FR{{\frak R}}
\def\FW{{\frak W}}

\def\SF{{\eusm F}}
\def\SH{{\eusm H}}
\def\SJ{{\eusm J}}
\def\SK{{\eusm K}}
\def\SL{{\eusm L}}
\def\SM{{\eusm M}}
\def\SN{{\eusm N}}

\def\SW{{\eusm W}}
\def\SX{{\eusm X}}

\def\SV{{\eusm V}}

\def\Ra{{\eurm a}}
\def\Rb{{\eurm b}}
\def\Rc{{\eurm c}}
\def\Rd{{\eurm d}}
\def\Re{{\eurm e}}
\def\Rf{{\eurm f}}

\def\Rl{{\eurm l}}
\def\Rm{{\eurm m}}

\def\Rq{{\eurm q}}

\def\Rs{{\eurm s}}
\def\Rt{{\eurm t}}
\def\Ru{{\eurm u}}
\def\Rx{{\eurm x}}
\def\Ry{{\eurm y}}

\def\RC{{\eurm C}}
\def\RD{{\eurm D}}

\def\RL{{\eurm L}}

\def\RP{{\eurm P}}
\def\RR{{\eurm R}}
\def\RS{{\eurm S}}

\def\RV{{\eurm V}}

\def\RX{{\eurm X}}
\def\RY{{\eurm Y}}
\def\RZ{{\eurm Z}}

\def\CMj{{\cmmib j}}

\noblackbox

\def\urlfont{\hyphenpenalty=10000 \hyphenchar\tentt='057 \tt}

\newbox\tmpbox\setbox\tmpbox\hbox{\abstractfont YITP-SB-09-26}
\Title{\vbox{\baselineskip12pt\hbox{\urlfont
arXiv:0911.2687}\hbox{YITP-SB-09-26}}}
{\vbox{
\centerline{Localization For Wilson Loops}
\smallskip
\centerline{In Chern-Simons Theory}}}
\smallskip
\centerline{Chris Beasley}
\medskip
\centerline{\it{Simons Center for Geometry and Physics}}
\centerline{\it{Stony Brook University}}
\centerline{\it{Stony Brook, NY 11794-3636}}
\bigskip\bigskip

We reconsider Chern-Simons gauge theory on a Seifert manifold $M$, which
is the total space of a nontrivial circle bundle over a Riemann surface
$\Sigma$, possibly with orbifold points.  As shown in previous work
with Witten, the path integral technique of non-abelian localization
can be used to express the partition function of Chern-Simons theory
in terms of the equivariant cohomology of the moduli space of flat
connections on $M$.  Here we extend this result to apply to the
expectation values of Wilson loop operators which wrap the circle
fibers of $M$ over $\Sigma$.  Under localization, such a Wilson loop
operator reduces naturally to the Chern character of an associated
universal bundle over the moduli space.  Along the way, we demonstrate
that the stationary-phase approximation to the Wilson loop path
integral is exact for torus knots in $S^3$, an observation made
empirically by Lawrence and Rozansky prior to this work.

\Date{November 2009}

\lref\AdamsZC{
D.~H.~Adams, ``The Semi-Classical Approximation for the Chern-Simons Partition  Function,'' Phys.\ Lett.\  B {\bf 417} (1998) 53--60, {\urlfont hep-th/9709147}.}

\lref\AganagicDH{
M.~Aganagic, A.~Neitzke, and C.~Vafa, ``BPS Microstates and the Open
Topological String Wave Function,'' {\urlfont hep-th/0504054}.}

\lref\AganagicJS{
M.~Aganagic, H.~Ooguri, N.~Saulina and C.~Vafa,
``Black Holes, q-Deformed 2d Yang-Mills, and Non-Perturbative
Topological Strings,'' {\urlfont hep-th/0411280}.}

\lref\AlekseevVX{
A.~Alekseev, L.~D.~Faddeev, and S.~L.~Shatashvili, ``Quantization of
Symplectic Orbits of Compact Lie Groups by Means of the Functional
Integral,'' J.\ Geom.\ Phys.\ {\bf 5} (1989) 391--406.}

\lref\AlvarezZV{
O.~Alvarez, I.~M.~Singer, and P.~Windey, ``Quantum Mechanics and the
Geometry of the Weyl Character Formula,'' Nucl.\ Phys.\  B {\bf 337}
(1990) 467--486.}

\lref\Apostol{
T.~Apostol, {\it Modular Functions and Dirichlet Series in Number
Theory}, Springer-Verlag, New York, 1976.}

\lref\ApostolII{
T.~Apostol, {\it Introduction to Analytic Number Theory},
Springer-Verlag, New York, 1976.}

\lref\ArsiwallaJB{
X.~Arsiwalla, R.~Boels, M.~Mari\~no, and A.~Sinkovics, ``Phase
Transitions in $q$-deformed 2d Yang-Mills Theory and Topological
Strings,'' Phys.\ Rev.\ D {\bf 73} (2006) 026005, {\urlfont
hep-th/0509002}.}

\lref\AtiyahCS{
M.~F. Atiyah, ``Circular Symmetry and Stationary-Phase
Approximation,'' in {\it Colloquium in Honor of Laurent Schwartz},
Vol. 1, Ast\'erisque {\bf 131} (1985) 43--59.}

\lref\AtiyahKN{
M.~Atiyah, {\it The Geometry and Physics of Knots}, Cambridge
University Press, Cambridge, 1990.}

\lref\AtiyahRB{
M.~Atiyah and R.~Bott,
``The Moment Map and Equivariant Cohomology,''
Topology {\bf 23} (1984) 1--28.}

\lref\AtiyahYM{
M.~Atiyah and R.~Bott,
``Yang-Mills Equations Over Riemann Surfaces,''
Phil. Trans. R. Soc. Lond. {\bf A308} (1982) 523--615.}

\lref\AtiyahFM{
M.~Atiyah, ``On Framings of Three-Manifolds'', Topology {\bf 29}
(1990) 1--7.}

\lref\APS{
M.~F.~Atiyah, V.~Patodi, and I.~Singer, ``Spectral Asymmetry and
Riemannian Geometry, I, II, III,''  Math. Proc. Camb. Phil. Soc. {\bf
77} (1975) 43--69;  {\bf 78} (1975) 405--432;  {\bf 79} (1976)
71--99.}

\lref\AtiyahRBII{
M.~F.~Atiyah and R.~Bott,
``A Lefschetz Fixed Point Formula for Elliptic Complexes, II,'' 
Ann. of Math. {\bf 88} (1968) 451--491.}

\lref\BalachandranUB{
 A.~P.~Balachandran, S.~Borchardt, and A.~Stern,
 ``Lagrangian And Hamiltonian Descriptions of Yang-Mills Particles,''
 Phys.\ Rev.\ D {\bf 17} (1978) 3247--3256.}
 
\lref\BaulieuNJ{
L.~Baulieu, A.~Losev, and N.~Nekrasov, ``Chern-Simons and Twisted Supersymmetry in Various Dimensions,'' Nucl.\ Phys.\  B {\bf 522} (1998) 82--104, {\urlfont hep-th/9707174}.}

\lref\BeasleyVF{
C.~Beasley and E.~Witten, ``Non-Abelian Localization For Chern-Simons
Theory,'' J.\ Differential Geom.\ {\bf 70} (2005) 183--323, {\urlfont
hep-th/0503126}.}

\lref\Blair{
D.~E.~Blair, {\it Riemannian Geometry of Contact and Symplectic
Manifolds}, Birkh\"auser, Boston, 2002.}

\lref\BlauGH{
M.~Blau and G.~Thompson, ``Chern-Simons Theory on $S^1$-bundles:
Abelianisation and q-Deformed Yang-Mills Theory,'' JHEP {\bf 0605}
(2006) 003, {\urlfont hep-th/0601068}.}

\lref\BlauHJ{
M.~Blau and G.~Thompson, ``Lectures on 2-d Gauge Theories: Topological
Aspects and Path Integral Techniques,'' in {\it Proceedings of the
1993 Trieste Summer School in High Energy Physics and Cosmology}, Ed. by
E.~Gava et al., ICTP Series in Theoretical Physics, vol.~10, World
Scientific, Hackensack, NJ, 1994, {\urlfont hep-th/9310144}.}

\lref\BlauRS{
M.~Blau and G.~Thompson,
``Localization and Diagonalization: A Review of Functional Integral
Techniques For Low Dimensional Gauge Theories and Topological Field
Theories,'' J.\ Math.\ Phys.\  {\bf 36} (1995) 2192--2236, {\urlfont
hep-th/9501075}.}

\lref\Boden{
H.~Boden, C.~Herald, P.~Kirk, and E.~Klassen, ``Gauge Theoretic
Invariants of Dehn Surgeries on Knots,'' Geom.\ Topol.\ {\bf 5} (2001)
143--226, {\urlfont math.GT/9908020}.}

\lref\BottBW{
R.~Bott, ``Homogeneous Vector Bundles,'' Ann.\ of Math.\ {\bf 66}
(1957) 203--248.}

\lref\BottT{
R.~Bott and L.~Tu, {\it Differential Forms in Algebraic Topology},
Springer-Verlag, New York, 1982.}

\lref\BuscherT{
T.~Buscher, ``Path Integral Derivation of Quantum Duality in Nonlinear
Sigma Models,''  Phys. Lett. {\bf B201} (1988) 466--472.}

\lref\CaporasoFP{
N.~Caporaso, M.~Cirafici, L.~Griguolo, S.~Pasquetti, D.~Seminara, and
R.~J.~Szabo, ``Topological Strings and Large $N$ Phase Transitions,
II: Chiral Expansion of $q$-deformed Yang-Mills Theory,'' JHEP {\bf
0601} (2006) 036, {\urlfont hep-th/0511043}.}

\lref\CaporasoTA{
N.~Caporaso, M.~Cirafici, L.~Griguolo, S.~Pasquetti, D.~Seminara, and
R.~J.~Szabo, ``Topological Strings and Large $N$ Phase Transitions, I:
Nonchiral Expansion of $q$-deformed Yang-Mills Theory,'' JHEP {\bf
0601} (2006) 035, {\urlfont hep-th/0509041}.}

\lref\CordesFC{
S.~Cordes, G.~W.~Moore and S.~Ramgoolam,
``Lectures on 2-d Yang-Mills Theory, Equivariant Cohomology and
Topological Field Theories,''
Nucl.\ Phys.\ Proc.\ Suppl.\  {\bf 41} (1995) 184--244,
{\urlfont hep-th/9411210}.}

\lref\DaskW{
G.~Daskalopoulos and R.~Wentworth, ``Geometric Quantization for the
Moduli Space of Vector Bundles with Parabolic Structure,'' in {\it
Geometry, Topology, and Physics (Campinas, 1996)}, Ed. B.~Apanasov et
al., pp.~119--155, de Gruyter, Berlin, 1997.}

\lref\deHaroRN{
S.~de Haro, ``A Note on Knot Invariants and $q$-deformed 2d
Yang-Mills,'' Phys.\ Lett.\ B {\bf 634} (2006) 78--83, {\urlfont
hep-th/0509167}.}

\lref\deHaroRZ{
S.~de Haro and M.~Tierz, ``Discrete and Oscillatory
Matrix Models in Chern-Simons Theory,'' Nucl.\ Phys.\ {\bf B} 731
(2005) 225--241, {\urlfont hep-th/0501123}.}

\lref\deHaroWN{
S.~de Haro, ``Chern-Simons Theory, 2d Yang-Mills, and
Lie Algebra Wanderers,'' Nucl.\ Phys.\ {\bf B} 730 (2005) 312--351,
{\urlfont hep-th/0412110}.}

\lref\deHaroUZ{
S.~de Haro, ``Chern-Simons Theory in Lens Spaces From
2d Yang-Mills on the Cylinder,'' JHEP {\bf 0408} (2004) 041, {\urlfont
hep-th/0407139}.}

\lref\deHaroID{
S.~de Haro and M.~Tierz, ``Brownian Motion, Chern-Simons Theory, and
2d Yang-Mills,'' Phys.\ Lett.\ {\bf B601} (2004) 201--208, {\urlfont
hep-th/0406093}.}

\lref\deHaroWQ{
S.~de Haro, S.~Ramgoolam, and A.~Torrielli, ``Large-$N$ Expansion of
$q$-Deformed Two-Dimensional Yang-Mills Theory and Hecke Algebras,''
Commun.\ Math.\ Phys.\ {\bf 273} (2007) 317--355, {\urlfont
hep-th/0603056}.}

\lref\Deligne{
E.~Witten, ``Dynamics of Quantum Field Theory,'' in {\it Quantum Fields
and Strings: A Course for Mathematicians, Vol. 2}, Ed. by P. Deligne et al.,
American Mathematical Society, Providence, Rhode Island, 1999.}

\lref\DeligneII{
E.~Witten, ``Index of Dirac Operators,'' in {\it Quantum Fields and
Strings: A Course for Mathematicians, Vol. 1}, Ed. by P. Deligne et al.,
American Mathematical Society, Providence, Rhode Island, 1999.}

\lref\DiakonovFC{
D.~Diakonov and V.~Y.~Petrov, ``A Formula for the Wilson Loop,''
Phys.\ Lett.\ B {\bf 224} (1989) 131--135.}

\lref\DiFrancesco{
P.~Di~Francesco, P.~Mathieu, and D.~S\'en\'echal, {\it Conformal Field
Theory}, Springer-Verlag, New York, 1997.}

\lref\DijkgraafSB{
R.~Dijkgraaf and H.~Fuji, ``The Volume Conjecture and Topological
Strings,'' {\urlfont arXiv:0903.2084 [hep-th]}.}

\lref\Donaldson{
S.~K.~Donaldson, ``Moment Maps and Diffeomorphisms,'' in {\it Sir
Michael Atiyah: A Great Mathematician of the Twentieth Century}, Asian
J.\ Math.\ {\bf 3} (1999) 1--15.}

\lref\DonaldsonII{
S.~K.~Donaldson, ``Gluing Techniques in the Cohomology of Moduli
Spaces,'' in {\it Topological Methods in Modern Mathematics}, Ed. by
L.~Goldberg and A.~Phillips, pp. 137--170, Publish or Perish, Houston,
Texas, 1993.}

\lref\Drezet{
J.-M. Drezet and M.~S.~Narasimhan, ``Groupe de Picard des vari\'et\'es
de modules de fibr\'es semi-stables sur les courbes alg\'ebriques,''
Invent.\ Math.\ {\bf 97} (1989) 53--94.}

\lref\Dubois{
J.~Dubois and R.~Kashaev, ``On the Asymptotic Expansion of the Colored
Jones Polynomial for Torus Knots,'' Math.\ Ann.\ {\bf 339} (2007)
757--782, {\urlfont math.GT/0510607}.}

\lref\Duistermaat{
J.~J.~Duistermaat and G.~J.~Heckman, ``On the Variation in the
Cohomology of the Symplectic Form of the Reduced Phase Space,''
Invent.\ Math.\ {\bf 69} (1982) 259--268; Addendum, Invent.\ Math.\
{\bf 72} (1983) 153--158.}

\lref\ElitzurNR{
S.~Elitzur, G.~Moore, A.~Schwimmer, and N.~Seiberg,  ``Remarks on
the Canonical Quantization of the Chern-Simons-Witten Theory,''
Nucl.\ Phys.\ B {\bf 326} (1989) 108--134.}

\lref\Etnyre{
J.~B.~Etnyre,
``Introductory Lectures on Contact Geometry,'' in {\it Topology and
Geometry of Manifolds (Athens, GA 2001)}, Proc.\ Sympos.\ Pure Math.\
{\bf 71}, Amer.\ Math.\ Soc., Providence, RI, 2003,
{\urlfont math.SG/0111118}.}

\lref\EtnyreII{
J.~B.~Etnyre,
``Legendrian and Transversal Knots,'' in {\it Handbook of Knot
Theory}, Ed. W.~Menasco and M.~Thistlethwaite, pp.~105--185,
Elsevier, Amsterdam, 2005, {\urlfont math.SG/0306256}.}

\lref\FreedJQ{
D.~S.~Freed, ``Remarks on Chern-Simons Theory,'' Bull.\ Amer.\ Math.\
Soc.\ {\bf 46} (2009) 221--254, {\urlfont arXiv:0808.2507~[math.AT]}.}

\lref\FreedRG{
D.~Freed and R.~Gompf, ``Computer Calculation of Witten's 3-Manifold
Invariant,'' Commun.\ Math.\ Phys.\  {\bf 141} (1991) 79--117.}

\lref\Fulton{
W.~Fulton and J.~Harris, {\it Representation Theory: A First Course},
Springer, New York, 1991.}

\lref\Furuta{
M.~Furuta and B. Steer, ``Seifert Fibred Homology 3-Spheres and the
Yang-Mills Equations on Riemann Surfaces with Marked Points,'' Adv. in
Math. {\bf 96} (1992) 38--102.}

\lref\GeigesHJ{
H.~Geiges, ``Contact Geometry,'' in {\it Handbook of Differential
Geometry, Vol. II}, Ed. by F.~Dillen and L.~Verstraelen, pp.~315--382,
Elsevier, Amsterdam, 2006, {\urlfont math.SG/0307242}.}

\lref\Griffiths{
P.~Griffiths and J.~Harris, {\it Principles of Algebraic Geometry},
John Wiley and Sons, Inc., New York, 1978.}

\lref\GuilleminS{
V.~Guillemin and S.~Sternberg, {\it Symplectic Techniques in Physics},
Cambridge University Press, Cambridge, 1984.}

\lref\GuilleminSII{
V.~Guillemin and S.~Sternberg, {\it Supersymmetry and Equivariant de
Rham Theory}, Springer, Berlin, 1999.}

\lref\HahnAT{
A.~Hahn, ``Chern-Simons Models on $S^2\times S^1$, Torus Gauge Fixing,
and Link Invariants I,II'' J.\ Geom.\ Phys.\ {\bf 53} (2005) 275--314;
J.\ Geom.\ Phys.\ {\bf 58} (2008) 1124--1136.}

\lref\Hamilton{
M.~Hamilton and L.~Jeffrey, ``Symplectic Fibrations and Riemann-Roch
Numbers of Reduced Spaces,'' Q.\ J.\ Math {\bf 56} (2005) 541--552, 
{\urlfont math.SG/0403004}.}

\lref\Hirzebruch{
F.~Hirzebruch, {\it Topological Methods in Algebraic Geometry}, 3rd
ed., Springer-Verlag, New York, 1966.}

\lref\ItzhakiRC{
N.~Itzhaki, ``Anyons, 't Hooft Loops, and a Generalized Connection in
Three Dimensions,'' Phys.\ Rev.\ D {\bf 67} (2003) 065008, {\urlfont
hep-th/0211140}.}

\lref\JafferisJD{
D.~Jafferis and J.~Marsano, ``A DK Phase Transition in $q$-deformed
Yang-Mills on $S^2$ and Topological Strings,'' {\urlfont
hep-th/0509004}.}

\lref\JeffreyLC{
L.~Jeffrey, ``Chern-Simons-Witten Invariants of Lens Spaces and Torus
Bundles, and the Semiclassical Approximation,'' Commun.\ Math.\ Phys.\
{\bf 147} (1992) 563--604.}

\lref\JeffreyLCII{
L.~Jeffrey, ``Extended Moduli Spaces of Flat Connections on Riemann
Surfaces,'' Math.~Ann. {\bf 298} (1994) 667--692.}

\lref\JeffreyKW{
L.~Jeffrey and F.~Kirwan, ``Intersection Theory on Moduli Spaces of
Holomorphic Bundles of Arbitrary Rank on a Riemann Surface,'' Ann. of
Math. {\bf 148} (1998) 109--196, {\urlfont alg-geom/9608029}.}

\lref\JeffreyPB{
L.~Jeffrey, ``The Verlinde Formula For Parabolic Bundles,'' J. London
Math. Soc. 63 (2001) 754--768, {\urlfont math.AG/0003150}.}

\lref\JeffreyTH{
L.~Jeffrey, {\it On Some Aspects of Chern-Simons Gauge Theory},
D.Phil. thesis, University of Oxford, 1991.}

\lref\JeffreyKY{
L.~Jeffrey, Y-H.~Kiem, F.~Kirwan, and J.~Woolf, ``Intersection
Pairings on Singular Moduli Spaces of Bundles over a Riemann Surface
and Their Partial Desingularisations,'' Transform.\ Groups {\bf 11}
(2006) 439--494, {\urlfont math.AG/0505362}.}

\lref\JeffreyMC{
L.~Jeffrey and B.~McLellan, ``Non-Abelian Localization for $U(1)$
Chern-Simons Theory,'' {\urlfont arXiv:0903.5093 [math.DG]}.}

\lref\JonesVFR{
V.~F.~R.~Jones, ``Hecke Algebra Representations of Braid Groups and
Link Polynomials,'' Ann. of Math.  {\bf 126} (1987) 335-388.}

\lref\Garoufalidis{
S.~Garoufalidis, {\it Relations Among $3$-Manifold Invariants},
Ph.D. thesis, University of Chicago, 1992.}

\lref\GopakumarII{
R.~Gopakumar and C.~Vafa,
``M-theory and Topological Strings. I,''
{\urlfont hep-th/9809187}.}

\lref\GopakumarKI{
R.~Gopakumar and C.~Vafa,
``On the Gauge Theory/Geometry Correspondence,''
Adv.\ Theor.\ Math.\ Phys.\  {\bf 3} (1999) 1415--1443,
{\urlfont hep-th/9811131}.}

\lref\GukovJK{
S.~Gukov and E.~Witten, ``Gauge Theory, Ramification, and the
Geometric Langlands Program,'' in {\it Current Developments in
Mathematics, 2006}, Ed. by D.~Jerison et. al., pp. 35--180, Int.\
Press, Somerville, Massachusetts, 2008, {\urlfont hep-th/0612073}.}

\lref\Hatcher{
A.~Hatcher, {\it Algebraic Topology}, Cambridge University Press, 
Cambridge, 2002.}

\lref\Helgason{
S.~Helgason, {\it Differential Geometry, Lie Groups, and Symmetric
Spaces}, Academic Press, San Diego, 1978.}

\lref\IsidroFZ{
J.~Isidro, J.~Labastida, and A.~Ramallo, ``Polynomials for Torus Links
from Chern-Simons Gauge Theories,'' Nucl.\ Phys.\ B {\bf 398} (1993)
187--236, {\urlfont hep-th/9210124}.}

\lref\LawrenceRZ{
R.~Lawrence and L.~Rozansky,
``Witten-Reshetikhin-Turaev Invariants of Seifert Manifolds,''
Commun.\ Math.\ Phys.\  {\bf 205} (1999) 287--314.}

\lref\KapustinWY{
A.~Kapustin, B.~Willett, and I.~Yaakov, ``Exact Results for Wilson
Loops in Superconformal Chern-Simons Theories with Matter,'' {\urlfont
arXiv:0909.4559 [hep-th]}.}

\lref\KashaevT{
R.~M.~Kashaev and O.~Tirkkonen, ``A Proof of the Volume Conjecture for
Torus Knots,'' J.\ Math.\ Sci\ (N.~Y.) {\bf 115} (2003) 2033--2036,
{\urlfont math.GT/9912210}.}

\lref\Kawasaki{
T.~Kawasaki, ``The Riemann-Roch Theorem for Complex $V$-manifolds,''
Osaka J. Math. {\bf 16} (1979) 151--159.}

\lref\KirillovAA{
A.~A.~Kirillov, {\it Elements of the Theory of Representations},
Trans. by E.~Hewitt, Springer-Verlag, Berlin, 1976.}

\lref\Klassen{
E.~Klassen, ``Representations of Knot Groups in $SU(2)$,''
Trans.\ Amer.\ Math.\ Soc.\ {\bf 326} (1991) 795--828.}

\lref\Labastida{
J.~Labastida, P.~Llatas, and A.~Ramallo, ``Knot Operators in
Chern-Simons Gauge Theory,'' Nucl.\ Phys.\ B {\bf 348} (1991)
651--692.}

\lref\MarinoFK{
M.~Mari\~no,
``Chern-Simons Theory, Matrix Integrals, and Perturbative Three-Manifold
Invariants,''
Commun.\ Math.\ Phys.\ {\bf 253} (2004) 25--49,
{\urlfont hep-th/0207096}.}

\lref\Martinet{
J.~Martinet, ``Formes de contact sur les variet\'et\'es de dimension
3,'' Springer Lecture Notes in Math {\bf 209} (1971) 142--163.}

\lref\MehtaSH{
V.~B.~Mehta and C.~S.~Seshadri, ``Moduli of Vector Bundles on Curves
with Parabolic Structures,'' Math.\ Ann.\ {\bf 248} (1980) 205--239.} 

\lref\MigdalZG{
A.~A.~Migdal,
``Recursion Equations In Gauge Field Theories,''
Zh.\ Eksp.\ Teor.\ Fiz.\  {\bf 69} (1975) 810--822
[Sov.\ Phys.\ JETP {\bf 42} (1975) 413--418].}

\lref\Moser{
L.~Moser, ``Elementary Surgery Along a Torus Knot,'' Pacific J.\ Math
{\bf 38} (1971) 737--745.}

\lref\Murakami{
H.~Murakami, ``An Introduction to the Volume Conjecture and Its
Generalizations,'' Acta Math.\ Vietnam.\ {\bf 33} (2008) 219--253,
{\urlfont arXiv:0802.0039 [math.GT]}.}

\lref\Narasimhan{
M.~Narasimhan and C.~Seshadri, ``Stable and Unitary Vector Bundles on
a Compact Riemann Surface,'' Ann.\ of Math.\ {\bf 82} (1965) 540--567.}

\lref\NekrasovNA{
N.~Nekrasov, {\it Four-Dimensional Holomorphic Theories}, Ph.D. thesis, Princeton University, 1996.}

\lref\Nicolaescu{
L.I.~Nicolaescu, ``Finite Energy Seiberg-Witten Moduli Spaces on
Four-Manifolds Bounding Seifert Fibrations,''  Comm.\ Anal.\
Geom. {\bf 8} (2000) 1027--1096, {\urlfont dg-ga/9711006}.}

\lref\OrlikPK{
P.~Orlik, {\it Seifert Manifolds},
Lecture Notes in Mathematics {\bf 291}, Ed. by A. Dold and B. Eckmann,
Springer-Verlag, Berlin, 1972.}

\lref\Ohtsuki{
``Problems on Invariants of Knots and 3-Manifolds,'' Ed. by T.~Ohtsuki
with an introduction by J.~Roberts, in {\it Invariants of Knots and
3-Manifolds (Kyoto, 2001)}, pp. 377--572, Geom. Topol. Monogr. 4,
Geom. Topol. Publ., Coventry, 2002, {\urlfont math.GT/0406190}.}

\lref\ParadanPE{
P.-E.~Paradan, ``The Moment Map and Equivariant Cohomology with
Generalized Coefficients,'' Topology {\bf 39} (2000) 401--444.}

\lref\PestunRZ{
V.~Pestun, ``Localization of Gauge Theory on a Four-Sphere and
Supersymmetric Wilson Loops,'' {\urlfont arXiv:0712.2824 [hep-th]}.}

\lref\PressleySG{
A.~Pressley and G.~Segal, {\it Loop Groups}, Clarendon Press, Oxford,
1986.}

\lref\ReshetikhinTU{
N.~Reshetikhin and V.~G.~Turaev, ``Invariants of $3$-Manifolds via
Link Polynomials and Quantum Groups,'' Invent.\ Math.\ {\bf 103} (1991)
547--597.}

\lref\RocekPS{
M.~Rocek and E.~Verlinde, ``Duality, Quotients, and Currents,''
Nucl.\ Phys.\ B {\bf 373} (1992) 630--646,
{\urlfont hep-th/9110053}.}

\lref\RozanskyL{
L. Rozansky, ``A Large $k$ Asymptotics of Witten's Invariant of Seifert
Manifolds,'' Comm.\ Math.\ Phys.\ {\bf 171} (1995) 279--322, {\urlfont
hep-th/9303099}.}

\lref\RozanskyWV{
L.~Rozansky,
``Residue Formulas for the Large $k$ Asymptotics of Witten's Invariants of
Seifert Manifolds: The Case of SU(2),''
Commun.\ Math.\ Phys.\  {\bf 178} (1996) 27--60, {\urlfont
hep-th/9412075}.}

\lref\SatakeI{
I.~Satake, ``On a Generalization of the Notion of Manifold,''
Proc. Nat. Acad. Sci. USA {\bf 42} (1956) 359--363.}

\lref\SatakeII{
I.~Satake, ``The Gauss-Bonnet Theorem for V-manifolds,''
J. Math. Soc. Japan {\bf 9} (1957) 464--492.}

\lref\StoneFU{
M.~Stone, ``Supersymmetry and the Quantum Mechanics of Spin,'' 
Nucl.\ Phys.\  B {\bf 314} (1989) 557--586.}

\lref\TelemanC{
C.~Teleman, ``$K$-theory of the Moduli Space of Bundles on a Surface
and Deformations of the Verlinde Algebra,'' in {\it Topology,
Geometry, and Quantum Field Theory: Proceedings of the Symposium in
Honour of the 60th Birthday of Graeme Segal}, Ed. by U.~Tillmann, LMS
Lecture Notes 308, Cambridge University Press, Cambridge, 2004.}

\lref\TelemanCW{
C.~Teleman and C.~T.~Woodward, ``The Index Formula for the Moduli of
$G$-bundles,'' {\urlfont math.AG/0312154}.}

\lref\TelemanQC{
C.~Teleman, ``The Quantization Conjecture Revisited,'' Ann.\ of Math.\
{\bf 152} (2000) 1--43, {\urlfont math.AG/9808029}.}

\lref\Weitsman{
J.~Weitsman, ``Fermionization and Convergent Perturbation Expansions
in Chern-Simons Gauge Theory,'' {\urlfont arXiv:0902.0097 [math-ph]}.}

\lref\WittenFB{
E.~Witten,
``Chern-Simons Gauge Theory as a String Theory,''
Prog.\ Math.\  {\bf 133} (1995) 637--678,
{\urlfont hep-th/9207094}.}

\lref\WittenGF{
E.~Witten, ``On S-Duality in Abelian Gauge Theory,''
Selecta Math.\  {\bf 1} (1995) 383--410,
{\urlfont hep-th/9505186}.}

\lref\WittenHC{
E.~Witten,
``(2+1)-Dimensional Gravity As An Exactly Soluble System,''
Nucl.\ Phys.\ B {\bf 311} (1988) 46--78.}

\lref\WittenHF{
E.~Witten,
``Quantum Field Theory and the Jones Polynomial,''
Commun.\ Math.\ Phys.\  {\bf 121} (1989) 351--399.}

\lref\WittenSX{
E.~Witten,
``Topology Changing Amplitudes In (2+1)-Dimensional Gravity,''
Nucl.\ Phys.\ B {\bf 323} (1989) 113--140.}

\lref\WittenWE{
E.~Witten,
``On Quantum Gauge Theories in Two-Dimensions,''
Commun.\ Math.\ Phys.\  {\bf 141} (1991) 153--209.}

\lref\WittenWF{
E.~Witten,
``Gauge Theories and Integrable Lattice Models,'' 
Nucl.\ Phys.\  B {\bf 322} (1989) 629--697.}

\lref\WittenXU{
E.~Witten,
``Two-dimensional Gauge Theories Revisited,''
J.\ Geom.\ Phys.\  {\bf 9} (1992) 303--368,
{\urlfont hep-th/9204083}.}

\lref\WongVW{
S.~K.~Wong,
``Field And Particle Equations For The Classical Yang-Mills Field And
Particles With Isotopic Spin,'' Nuovo Cim.\ A {\bf 65} (1970)
689--694.}

\lref\Woodhouse{N.~M.~J. Woodhouse, {\it Geometric Quantization,
Second Ed.}, Clarendon Press, Oxford, 1992.}

\lref\WoodwardCT{
C.~T.~Woodward,
``Localization for the Norm-Square of the Moment Map and the
Two-Dimensional Yang-Mills Integral,'' J.\ Symplectic Geom.\ {\bf 3}
(2005) 17--54, {\urlfont math.SG/0404413}.}

\listtoc\writetoc

\newsec{Introduction}

An outstanding issue in the study of Chern-Simons gauge theory has
long been to understand more precisely the Lagrangian formulation of
this theory via the Feynman path integral, in terms of which the
partition function $Z$ at level $k$ is described formally as an
integral over the space of all connections $A$ on a given
three-manifold
$M$,\countdef\M=10\M=\pageno \countdef\AA=11\AA=\pageno
\countdef\k=12\k=\pageno \countdef\Zk=13\Zk=\pageno
\eqn\CSZ{ Z(k) \,=\, \int \! \CD \! A \; \exp{\left[i {k \over {4
\pi}} \int_M \! \Tr\!\left( A \^ d A + {2 \over 3} A \^ A \^ A
\right)\right]}\,.}
Of course, as explained by Witten \WittenHF\ in his foundational work
on Chern-Simons theory, the Hamiltonian formalism can be
alternatively applied to give a completely rigorous \ReshetikhinTU\
and explicitly computable description of the Chern-Simons partition
function in terms of two-dimensional rational conformal field theory
and a ``cut-and-paste'' presentation of $M$ via surgery on links in
$S^3$.  Yet despite its computability, this algebraic definition of
$Z(k)$ as a quantum three-manifold invariant obscures many features
which are manifest in the preceding path integral and which one would
like to understand more deeply.

As a simple example, in the semi-classical limit that $k$ is large, a
naive stationary-phase approximation can be applied to the path
integral, and this approximation implies asymptotic behavior for
$Z(k)$ that is far from evident in the complicated, exact expressions
that arise from conformal field theory --- an observation recently
emphasized by Freed in \FreedJQ.  Nonetheless, the predicted asymptotic
behavior can be checked in examples, as was done early on by Freed and
Gompf \FreedRG, Jeffrey \JeffreyLC, and Garoufalidis \Garoufalidis.  See  \AdamsZC\ for related analysis, and see \S $7$ in \Ohtsuki\ for a survey of continuing work in this area.

For many topological quantum field theories, including most notably 
topological Yang-Mills theory in dimensions two and four (reviewed
nicely in \CordesFC), the semi-classical approximation to the path
integral is actually exact, due to the presence of a conserved,
nilpotent scalar supercharge which is interpreted formally as a BRST
operator in the theory.   As a result, the path integral in those
examples reduces to the integral of an appropriate cohomology class
over a finite-dimensional moduli space of supersymmetric solutions.

In the happy circumstance above, the partition function can be
described not only algebraically in the Hamiltonian formalism, as for
any topological quantum field theory, but also cohomologically in the
Lagrangian formalism, and one might hope to benefit by comparing the
two descriptions.  Indeed, in a spectacular application, Witten
\WittenXU\ used a Hamiltonian computation, originally due to Migdal
\MigdalZG, of the two-dimensional Yang-Mills partition function to
deduce via its path integral interpretation very general results about
the cohomology ring of the moduli space of flat connections on a
Riemann surface.

From this perspective one might wonder to what extent, if any, the
topological observables in Chern-Simons theory also admit a
cohomological interpretation of the sort that arises naturally in
Yang-Mills theory.

The present paper is a sequel to our earlier work \BeasleyVF\ with
Witten in which we gave a partial answer to the preceding question.
There we demonstrated that the partition function of Chern-Simons
theory on a three-manifold $M$ does admit a cohomological
interpretation in the special case that $M$ is a Seifert manifold.
Such a three-manifold can be described succinctly as the total space of a
nontrivial circle bundle over a Riemann surface
$\Sigma$, \countdef\Sig=14\Sig=\pageno
\eqn\SFRT{\matrix{
&S^1\,\longrightarrow\,M\cr
&\mskip 75mu\big\downarrow\lower 0.5ex\hbox{$^\pi$}\cr
&\mskip 65mu\Sigma\cr}\,,}
where $\Sigma$ is generally allowed to have orbifold points and the
circle bundle is allowed to be a corresponding orbifold bundle.

Specifically, the results in \BeasleyVF\ were based upon the technique
of non-abelian localization as applied to the Chern-Simons path
integral in \CSZ.  Very briefly, non-abelian localization
provides a cohomological interpretation for a special class of
symplectic integrals which are intimately related to symmetries.
These integrals take the canonical form 
\eqn\PZSM{ Z(\epsilon) \,=\, \int_X \!
\exp{\!\left[\Omega - {1 \over {2 \epsilon}} \left(\mu,\mu\right)
\right]}\,.}
Here $X$ is an arbitrary symplectic manifold with symplectic
form $\Omega$.  We assume that a Lie group $H$ acts on $X$ in a
Hamiltonian fashion with moment map ${\mu: X \to \Fh^*}$, where $\Fh^*$
is the dual of the Lie algebra $\Fh$ of $H$.  We also introduce an 
invariant quadratic form $(\,\cdot\,,\,\cdot\,)$ on $\Fh$ and dually on
$\Fh^*$ to define the function ${S = \ha (\mu,\mu)}$ appearing in the
integrand of $Z(\epsilon)$.  Finally, $\epsilon$ is a coupling
parameter.

Although it is far from evident that the path integral in \CSZ\ bears
any relation to the canonical symplectic integral in \PZSM, we 
nevertheless explained in \BeasleyVF\ how to recast the Chern-Simons
path integral as such a symplectic integral when $M$ is a Seifert
manifold.  Given this initial and somewhat miraculous step, general
facts about non-abelian localization were then enough to imply the
exactness of the semi-classical approximation to the Chern-Simons path
integral on a Seifert manifold, a result obtained empirically from
known formulae for $Z(k)$ by Lawrence and Rozansky \LawrenceRZ\ and 
Mari\~no \MarinoFK\ prior to our work.  Finally, using localization
we obtained a precise description of the Chern-Simons partition
function in terms of the equivariant cohomology of the moduli space of
flat connections on $M$.

At this point, an obvious further question to ask is whether
non-abelian localization can be applied to give an exact, cohomological 
description for any {\sl other} quantities in Chern-Simons theory
beyond the partition function.  Of course, the other quantities in 
question must be the expectation values of Wilson loop operators, and
the purpose of this paper is to explain how to apply non-abelian
localization to analyze the Chern-Simons path integral including
Wilson loop insertions.

We recall that a Wilson loop operator $W_R(C)$ in any gauge theory
on a manifold $M$ is described by the data of an oriented, closed
curve $C$ \countdef\C=15\C=\pageno which is smoothly embedded in $M$
and which is decorated by an irreducible representation $R$
\countdef\R=17\R=\pageno of the gauge group
$G$. \countdef\G=16\G=\pageno As a classical functional of the
connection $A$, the Wilson loop operator is then 
given simply by the trace in $R$ of the holonomy\foot{Because we work
in conventions for which ${d_A = d + A}$ is the covariant derivative, 
the holonomy of $A$ around $C$ is given by ${P\exp{\!\left(-\oint_C
A\right)}}$, with the minus sign as above.} of $A$ around $C$,
\eqn\DEFWVC{ W_R(C) \,=\, \Tr_R \, P\exp{\!\left(-\oint_C\!
A\right)}\,.}
\countdef\WRC=18\WRC=\pageno

To describe the expectation value of $W_R(C)$ in the Lagrangian
formulation of Chern-Simons theory, we introduce the
absolutely-normalized Wilson loop path integral
\countdef\ZkCR=19\ZkCR=\pageno 
\eqn\WLPI{ Z(k; C, R) \,=\, \int \! \CD \! A
\;\; W_R(C) \; \exp{\left[i {k \over {4 \pi}} \int_M \! \Tr\left( A \^
d A + {2 \over 3} A \^ A \^ A \right)\right]}\,,}
in terms of which the Wilson loop expectation value is given by the
ratio 
\eqn\VEVWVC{ \Big\langle W_R(C)\Big\rangle \,=\, {{Z(k; C,
R)}\over{Z(k)}}\,.}
Just as for the partition function, the Wilson loop path integral
\WLPI\ can be computed exactly using the Hamiltonian formulation of
Chern-Simons theory and its relation to rational conformal field
theory.  In that approach, the expectation values of Wilson loop
operators lead naturally \WittenHF\ to knot invariants such as the
celebrated Jones polynomial.  

In this paper, our perspective on the Wilson loop operator is rather 
different.  Here we wish to apply non-abelian localization to the
Wilson loop path integral in \WLPI\ to obtain a new, complementary
description of $W_R(C)$ as a cohomology class on the moduli space of
flat connections on $M$.

\bigskip\noindent{\it Some Experimental Evidence}\smallskip

At the outset, it is again far from clear that the Wilson loop
path integral bears any relation to the canonical symplectic integral
in \PZSM\ to which non-abelian localization applies.  To present one
suggestive piece of evidence, let us consider the simplest 
Wilson loop --- namely, the unknot Wilson loop --- in Chern-Simons
theory on $S^3$ with gauge group $SU(2)$.  Irreducible representations
of $SU(2)$ are uniquely labelled by their dimension, and we let $\CMj$
denote the irreducible representation of $SU(2)$ with dimension $j$.

For the unknot, the absolutely-normalized Wilson loop path integral in
\WLPI\ is given exactly by 
\eqn\UNKNOTW{ Z\big(k; \bigcirc, \CMj\big) \,=\, \sqrt{2 \over
{k+2}} \sin{\left({{\pi \, j} \over {k+2}}\right)}\,,\qquad\qquad
j=1,\ldots,k+1\,.}
As indicated, $\CMj$ runs without loss over the finite set of
irreducible representations which are integrable in the $SU(2)$
current algebra at level $k$.  This simple result was first obtained
by Witten in \WittenHF\ using the Hamiltonian formulation of
Chern-Simons theory, and as a special case, when ${\CMj={\bf 1}}$ is
trivial, the general formula for $Z(k; \bigcirc, \CMj)$ reduces to the
standard expression for the $SU(2)$ partition function $Z(k)$ of
Chern-Simons theory on $S^3$.

From the semi-classical perspective, we can gain greater insight into
the exact formula for $Z\big(k;\bigcirc, \CMj\big)$ by rewriting
\UNKNOTW\ as a contour integral over the real axis,
\eqn\UNKNOTWII{ Z\big(k; \bigcirc, \CMj\big) \,=\, {1 \over {2 \pi
i}} \, \e{-{{i \pi (1 + j^2)} \over {2(k+2)}}} \int_{-\infty}^{+\infty}
\!\! dx \;\; \ch_\CMj\!\biggr(\e{{i \pi} \over 4} \, {x\over 2}\biggr) \,
\sinh^2{\!\left(\e{{i \pi} \over 4} \, {x\over 2}\right)} \>
\exp{\!\left(-{{(k+2)} \over {8 \pi}} x^2\right)}\,.} 
Here $\ch_\CMj$ is the character of $SU(2)$ associated to the
representation $\CMj$,
\eqn\CHN{ \ch_\CMj(y) \,=\, {\sinh(j \, y) \over \sinh(y)} \,=\, \e{(j-1)
y} \,+\, \e{(j-3) y} \,+\, \cdots \,+\, \e{-(j-3) y} \,+\, \e{-(j-1)
y}\,,}
and the equality between the expressions in \UNKNOTW\ and \UNKNOTWII\
follows by evaluating \UNKNOTWII\ as a sum of elementary Gaussian
integrals.

Now, the only flat connection on $S^3$ is the trivial connection, and
for the case of the partition function $Z(k)$, we explained previously
\BeasleyVF\ how the contour integral over $x$ in \UNKNOTWII\ can be
interpreted very precisely as the stationary-phase contribution from
the trivial connection to the Chern-Simons path integral.
From a geometric perspective, the contour integral arises as an
integral over the Cartan subalgebra of $SU(2)$, regarded as the group
of constant gauge transformations on $S^3$.  The constant gauge
transformations are the stabilizer of the trivial connection in the
group of all gauge transformations, and the presence of this
stabilizer group plays an important role in the semi-classical
analysis of the Chern-Simons path integral.  In any event, as evident
from \UNKNOTWII, the stationary-phase approximation to the
Chern-Simons path integral on $S^3$ is exact.

Given this interpretation of $Z(k; \bigcirc, \CMj)$ in the
special case ${\CMj = {\bf 1}}$, it is very tempting to apply the same
interpretation for arbitrary $\CMj$, so that the contour integral
over $x$ in \UNKNOTWII\ more generally represents the stationary-phase
contribution from the trivial connection to the Wilson loop path
integral in \WLPI.  Moreover, since all dependence on the
representation $\CMj$ enters the integrand of \UNKNOTWII\ through the
$SU(2)$ character $\ch_\CMj$, we naturally identify $\ch_\CMj$ as
the avatar of the unknot Wilson loop operator itself when the path
integral in \WLPI\ is reduced to the contour integral in \UNKNOTWII.  

The semi-classical identification of the unknot Wilson loop operator
with a character is quite elegant, and one of our eventual results
will be to obtain this identification directly from the path integral
via non-abelian localization.  However, the unknot Wilson loop in
$S^3$ is also very special.  For instance, if we regard $S^3$ as a
Seifert manifold by virtue of the Hopf fibration over $S^2$, 
\eqn\SFRTSTHR{\matrix{
&S^1\,\longrightarrow\,S^3\cr
&\mskip 65mu \big\downarrow\lower 0.5ex\hbox{$^\pi$}\cr
&\mskip 60mu S^2\cr}\,,}
then the unknot can be represented by one of the $S^1$ fibers in
\SFRTSTHR.  As a result, the unknot Wilson loop is the unique
Wilson loop in $S^3$ which respects the distinguished $U(1)$ action
rotating the fibers of \SFRTSTHR.

To apply non-abelian localization to the Wilson loop path integral on
a general Seifert manifold $M$, we must similarly assume that the
Wilson loop operator respects the $U(1)$ action on $M$ which rotates
the fibers in \SFRT.  As for the unknot, such a Wilson loop operator
necessarily wraps a Seifert fiber of $M$.  To avoid potential
confusion later, we refer to the Wilson loops wrapping Seifert fibers of
$M$ as ``Seifert loops'' to distinguish them from arbitrary Wilson
loops in $M$, about which we will also have some things to say.

Perhaps surprisingly, even for $S^3$ the unknot is not the only knot
which a Seifert loop operator can wrap.  As we recall in Section $7.1$,
$S^3$ admits infinitely-many distinct Seifert presentations as the
total space of a nontrivial circle bundle over a Riemann surface of
genus zero with two orbifold points, of relatively-prime orders ${\bf
p}$ and ${\bf q}$.  The associated Seifert loop operator then wraps a
$({\bf p},{\bf q})$-torus knot in $S^3$.  Precisely for this class of
knots, Lawrence and  Rozansky \LawrenceRZ\ have again observed on the basis 
of empirical formulae generalizing \UNKNOTW\ and \UNKNOTWII\ that the
stationary-phase approximation to the Wilson loop path integral is
exact.  So like our previous work in  \BeasleyVF, one very specific
motivation for the present paper is to offer a theoretical explanation
of the remarkable results in \LawrenceRZ.

The special nature of the Seifert loop operators in $M$ has been also
pointed out by Aganagic, Neitzke, and Vafa \AganagicDH, who consider such
Wilson loops and others in the context of $q$-deformed Yang-Mills 
theory on the Riemann surface $\Sigma$.  See \refs{\deHaroID\deHaroUZ\AganagicJS\deHaroWN\deHaroRZ\AganagicDH\ArsiwallaJB\JafferisJD\CaporasoTA\deHaroRN\CaporasoFP\BlauGH\deHaroWQ{--}\JeffreyMC}
for a variety of additional papers which discuss Chern-Simons theory on a
Seifert manifold and the closely related subject of $q$-deformed
Yang-Mills theory on a Riemann surface.  We also mention attempts by
Hahn \HahnAT\ to make sense of the Wilson loop path integral in
Chern-Simons theory in a more formal mathematical framework.  See
\Weitsman\ for other work in this direction.  Finally,
we refer the interested reader to the very beautiful and roughly
analogous work by Pestun \PestunRZ\ on path integral localization for
supersymmetric circular Wilson loop operators in four-dimensional
Yang-Mills theory.  See also the work of Kapustin and collaborators
\KapustinWY\ (which has a certain degree of overlap with this paper)
for a more recent application of Pestun's techniques to Wilson loops
in superconformal Chern-Simons theories.  Last but not least, related ideas 
have been pursued by Nekrasov and collaborators in \refs{\NekrasovNA, \BaulieuNJ}.

\bigskip\noindent{\it The Plan of the Paper}\smallskip

In order to apply non-abelian localization to the general Seifert loop
path integral in \WLPI, we must first recast this path integral as a
symplectic integral of the canonical form \PZSM.  Although not
immediately obvious, the symplectic description of the Seifert loop 
path integral turns out to be a wonderfully natural extension of our
prior results for the Chern-Simons path integral in \CSZ.  To make this
fact apparent and also for sake of readability, we have endeavored to
keep the presentation self-contained --- a goal which at least partially
accounts for the length of the present paper.

With the ambition above in mind, we begin in Section $2$ by quickly
recalling how the path integral of Yang-Mills theory on a Riemann
surface $\Sigma$ furnishes the basic example of a symplectic 
integral of the canonical form \PZSM.  Along the way, we establish
some standard notation and conventions for the paper.

Next, in Section $3$ we review how the path integral which describes
the partition function of Chern-Simons theory on a Seifert manifold
can also be put into the canonical symplectic form.  The material here
closely follows \S $3$ of \BeasleyVF.

Finally, in Section $4$ we generalize the results in Section $3$ to the
Seifert loop path integral.

At the heart of Section $4$ lies a very old and very general piece of
gauge theory lore, which is perhaps worth mentioning now.  As
suggested by Witten in one of the small gems of \WittenHF, if we are
to analyze a Wilson loop expectation value semi-classically, we must
use a corresponding semi-classical description for the Wilson loop
operator itself.  That is, rather than applying the conventional
definition of $W_R(C)$ in \DEFWVC, we represent the Wilson loop
operator by a path integral over an auxiliary bosonic field $U$
attached to the curve $C$ and coupled to the restriction of $A$ to
$C$.  In these terms, we schematically write 
\eqn\NEWWC{ W_R(C) \,=\, \int\!\!\CD\!U \, \exp{\!\Big[
i\,{\Rc\Rs}_\alpha\big(U; A|_C\big)\Big]}\,.}
A bit more precisely, the field $U$ describes a one-dimensional
sigma model on $C$ whose target space is the coadjoint orbit
$\CO_\alpha$ of $G$ which passes through the highest weight $\alpha$
of $R$, and ${\Rc\Rs}_\alpha\big(U; A|_C\big)$ is a local,
gauge-invariant, and indeed topological action that specifies the
coupling of $U$ to the background gauge field $A$.  Because the
semi-classical description \NEWWC\ of $W_R(C)$ proves to be an
essential ingredient for our analysis, we review this description in
detail in Section $4$.

Athough Chern-Simons theory might seem to be characterized as an
intrinsically three-dimensional gauge theory, an important outcome of
our work in both Sections $3$ and $4$ is to provide a general
reformulation of Chern-Simons theory on an arbitrary three-manifold 
$M$ in such a way that one of the three components of the connection
$A$ completely decouples.  This construction fundamentally underlies
our work on non-abelian localization, but it may have other
applications as well.  At the moment, one tantalizing geometric aspect
of this reformulation of Chern-Simons theory is that it relies upon
the choice of a contact structure on $M$.

Although perhaps anticlimactic, in Section $5$ we return to two
dimensions and consider the analogue for the Seifert loop operator in
Yang-Mills theory on $\Sigma$.  As pointed out long ago by Witten 
\WittenWE, the two-dimensional analogue of the Seifert loop operator
is a local ``monodromy'' operator which inserts a classical
singularity into the gauge field $A$ at a marked point of $\Sigma$.
Like the Seifert loop operators, the monodromy operators in
two-dimensional Yang-Mills theory are described by a path integral of
the canonical symplectic form \PZSM.  This beautiful observation,
which we review in some detail, can be traced back to remarks of Atiyah in
\S $5.2$ of \AtiyahKN.  The monodromy operators in two-dimensional
Yang-Mills theory are also related to the surface operators
considered more recently by Gukov and Witten \GukovJK\ in the context
of four-dimensional gauge theory.  Indeed, portions of the exposition
in Section $5$ have a basic overlap with material in \GukovJK.

In preparation for actual computations, we discuss in Section $6$
general aspects of non-abelian localization.  In particular, we
explain how non-abelian localization provides a cohomological
interpretation for symplectic integrals such as \PZSM, and we recall a
general non-abelian localization formula derived in \BeasleyVF.
Thankfully, this localization formula is again applicable, so we do
not need to extend the lengthy technical analysis of \BeasleyVF.  At
the end of Section $6$, we also review two elementary,
finite-dimensional examples of non-abelian localization.

Finally, in Section $7$ we perform explicit computations of Seifert
loop path integrals using non-abelian localization.  As in our previous
study of the Chern-Simons partition function, we focus on two extreme
cases.

First, in the case that $M$ is a Seifert homology sphere, we evaluate
the local contribution from the trivial connection to the 
Seifert loop path integral.  Because the first homology group of $M$
is by assumption zero, ${H_1(M;\BZ) = 0}$, the trivial connection is
an isolated flat connection.  However, the trivial connection is also
fixed by constant gauge transformations on $M$.  So just as in
\BeasleyVF, the contribution of the trivial connection to the Seifert
loop path integral reduces to an integral over the Cartan subalgebra $\Ft$
of the gauge group $G$ itself.  As we found experimentally in 
\UNKNOTWII, the Seifert loop operator for a given irreducible
representation $R$ of $G$ then appears in the integral over $\Ft$ as
the corresponding character $\ch_R$.  

Among other results, we thus provide an exact path integral calculation for
the expectation value of a Wilson loop operator wrapping an arbitrary
torus knot in $S^3$.  As a special case, for gauge group ${G=SU(2)}$ we
recover the well-known expression for the Jones polynomial of a torus
knot.  Another satisfying aspect of this calculation is that we
observe the renowned Weyl character formula to emerge naturally from
the Chern-Simons path integral, in a manner reminiscent of its classic
derivation by Atiyah and Bott \AtiyahRBII\ from index theory.

At the opposite extreme, in the case that $M$ is a smooth circle
bundle over a Riemann surface $\Sigma$ of genus ${h \ge 1}$, we
compute the cohomology class that represents the Seifert loop operator
on a smooth component of the moduli space of flat connections on $M$.
Such a component consists of irreducible connections, whose
stabilizers in the group of all gauge transformations arise solely
from the center of $G$.  In this setting, the cohomology class that
describes the Seifert loop operator turns out to be extremely
natural.  Namely, the Seifert loop class is given by the Chern
character\foot{This description of the Seifert loop class was
presciently suggested to me by E.~Witten as I was in the midst of this
work.} of a universal bundle associated to the representation $R$.  In
fact, the Seifert loop class already appears in related work of
Teleman and Woodward \refs{\TelemanC,\TelemanCW}, where it derives
from a tautological Atiyah-Bott generator in the $K$-theory of the
moduli stack of holomorphic bundles over $\Sigma$.  As we discuss, our
computation of the Seifert loop class is also strongly suggested by
prior work of Jeffrey \JeffreyPB\ on the Verlinde formula.

For the convenience of the reader, we include in Appendix A an index
of commonly-used notation.  The remaining appendices contain a few 
technical calculations associated to our work in Section $7$.

\newsec{The Symplectic Geometry of Yang-Mills Theory on a Riemann
Surface}

A central theme throughout this work is the relationship between
Chern-Simons theory on a Seifert manifold $M$ and Yang-Mills theory on
the associated Riemann surface $\Sigma$.  Thus as a warm-up for our
discussion of path integrals in Chern-Simons theory, let us quickly
recall the much simpler symplectic interpretation for the path
integral of two-dimensional Yang-Mills theory.

We start by considering the usual path integral which describes the
partition function of Yang-Mills theory on a Riemann surface $\Sigma$,
\eqn\YMD{\eqalign{
&Z(\epsilon) \,=\, {1 \over {\Vol(\CG(P))}} \,
\left({1 \over {2 \pi \epsilon}}\right)^{\Delta_{\CG(P)}/2} \,
\int_{\CA(P)} \! \CD \! A \; \exp{\left[ {1 \over {2 \epsilon}}
\int_\Sigma \! \Tr\left(F_A \^ \* F_A\right)\right]}\,,\cr
&\Delta_{\CG(P)} \,=\, \dim \CG(P)\,.\cr}}
Here we assume that the Yang-Mills gauge group $G$ is compact,
connected, and simple.  At times, especially for our discussion of
Chern-Simons theory, we will further specialize to the case that $G$
is simply-connected as well.

As is standard in gauge theory, we have introduced in \YMD\ 
an invariant quadratic form `$\Tr$' on the Lie algebra $\Fg$
\countdef\LieG=23\LieG=\pageno of $G$.  Our conventions for the form
`$\Tr$' \countdef\TR=24\TR=\pageno are as follows.  If ${G = 
SU(r+1)}$, then `$\Tr$' denotes the trace in the fundamental
representation.  Because the generators of the Lie algebra of
$SU(r+1)$ are anti-hermitian, the trace then determines a {\sl 
negative}-definite quadratic form on the Lie algebra.  For other
simple Lie groups, `$\Tr$' denotes the unique invariant,
negative-definite quadratic form on $\Fg$ which is normalized so that,
for simply-connected $G$, the Chern-Simons level $k$ in \CSZ\ obeys
the conventional integral quantization.  

With these conventions, the gauge field $A$ is anti-hermitian, and the
covariant derivative defined by $A$ is ${d_A = d + A}$.
\countdef\dA=25\dA=\pageno The curvature
of $A$ is then ${F_A = dA + A\^A}$, \countdef\FA=26\FA=\pageno as
appears in the Yang-Mills action in \YMD.  Of course, the parameter
$\epsilon$ appearing there is related to the conventional Yang-Mills
coupling $g_{\rm ym}$ via ${\epsilon = g_{\rm ym}^2}$.

In order to define $Z$ formally, we fix a principal $G$-bundle $P$
\countdef\P=20\P=\pageno over $\Sigma$.  Then the space $\CA(P)$
\countdef\AP=21\AP=\pageno over which we integrate in
\YMD\ is the space of connections on $P$.  The group $\CG(P)$
\countdef\GP=22\GP=\pageno of gauge transformations acts on $\CA(P)$,
and we have normalized $Z$ in \YMD\ by dividing by the volume of
$\CG(P)$ and a formal power of $\epsilon$.  As we explained in
\BeasleyVF, this normalization of $Z$ is the natural normalization
when we apply non-abelian localization to compute the two-dimensional
Yang-Mills path integral.

The space $\CA(P)$ is an affine space, which means that, if we choose
a particular basepoint $A_0$ in $\CA(P)$, then we can identify
$\CA(P)$ with its tangent space at $A_0$.  This tangent space is the
vector space of sections of the bundle $\Omega^1_\Sigma \otimes
\ad(P)$ of one-forms on $\Sigma$ taking values in the adjoint bundle
associated to $P$.  In other words, an arbitrary connection $A$ on $P$
can be written as $A = A_0 + \eta$ for some section $\eta$ of
$\Omega^1_\Sigma \otimes \ad(P)$.  Because the space $\CA(P)$ is
affine, we define the path-integral measure $\CD\!A$ up to an overall
multiplicative constant by taking any translation-invariant measure on
$\CA(P)$.

To define the Yang-Mills action in \YMD, we must introduce a
duality operator $\*$ on $\Sigma$.  For two-dimensional Yang-Mills
theory, we only require that the operator $\*$ relates zero-forms to
two-forms, and to obtain such an operator we only need a symplectic
structure, as opposed to a metric, on $\Sigma$.  Given a
\countdef\SmOm=35\SmOm=\pageno symplectic form $\omega$ on $\Sigma$,
we define $\*$ by the condition ${\* 1 = \omega}$.  The symplectic
form $\omega$ is invariant under all area-preserving diffeomorphisms
of $\Sigma$, and this large group acts as a symmetry of
two-dimensional Yang-Mills theory.  More precisely, this symmetry
group is ``large'' in the sense that its complexification is the full
group of orientation-preserving diffeomorphisms of $\Sigma$
\Donaldson.  This fact is fundamentally responsible for the
topological nature of two-dimensional Yang-Mills theory.

\bigskip\noindent{\it The Yang-Mills Path Integral as a Symplectic
Integral}\smallskip

To interpret the two-dimensional Yang-Mills path integral as a
symplectic integral of the canonical form in \PZSM, we just need to
identify the gauge theory counterparts of the abstract geometric data
which enter the symplectic integral.  These data consist of a
symplectic manifold $X$ with symplectic form $\Omega$, a Lie group $H$
with Hamiltonian action on $X$, and an invariant quadratic form
$(\,\cdot\,,\,\cdot\,)$ on the Lie algebra $\Fh$ of $H$.  The
corresponding quantities in two-dimensional Yang-Mills theory are very
easy to guess.

First, the affine space $\CA(P)$ carries a natural symplectic form
$\Omega$ determined by the intersection pairing on $\Sigma$ itself.
Explicitly, if $\eta$ and $\xi$ are any two tangent vectors to
$\CA(P)$, represented on $\Sigma$ by sections of $\Omega^1_\Sigma
\otimes \ad(P)$, then $\Omega$ is defined by 
\eqn\OMYM{ \Omega(\eta, \xi) = - \int_\Sigma \Tr\big( \eta \^ \xi
\big)\,.}\countdef\Om=26\Om=\pageno
Clearly $\Omega$ is closed, non-degenerate, and invariant under
both gauge transformations and translations on $\CA(P)$.  Thus
$\CA(P)$ plays the role of the abstract symplectic manifold $X$ 
in \PZSM.  

Similarly, we identify the translation-invariant path-integral measure
$\CD\!A$ with the symplectic measure on $\CA(P)$ determined by
$\Omega$.  Quite generally, if $X$ is a symplectic manifold of
dimension $2n$ with symplectic form $\Omega$, then the symplectic
measure on $X$ is given by the top-form $\Omega^n / n!$.  This measure
can be represented uniformly for $X$ of arbitrary dimension by the
expression $\exp{\!(\Omega)}$, where we implicitly  pick out from the
series expansion of the exponential the term which is of top degree on
$X$.  So we formally write ${\CD\!A = \exp{\!(\Omega)}}$.

As for the Hamiltonian group $H$ acting on $X$, the obvious candidate
for this role in two-dimensional Yang-Mills theory is the group
$\CG(P)$ of gauge transformations acting on $\CA(P)$.  Indeed, as was
observed long ago by Atiyah and Bott \AtiyahYM, the action of $\CG(P)$
on $\CA(P)$ is Hamiltonian with respect to the symplectic form $\Omega$.

To recall what the Hamiltonian condition implies, we consider the
general situation that a connected Lie group $H$ with Lie algebra
$\Fh$ acts on a symplectic manifold $X$ preserving the symplectic form
$\Omega$.   The action of $H$ on $X$ is then Hamiltonian when there
exists an algebra homomorphism from $\Fh$ to the algebra of functions
on $X$ under the Poisson bracket.  The Poisson bracket of functions
$f$ and $g$ on $X$ is given by $\{f,g\} = -V_f(g)$, where $V_f$ is the
Hamiltonian vector field associated to $f$.  This vector field is
determined by the relation ${d f = \iota_{V_f} \Omega}$, where
$\iota_{V_f}$ is the interior product with $V_f$.  More explicitly, in
local canonical coordinates on $X$, the components of $V_f$ are
determined by $f$ as ${V_f^m = - (\Omega^{-1})^{m n} \, \partial_n
f}$, where $\Omega^{-1}$ is an ``inverse'' to $\Omega$ that arises by
considering the symplectic form as an isomorphism ${\Omega: TM
\rightarrow T^*M}$ with inverse ${\Omega^{-1}:T^*M \rightarrow TM}$.  In
coordinates, $\Omega^{-1}$ is defined by $(\Omega^{-1})^{l m} \,
\Omega_{m n} \,=\, \delta^l_n$, and  ${\{f, g\} = \Omega_{m n} V^m_f
V^n_g}$.  

The algebra homomorphism from the Lie algebra $\Fh$ to the algebra of
functions on $X$ under the Poisson bracket is then specified by a
moment map ${\mu: X \rightarrow \Fh^*}$, \countdef\Mom=33\Mom=\pageno
under which an element $\phi$ of $\Fh$ is sent to the function
$\langle\mu,\phi\rangle$ on $X$, where
$\langle\,\cdot\,,\,\cdot\,\rangle$ is the dual pairing 
between $\Fh$ and $\Fh^*$.  More generally, we use
$\langle\,\cdot\,,\,\cdot\,\rangle$ \countdef\BRC=32\BRC=\pageno 
throughout this paper to denote the canonical pairing between any
vector space and its dual.  The moment map by definition satisfies the
relation 
\eqn\MOMMAPEQ{d\langle\mu,\phi\rangle = \iota_{V(\phi)} \Omega\,,}
where $V(\phi)$ is the vector field on $X$ which is generated by the
infinitesimal action of $\phi$.  In terms of $\mu$, the Hamiltonian
condition then becomes the condition that $\mu$ also satisfy
\eqn\HOMMUII{ \left\{ \langle\mu,\phi\rangle,\langle\mu,\psi\rangle
\right\} = \langle\mu,[\phi,\psi]\rangle\,.}
Geometrically, the equation \HOMMUII\ is an infinitesimal
expression of the condition that the moment map $\mu$ commute with the
action of $H$ on $X$ and the coadjoint action of $H$ on $\Fh^*$.

Returning from this abstract discussion to the case of Yang-Mills
theory on $\Sigma$, let us consider the moment map for the action of
$\CG(P)$ on $\CA(P)$.  If $\phi$ is an element of the Lie algebra of
$\CG(P)$ and hence is represented on $\Sigma$ by a section of
$\ad(P)$, the corresponding vector field $V(\phi)$ on $\CA(P)$ is
given as usual by\foot{In order to make some conventions for our
discussion of Wilson loops more natural, we have made the opposite
choice for the sign of $V(\phi)$ as compared to that in \BeasleyVF.
As a result, certain formulae in this paper differ by signs from the
corresponding expressions in \BeasleyVF.}
\eqn\DGA{ V(\phi) \,=\, -d_A \phi\,,\qquad d_A \phi \,=\, d\phi \,+\,
\left[A,\phi\right].}
We then compute directly using \OMYM,
\eqn\DOMII{ \iota_{V(\phi)} \Omega \,=\, \int_\Sigma \Tr\!\left( d_A
\phi \^ \delta A \right) \,=\, -\int_\Sigma \Tr\!\left(\phi \> d_A \delta
A\right) \, = \, -\delta \! \int_\Sigma \Tr\!\left(F_A \, \phi\right)\,.}
Here we write $\delta$ for the exterior derivative acting on $\CA(P)$,
so that, for instance, $\delta A$ is regarded as a one form on
$\CA(P)$.  Thus the relation \MOMMAPEQ\ determines, up to an additive
constant, that the moment map $\mu$ for the action of $\CG(P)$ on
$\CA(P)$ is given by
\eqn\YMMO{ \langle\mu\,,\phi\rangle \,=\, -\int_\Sigma \Tr\big(F_A
\, \phi\big)\,.}
One can then check directly that $\mu$ in \YMMO\ satisfies the
condition \HOMMUII\ that it arise from a Lie algebra homomorphism, and
this condition fixes the arbitrary additive constant that could
otherwise appear in $\mu$ to be zero.  

Thus $\CG(P)$ acts in a Hamiltonian fashion on $\CA(P)$ with moment
map ${\mu = F_A}$.  Here we regard the curvature $F_A$, transforming
on $\Sigma$ as a section of $\Omega^2_\Sigma \otimes \ad(P)$, more
abstractly as an element of the dual of the Lie algebra of $\CG(P)$.

Finally, to define the canonical symplectic integral in \PZSM, the Lie
algebra $\Fh$ of the Hamiltonian group $H$ must carry an
\countdef\PR=34\PR=\pageno 
invariant quadratic form
$(\,\cdot\,,\,\cdot\,)$, which we use to define the 
invariant function ${S = \ha (\mu,\mu)}$.  In the case of Yang-Mills
theory on $\Sigma$, we use the duality operator $\*$ and the invariant
form `$\Tr$' to introduce the obvious positive-definite, invariant
quadratic form on the Lie algebra of $\CG(P)$, given explicitly by 
\eqn\QFG{ (\phi,\phi) \,=\, - \int_\Sigma \! \Tr\big(\phi \^ \*
\phi\big)\,.}
We formally define the volume of $\CG(P)$ as appears in \YMD\ using
the quadratic form \QFG.  With respect to this quadratic form, the
Yang-Mills action is precisely the square of the moment map in \YMMO,
\eqn\YMS{ S \,=\, \ha (\mu,\mu) \,=\, - \ha \int_\Sigma \!
\Tr\big(F_A \^ \* F_A\big)\,.}

As a result, the path integral \YMD\ of Yang-Mills theory on $\Sigma$
can be recast completely in terms of the symplectic data associated to
the Hamiltonian action of $\CG(P)$ on $\CA(P)$,
\eqn\PZYMII{ Z(\epsilon) \,=\,  {1 \over {\Vol(\CG(P))}} \,
\left({1 \over {2 \pi \epsilon}}\right)^{\Delta_{\CG(P)}/2} \,
\int_{\CA(P)} \! \exp{\!\left[ \Omega - {1 \over {2 \epsilon}}
\left(\mu,\mu\right) \right]}\,,}
just as in \PZSM.

\newsec{The Symplectic Geometry of Chern-Simons Theory on a Seifert
Manifold}

Following \S $3$ of \BeasleyVF, our goal in this section is to review 
how the path integral which describes the partition function of
Chern-Simons theory on a Seifert manifold can be recast as a
symplectic integral of the canonical form \PZSM.  For the convenience 
of the reader, the treatment below is both self-contained and
reasonably complete, though not so exhaustive as that in \BeasleyVF.

To setup notation, we consider Chern-Simons gauge theory on a
three-manifold $M$ with compact, connected, simply-connected, and
simple gauge group $G$.  With these assumptions, any principal
$G$-bundle $P$ on $M$ is topologically trivial, and we denote by $\CA$
the affine space of connections on the trivial bundle.
\countdef\CurlyA=36\CurlyA=\pageno  We denote by
$\CG$ \countdef\CurlyG=37\CurlyG=\pageno the group of gauge
transformations acting on $\CA$. 

We begin with the Chern-Simons path integral
\countdef\Eps=40\Eps=\pageno 
\eqn\PZCS{\eqalign{
Z(\epsilon) \,&=\, {1 \over {\Vol(\CG)}} \, \left({1 \over {2 \pi
\epsilon}}\right)^{\Delta_{\CG}} \, \int_\CA \! \CD \! A \;
\exp{\left[{i \over {2 \epsilon}} \, \int_M \! \Tr\!\left( A \^ d A + {2
\over 3} A \^ A \^ A \right)\right]}\,,\cr
\epsilon &= {{2 \pi} \over k}\,,\qquad \Delta_{\CG} = \dim \CG\,.}}
Here we have introduced a coupling parameter $\epsilon$ by analogy to the
canonical symplectic integral in \PZSM, and we have included a number
of formal factors in $Z$.  First, we have the measure $\CD A$ on
$\CA$, which we again define up to norm as a translation-invariant
measure on $\CA$.  As usual, we have also divided the path integral by
the volume of the group of gauge transformations $\CG$.  Finally, to
be fastidious, we have normalized $Z$ by a formal power of $\epsilon$
which as in \YMD\ is natural when $Z$ is defined by localization.

At the moment, we make no assumption about the three-manifold $M$.
However, if $M$ is a Seifert manifold, then to interpret the
Chern-Simons path integral symplectically we must eventually
decouple one of the three components of the gauge field $A$.  This
observation motivates the following reformulation of Chern-Simons
theory, which proves to be key to the rest of the paper.

\subsec{A New Formulation of Chern-Simons Theory, Part I}

In order to decouple one of the components of $A$, we begin by
choosing a one-dimensional subbundle of the cotangent bundle $T^* M$
of $M$.  Locally on $M$, this choice can be represented by the choice
of an everywhere non-zero one-form $\kappa$, so that the subbundle of
$T^* M$ consists of all one-forms proportional to $\kappa$.  However,
if $t$ is any non-zero function, then clearly $\kappa$ and $t \,
\kappa$ generate the same subbundle in $T^* M$.  Thus, our choice of a
one-dimensional subbundle of $T^* M$ corresponds locally to the choice
of an equivalence class of one-forms under the relation
\eqn\EQVA{ \kappa \sim t \, \kappa\,.}
We note that the representative one-form $\kappa$ which generates the
subbundle need only be defined locally on $M$.  Globally, the
subbundle might or might not be generated by a non-zero one-form which is
defined everywhere on $M$; this condition depends upon whether the
sign of $\kappa$ can be consistently defined under \EQVA\ and thus
whether the subbundle is orientable or not.

We now attempt to decouple one of the three components of $A$.
Specifically, our goal is to reformulate Chern-Simons theory on $M$ as
a theory which respects a new local symmetry under which $A$ varies as
\eqn\SHFTA{ \delta A \,=\, \sigma \kappa\,.}
Here $\sigma$ is an arbitrary section of the bundle $\Omega^0_M
\otimes \Fg$ of Lie algebra-valued functions on $M$.

The Chern-Simons action certainly does not respect the local
``shift'' symmetry in \SHFTA.  However, we can trivially introduce this
shift symmetry into Chern-Simons theory if we simultaneously introduce
a new scalar field $\Phi$ on $M$ which transforms like $A$ in the adjoint
representation of the gauge group.  Under the shift symmetry, $\Phi$
transforms as
\eqn\SHFTP{ \delta \Phi \,=\, \sigma\,.}
For future reference, we denote the infinite-dimensional group of
shift symmetries parametrized by $\sigma$ as
$\CS$. \countdef\CurlyS=38\CurlyS=\pageno

Now, if $\kappa$ in \SHFTA\ is scaled by a non-zero function $t$ so
that ${\kappa \to t \, \kappa}$, then this rescaling can be absorbed into
the arbitrary section $\sigma$ which also appears in \SHFTA\ so that
the transformation law for $A$ is well-defined.  However, from the
transformation \SHFTP\ of $\Phi$ under the same symmetry, we see that
because we absorb $t$ into $\sigma$ we must postulate an inverse
scaling of $\Phi$, so that ${\Phi \to t^{-1} \Phi}$.  As a result,
although $\kappa$ is only locally defined up to scale, the product
$\kappa \, \Phi$ is well-defined on $M$.

The only extension of the Chern-Simons action which now incorporates
both $\Phi$ and the shift symmetry is the Chern-Simons functional
$\RC\RS(\,\cdot\,)$ of the shift-invariant combination ${A - \kappa \,
\Phi}$.  Thus, we consider the theory with action
\countdef\CSFun=39\CSFun=\pageno 
\eqn\SAPI{ S(A,\Phi) \,=\, \RC\RS(A - \kappa \, \Phi)\,,}
or more explicitly,
\eqn\SAPII{ S(A, \Phi) \,=\, \RC\RS(A) - \int_M \Big[ 2 \kappa \^ \Tr(\Phi
F_A) - \kappa \^ d \kappa \, \Tr(\Phi^2)\Big]\,.}

To proceed, we play the usual game used to derive field theory
dualities by path integral manipulations, as for $T$-duality in two
dimensions \refs{\BuscherT,\RocekPS} or abelian $S$-duality in four
dimensions \WittenGF.  We have introduced a new degree of freedom,
namely $\Phi$, into Chern-Simons theory, and we have simultaneously
enlarged the symmetry group of the theory so that this degree of
freedom is completely gauge trivial.  As a result, we can either use
the shift symmetry \SHFTP\ to gauge $\Phi$ away, in which case we
recover the usual description of Chern-Simons theory, or we can
integrate $\Phi$ out, in which case we obtain a new description of
Chern-Simons theory which respects the action of the shift symmetry
\SHFTA\ on $A$.

\bigskip\noindent{\it A Contact Structure on $M$}\smallskip

Hitherto, we have supposed that the one-dimensional subbundle of $T^*
M$ represented by $\kappa$ is arbitrary, but at this point we must
impose an important geometric condition on this subbundle.  From the
action $S(A,\Phi)$ in \SAPII, we see that the term quadratic in $\Phi$ is
multiplied by the local three-form $\kappa \^ d\kappa$.  In order for
this quadratic term to be everywhere non-degenerate on $M$, so that we can
easily perform the path integral over $\Phi$, we require that $\kappa
\^ d\kappa$ is also everywhere non-zero on $M$.

Although $\kappa$ itself is only defined locally and up to rescaling
by a non-zero function $t$, the condition that $\kappa \^ d\kappa \neq
0$ pointwise on $M$ is a globally well-defined condition on the
subbundle generated by $\kappa$.  For when $\kappa$ scales as $\kappa
\to t\,\kappa$ for any non-zero function $t$, we easily see that
$\kappa \^ d\kappa$ also scales as $\kappa \^ d\kappa \to t^2 \,
\kappa \^ d\kappa$.  Thus, the condition that $\kappa \^ d\kappa \neq
0$ is preserved under arbitrary rescalings of $\kappa$.

The structure which we thus introduce on $M$ is the choice of a
one-dimensional subbundle of $T^* M$ for which any local generator
$\kappa$ satisfies $\kappa \^ d\kappa \neq 0$ at each point of $M$.
This geometric structure, which appears so naturally here, is known as a
contact structure \refs{\Etnyre\GeigesHJ{--}\Blair}.  More generally, on
an arbitrary manifold $M$ of odd dimension $2n+1$, a contact
structure on $M$ is defined as a one-dimensional subbundle of $T^* M$
for which the local generator $\kappa$ satisfies $\kappa \^
(d\kappa)^n \neq 0$ everywhere on
$M$. \countdef\Contact=41\Contact=\pageno

In many ways, a contact structure is the analogue of a symplectic
structure for manifolds of odd dimension.  The fact that we must
choose a contact structure on $M$ for our reformulation of
Chern-Simons theory is thus closely related to the fact, mentioned
previously, that we must choose a symplectic structure on the Riemann
surface $\Sigma$ in order to define Yang-Mills theory on $\Sigma$.

We will say a bit more about contact structures on Seifert manifolds
later, but for now, we just observe that, by a classic theorem of Martinet
\Martinet, any compact, orientable\foot{We note that, because $\kappa
\^ d\kappa \to t^2 \, \kappa \^ d\kappa$ under a local rescaling of
$\kappa$ and because $t^2$ is always positive, the sign of the local
three-form $\kappa \^ d\kappa$ is well-defined.  So any three-manifold
with a contact structure is necessarily orientable.} three-manifold
possesses a contact structure.

\bigskip\noindent{\it Path Integral Manipulations}\smallskip

Without loss of generality, we choose a contact structure on the
three-manifold $M$, and we consider the theory defined by the path
integral 
\eqn\PZCSII{\eqalign{
Z(\epsilon) \,&=\, {1 \over {\Vol(\CG)}} \, {1 \over {\Vol(\CS)}}
\, \left({1 \over {2 \pi \epsilon}}\right)^{\Delta_{\CG}} \, \times
\cr
&\times \, \int \CD \! A \, \CD \! \Phi \; \exp{\!\left[ {i \over {2
\epsilon}} \left( \RC\RS\!\left(A\right) - \int_M \! 2 \kappa \^
\Tr\!\left(\Phi F_A\right) \,+\, \int_M \! \kappa \^ d \kappa \,
\Tr\!\left(\Phi^2\right)\right)\right]}\,.\cr}}
Here the measure $\CD \! \Phi$ is defined independently of any metric on
$M$ by the invariant, positive-definite quadratic form
\eqn\DPHI{ \left(\Phi,\Phi\right) = - \int_M \kappa \^
d\kappa \, \Tr\!\left(\Phi^2\right)\,,}
which is invariant under the scaling ${\kappa \to t\, \kappa}$,
${\Phi \to t^{-1} \, \Phi}$.  We similarly use this quadratic form
to define formally the volume of the group $\CS$ of shift symmetries,
as appears in the normalization of \PZCSII.

Using the shift symmetry \SHFTP, we can fix $\Phi = 0$ trivially, with
unit Jacobian, and the resulting group integral over $\CS$ produces a
factor of $\Vol(\CS)$ to cancel the corresponding factor in the
normalization of $Z(\epsilon)$.  Hence, the new theory defined by
\PZCSII\ is fully equivalent to Chern-Simons theory.

On the other hand, because the field $\Phi$ appears only quadratically
in the action \SAPII, we can also perform the path integral over
$\Phi$ directly.  Upon integrating out $\Phi$, the new action
$S(A)$ for the gauge field becomes
\eqn\SAA{ S(A) \,=\, \int_M \! \Tr \!\left( A \^ d A + {2 \over 3} A \^ A
\^ A \right) \,-\, \int_M {1 \over {\kappa \^ d \kappa}}
\Tr\Big[ (\kappa \^ F_A)^2 \Big]\,.}
We find it convenient to abuse notation slightly by writing ``$1 /
\kappa \^ d\kappa$'' in \SAA.  To explain this notation precisely, we
observe that, as $\kappa \^ d\kappa$ is nonvanishing, we can always
write ${\kappa \^ F_A = \varphi \, \kappa \^ d\kappa}$ for some function
$\varphi$ on $M$ taking values in the Lie algebra $\Fg$.  Thus, we set
${\kappa \^ F_A / \kappa \^ d\kappa = \varphi}$, and the second term in
$S(A)$ becomes $\int_M \kappa \^ \Tr\!\left(F_A \varphi\right)$.  As our
notation in \SAA\ suggests, this term is invariant under the
transformation $\kappa \to t\,\kappa$, since $\varphi$ transforms as
${\varphi \to t^{-1} \, \varphi}$.

By construction, the new action $S(A)$ in \SAA\ is invariant under the
action of the shift symmetry \SHFTA\ on $A$.  Alternatively, one can
directly check the shift-invariance of $S(A)$, for which one notes
that the expression $\kappa \^ F_A$ transforms under the shift symmetry as 
\eqn\SHFTF{ \kappa \^ F_A \,\longrightarrow\, \kappa \^ F_A + \sigma \,
\kappa\^d\kappa\,.}

The partition function $Z(\epsilon)$ now takes the form
\eqn\PZCSIII{\eqalign{Z(\epsilon) \,&=\, {1 \over {\Vol(\CG)}} \, {1
\over {\Vol(\CS)}} \, \left({{-i} \over {2 \pi
\epsilon}}\right)^{\Delta_{\CG}/2} \, \times\cr
&\times \, \int_\CA \CD \! A \; \exp{\!\left[{i \over {2
\epsilon}}\left( \int_M \! \Tr\!\left( A \^ d A + {2 \over 3} A \^ A
\^ A \right) \,-\, \int_M {1 \over {\kappa \^ d \kappa}} \Tr\!\left[
\left(\kappa \^ F_A\right)^2 \right]\right)\right] }\,,\cr}}
where the Gaussian integral over $\Phi$ cancels some factors of
$2 \pi \epsilon$ in the normalization of $Z$.  As is standard, in
integrating over $\Phi$ we assume that the integration contour has
been slightly rotated off the real axis, effectively giving $\epsilon$
a small imaginary part, to regulate the oscillatory Gaussian integral.
Thus, the theory described by the path integral \PZCSIII\ is fully
equivalent to Chern-Simons theory, but now one component of $A$
manifestly decouples.

\subsec{The Chern-Simons Path Integral as a Symplectic Integral}

Our reformulation of the Chern-Simons partition function in \PZCSIII\
applies to any three-manifold $M$ with a specified contact structure.
However, in order to apply non-abelian localization to Chern-Simons
theory on $M$, we require that $M$ possesses additional symmetry.

Specifically, we require that $M$ admits a locally-free $U(1)$ action,
which means that the generating vector field on $M$ associated to the
infinitesimal action of $U(1)$ is nowhere vanishing.  A free $U(1)$
action on $M$ clearly satisfies this condition, but more generally it
is satisfied by any $U(1)$ action such that no point on $M$ is fixed
by all of $U(1)$ (at such a point the generating vector field would
vanish).  Such an action need not be free, since some points on $M$
could be fixed by a cyclic subgroup of $U(1)$.  The class of
three-manifolds which admit a $U(1)$ action of this sort are precisely
the Seifert manifolds.

To proceed further to a symplectic description of the Chern-Simons
path integral, we now restrict attention to the case that $M$ is a
Seifert manifold.  We first review a few basic facts about such
manifolds, for which a complete reference is \OrlikPK.

\bigskip\noindent{\it Contact Structures on Seifert
Manifolds}\smallskip

For simplicity, we begin by assuming that the three-manifold $M$
admits a free $U(1)$ action.  In this case, $M$ is the total
space of a circle bundle over a Riemann surface $\Sigma$,
\eqn\FBII{\matrix{
&S^1\,{\buildrel n\over\longrightarrow}\,M\cr
&\mskip 65mu\big\downarrow\lower 0.5ex\hbox{$^\pi$}\cr
&\mskip 55mu\Sigma\cr}\,,}
and the free $U(1)$ action simply arises from rotations in the
fiber of \FBII.  The topology of $M$ is completely determined
by the genus $h$ of $\Sigma$ \countdef\genus=43\genus=\pageno and the
degree $n$ of the bundle. \countdef\degree=44\degree=\pageno Assuming
that the bundle is nontrivial, we can always arrange by a suitable
choice of orientation for $M$ that $n \ge 1$.

At this point, one might wonder why we restrict attention to the case
of nontrivial bundles over $\Sigma$.  As we now explain, in this case
$M$ admits a natural contact structure which is invariant under the
action of $U(1)$.  As a result, our reformulation of Chern-Simons
theory in \PZCSIII\ still respects this crucial symmetry of $M$.

To describe the $U(1)$ invariant contact structure on $M$, we
simply exhibit an invariant one-form $\kappa$, defined globally on
$M$, which satisfies the contact condition that $\kappa \^
d\kappa$ is nowhere vanishing.  To describe $\kappa$, we begin by
choosing a symplectic form $\omega$ on $\Sigma$ which is
normalized so that \eqn\INTKII{ \int_\Sigma \omega \,=\, 1\,.}
Regarding $M$ as the total space of a principal $U(1)$-bundle, we
take $\kappa$ to be a connection on this bundle (and hence a
real-valued one-form on $M$) whose curvature satisfies \eqn\CONII{
d\kappa = n \, \pi^* \omega\,,} where we recall that $n \ge 1$ is
the degree of the bundle.  For a nice, explicit description of
$\kappa$ in this situation, see the description of the angular
form in \S $6$ of \BottT.

We let $\RR$ (for ``rotation'') be the non-vanishing vector field on $M$
which generates the $U(1)$ action and which is normalized so that its
orbits have unit period.  \countdef\BoldR=55\BoldR=\pageno By the
fundamental properties of a connection, $\kappa$ is invariant under
the $U(1)$ action and satisfies $\langle\kappa, \RR\rangle = 1$.  Here
$\langle\,\cdot\,,\,\cdot\,\rangle$ again denotes the canonical
dual pairing.  Thus, $\kappa$ pulls back to a non-zero one-form which
generates the integral cohomology of each $S^1$ fiber of $M$, and we
immediately see from \CONII\ that $\kappa \^ d\kappa$ is everywhere
non-vanishing on $M$ so long as the bundle is nontrivial.

Of course, in the above construction we have assumed that $M$ admits a
free $U(1)$ action, which is a more stringent requirement than the
condition that no point of $M$ is completely fixed by the $U(1)$
action.  However, an arbitrary Seifert manifold does admit an orbifold
description precisely analogous to the description of $M$ as a principal
$U(1)$-bundle over a Riemann surface.  We simply replace the smooth
Riemann surface $\Sigma$ with an orbifold $\widehat\Sigma$, and we replace
the principal $U(1)$-bundle over $\Sigma$ with its orbifold counterpart,
in such a way that the total space is a smooth three-manifold.

Concretely, the orbifold base $\widehat\Sigma$ of $M$ is now described by
a Riemann surface of genus $h$ with marked points $p_j$ for 
${j=1,\ldots,N}$ at which the coordinate neighborhoods are modeled not
on $\BC$ but on $\BC / \BZ_{a_j}$ for some cyclic group
$\BZ_{a_j}$, which acts on the local coordinate $z$ at $p_j$ as 
\eqn\ZAORB{ z \mapsto \zeta \cdot z\,,\qquad \zeta = \e{2 \pi i /
a_j}\,.}
The choice of the particular orbifold points $p_j$ is topologically
irrelevant, and the orbifold base $\widehat\Sigma$ can be completely specified
by the genus $h$ and the set of integers $\{a_1,\ldots,a_N\}$.

We now consider a line $V$-bundle over $\widehat\Sigma$.  Such an object
is precisely analogous to a complex line bundle, except that the local
trivialization over each orbifold point $p_j$ of $\widehat\Sigma$ is
now modeled on $\BC \times \BC/\BZ_{a_j}$, where $\BZ_{a_j}$
acts on the local coordinates $(z,s)$ of the base and fiber as
\eqn\ZAORBF{  z \mapsto \zeta \cdot z\,,\qquad s \mapsto
\zeta^{b_j} \cdot s\,,\qquad \zeta = \e{2 \pi i / a_j}\,,}
for some integers $0 \le b_j < a_j$.

Given such a line $V$-bundle over $\widehat\Sigma$, an arbitrary Seifert
manifold $M$ can be described as the total space of the associated
$S^1$ fibration.  Of course, we require that $M$ itself be smooth.
This condition implies that each pair of integers $(a_j,b_j)$
above must be relatively-prime, so that the local action \ZAORBF\ of the
orbifold group $\BZ_{a_j}$ on ${\BC \times S^1}$ is free.  In
particular, we require ${b_j \neq 0}$ above.

The $U(1)$ action on $M$ again arises from rotations in the fibers
over $\widehat\Sigma$, but this action is no longer free.  Rather, the points
in the $S^1$ fiber over each ramification point $p_j$ of $\widehat\Sigma$
are fixed by the cyclic subgroup $\BZ_{a_j}$ of $U(1)$, due to
the orbifold identification in \ZAORBF.

Once the integers $\{b_1,\ldots,b_N\}$ are fixed, the
topological isomorphism class of a line $V$-bundle on $\widehat\Sigma$
is specified by a single integer $n$, the degree.  Thus, in total, the
description of an arbitrary Seifert manifold $M$ is given by the
Seifert invariants \countdef\Seifert=42\Seifert=\pageno
\eqn\SFRT{\Big[h;n;(a_1,b_1), \ldots, (a_N,b_N)\Big]\,,\qquad
\gcd(a_j,b_j) = 1\,.}

Because the basic notions of bundles, connections, curvatures, and
(rational) characteristic classes generalize immediately from smooth
manifolds to orbifolds \refs{\SatakeI, \SatakeII}, our previous
construction of an invariant contact form $\kappa$ as a connection on
a principal $U(1)$-bundle immediately generalizes to the orbifold
situation here.   In the orbifold case, if $\widehat\CL$ denotes the line
$V$-bundle over $\widehat\Sigma$ which describes $M$ with Seifert
invariants \SFRT, the Chern class of $\widehat\CL$ is given by 
\eqn\CHRNCL{ c_1(\widehat\CL) \,=\, n + \sum_{j=1}^N {{b_j} \over
{a_j}}\,.} \countdef\CurlyL=45\CurlyL=\pageno
So long as $c_1(\widehat\CL)$ is non-zero (and positive by convention),
then $\widehat\CL$ is non-trivial, generalizing the previous condition
that ${n\ge 1}$.  In particular, $n$ can now be any integer such that
${c_1(\widehat\CL) > 0}$.  To define a contact structure on $M$ by analogy
to \CONII, we then choose the connection $\kappa$ so that its
curvature is given by 
\eqn\CHRNCLII{ d\kappa \,=\, \left(n + \sum_{j=1}^N {{b_j} \over
{a_j}}\right) \pi^* \widehat\omega\,,}
where $\widehat\omega$ is a symplectic form on $\widehat\Sigma$ of unit
volume, as in \INTKII. 

In the course of our discussion, we have distinguished the orbifold
$\widehat\Sigma$ from the smooth Riemann surface $\Sigma$.  In the future,
we will not make this artificial distinction, and for our discussion of
Chern-Simons theory, we use $\Sigma$ to denote an arbitrary Riemann
surface, possibly with orbifold points.

\bigskip\noindent{\it A Symplectic Structure for Chern-Simons
Theory}\smallskip

We now specialize to the case of Chern-Simons theory on a Seifert
manifold $M$, which carries a distinguished $U(1)$ action and an
invariant contact form $\kappa$.  Our first task is to identify the
symplectic space in Chern-Simons theory on $M$ which is to play the
role of $X$ in the canonical symplectic integral \PZSM.

Initially, the path integral of Chern-Simons theory is an integral
over the affine space $\CA$ of all connections on $M$, and unlike the
case for two-dimensional Yang-Mills theory, $\CA$ is not naturally
symplectic.  However, we now reap the reward of our reformulation of
Chern-Simons theory to decouple one component of $A$.  Specifically,
we consider the following two-form $\Omega$ on $\CA$.  If $\eta$ and
$\xi$ are any two tangent vectors to $\CA$, and hence are represented
by sections of the bundle $\Omega^1_M \otimes \Fg$ on $M$, we define
$\Omega$ by \countdef\OmBarA=47\OmBarA=\pageno
\eqn\BO{\Omega(\eta,\xi) = - \int_M \kappa\^\Tr\!\left( \eta \^ \xi
\right)\,.}

Because $\kappa$ is a globally-defined one-form on $M$, the 
expression for $\Omega$ in \BO\ is also well-defined.  Further,
$\Omega$ is manifestly closed and invariant under all symmetries.  In
particular, $\Omega$ is invariant under the group $\CS$ of shift
symmetries, and by virtue of this shift invariance, $\Omega$ is
degenerate along tangent vectors to $\CA$ of the form $\sigma \kappa$,
where $\sigma$ is an arbitrary section of $\Omega^0_M \otimes \Fg$.

Unlike the gauge symmetry $\CG$, which acts nonlinearly on $\CA$, the
shift symmetry $\CS$ acts in a simple, linear fashion on $\CA$.  Thus
we can trivially take the quotient of $\CA$ by the action of $\CS$,
which we denote as \countdef\BarA=46\BarA=\pageno 
\eqn\CMI{ \bar\CA \,=\, \CA / \CS\,.}
Under this quotient, the pre-symplectic form $\Omega$ on $\CA$
descends immediately to a symplectic form on $\bar\CA$, which becomes
a symplectic space naturally associated to Chern-Simons theory on
$M$.  So $\bar\CA$ plays the role of the abstract symplectic manifold
$X$ in \PZSM.

Our reformulation of the Chern-Simons action $S(A)$ in \SAA\ is invariant
under the shift symmetry $\CS$, so $S(A)$ immediately descends to the
quotient $\bar\CA$.  But we should also think (at least formally)
about the path integral measure $\CD \! A$.  

As we explained in some detail in \S $3.3$ of \BeasleyVF, the
translation-invariant measure $\CD A$ on $\CA$ pushes down to the
symplectic measure defined by $\Omega$ on $\bar\CA$.  Along the way,
the formal integral over the orbits of $\CS$ contributes a factor of
the volume $\Vol(\CS)$ to cancel the prefactor in \PZCSIII.
Consequently, the Chern-Simons path integral reduces to the following 
integral over $\bar\CA$,
\eqn\PZCSIV{ Z(\epsilon) \,=\, {1 \over {\Vol(\CG)}} \, \left({{-i} \over
{2 \pi \epsilon}}\right)^{\Delta_{\CG}/2} \, \int_{\bar\CA}
\exp{\!\left[\Omega + {i \over {2 \epsilon}} \,
S\!\left(A\right)\right]}\,,}
with 
\eqn\SAAII{
S(A) \,=\, \int_M \! \Tr{\!\left( A \^ d A + {2 \over 3} A \^ A \^ A
\right)} - \int_M {1 \over {\kappa \^ d \kappa}}
\Tr\Big[ (\kappa \^ F_A)^2 \Big]\,.}

\subsec{Hamiltonian Symmetries}

To complete our symplectic description of the Chern-Simons path
integral on $M$, we must show that the action $S(A)$ in \SAAII\ is
the square of a moment map $\mu$ for the Hamiltonian action of some
symmetry group $\CH$ on the symplectic space $\bar\CA$.

By analogy to the case of Yang-Mills theory on $\Sigma$, one might
naively guess that the relevant symmetry group for Chern-Simons theory
would also be the group $\CG$ of gauge transformations.  One can
easily check that the action of $\CG$ on $\CA$ descends under the 
quotient to a well-defined action on $\bar\CA$, and clearly the
symplectic form $\Omega$ on $\bar\CA$ is invariant under $\CG$.
However, one interesting aspect of non-abelian localization for
Chern-Simons theory is the fact that the group $\CH$ which we use for
localization must be somewhat more complicated than $\CG$ itself.

A trivial objection to using $\CG$ for localization is that, by
construction, the square of the moment map $\mu$ for any Hamiltonian
action on $\bar\CA$ defines an invariant function on $\bar\CA$, but
the action $S(A)$ is not invariant under the group $\CG$.  Instead,
the action $S(A)$ is the sum of a manifestly gauge-invariant term and
the usual Chern-Simons action, which shifts by integral multiples of
$8 \pi^2$ under ``large'' gauge transformations, those not continuously
connected to the identity in $\CG$.

This trivial objection is easily overcome.  We consider not the
disconnected group $\CG$ of all gauge transformations but only the
identity component $\CG_0$ of this group, under which $S(A)$ is
invariant.\countdef\CurlyGZero=48\CurlyGZero=\pageno

We now consider the action of $\CG_0$ on $\bar\CA$, and our first task
is to determine the corresponding moment map $\mu$.  If $\phi$ is an
element of the Lie algebra of $\CG_0$, described by a section of the
bundle $\Omega^0_M \otimes \Fg$ on $M$, then the corresponding vector
field $V(\phi)$ generated by $\phi$ on $\CA$ is given\foot{Once again,
we warn the reader that our convention for the sign of $V(\phi)$ is
opposite from that in \BeasleyVF.} by $V(\phi) = -d_A
\phi$.  Thus, from our expression for the symplectic form $\Omega$ in
\BO\ we see that \eqn\BOII{ \iota_{V(\phi)} \Omega \,=\, \int_M
\kappa \^ \Tr\!\left(d_A \phi \^ \delta A\right)\,.}
Integrating by parts with respect to $d_A$, we can rewrite \BOII\ in
the form $\delta\langle\mu,\phi\rangle$, where
\eqn\MU{\langle\mu,\phi\rangle = -\int_M \kappa\^\Tr\Big(\phi
F_A\Big) + \int_M d\kappa\^\Tr\Big(\phi (A - A_0)\Big)\,.}
Here $A_0$ is an arbitrary connection, corresponding to a basepoint in
$\CA$, which we must choose so that the second term in \MU\ can be
honestly interpreted as the integral of a differential form on $M$.
In the case that the gauge group $G$ is simply-connected, so that the 
principal $G$-bundle over $M$ is necessarily trivial, the choice of a
basepoint connection $A_0$ corresponds geometrically to the choice of a
trivialization for the bundle on $M$.  We will say more about this
choice momentarily, but we first observe that the expression for $\mu$
in \MU\ is invariant under the shift symmetry and immediately descends
to a moment map for the action of $\CG$ on $\bar\CA$.

The fact that we must choose a basepoint $A_0$ in $\CA$ to define the
moment map is very important in the following, and it is fundamentally
a reflection of the affine structure of $\CA$.  In general, an affine
space is a space which can be identified with a vector space only
after some basepoint is chosen to represent the origin.  In the case
at hand, once $A_0$ is chosen, we can identify $\CA$ with the vector
space of sections $\eta$ of the bundle $\Omega^1_M \otimes \Fg$ on
$M$, via $A = A_0 + \eta$, as we used in \MU.  However, $\CA$ is not
naturally itself a vector space, since $\CA$ does not intrinsically
possess a distinguished origin.  This statement corresponds to the
geometric statement that, though our principal $G$-bundle on $M$ is
trivial, it does not possess a canonical trivialization.

In terms of the moment map $\mu$, the choice of $A_0$ simply
represents the possibility of adding an arbitrary constant to $\mu$.
In general, our ability to add a constant to $\mu$ means that
$\mu$ need {\it not} determine a Hamiltonian action of $\CG_0$ on
$\bar\CA$.  Indeed, as we show below, the action of $\CG_0$ on
$\bar\CA$ is not Hamiltonian and we cannot simply use $\CG_0$ to
perform localization.

In order not to clutter the expressions below, we assume henceforth
that we have fixed a trivialization of the $G$-bundle on $M$ and we
simply set $A_0 = 0$.

To determine whether the action of $\CG_0$ on $\bar\CA$ is
Hamiltonian, we must check the condition \HOMMUII\ that $\mu$
determine a homomorphism from the Lie algebra of $\CG_0$ to the
algebra of functions on $\bar\CA$ under the Poisson bracket.  So we
directly compute
\eqn\HAM{\eqalign{
\Big\{ \langle\mu,\phi\rangle, \langle\mu,\psi\rangle \Big\} \,&=\,
\Omega\Big(d_A \phi, d_A \psi\Big)\,=\, - \int_M \kappa \^
\Tr\!\left(d_A \phi \^ d_A \psi\right)\,,\cr
\,&=\, -\int_M \kappa\^\Tr\Big([\phi,\psi] F_A\Big) -
\int_M d \kappa\^\Tr\Big(\phi \, d_A \psi\Big)\,,\cr
\,&=\, \langle\mu,[\phi,\psi]\rangle -
\int_M d\kappa \^ \Tr\Big(\phi \, d \psi\Big)\,.\cr}}

Thus, the failure of $\mu$ to determine an algebra homomorphism is
measured by the cohomology class of the Lie algebra cocycle
\countdef\LieCoc=49\LieCoc=\pageno
\eqn\COC{\eqalign{ c(\phi,\psi) \,&=\, \Big\{ \langle\mu,\phi\rangle,
\langle\mu,\psi\rangle \Big\} - \langle\mu,[\phi,\psi]\rangle\,,\cr
&\,=\, - \int_M d\kappa \^ \Tr\Big(\phi \, d \psi\Big) \,=\, - \int_M
\kappa \^ d\kappa \Tr\Big(\phi \lie_\RR \psi\Big)\,.\cr}}
In the second line of \COC, we have rewritten the cocycle more
suggestively by using the Lie derivative $\lie_\RR$ along the vector
field $\RR$ on $M$ which generates the $U(1)$ action.
\countdef\PoundR=56\PoundR=\pageno The class of this
cocycle is not zero, and no Hamiltonian action on $\bar\CA$ exists for the
group $\CG_0$.

\bigskip\noindent{\it Some Facts About Loop Groups}\smallskip

The cocycle appearing in \COC\ has a very close relationship to a
similar cocycle that arises in the theory of loop groups, and some
well-known loop group constructions feature heavily in our study of
Chern-Simons theory.  We briefly review these ideas, for which a
general reference is \PressleySG.

When $G$ is a finite-dimensional Lie group, we recall that the loop
group $LG$ is defined as the group of smooth maps $\Map(S^1,G)$ from
$S^1$ to $G$. \countdef\LoopG=50\LoopG=\pageno Similarly, the Lie
algebra $L\Fg$ of $LG$ is the algebra $\Map(S^1,\Fg)$ of smooth maps
from $S^1$ to $\Fg$. \countdef\Loopg=51\Loopg=\pageno When $\Fg$ is
simple, then the Lie algebra $L\Fg$ admits a unique, 
$G$-invariant cocycle up to scale, and this cocycle is directly
analogous to the cocycle we discovered in \COC.  If $\phi$ and $\psi$
are elements in the Lie algebra $L\Fg$, then this cocycle is defined by
\eqn\LGCOC{ c(\phi,\psi) = -\int_{S^1} \Tr\Big(\phi \, d \psi\Big) \,=\,
-\int_{S^1} \, d\tau \, \Tr\Big(\phi \lie_\RR \psi\Big)\,.}
In passing to the last expression, we have by analogy to \COC\
introduced a unit-length parameter $\tau$ on $S^1$, so that $\int_{S^1}
d\tau = 1$, and we have introduced the dual vector field $\RR =
\partial/\partial\tau$ which generates rotations of $S^1$.

In general, if $\Fg$ is any Lie algebra and $c$ is a nontrivial cocycle,
then $c$ determines a corresponding central extension $\wt\Fg$ of $\Fg$,
\eqn\ZNTRL{ \BR \longrightarrow \wt\Fg \longrightarrow \Fg\,.}
As a vector space, $\wt\Fg = \Fg \oplus \BR$, and the Lie algebra of
$\wt\Fg$ is given by the bracket
\eqn\BRAK{  \Big[ (\phi, a), (\psi, b) \Big] = \Big(
[\phi,\psi],\, c(\phi,\psi) \Big)\,,}
where $\phi$ and $\psi$ are elements of $\Fg$, and $a$ and $b$ are
elements of $\BR$.

In the case of the Lie algebra $L\Fg$, the cocycle $c$ appearing in
\LGCOC\ consequently determines a central extension $\wt{{L\Fg}}$ of
$L\Fg$.  When $G$ is simply connected, the extension determined
by $c$ or any integral multiple of $c$ lifts to a corresponding
extension of $LG$ by $U(1)$,
\eqn\LGE{ U(1)_\RZ \longrightarrow \wt{{LG}} \longrightarrow LG\,.}
Here we use the subscript `$\RZ$' to distinguish the central $U(1)$ in
$\wt{{LG}}$ from another $U(1)$ that will appear shortly.
Topologically, the extension $\wt{{LG}}$ is the total space of the
$S^1$ bundle over $LG$ whose Euler class is represented by the cocyle
of the extension, interpreted as an invariant two-form on $LG$.  The fact
that the Euler class must be integral is responsible for the
corresponding quantization condition on the cocycle of the extension.

When $\Fg$ is simple, the algebra $L\Fg$ has a non-degenerate,
invariant inner product which is unique up to scale and is given by
\eqn\INVLG{ \left(\phi,\psi\right) \,=\, -\int_{S^1} d\tau \,
\Tr\!\left(\phi\psi\right)\,.}
On the other hand, the corresponding extension $\wt{{L\Fg}}$ does {\it
not} possess a non-degenerate, invariant inner product, since any
element of $\wt{{L\Fg}}$ can be expressed as a commutator, so that
$[\wt{{L\Fg}},\wt{{L\Fg}}] = \wt{{L\Fg}}$, and the center of
$\wt{{L\Fg}}$ is necessarily orthogonal to every commutator under an
invariant inner product.

However, we can also consider the semidirect product $U(1)_\RR \ltimes
\wt{{LG}}$.  Here $U(1)_\RR$ is the group acting on $S^1$ by rigid
rotations, inducing a natural action on $\wt{{LG}}$ by which we define
the product.  The important observation about the group $U(1)_\RR
\ltimes \wt{{LG}}$ is that it does admit an invariant, non-degenerate
inner product on its Lie algebra.

Explicitly, the Lie algebra of $U(1)_\RR \ltimes \wt{{LG}}$ is identified
with $\BR \oplus \wt{{L\Fg}} = \BR \oplus L\Fg \oplus \BR$ as a
vector space, and the Lie algebra is given by the bracket
\eqn\BRAKII{ \Big[(p,\phi,a),(q,\psi,b)\Big] \,=\,
\Big(0,\,[\phi,\psi] + p \lie_\RR \psi - q \lie_\RR
\phi,\,c(\phi,\psi)\Big)\,,}
where $\lie_\RR$ is the Lie derivative with respect to the vector
field $\RR$ generating rotations of $S^1$.  We then consider the manifestly
non-degenerate inner product on $\BR \oplus \wt{{L\Fg}}$ which is
given by 
\eqn\FRMLG{ \Big((p,\phi,a),(q,\psi,b)\Big) \,=\, - \int_M d\tau
\, \Tr(\phi \psi) - p b - q a\,.}  
One can directly check that this inner product is invariant under the
adjoint action determined by \BRAKII.  We note that although this
inner product is non-degenerate, it is not positive-definite because
of the last two terms in \FRMLG.

\bigskip\noindent{\it Extension to Chern-Simons Theory}\smallskip

We now return to our original problem, which is to find a Hamiltonian
action of a group $\CH$ on $\bar\CA$ to use for localization.  The
natural guess to consider the identity component $\CG_0$ of the gauge
group does not work, because the cocycle $c$ in \COC\ obstructs the
action of $\CG_0$ on $\bar\CA$ from being Hamiltonian.

However, motivated by the loop group constructions, we consider now
the central extension ${\wt\CG}_0$ of $\CG_0$ by $U(1)$ which is
determined by the cocycle $c$ in \COC,
\countdef\TildG=52\TildG=\pageno 
\eqn\ZNTRLG{ U(1)_\RZ \longrightarrow {\wt\CG}_0 \longrightarrow
\CG_0\,.}
Again we use the subscript `$\RZ$' to distinguish the central
$U(1)_\RZ$ in ${\wt\CG}_0$ from the geometric $U(1)_\RR$ that acts on
the Seifert manifold
$M$.\countdef\UoneZ=53\UoneZ=\pageno\countdef\UoneR=54\UoneR=\pageno

We assume that the central $U(1)_\RZ$ subgroup of $\wt{\CG}_0$ acts
trivially on $\bar\CA$, so that the moment map for the central
generator $(0,a)$ of the Lie algebra is constant.  Then by
construction, we see from \COC\ and \BRAK\ that the new moment map for
the action of ${\wt\CG}_0$ on $\bar\CA$ is given by the sum 
\eqn\MUII{ \left\langle\mu,\left(\phi,a\right)\right\rangle \,=\,
-\int_M \kappa\^\Tr\!\left(\phi F_A\right) + \int_M
d\kappa\^\Tr\!\left(\phi A \right) + a\,,} 
and this moment map does satisfy the Hamiltonian condition
\eqn\HAMII{ \Big\{ \big\langle\mu,(\phi,a)\big\rangle,
\big\langle\mu,(\psi,b)\big\rangle \Big\} =
\Big\langle\mu,\big[(\phi,a),(\psi,b)\big]\Big\rangle\,.}
The action of the extended group $\wt{\CG_0}$ on $\bar\CA$ is thus
Hamiltonian with moment map in \MUII.

But ${\wt\CG}_0$ is still not the group $\CH$ which we must use to
perform non-abelian localization in Chern-Simons theory!  In order to
realize the action $S(A)$ as the square of the moment map $\mu$ for some
Hamiltonian group action on $\bar\CA$, the Lie algebra of the group
must first possess a non-degenerate, invariant inner product.  Just
as for the loop group extension $\wt{{LG}}$, the group ${\wt\CG}_0$
does not possess such an inner product.

However, we can elegantly remedy this problem, just as it was remedied
for the loop group, by also considering the geometric action of
$U(1)_\RR$ on $M$.  The $U(1)_\RR$ action on $M$ induces an action of
$U(1)_\RR$ on ${\wt\CG}_0$, so we consider the associated semidirect
product $U(1)_\RR \ltimes {\wt\CG}_0$.  A non-degenerate,
invariant inner product on the Lie algebra of $U(1)_\RR \ltimes
{\wt\CG}_0$ is given by 
\eqn\FRM{ \Big((p,\phi,a),(q,\psi,b)\Big) =
- \int_M \kappa\^d\kappa \, \Tr(\phi \psi) - p b - q a\,,}
in direct correspondence with \FRMLG.  As for the loop group, this
quadratic form is of indefinite signature, due to the hyperbolic form
of the last two terms in \FRM.

Finally, the $U(1)_\RR$ action on $M$ immediately induces a
corresponding action on $\CA$.  Since the contact form $\kappa$ is invariant
under this action, the induced $U(1)_\RR$ action on $\CA$ descends to a
corresponding action on the quotient $\bar\CA$.  In general, the
vector field upstairs on $\CA$ which is generated by an arbitrary
element $(p,\phi,a)$ of the Lie algebra of $U(1)_\RR \ltimes {\wt\CG}_0$ is
then given by 
\eqn\DA{ \delta A = -d_A \phi + p \, \lie_\RR A\,,}
where we recall that $\RR$ is the vector field on $M$ generating the
action of $U(1)_\RR$.  Clearly the moment for the new generator
$(p,0,0)$ is given by 
\eqn\MUIII{
\Big\langle\mu,(p,0,0)\Big\rangle \,=\, - \ha \, p \int_M
\kappa\^\Tr\!\left(\lie_\RR A\^A\right)\,.}
This moment is manifestly invariant under the shift symmetry and
descends to $\bar\CA$.

In fact, the action of $U(1)_\RR \ltimes {\wt\CG}_0$ on $\bar\CA$ is
Hamiltonian, with moment map
\eqn\MUIV{\Big\langle\mu,(p,\phi,a)\Big\rangle \,=\,
-\ha p \int_M \kappa\^\Tr\!\left(\lie_\RR A\^A\right) \,-\, \int_M
\kappa\^\Tr\!\left(\phi F_A\right) \,+\, \int_M d\kappa\^\Tr\!\left(\phi
A\right) \,+\, a\,.}
To check this statement, it suffices to compute
$\Big\{\langle\mu,(p,0,0)\rangle,\langle\mu,(0,\psi,0)\rangle\Big\}$,
which is the only nontrivial Poisson bracket that we have not already
computed.  Thus,
\eqn\HAMIII{\eqalign{
\Big\{\big\langle\mu,(p,0,0)\big\rangle,\big\langle\mu,(0,\psi,0)
\big\rangle\Big\}
\,&=\, \Omega\Big(p \, \lie_\RR A,-d_A\psi\Big)\,=\, p \int_M
\kappa\^\Tr\!\left(\lie_\RR A \^ d_A \psi\right)\,,\cr
\,&=\, -p \int_M \kappa\^\Tr\!\left(\lie_\RR\psi \, F_A\right) + p \int_M
d\kappa\^\Tr\!\left(\lie_\RR\psi \, A\right)\,,\cr
\,&=\, \Big\langle\mu,(0,p\,\lie_\RR\psi,0)\Big\rangle\,,}}
as required by the Lie bracket \BRAKII.  So we identify
\eqn\BIGCH{ \CH \,=\, U(1)_\RR \ltimes {\wt\CG}_0}
as the relevant group of Hamiltonian symmetries to use for
localization in Chern-Simons theory.\countdef\CurlyH=57\CurlyH=\pageno 

\subsec{The Shift-Invariant Action as the Square of the Moment Map}

By construction, the square $(\mu,\mu)$ of the moment map $\mu$ in
\MUIV\ for the Hamiltonian action of $\CH$ on $\bar\CA$ is a function on
$\bar\CA$ invariant under $\CH$.  The new Chern-Simons action
$S(A)$ in \SAA\ is also a function on $\bar\CA$ invariant under
$\CH$.  Given the high degree of symmetry, we certainly expect
that $(\mu,\mu)$ and $S(A)$ agree up to normalization.  We now check
this fact and fix the relative normalization.

From \FRM\ and \MUIV, we see immediately that 
\eqn\MUSQR{ (\mu,\mu) \,=\, \int_M\kappa\^\Tr\Big(\lie_\RR A\^A\Big) \,-\, 
\int_M\kappa\^d\kappa \, \Tr\!\left[\left({{\kappa\^F_A - d\kappa\^A} \over
{\kappa\^d\kappa}}\right)^2\right].}
Using the identity 
\eqn\IXA{ \iota_\RR A = {{d\kappa\^ A} \over {d\kappa\^\kappa}}\,,}
let us rewrite \MUSQR\ as 
\eqn\MUSQRII{\eqalign{ 
(\mu,\mu) \,&=\, \int_M \kappa\^\Tr\Big(\lie_\RR
A\^A\Big) \,+\, 2 \int_M \kappa \^ \Tr\Big[(\iota_\RR A) \, F_A\Big] \,-\, 
\int_M \kappa \^d\kappa \, \Tr\Big[(\iota_\RR A)^2\Big] \,-\,\cr
&-\,\int_M {1 \over {\kappa \^ d\kappa}} \,
\Tr\Big[\left(\kappa\^F_A\right)^2\Big]\,.\cr}}

We also require a somewhat more baroque identity,
\eqn\CSID{\eqalign{ 
\RC\RS(A) \,&=\, \int_M \Tr\!\left(A \^ dA + {2 \over 3}
A\^A\^A\right)\,,\cr
&=\, \int_M \kappa\^\Tr\Big(\lie_\RR A\^A\Big) \,+\,
2 \int_M \kappa \^ \Tr\Big[(\iota_\RR A) \, F_A\Big] \,-\, 
\int_M \kappa \^d\kappa \, \Tr\Big[(\iota_\RR A)^2\Big]\,.\cr}}
At first glance, the identity in \CSID\ may not be immediately
obvious, but it can be checked by elementary means.  See \S $3.5$ of
\BeasleyVF, especially $(3.56)$ and $(3.58)$ therein\foot{A slight
discrepancy exists between the numbering of equations in the arXived
and published versions of \BeasleyVF.  We refer throughout to the
version of \BeasleyVF\ on the electronic arXiv.}, for a very explicit
demonstration of \CSID.

From \MUSQRII\ and \CSID, we then obtain the beautiful result,
\eqn\CSMU{\eqalign{
(\mu,\mu)\,&=\, \RC\RS(A) \,-\, \int_M {1 \over {\kappa \^ d \kappa}}
\Tr\Big[ (\kappa\^F_A)^2\Big]\,,\cr
&=\, S(A)\,.\cr}}
So we finally write the Chern-Simons partition function as a symplectic
integral over $\bar\CA$ of the canonical form,
\eqn\PZCSV{ Z(\epsilon) \,=\, {1 \over {\Vol(\CG)}} \, \left({{-i} \over
{2 \pi \epsilon}}\right)^{\Delta_{\CG}/2} \, \int_{\bar\CA} \,
\exp{\!\left[\Omega + {i \over {2 \epsilon}}
\left(\mu,\mu\right)\right]}\,.}

\newsec{On Wilson Loops and Seifert Loops in Chern-Simons Theory}

Our goal is now to extend the results in Section $3$ concerning the
partition function of Chern-Simons theory to corresponding results for
the expectation values of Wilson loop operators.

As in Section $3$, we do not require the Seifert condition initially.
So we consider Chern-Simons theory on an arbitrary three-manifold $M$, 
endowed with a contact structure represented locally by a one-form
$\kappa$.  We similarly consider a general Wilson loop
operator 
\eqn\DEFWVCII{ W_R(C) \,=\, \Tr_R \, P\exp{\!\left(-\oint_C\!
A\right)}\,,}
where $C$ is an oriented closed curve smoothly embedded\foot{The
condition that $C$ be smoothly embedded in $M$ is not strictly
required to define $W_R(C)$ as a sensible operator in gauge theory.
Indeed, the Wilson loop expectation value in Chern-Simons theory can
be computed exactly even for the case that $C$ is an arbitrary closed
graph \WittenWF\ in $M$.} in $M$, and $R$ is an irreducible
representation of the simply-connected gauge group $G$.  

Throughout this paper, we find it useful to characterize the
representation $R$ in terms of its highest weight.  So we pick a
decomposition of the set $\FR$ \countdef\Roots=58\Roots=\pageno of
roots of $G$ into positive and negative subsets, ${\FR = \FR_{+}
\!\cup \FR_{-}}$.  \countdef\PosNegRoots=59\PosNegRoots=\pageno With
respect to that decomposition, we then take ${\alpha \ge 0}$ to be the
highest weight of $R$. \countdef\Aleph=60\Aleph=\pageno

We begin with the Wilson loop path integral,
\eqn\PZCSWL{\eqalign{
Z(\epsilon; C, R) \,&=\, {1 \over {\Vol(\CG)}} \, \left({1 \over {2 \pi
\epsilon}}\right)^{\Delta_{\CG}} \, \int \! \CD \! A
\;\; W_R(C) \; \exp{\left[{i \over {2 \epsilon}} \int_M \! \Tr\!\left( A \^
d A + {2 \over 3} A \^ A \^ A \right)\right]}\,,\cr
\epsilon &= {{2 \pi} \over k}\,,\qquad \Delta_{\CG} = \dim \CG\,.}}
Here we have been careful to normalize $Z(\epsilon; C, R)$ precisely
as we normalized the basic Chern-Simons path integral in \PZCS.

\subsec{A Semi-Classical Description of the Wilson Loop Operator}

This paper relies on only one good idea, to which we now come.

We clearly need a good idea, because a naive attempt to reapply
the path integral manipulations of Section $3$ to the Wilson loop path
integral in \PZCSWL\ runs immediately aground.  To illustrate the
difficulty with the direct approach, let us consider the obvious way
to rewrite the Wilson loop path integral in a shift-invariant form,
\eqn\PZCSWLII{ Z(\epsilon; C, R) \,=\, {1 \over
{\Vol(\CG)}} \, {1 \over {\Vol(\CS)}} \, \left({1 \over {2 \pi
\epsilon}}\right)^{\Delta_{\CG}} \,\int \CD \! A \, \CD \! \Phi
\,\, \SW_{\!R}(C) \,\exp{\left[ {i \over {2
\epsilon}} \RC\RS(A - \kappa \, \Phi)\right]}\,.}
Here $\SW_{\!R}(C)$ denotes the generalized Wilson loop operator defined
not using $A$ but using the shift-invariant combination $A - \kappa
\, \Phi$, so that 
\eqn\SHFTWL{ \SW_{\!R}(C) \,=\, \Tr_R \, P\exp{\!\left[-\oint_C \big(A -
\kappa \, \Phi\big)\right]}\,.}
Exactly as for our discussion of the analogous path integral in
\PZCSII, we can use the shift symmetry to fix ${\Phi = 0}$, after
which the path integral in \PZCSWLII\ reduces trivially to the Wilson loop
path integral in \PZCSWL.  

However, to learn something useful from \PZCSWLII\ we need to perform
the path integral over $\Phi$, and as it stands, this integral is not
easy to do.  Because the generalized Wilson loop operator $\SW_{\!R}(C)$
is expressed in \SHFTWL\ as a complicated, non-local functional of
$\Phi$, the path integral over $\Phi$ in \PZCSWLII\ is not a
Gaussian integral that we can trivially evaluate as we did for the
Chern-Simons path integral in \PZCSII.  

Very concretely, if we expand the path-ordered exponential in \SHFTWL,
we immediately encounter an awkward series of multiple integrals over
$C$ extending to arbitrary order in $\Phi$,
\eqn\BADPHI{\eqalign{
\SW_{\!R}(C) \,&=\, \dim R \,+\, \ha \oint_{C \times C} \!\! \Tr_R \,
P\Big[(A - \kappa \, \Phi)(\tau) \cdot (A - \kappa \, \Phi)(\tau')\Big] \,-\,\cr
&-\, {1\over 6} \oint_{C \times C \times C} \mskip -10mu \Tr_R \,
P\Big[(A - \kappa \, \Phi)(\tau) \cdot (A - \kappa \, \Phi)(\tau')
\cdot (A - \kappa \, \Phi)(\tau'') \Big] \,+\, \cdots\,.\cr}}
Here we have introduced separate parameters $\tau$, $\tau'$, and
$\tau''$ on $C$ to make the structure of the double and triple
integrals over $C$ manifest, and we recall that the path-ordering
symbol $P$ implies that the factors in the multiple integrals are
ordered so that ${\tau \ge \tau' \ge \tau''}$.  We indicate
similar terms of quartic and higher order by `$\cdots$'.\foot{In
principle, a term linear in $(A-\kappa \, \Phi)$ also appears in the
series expansion, but this term vanishes identically when $G$ is
simply-connected and simple, as we assume.}  As is hopefully clear,
any direct attempt\foot{Though we will not make use 
of the following observation in this paper, the path integral over
$\Phi$ in \PZCSWLII\ can be performed directly, just as in Section
$3$, in the very special case that $C$ is a Legendrian curve
\EtnyreII\ in $M$.  By definition, if $C$ is a Legendrian curve, the
tangent vector to $C$ lies everywhere in the kernel of $\kappa$.  In
this situation, $\Phi$ completely decouples from the generalized
Wilson loop operator $\SW_{\!R}(C)$ in \SHFTWL, and the ordinary
Wilson loop operator $W_R(C)$ is automatically shift-invariant.

However, we do not wish to make any special assumptions about $C$ at
the moment.  For one reason, if $M$ is a Seifert manifold with the
$U(1)$ invariant contact structure introduced in Section $3.2$, the
pullback of $\kappa$ to each Seifert fiber is non-vanishing by
construction, so the Seifert fibers are not Legendrian.} to perform
the path integral over $\Phi$ in \PZCSWLII\ would be painful, to say
the least.

A more fundamental perspective on our problem is the following.  Let
us return to the description of the ordinary Wilson loop operator  
$W_R(C)$ as the trace in the representation $R$ of the holonomy of
$A$ around $C$,
\eqn\DEFWVCIII{ W_R(C) \,=\, \Tr_R \, P\exp{\left(-\oint_C\!
A\right)}\,.}
As observed by Witten in \S $3.3$ of \WittenHF, this description of
$W_R(C)$ should be regarded as intrinsically quantum mechanical, for
the simple reason that $W_R(C)$ can be naturally interpreted in 
\DEFWVCIII\ as the partition function of an auxiliary quantum system
attached to the curve $C$.  We will eventually make this interpretation
very precise, but briefly, the representation $R$ is to be
identified with the Hilbert space of the system, the holonomy of $A$
is to be identified with the time-evolution operator around $C$, and
the trace over $R$ is the usual trace over the Hilbert space 
that defines the partition function in the Hamiltonian formalism.

Because the notion of tracing over a Hilbert space is inherently
quantum mechanical, any attempts to perform essentially classical path
integral manipulations involving the expressions in \SHFTWL\ or
\DEFWVCIII\ are misguided at best.  Rather, if we hope to generalize 
the simple, semi-classical path integral manipulations of Section
$3$ to apply to the Wilson loop path integral in \PZCSWL, we need to
use an alternative description for the Wilson loop operator that is itself
semi-classical.

More precisely, we want to replace the quantum mechanical trace over
$R$ in \DEFWVCIII\ by a path integral over an auxiliary bosonic field
$U$ which is attached to the curve $C$ and coupled to the connection
$A$ as a background field, so that schematically  
\eqn\NEWWC{ W_R(C) \,=\, \int \!\CD\!U \, \exp{\Big[
i\,{\Rc\Rs}_\alpha\big(U; A|_C\big)\Big]}\,.}
Here ${\Rc\Rs}_\alpha\big(U; A|_C\big)$ is an action, depending upon the
representation $R$ through its highest weight $\alpha$, which is a
local, gauge-invariant functional of the auxiliary field $U$ and the
restriction of $A$ to $C$.  Not surprisingly, this semi-classical
description \NEWWC\ of $W_R(C)$ turns  out to be the key ingredient
required to reformulate the Wilson loop path integral in a
shift-invariant fashion.

The idea of representing the Wilson loop operator by a path integral
as in \NEWWC\ is far from new.  In the context of Chern-Simons
theory, this device has already been applied to the canonical
quantization of the theory by Elitzur and collaborators in \ElitzurNR.
As will be clear, though, the path integral description \NEWWC\ of the
Wilson loop operator holds much more generally for any gauge theory in
any dimension.  In the context of four-dimensional Yang-Mills theory,
the semi-classical description of $W_R(C)$ is then much older, going back
(at least) to work of Balachandran, Borchardt, and Stern
\BalachandranUB\ in the 1970's.  See
\refs{\AlekseevVX\DiakonovFC{--}\ItzhakiRC} for other
appearances of this idea, including a recent application to the volume
conjecture for the colored Jones polynomial in \DijkgraafSB.

Despite such history, we now review in some detail how the path
integral description \NEWWC\ of the Wilson loop operator works.  We
do so both for sake of completeness and to introduce a few geometric
ideas which become necessary later.  See also \S $7.7$ of \Deligne\
for a very clear and somewhat more concise exposition of the following
material.

As we have indicated, the basic idea behind the path integral
description \NEWWC\ of $W_R(C)$ is very simple.  We interpret the
closed curve $C$ as a periodic ``time'' for the field $U$, and we 
apply the Hamiltonian formalism to rewrite the path integral over $U$
axiomatically as the quantum mechanical trace of the corresponding
time-evolution operator around $C$,
\eqn\ZC{ W_R(C) \,=\, \Tr_\SH \, P\exp{\!\left(-i
\oint_C\!{\bf H}\right)}\,.} 
Here $\SH$ is the Hilbert space which we obtain by quantizing the
field $U$, and ${\bf H}$ is the Hamiltonian which acts upon $\SH$ to
generate infinitesimal translations along $C$.  

Comparing the conventional description of the Wilson loop operator in
\DEFWVCIII\ to the axiomatic expression in \ZC, we see that the two
agree if we identify\foot{We follow the standard physical definition
according to which ${\bf H}$ is a hermitian operator, accounting for
the `$-i$' in \ZC.  We also recall that the gauge field $A$ is valued
in the Lie algebra $\Fg$, so $A$ is anti-hermitian and no `$i$'
appears in the holonomy.}
\eqn\ISQM{\eqalign{
R \,&\longleftrightarrow\,\SH\,,\cr
P\exp{\!\left(-\oint_C\! A\right)}\,&\longleftrightarrow\, P
\exp{\!\left(-i \oint_C\!{\bf H}\right)}\,.\cr}}
Hence to make the Wilson loop path integral in \NEWWC\ precise, we
need only exhibit a classical theory on $C$, for which the gauge group
$G$ acts as a symmetry, such that upon quantization we obtain a
Hilbert space $\SH$ isomorphic to $R$ and for which the
time-evolution operator around $C$ is given by the holonomy of $A$,
acting as an element of $G$ on $R$. 

\bigskip\noindent{\it Coadjoint Orbits of $G$}\smallskip

Now, of the two identifications in \ISQM, the more fundamental by far
is the identification of the irreducible representation $R$ with a
Hilbert space, obtained by quantizing some classical phase space upon
which $G$ acts as a symmetry.  So before we even consider what
classical theory must live on $C$ to describe the Wilson loop
operator, we can ask the simpler and more basic question ---
what classical phase space must we quantize to obtain $R$ as a
Hilbert space?

This question is beautifully answered by the Borel-Weil-Bott theorem
\BottBW, which explains how to obtain each irreducible representation
of $G$ by quantizing a corresponding coadjoint orbit.  Though the
quantum interpretation of the Borel-Weil-Bott theorem is quite
standard (see for instance \S $15$ of \KirillovAA), we nonetheless
review it now.  Once we do so, we will find it very easy to exhibit
the classical defect theory that must live on $C$ to describe the
Wilson loop operator.

More or less as a means to establish notation and conventions,
let us first recall some elementary facts about the geometry of the
coadjoint orbits of $G$.  We first fix a maximal torus ${T \subset
G}$, for which ${\Ft \subset \Fg}$ is the associated Cartan
subalgebra.\countdef\MaxT=61\MaxT=\pageno\countdef\Maxt=62\Maxt=\pageno 
By definition, the coadjoint orbits of $G$ are embedded in the dual
$\Fg^*$ of the Lie algebra $\Fg$.  However, given the invariant metric
on $\Fg$ defined by the pairing 
\eqn\METG{ (x,y) \,=\, -\Tr(x y)\,,\qquad x,y \in \Fg\,,}
we are free to identify ${\Fg \cong \Fg^*}$ and hence equivalently to 
consider the adjoint orbits of $G$ embedded in $\Fg$ itself.  Though
perhaps slightly unnatural from a purely mathematical perspective, we
find the latter convention convenient.  Thus, given an element
${\lambda \in \Ft}$, we let ${\CO_\lambda \subset \Fg}$ be the orbit
through $\lambda$ under the adjoint action of
$G$. \countdef\Littlelam=63\Littlelam=\pageno\countdef\BigOl=64\BigOl=\pageno

Alternatively, $\CO_\lambda$ can be regarded as a quotient $G/G_\lambda$,
where $G_\lambda$ is the subgroup of $G$ which fixes $\lambda$ under the
adjoint action. \countdef\BigGl=65\BigGl=\pageno  If $\lambda$ is a
generic element of $\Ft$, then ${G_\lambda = T}$, in which case
$\lambda$ is said to be {\sl regular}.  However, $G_\lambda$ can be
strictly larger than $T$, culminating in the extreme case that
${\lambda = 0}$ and ${G_\lambda = G}$.  The distinction between
$\lambda$ regular and irregular will at times be important, but for
the moment we treat these cases uniformly.

As our notation suggests, we assume that $G_\lambda$ acts on $G$ from
the right, so that the quotient $G/G_\lambda$ is defined by the relation 
\eqn\QUOTR{ g \sim g\,h^{-1}\,,\qquad g \in G\,,\qquad h \in
G_\lambda\,.} 
With this convention, we identify $G/G_\lambda$ with $\CO_\lambda$ via
the map 
\eqn\QUOTO{ g \, G_\lambda \,\longmapsto\, g\,\lambda\,g^{-1}\,,}
and  under \QUOTO, the left-action of $G$ on itself descends to a
transitive action of $G$ on $\CO_\lambda$.

In order to discuss the quantization of $\CO_\lambda$ as a classical
phase space, we must endow $\CO_\lambda$ with additional geometric
structure, starting with a coadjoint symplectic form $\nu_\lambda$.
To describe the coadjoint symplectic form explicitly, we introduce
the canonical left-invariant one-form $\theta$ on $G$ which is valued
in $\Fg$,  
\eqn\CANL{ \theta \,=\, g^{-1} \, dg\,,\qquad g \in
G\,.}\countdef\CartanForm=66\CartanForm=\pageno
Using the form `$\Tr$', we then define a real-valued, pre-symplectic
one-form $\Theta_\lambda$ on $G$,
\eqn\BIGTH{ \Theta_\lambda \,=\, -\big(\lambda\,,\theta\big) \,=\,
\Tr\big(\lambda\,\theta\big)\,,}\countdef\PreSym=67\PreSym=\pageno
in terms of which we set 
\eqn\COADJ{ \nu_\lambda \,=\, d\Theta_\lambda \,=\,
\ha\big(\lambda\,,[\theta\,,\theta]\big) \,=\,
-\ha\big(\theta\,,[\lambda\,,
\theta]\big)\,.}\countdef\Coadj=68\Coadj=\pageno
In passing from \BIGTH\ to \COADJ, we use that ${d\theta = -\theta \^
\theta = -\ha\big[\theta, \theta\big]}$.

Quite literally, we have written $\nu_\lambda$ in \COADJ\ as a
two-form on $G$.  However, since $G_\lambda$ preserves $\lambda$ under
the adjoint action, $\nu_\lambda$ is invariant under both the
left-action of $G$ and the right-action of $G_\lambda$ and vanishes upon
contraction with any vector tangent to $G_\lambda$, so this two-form
descends to an invariant two-form on $\CO_\lambda$.  

As a two-form on $\CO_\lambda$, clearly $\nu_\lambda$ is closed.
Furthermore, if we let $\Fg_\lambda$ \countdef\Biggl=69\Biggl=\pageno
denote the Lie algebra of the stabilizer group $G_\lambda$ and
${\Fg\ominus\Fg_\lambda}$ denote the orthocomplement to $\Fg_\lambda$
in $\Fg$, then the adjoint action of $\lambda$ on
${\Fg\ominus\Fg_\lambda}$ is by definition non-degenerate, with 
non-zero eigenvalues.  Identifying ${\Fg\ominus\Fg_\lambda}$ geometrically
with the tangent space to ${\CO_\lambda = G/G_\lambda}$ at the identity
coset, we see from \COADJ\ that $\nu_\lambda$ is thus non-degenerate
as a symplectic form on $\CO_\lambda$.

\bigskip\noindent{\it The Coadjoint Moment Map}\smallskip

One goal of the present discussion is to keep careful track of the
various signs and orientations which eventually enter our localization
computations.  We have already chosen a particular sign in the
definition \BIGTH\ of the pre-symplectic one-form $\Theta_\lambda$, and
hence in the definition of the invariant symplectic form
$\nu_\lambda$.  This choice of sign has a particularly felicitous 
consequence for the moment map which describes the infinitesimal
action of $G$ on $\CO_\lambda$.

To compute the moment map for the action of $G$ on $\CO_\lambda$,
we identify $\CO_\lambda$ with the quotient $G/G_\lambda$ under the
map in \QUOTO.  The vector field $V(\phi)$ on $\CO_\lambda$ generated
by an element ${\phi\in\Fg}$ is then given at the coset ${g \,
G_\lambda}$ simply by 
\eqn\DELTG{ \delta g = \phi \cdot g\,.}  
From \COADJ, we also see that 
\eqn\IOTVP{ \iota_{V(\phi)} \nu_\lambda \,=\, \iota_{V(\phi)}
d\Theta_\lambda \,=\, -d\left(\iota_{V(\phi)}\Theta_\lambda\right).} 
In passing to the second equality in \IOTVP, we use that the Lie
derivative ${\lie_{V(\phi)} = \{d\,,\iota_{V(\phi)}\}}$ along
$V(\phi)$ annihilates the invariant one-form $\Theta_\lambda$.  Thus
via \MOMMAPEQ\ and \BIGTH, the moment map
${\mu:\CO_\lambda\rightarrow\Fg^*}$ for the action of $G$ on
$\CO_\lambda$ is given by 
\eqn\COADJMOM{ \langle\mu,\phi\rangle \,=\, -\iota_{V(\phi)}
\Theta_\lambda \,=\, \big(g\,\lambda\,g^{-1}, \phi\big)\,=\,
-\Tr\big[(g\,\lambda\,g^{-1}) \cdot \phi\big]\,.}

An arbitrary constant could {\it a priori} appear in \COADJMOM, but
only if we set this constant to zero does the moment map $\mu$ satisfy
the Hamiltonian condition in \HOMMUII.  Hence $G$ acts in a Hamiltonian
fashion on $\CO_\lambda$, with moment map given by the natural
embedding ${\CO_\lambda \subset \Fg \cong \Fg^*}$ in \QUOTO.

\bigskip\noindent{\it $\CO_\lambda$ as a K\"ahler Manifold}\smallskip

In a nutshell, the Borel-Weil-Bott theorem concerns the algebraic
geometry of the orbit $\CO_\lambda$.  So we must also introduce a
complex structure $\SJ$ on
$\CO_\lambda$. \countdef\CurlyJ=70\CurlyJ=\pageno  Like the coadjoint 
symplectic form, $\SJ$ will be invariant under $G$.  Additionally,
$\SJ$ will be compatible with $\nu_\lambda$ in the sense that
$\nu_\lambda(\,\cdot\,,\SJ\,\cdot\,)$ defines a homogeneous 
K\"ahler metric on $\CO_\lambda$ for which $\nu_\lambda$ is the
K\"ahler form.  The existence of such a complex structure can be
understood in various ways.  Here we take a rather down-to-earth
approach and exhibit $\SJ$ directly as an invariant tensor on $\CO_\lambda$.

At the outset, we find it convenient to introduce the complexified Lie 
algebras ${\Ft_\BC \!= \Ft \otimes \BC}$, ${\Fg_\BC \!= \Fg \otimes
\BC}$, and ${\Fg_{\lambda,\BC} = \Fg_\lambda^{} \otimes \BC}$.
\countdef\LieGc=71\LieGc=\pageno\countdef\Maxtc=72\Maxtc=\pageno  The
invariant metric \METG\ on $\Fg$ extends immediately to a hermitian
metric on $\Fg_\BC$, and we identify the hermitian complement
${\Fg_\BC\!\ominus\Fg_{\lambda,\BC}}$ with the complexified tangent space
to $\CO_\lambda$ at the identity coset.

Any invariant tensor on $\CO_\lambda$ is determined by its value at a
single point.  Hence an invariant complex structure $\SJ$ on
$\CO_\lambda$ can be described algebraically as a splitting of the
complexified tangent space at the identity coset into two
complementary, half-dimensional subspaces $\Fg^{(1,0)}$ and $\Fg^{(0,1)}$,
\eqn\SPLTGC{ \Fg_\BC\!\ominus\Fg_{\lambda,\BC} \,=\,
\Fg^{(1,0)}\oplus\Fg^{(0,1)}\,,}  
which we declare to consist of the respective holomorphic and
anti-holomorphic tangent vectors at that point and upon which
$\SJ$ acts with eigenvalues $\pm i$.

The splitting in \SPLTGC\ can be obtained from any decomposition 
of the root system $\FR$ of $G$ into positive and negative subsets
$\FR_\pm$, so that ${\FR = \FR_{+} \!\cup \FR_{-}}$.  As standard, we
often write ${\beta > 0}$ for positive roots ${\beta \in \FR_+}$, and
similarly ${\beta < 0}$ for negative roots
${\beta\in\FR_-}$. \countdef\Betas=73\Betas=\pageno  

Of course, the decomposition ${\FR = \FR_{+} \!\cup \FR_{-}}$ is not 
unique, and different choices are related by the action of the Weyl
group $\FW$ of $G$.  However, in the generic case that ${\lambda \in
\Ft}$ is regular, we can fix this ambiguity by requiring that
$\lambda$ itself lies in the positive Weyl chamber ${\RC_+ \subset
\Ft}$.  We recall that $\RC_+$ is the polyhedral cone, with vertex at
the origin, defined by the inequalities 
\eqn\POSWEYL{ \RC_+ \,=\, \big\{ \xi \in \Ft\,\big|\,
\langle\beta,\xi\rangle \ge 0 \hbox{ for all } \beta > 0\big\}\,.}
\countdef\PosWyl=74\PosWyl=\pageno  Thus, if $\lambda$ is regular, we
simply take $\FR_+$ to be the subset of ${\beta\in\FR}$ such that
${\langle\beta,\lambda\rangle > 0}$.  Again, we often write ${\lambda
\ge 0}$ to indicate that $\lambda$ lies in $\RC_+$, with strict
positivity when $\lambda$ is regular.

If $\lambda$ is not regular, then inevitably some non-zero roots
${\beta_\perp \in \FR}$ satisfy ${\langle\beta_\perp,\lambda\rangle =
0}$.  Such ${\beta_\perp\in\FR}$ can be identified as roots of the
stabilizer group $G_\lambda$.  If $G_\lambda$ is non-trivial, we
simply pick $\FR_\pm$ so that $\lambda$ lies on a boundary wall of the
positive Weyl chamber $\RC_+$.  This choice is determined up to the action of
the Weyl group $\FW_\lambda$ of $G_\lambda$.  By definition, $\lambda$
is fixed by $\FW_\lambda$, and as will be clear, the ambiguity under
$\FW_\lambda$ is an unbroken gauge symmetry which eventually factors
out of our computations.

Given the decomposition ${\FR = \FR_+ \!\cup \FR_-}$, we obtain an
associated decomposition of ${\Fg_\BC \!\ominus \Ft_\BC}$ into
positive and negative rootspaces $\Fg_\pm$, so that ${\Fg_\BC
\!\ominus \Ft_\BC = \Fg_+ \!\oplus \Fg_-}$.  Briefly, to define
$\Fg_\pm$ themselves, we recall that the rootspace $\Fe_\beta$
\countdef\eRtspace=75\eRtspace=\pageno associated to any non-zero root
$\beta$ is the one-dimensional eigenspace in ${\Fg_\BC \!\ominus
\Ft_\BC}$ upon which elements ${\xi\in\Ft}$ act via the Lie bracket
with eigenvalue $+i \, \langle \beta, \xi \rangle$. That is, if
$x_\beta$ is an element of $\Fe_\beta$,
\eqn\ROOTSP{ \left[\xi\,, x_\beta\right] \,=\, +i \, \langle \beta,
\xi \rangle \, x_\beta\,,\qquad\qquad x_\beta \,\in\,\Fe_\beta\,.}
The rootspaces $\Fg_\pm$ are then given by the direct sums of the
eigenspaces $\Fe_\beta$ for positive and negative $\beta$,
\eqn\POSROOTSP{\Fg_{+} \,=\,\bigoplus_{\beta\in\FR_+} \,
\Fe_\beta\,, \qquad\qquad 
\Fg_{-} \,=\, \bigoplus_{\beta\in\FR_-} \,
\Fe_\beta\,.}

We now meet a crucial sign.  To define the holomorphic and
anti-holomorphic tangent spaces $\Fg^{(1,0)}$ and $\Fg^{(0,1)}$ in
\SPLTGC, we naturally use the positive and negative rootspaces
$\Fg_\pm$.  But which of $\Fg_+$ and $\Fg_-$ is to be holomorphic?  

The answer to this question is determined by the requirement that
$\nu_\lambda(\,\cdot\,,\SJ\,\cdot\,)$ defines a positive-definite,
as opposed to negative-definite, hermitian form on the tangent space
to $\CO_\lambda$.  Using the definition of $\nu_\lambda$ in \COADJ\
and the convention that ${\lambda\ge0}$, one can check that the correct
assignment is 
\eqn\HOLTN{ \Fg^{(1,0)} \,=\, (\Fg_\BC \ominus \Fg_{\lambda,\BC}) \cap
\Fg_{+},\qquad\qquad \Fg^{(0,1)} \,=\, (\Fg_\BC \ominus
\Fg_{\lambda,\BC}) \cap \Fg_{-}\,.}
\countdef\LieGplus=76\LieGplus=\pageno  In taking the intersection of
$\Fg_\BC \ominus \Fg_{\lambda,\BC}$ with $\Fg_+$ to define the
holomorphic tangent space $\Fg^{(1,0)}$, we allow for the possibility
that $\lambda$ is irregular.  If $\lambda$ is regular and
${\Fg_\lambda = \Ft}$, then the assignment in \HOLTN\ 
merely reduces to\foot{At this stage, a warning is in order.  In more
algebraic approaches to the Borel-Weil-Bott theorem, for which ${G/T =
G_\BC/B}$ is presented as the quotient of the associated complex group $G_\BC$
by a Borel subgroup $B$, the standard convention is to take the 
roots of $B$ to be positive roots of $G$.  Hence in much of the
literature, the holomorphic tangent space $\Fg^{(1,0)}$ is identified with
the negative rootspace $\Fg_-$, opposite to \HOLTN.  That convention
ultimately leads to a characterization of $R$ via lowest (as opposed
to highest) weights, which we prefer to avoid.}
\eqn\HOLTNRG{ \Fg^{(1,0)} \!= \Fg_+\,,\qquad\qquad \Fg^{(0,1)} \!=
\Fg_-\,,\qquad\qquad \lambda > 0 \quad \hbox{regular}\,.}

With $\Fg^{(1,0)}$ and $\Fg^{(0,1)}$ in hand, we immediately define
the tensor $\SJ$ as an invariant almost-complex structure on
$\CO_\lambda$.  By virtue of the elementary Lie algebra relation
${[\Fg_{+}, \Fg_{+}] \subseteq \Fg_{+}}$, the Nijenhuis tensor
associated to this almost-complex structure vanishes, so that $\SJ$ in
fact defines an honest, integrable complex structure on
$\CO_\lambda$.  Finally, because the hermitian metric on $\Fg_\BC$
only pairs elements of $\Fg_{+}$ with elements of $\Fg_{-}$, and
because $\lambda$ preserves these spaces under the adjoint action, we
see directly from \COADJ\ that $\nu_\lambda$ has
holomorphic/anti-holomorphic type $(1,1)$ with respect to $\SJ$.  Thus 
$\nu_\lambda$ and $\SJ$ together endow the orbit $\CO_\lambda$ with a
homogeneous K\"ahler structure.

\bigskip\noindent{\it A Prequantum Line Bundle on $\CO_\lambda$}\smallskip

From a physical perspective, the Borel-Weil-Bott theorem explains how
to quantize $\CO_\lambda$ as a K\"ahler manifold, so that is what
we shall now do.  We follow the recipe of geometric quantization,
for which a nice reference is \Woodhouse.  Admittedly, the following
exposition puts the cart before the horse, since the notions of
geometric quantization were developed partly based upon this example.

In general, quantization is a delicate procedure.  Athough $\lambda$
parametrizes a continuous family of K\"ahler orbits in $\Fg$,  only a
discrete subset of this family can actually be quantized.
Specifically, according to the usual Bohr-Sommerfeld condition, the
symplectic form $\nu_\lambda$ must derive from a corresponding prequantum line
bundle over $\CO_\lambda$.  By definition, a prequantum line bundle on
any symplectic manifold $X$ with symplectic form $\Omega$ is a unitary
line bundle with connection whose curvature is given by ${+i\,\Omega}$.
In the present setting, as we now explain, the coadjoint symplectic
form $\nu_\lambda$ on $\CO_\lambda$ derives from a prequantum line
bundle precisely when ${\lambda\in\Ft}$ is quantized as a weight of $G$.

The last statement may require a minor clarification.  The
weight lattice ${\Gamma_{\rm wt} \subset \Ft^*}$ of $G$ is defined
intrinsically as a lattice in the dual $\Ft^*$ of the Cartan
subalgebra.  However, we once again use the invariant metric \METG\ on
$\Fg$ to identify ${\Ft \cong \Ft^*}$, under which the dual
${\lambda^*\!\in\Ft^*}$ of $\lambda$ is defined via the relation 
${\langle\lambda^*\,,\,\cdot\,\rangle \,=\,
(\lambda\,,\,\cdot\,) \,=\, -\Tr(\lambda\,\,\cdot\,)}$.
So by the quantization of $\lambda$ as a weight of $G$, we mean
literally that $\lambda^*$ is quantized as an element of $\Gamma_{\rm
wt}$.

Henceforth, to avoid cluttering the notation, we implicitly
apply the isomorphism ${\Ft\cong\Ft^*}$ induced by `${-\Tr}$'
and do not attempt to distinguish $\lambda$ from $\lambda^*$.  By the
same token, if ${\alpha\in\Gamma_{\rm wt}}$ is a weight of $G$, we do
not distinguish $\alpha$ from its dual in $\Ft$.

To explain why $\lambda$ must be quantized as a weight of $G$, we
classify the possible line bundles on the coadjoint orbits of $G$.  As
an immediate simplifying observation, all coadjoint orbits
$\CO_\lambda$ take the form of quotients $G/G_\lambda$, where the
stabilizer $G_\lambda$ necessarily contains the torus $T$.  Under the
quotient map from $G/T$ to $G/G_\lambda$, any line bundle on
$G/G_\lambda$ immediately pulls back to a line bundle on $G/T$.  So we
need only classify line bundles on $G/T$.

As usual, line bundles on $G/T$ are classified topologically by
elements of $H^2(G/T;\BZ)$.  On the other hand, $G/T$ sits
tautologically as the base of a principal $T$-bundle whose total space
is $G$,
\eqn\PGTBT{\matrix{
&T\,\longrightarrow\,G\cr
&\mskip 55mu\big\downarrow\cr
&\mskip 55mu G/T\cr}\,.}
The Leray spectral sequence\foot{For an excellent introduction to
spectral sequences, see \S $14$ of \BottT.} associated to the
fibration in \PGTBT\ implies that 
\eqn\ISOI{ H^2\big(G/T; \BZ\big) \,=\, H^1\!\big(T;\BZ\big) \,=\,
\Hom\!\big(T,\, U(1)\big)\,.}
For our application to Chern-Simons theory, we assume that the group
$G$ is simply-connected.  The lattice of homomorphisms from $T$ to
$U(1)$ is then isomorphic to the weight lattice of $G$, 
\eqn\WTG{ \Gamma_{\rm wt} \,\cong\, \Hom\!\big(T,\, U(1)\big)\,.}
\countdef\Gamwt=77\Gamwt=\pageno  Thus, at the level of topology, line
bundles on $G/T$ are in one-to-one correspondence with weights of $G$.

The relation between line bundles on $G/T$ and weights of $G$ can be
understood more concretely as follows.  For each weight
${\alpha\in\Gamma_{\rm wt}}$, we let $\varrho_\alpha$ be the corresponding 
homomorphism, 
\eqn\VRRHO{ \varrho_\alpha:T\longrightarrow U(1)\,.}
Explicitly, $\varrho_\alpha$ is given in terms of $\alpha$ by 
\eqn\HOMRHO{
\varrho_\alpha(t) \,=\,
\exp{\!\big[i\,\langle\alpha\,,\xi\rangle\big]},
\qquad\qquad t=\exp(\xi)\,,\qquad\qquad \xi\in\Ft\,.}
As indicated in \HOMRHO, $\xi$ is a logarithm of ${t\in T}$.  This
logarithm is only defined up to shifts ${\xi \mapsto \xi + 2
\pi y}$, where ${y\in\Ft}$ satisfies the integrality condition 
\eqn\CONDY{ \exp{\!(2\pi y)}\,=\,1\,.}
Equivalently, each element $y$ in \CONDY\ corresponds to a
homomorphism from $U(1)$ to $T$.  By definition, the lattice of such
homomorphisms is the cocharacter lattice of
$G$,\countdef\Gamch=78\Gamch=\pageno 
\eqn\COCHAR{ \Gamma_{\rm cochar} \,\cong\,
\Hom\!\big(U(1),\,T\big)\,,}
which for simply-connected $G$ reduces to the coroot lattice
$\Gamma_{\rm cort}$.  In any event, because the lattices in \WTG\ and
\COCHAR\ are canonically dual over $\BZ$, the weight $\alpha$
satisfies ${\langle\alpha,y\rangle\in\BZ}$ for all ${y\in\Gamma_{\rm
cochar}}$, and the homomorphism $\varrho_\alpha$ is well-defined.

Using the homomorphism in \HOMRHO, we now introduce the associated
line bundle $\FL(\alpha)$ over $G/T$,
\eqn\BGFLA{ \FL(\alpha) \,=\, G \times_{\varrho_\alpha} \BC\,.}
Here ``$\times_{\varrho_\alpha}$'' indicates that elements in the
product $G \times \BC$ are identified under the action of $T$ as 
\eqn\LRHO{ t\cdot\left(g\,, v\right)
\,=\,\left(g\,t^{-1},\,\varrho_\alpha(t) \cdot v\right)\,,\qquad g \in
G\,,\qquad v \in \BC\,,\qquad t \in T\,.}
Thus, each weight $\alpha$ determines a complex line bundle
$\FL(\alpha)$.  Conversely, since ${H^2(G;\BZ) = 0}$ under our
assumptions on $G$, any line bundle on $G/T$ pulls back to a
topologically-trivial line bundle on $G$ and hence can be represented 
by a quotient of ${G \times \BC}$ as in \LRHO.  Finally, if $\alpha$
happens to be an irregular weight, the homomorphism $\varrho_\alpha$
in \HOMRHO\ extends uniquely to a homorphism from $G_\alpha$ to
$U(1)$, and $\FL(\alpha)$ is the pullback from a corresponding line
bundle on $G/G_\alpha$.  \countdef\Gamrt=79\Gamrt=\pageno

To interpret $\FL(\alpha)$ as a prequantum line bundle, we still need
to endow $\FL(\alpha)$ with a hermitian metric and a compatible
unitary connection, both of which are invariant under $G$.
Thankfully, the metric and the connection are straightforward to 
describe.

For the metric, we note that sections of $\FL(\alpha)$ can be
identified with complex functions $f$ on $G$ which transform
equivariantly under the action of $T$,
\eqn\SLR{ f(g \, t^{-1}) \,=\, \varrho_\alpha(t) \cdot f(g)\,,\qquad 
g \in G\,,\qquad t \in T\,.}
Hence the invariant hermitian metric for complex functions on $G$,
unique up to norm, immediately descends to an invariant hermitian
metric for sections of $\FL(\alpha)$.

Because the metric on $\FL(\alpha)$ is invariant under $G$,
a compatible unitary connection on $\FL(\alpha)$ must be invariant
as well.  This observation, along with our explicit description of
$\varrho_\alpha$, suffices to fix the connection uniquely.  

Of course, one can describe a connection on a line bundle in many ways.  
As we have already defined $\FL(\alpha)$ in terms of the principal
$T$-bundle in \PGTBT, we also describe its connection in these terms. 
Globally, we take the invariant unitary connection on $\FL(\alpha)$
to be a left-invariant one-form $B$ on $G$ which is valued 
in the Lie algebra of $U(1)$, identified with $+i\,\BR$, and for which
${d_B = d + B}$ is the covariant derivative acting on
sections of $\FL(\alpha)$, identified as in \SLR\ with equivariant
functions on $G$.  Covariance of $d_B$ then implies that ${d_B
\varrho_\alpha(t) = 0}$, so $B$ must be given in terms of the
left-invariant Cartan form $\theta$ by 
\eqn\BIGB{ B \,=\, -i\,\langle\alpha\,,\theta\rangle \,=\,
i\,\Theta_{\alpha}\,.}
Since $\Theta_{\alpha}$ is the pre-symplectic one-form satisfying
${\nu_{\alpha} = d\Theta_{\alpha}}$, the coadjoint symplectic form
$\nu_\lambda$ on $\CO_\lambda$ derives from a prequantum line bundle
precisely when ${\lambda = \alpha}$ is quantized as a weight of $G$, 
in which case the prequantum line bundle is $\FL(\alpha)$.  Actually,
because $\lambda$ lies by convention in the positive Weyl chamber,
${\alpha \ge 0}$ is necessarily a {\sl highest} weight for $G$.

\bigskip\noindent{\it The Borel-Weil-Bott Theorem in Brief}\smallskip

At this point, the statement of the Borel-Weil-Bott theorem is
remarkably simple.\foot{For sake of time, we regrettably omit the
refinements due to Bott \BottBW.}  Because $\nu_{\alpha}$ is
the K\"ahler form for an invariant K\"ahler metric on
$\CO_{\alpha}$, the prequantum line bundle $\FL(\alpha)$ immediately
becomes a holomorphic line bundle of positive curvature over
$\CO_{\alpha}$.  As such, the prequantum line bundle naturally admits
global holomorphic sections, and via the usual recipe of geometric
quantization, we construct a Hilbert space $\SH$ for $\CO_{\alpha}$ as
the space of holomorphic sections of $\FL(\alpha)$.  That is, 
\eqn\GEOMHO{ \SH \,=\, H^0_{\bar\partial}\big(\CO_{\alpha},\,
\FL(\alpha)\big)\,,\qquad\qquad \alpha \ge 0\,.}

On the other hand, the holomorphic sections of $\FL(\alpha)$
automatically transform in a finite-dimensional representation of $G$
induced from its action on $\CO_{\alpha}$.  Perhaps not surprisingly,
the role of the Borel-Weil-Bott theorem is to identify this
representation as none other than the irreducible representation $R$
with highest weight $\alpha$,
\eqn\BWBT{ R \,\cong\, H^0_{\bar\partial}\big(\CO_{\alpha},\,
\FL(\alpha)\big)\,.}
In these terms, the Borel-Weil-Bott isomorphism \BWBT\ exhibits
$R$ as a Hilbert space obtained by quantizing $\CO_{\alpha}$ as  a
classical phase space.

Though we have introduced the general adjoint orbit $\CO_{\alpha}$,
the Borel-Weil-Bott theorem can be (and often is) stated purely in
terms of the maximal orbit described by $G/T$.  As mentioned earlier,
if $\alpha$ is not a regular weight of $G$, the prequantum line bundle
$\FL(\alpha)$ over ${\CO_{\alpha} = G/G_{\alpha}}$ pulls back to a
holomorphic line bundle over $G/T$.  The holomorphic sections of the
pullback also transform on $G/T$ in the representation $R$, so at
least from the perspective of algebraic geometry, the irreducible
representations of $G$ can be constructed more uniformly using
holomorphic line bundles on $G/T$ alone. 

However, if $\alpha$ is not a regular weight, the closed two-form
$\nu_{\alpha}$ is degenerate on $G/T$.  As an extreme example, if
${\alpha = 0}$, then $\FL(\alpha)$ is the trivial line bundle, and
${\nu_{\alpha} = 0}$.  Of course, in that case
$\CO_{\alpha}$ is simply a point.  So to interpret $\FL(\alpha)$
as a prequantum line bundle associated to a necessarily {\sl
non-degenerate} symplectic form, we prefer to state the
Borel-Weil-Bott theorem with $\FL(\alpha)$ being a line bundle over
the general adjoint orbit $\CO_{\alpha}$, as opposed to the maximal
orbit given by $G/T$.

\bigskip\noindent{\it An Elementary Example}\smallskip

Although we have given a careful description of the prequantum line
bundle $\FL(\alpha)$, it would unfortunately take us a bit too far
afield to demonstrate here the crucial Borel-Weil-Bott isomorphism in
\BWBT.  However, see for instance \S $23$ of \Fulton, particularly
Exercise $23.62$, for a nice proof.

In place of a proof, we end our geometric digression with an
elementary example of the Borel-Weil-Bott theorem in action.  Namely, 
we let $G$ be $SU(2)$, so that $T$ is $U(1)$, and we identify $G/T$
with $S^2$ via the Hopf fibration.  To endow $G/T$ with a complex
structure, we pick a positive Weyl chamber of $SU(2)$, corresponding
to an orientation for $S^2$, after which we regard $S^2$ as
$\BC\BP^1$.

To illustrate our conventions explicitly, we parametrize
elements $g$ of $SU(2)$ by means of complex variables ${(X,Y) \in
\BC^2}$, so that
\eqn\GSUTWO{ g \,=\, \pmatrix{\bar Y& X\cr -\bar X& Y}\,,\qquad |X|^2
\,+\, |Y|^2 \,=\, 1\,.}
Similarly, we parametrize elements $t$ in the maximal torus $T$ by
means of an angular variable $\theta$,
\eqn\TSUTWO{ t \,=\, \pmatrix{\e{i\,\theta}&0\cr 0&\e{-i\,\theta}}\,,\qquad
\theta \in [0,2\pi]\,.}
Under the action ${g\mapsto g \cdot t^{-1}}$, the variables
$(X,Y)$ then transform homogeneously with unit charge, ${(X,Y) \mapsto
\left(\e{i \theta} \, X,\e{i \theta} \, Y\right)}$.  As for the
complex structure on $G/T$, we pick the positive Weyl chamber so that the
positive rootspace $\Fg_+$ of $SU(2)$ is spanned by strictly
upper-triangular matrices, 
\eqn\POSRTSSU{ \Fg_+ \,=\, \BC \cdot \pmatrix{0&1\cr 0&0}\,.}
Conversely, $\Fg_-$ is spanned by lower triangular matrices.  With
this convention, ${[X:Y]}$ become homogeneous coordinates on
$\BC\BP^1$. 

The weights of $SU(2)$ are labelled by a single integer $m$.  If we
use the invariant metric `$-\Tr$' to dualize ${\Ft \cong \Ft^*}$, the
weights take the concrete form 
\eqn\WTSSUT{ \alpha \,=\, {{m}\over 2} \, \pmatrix{i&0\cr
0&-i}\,,\qquad m \in \BZ\,,}
such that $\alpha$ is positive when ${m \ge 0}$.  The associated
homomorphism ${\varrho_\alpha:T\rightarrow U(1)}$ in \HOMRHO\ then
becomes ${\varrho_\alpha(t) \,=\, \exp{\!(i\,m\,\theta)}}$, and the
holomorphic line bundle $\FL(\alpha)$ defined in \BGFLA\ and \LRHO\ 
is precisely the line bundle of degree, or monopole number,
$m$ on $\BC\BP^1$.

Via \SLR, we see that holomorphic sections of $\FL(\alpha)$ are given
by degree $m$ polynomials in the homogeneous coordinates ${[X:Y]}$ of
$\BC\BP^1$.  Under the action of $SU(2)$, these polynomials transform
naturally in the irreducible representation of dimension $m+1$,
realizing the Borel-Weil-Bott isomorphism in \BWBT.   Finally, the
K\"ahler metric on $\BC\BP^1$ compatible with the curvature of the
invariant connection on $\FL(\alpha)$ for ${m>0}$ is just a multiple
of the round, Fubini-Study metric.

\bigskip\noindent{\it More About Quantum Mechanics on Coadjoint
Orbits}\smallskip

Our discussion of the Borel-Weil-Bott theorem so far has been fairly 
abstract.  As the next step towards explaining the path integral
description of $W_R(C)$, let us place the isomorphism in \BWBT\ into its
proper physical context.

To start, we clearly want to consider the quantum mechanics of a
single particle moving on the orbit $\CO_\alpha$.  Formally, such a
particle is described by a one-dimensional sigma model of maps $U$
from $S^1$ to $\CO_\alpha$,\countdef\BigU=80\BigU=\pageno
\eqn\BIGU{ U\negthinspace: S^1 \longrightarrow \CO_\alpha\,.}
Though not particularly essential at the moment, we take the particle
worldline to be compact in anticipation of our application to the
Wilson loop operator, where the abstract $S^1$ will be identified with
the embedded curve ${C \subset M}$.

We now want to pick a classical action for $U$ such that the orbit
$\CO_\alpha$, with symplectic form $\nu_{\alpha}$, appears as the
corresponding classical phase space.  Once we do so, we can
immediately invoke the Borel-Weil-Bott isomorphism \BWBT\ to identify
the Hilbert space $\SH$ obtained by quantizing $U$ with the
representation $R$.  But what classical action for $U$ should we pick?

An immediate guess might be to consider the standard two-derivative
sigma model action associated to the invariant K\"ahler metric on
$\CO_\alpha$ defined by $\nu_{\alpha}$ and $\SJ$,
\eqn\STWODU{ S_\sigma(U) \,=\, \ha \, \oint_{S^1} \eta^{-\ha} \,\,
\nu_{\alpha}\big(dU,\,\SJ\cdot dU\big) \,=\, \ha \, \oint_{S^1}
\eta^{-\ha} \, (\gamma_\alpha)_{m n} \,\, {{dU}\over{d\tau}}^{\!m}
{{dU}\over{d\tau}}^{\!n}\,.}
Here $\tau$ is a coordinate along $S^1$, and ${\eta \equiv
\eta_{\tau \tau}}$ is a worldline metric on $S^1$.  Also,
${\gamma_\alpha = \nu_{\alpha}(\,\cdot\,,\SJ\,\cdot\,)}$ is the
invariant K\"ahler metric on $\CO_\alpha$.  Finally, for sake of
concreteness, we parametrize the map $U$ using local coordinates $u^m$
on $\CO_\alpha$, in terms of which we write the second expression in
\STWODU.

Though $S_\sigma(U)$ is a natural action to consider, it cannot be
correct for two reasons.  First and foremost, $S_\sigma(U)$ describes
a particle freely moving on $\CO_\alpha$, and the classical phase
space for this particle is not the orbit $\CO_\alpha$ but its 
cotangent bundle $T^*\CO_\alpha$, with the standard cotangent
symplectic structure.  The Hilbert space obtained by quantizing
$T^*\CO_\alpha$ would then be the infinite-dimensional space $L^2(\CO_\alpha)$
of square-integrable functions on $\CO_\alpha$, as opposed to the
finite-dimensional representation $R$.  Second, the sigma model action
$S_\sigma(U)$ depends upon the choice of a worldline metric $\eta$ on
$S^1$.  However, the Wilson loop operator $W_R(C)$ certainly does not
depend upon the choice of a metric on the corresponding curve $C$.

As these objections suggest, the correct action for $U$ should be of
first-order, not second-order, in the ``time'' derivative ${d/d\tau}$
along $S^1$, so that the classical phase space has a chance to be compact.
Furthermore, the action for $U$ should be topological in the sense that it 
does not depend upon the choice of a metric on $S^1$.

The requirements above are hallmarks of a Chern-Simons action.  Given
that we already possess an invariant unitary connection
$\Theta_{\alpha}$ on the line bundle $\FL(\alpha)$, let us consider the
following Chern-Simons-type action for
$U$,\countdef\CSalpha=81\CSalpha=\pageno
\eqn\SRU{ {\Rc\Rs}_\alpha(U) \,=\, \oint_{S^1}\!U^*(\Theta_{\alpha}) \,=\,
\oint_{S^1}\!(\Theta_{\alpha}){}_m \, {{dU}\over{d\tau}}^{\!m}\,.}
Here $U^*(\Theta_{\alpha})$ denotes the pullback of $\Theta_{\alpha}$ to
a connection over $S^1$, and ${\Rc\Rs}_\alpha(U)$ is quite literally the
one-dimensional Chern-Simons action for $U^*(\Theta_{\alpha})$.  Once
again, in the second expression of \SRU\ we write the pullback
$U^*(\Theta_{\alpha})$ using coordinates $u^m$ on $\CO_\alpha$, in terms
of which $\Theta_{\alpha}$ is represented locally by the one-form
${(\Theta_{\alpha}){}_m \, du^m}$.

As with any Chern-Simons action, the functional ${\Rc\Rs}_\alpha(U)$ is not
strictly invariant under ``large'', homotopically non-trivial gauge
transformations on $S^1$.  However, so long as ${\alpha\in\Gamma_{\rm wt}}$ is 
quantized as a weight of $G$ and hence $+i\,\Theta_{\alpha}$ is defined as
an honest $U(1)$-connection over $\CO_\alpha$, the value of
${\Rc\Rs}_\alpha(U)$ is well-defined modulo $2\pi$, a sufficient condition
to discuss a sensible path integral.

From a more physical perspective, the first-order action for $U$ in
\SRU\ specifies the minimal coupling of a charged particle on
$\CO_\alpha$ to a background magnetic field given by the coadjoint
symplectic form ${\nu_{\alpha} = d\Theta_{\alpha}}$.  From this 
perspective, the quantization of $\alpha$ as a weight of $G$ follows 
from the quantization of flux on a compact space.  The dynamics of
such a charged particle moving on $\CO_\alpha$ in the background
magnetic field $\nu_\alpha$ are then described by the total action 
\eqn\TOTS{ S_{\rm tot}(U) \,=\, S_\sigma(U) \,+\, {\Rc\Rs}_\alpha(U)\,.}
To pass from \TOTS\ to \SRU\ alone, we consider the low-energy limit,
for which ${\eta\rightarrow\infty}$.  In this limit, the 
two-derivative sigma model action $S_\sigma(U)$ in \TOTS\ becomes 
irrelevant, and the topological Chern-Simons-type term ${\Rc\Rs}_\alpha(U)$
provides an effective action for the Landau groundstates of the
particle.\foot{Even in the limit ${\eta\to\infty}$, the classical
action ${\Rc\Rs_\alpha(U)}$ may receive a non-trivial one-loop correction
when we perform the functional integral over non-zero Fourier modes
of $U$ in the sigma model with action $S_{\rm tot}(U)$.  Such
a quantum correction can at most shift the weight $\alpha$ to a new
weight $\alpha'$ and thus does not alter our general analysis of the
effective topological sigma model with action ${\Rc\Rs_\alpha(U)}$.
In Section $7$, we compute explicitly a related quantum shift in
$\alpha$.} 

To proceed with the analysis of the topological sigma model with
action ${\Rc\Rs}_\alpha(U)$, we first determine the classical phase space
for $U$.  Under a variation $\delta U$, the variation of the
first-order action ${\Rc\Rs_\alpha(U)}$ is given by 
\eqn\DLSRU{ \delta{\Rc\Rs}_\alpha(U) \,=\, \oint_{S^1}\!\nu_{\alpha}\big(\delta
U,\,d U\big) \,=\, \oint_{S^1}\!(\nu_{\alpha}){}_{m n} \,\, \delta
U^m \, {{dU}\over{d\tau}}^{\!n}\,.}
Once again, the first equality in \DLSRU\ derives from the relation
${\nu_{\alpha} = d\Theta_{\alpha}}$.  

Because $\nu_{\alpha}$ is non-degenerate as a symplectic form on
$\CO_\alpha$, the equations of motion which follow from \DLSRU\ imply
that ${dU^n/d\tau = 0}$.  Thus, classical solutions for $U$ are constant
maps, and the classical phase space for $U$ is the orbit $\CO_\alpha$
itself, as we originally required.  Furthermore, if we consider
$\delta U^n$ to represent a linearized coordinate on a neighborhood of
the identity coset in $\CO_\alpha$, then we see from \DLSRU\ that the
canonical momentum conjugate to $\delta U^n$ is ${\Pi_n = (\nu_{\alpha}){}_{m
n}\,\delta U^m}$.  Hence the classical Poisson bracket on $\CO_\alpha$
derives from the coadjoint symplectic form $\nu_{\alpha}$.

To quantize the compact phase space for $U$, we immediately
invoke our previous, more abstract discussion of the Borel-Weil-Bott
theorem.  Namely, because of the background magnetic field
$\nu_{\alpha}$, the wavefunctions for $U$ in the K\"ahler polarization
of $\CO_\alpha$ transform as holomorphic sections of the prequantum
line bundle $\FL(\alpha)$, where by convention ${\alpha \ge 0}$.  Thus
the Hilbert space for $U$ is again ${\SH = H^0_{\bar\partial}\big(\CO_\alpha,
\FL(\alpha)\big)}$, and in the more physical language above, $\SH$ can
be interpreted as the space of Landau levels of a single electron
moving on $\CO_\alpha$ in the magnetic field $\nu_{\alpha}$.  The
Borel-Weil-Bott theorem \BWBT\ then asserts that these Landau levels 
transform under $G$ in the representation $R$.

\bigskip\noindent{\it Coupling to the Bulk Gauge Field}\smallskip

The one-dimensional Chern-Simons sigma model with target space
$\CO_\alpha$ realizes the first identification in \ISQM\ required to
describe the Wilson loop operator $W_R(C)$ semi-classically.  To
realize the second identification in \ISQM, we simply attach the sigma
model to the curve ${C \subset M}$, and we couple the sigma model
field $U$ to the bulk connection $A$ by promoting the global action of
$G$ on $\CO_\alpha$ to a gauge symmetry.

The action of $G$ on $\CO_\alpha$ is perhaps most transparent when the
orbit is described as a quotient $G/G_\alpha$.  Even before we gauge the
Chern-Simons sigma model, let us apply this description of
$\CO_\alpha$ to rewrite the functional ${\Rc\Rs}_\alpha(U)$ in \SRU\ a bit
more concretely.  Given the sigma model map $U$ with target
${\CO_\alpha = G/G_\alpha}$, we lift $U$ to a map   
\eqn\MAPg{ g\negthinspace:S^1 \longrightarrow G\,.} 
Here $U$ and $g$ are related via ${U = g\,\alpha g^{-1}}$, just as in
\QUOTO.  In terms of $g$, we then rewrite ${\Rc\Rs}_\alpha(U)$ using
the explicit description \BIGTH\ of $\Theta_{\alpha}$ as a left-invariant
one-form on $G$,
\eqn\SRUII{ {\Rc\Rs}_\alpha(U) \,=\, \oint_{S^1}\!U^*(\Theta_{\alpha}) \,=\,
\oint_{S^1}\!\Tr\big(\alpha \cdot g^{-1} dg\big)\,.}

The expression for ${\Rc\Rs}_\alpha(U)$ in \SRUII\ is admirably
explicit, but let us quickly consider how it depends upon the choice of $g$.
First, a lift of $U$ to $g$ always exists over $S^1$.   The
obstruction to lifting $U$ is measured by a characteristic class of
degree two on $\CO_\alpha$, which vanishes for trivial reasons when
pulled back to $S^1$ under $U$.  But of course $g$ is only determined
by $U$ up to the local right-action of $G_\alpha$, under which a map
${h\negthinspace:S^1\rightarrow G_\alpha}$ acts on $g$ as ${g \mapsto
g \, h^{-1}}$.  If $h$ is homotopically trivial, one can easily check
that value of $\Rc\Rs_\alpha(U)$ in \SRUII\ is invariant under this
transformation.  Otherwise, under ``large'' gauge transformations by
homotopically non-trivial maps ${h\negthinspace:S^1\rightarrow
G_\alpha}$, the value of the functional in \SRUII\ generally shifts by
an  integral multiple of $2\pi$, reflecting the harmless ambiguity in
a Chern-Simons action.

In order to forestall potential confusion, let me emphasize that the
lift from $U$ to $g$ in \SRUII\ serves merely as a convenient device
to describe the functional ${\Rc\Rs}_\alpha(U)$ in a manifestly
$G$-invariant fashion.  All path integrals that we discuss will be
integrals over the space of sigma model maps 
${U\negthinspace:S^1 \rightarrow \CO_\alpha}$, as opposed to the space 
of maps ${g\negthinspace:S^1 \rightarrow G}$.

We now attach the sigma model with target space $\CO_\alpha$ to the
curve ${C \subset M}$, and we gauge the global action of $G$ on
$\CO_\alpha$.  The procedure of gauging the sigma model on $C$ is
entirely straightforward.  Nevertheless, at some risk of pedantry, we
provide a systematic discussion.

In order to keep track of the local action of $G$ in the sigma model,
we recall that the bulk gauge theory on $M$ is associated to a  
fixed principal $G$-bundle $P$ over $M$ on which $A$ is a connection,
\eqn\PGTM{\matrix{
&G \longrightarrow P\cr
&\mskip 50mu\big\downarrow\cr
&\mskip 47mu M\cr}\,\,.}
The restriction of $P$ to the embedded curve ${C\subset M}$ determines
a principal $G$-bundle $P|_C$ over $C$, and the map $g$ in \MAPg\ now
transforms as a section of $P|_C$.  Finally, the sigma model field $U$
itself transforms geometrically as a section of an associated bundle
$Q$ with fiber $\CO_\alpha$ over $C$,
\eqn\PGTC{\matrix{
&\CO_\alpha \longrightarrow Q\cr
&\mskip 63mu\big\downarrow\cr
&\mskip 63mu C\cr}\,\,.}
Explicitly, using the same notation as in \BGFLA, $Q$ is given by
${P|_C \times_G \CO_\alpha}$.

As a homogeneous bundle on $\CO_\alpha$, the prequantum line
bundle $\FL(\alpha)$ extends fiberwise in \PGTC\ to a line bundle over
$Q$.  This line bundle carries its own unitary connection, determined
by both $\Theta_{\alpha}$ and the bulk connection $A$ to be 
\eqn\WTBIGTH{ \Theta_{\alpha}(A) \,=\, \Tr\big(\alpha \cdot \, g^{-1}
d_A g\big)\,,}
where 
\eqn\COVDG{ d_A g = dg \,+\, A|_C \cdot g\,.}
As one can readily check, the covariant derivative in \COVDG\ behaves
correctly under a gauge transformation acting on $A$ and $g$ as 
\eqn\GT{\delta A \,=\, -d_A \phi\,,\qquad\qquad \delta g = \phi|_C \cdot
g\,,} 
where $\phi$ is a section of the adjoint bundle $\ad(P)$ on $M$.
Hence the expression for $\Theta_{\alpha}(A)$ in \WTBIGTH\ is invariant
under gauge transformations in $G$ on $M$ and otherwise transforms like
$\Theta_{\alpha}$ as a unitary connection over $Q$.

Using $\Theta_{\alpha}(A)$, we now write the gauge-invariant version of
the sigma model action on $C$ as 
\eqn\SRUIII{\eqalign{
{\Rc\Rs}_\alpha\big(U; A|_C\big) \,&=\, \oint_C
U^*\big(\Theta_{\alpha}(A)\big) \,=\, \oint_C
(\Theta_{\alpha}){}_m \, {{d_A U}\over{d\tau}}^{\!m}\,,\cr 
&=\, \oint_C \Tr\big(\alpha \cdot g^{-1} d_A g\big)\,.\cr}}
For concreteness, in the first line of \SRUIII\ we again write the
sigma model action in terms of local coordinates $u^m$ on
$\CO_\alpha$, and we have introduced there the covariant derivative
$d_A$ acting on $U$.  The action of $d_A$ on $U$ is inherited immediately
from its action on $g$ in \COVDG.

\bigskip\noindent{\it A Semi-Classical Description of the Wilson Loop
Operator}\smallskip

The gauged sigma model on $C$ with target space $\CO_\alpha$ finally
provides the promised semi-classical description for the Wilson loop
operator.  In short,
\eqn\NEWWCIII{\eqalign{ 
W_R(C) \,&=\, \epsilon^{\Delta_\alpha/2} \int_{L\CO_\alpha}\mskip -10mu
\CD\!U \, \exp{\Big[ i\,{\Rc\Rs}_\alpha\big(U;
A|_C\big)\Big]}\,,\cr
\epsilon \,&=\, {{2 \pi} \over k}\,,\qquad \Delta_\alpha \,=\,
\dim\,L\CO_\alpha\,.\cr}}

Formally, the sigma model path integral in \NEWWCIII\ is an integral
over the space of sections of the bundle $Q$ in \PGTC.  If we are
willing to forget that this space carries an action by the group $\CG$
of gauge transformations and thence to trivialize $Q$, the space
of sections of $Q$ can be equivalently regarded as the free loopspace
$L\CO_\alpha$ of the orbit $\CO_\alpha$.  We find the latter notation
convenient in \NEWWCIII\ and will phrase the discussion in terms of
$L\CO_\alpha$ throughout the remainder of the
paper.\countdef\LoopOalpha=82\LoopOalpha=\pageno

As is standard for sigma models, the loopspace $L\CO_\alpha$ carries
an invariant metric induced from the invariant K\"ahler metric on
$\CO_\alpha$, and the path integral measure $\CD\!U$ is the associated
Riemannian measure on $L\CO_\alpha$.

For our particular application to Chern-Simons theory, we have also
multiplied the path integral in \NEWWCIII\ by a formal power of the
Chern-Simons coupling $\epsilon$.  Because the relevant power is
half the dimension of $L\CO_\alpha$, this prefactor can be
equivalently absorbed into the Riemannian measure ${\CD\!U}$ on
$L\CO_\alpha$ if the sigma model metric on $L\CO_\alpha$ is rescaled
by $\epsilon$.  Indeed, this rescaling is the ultimate reason we
introduce the prefactor at all.  As will prove convenient, we write 
${\epsilon L\CO_\alpha}$ to indicate the loopspace $L\CO_\alpha$ 
equipped with the sigma model metric induced from the invariant
K\"ahler metric on ${\epsilon\,\CO_\alpha \equiv 
\CO_{\epsilon \, \alpha}}$.  In that abbreviated notation,
\eqn\NEWWCIIIRES{ W_R(C) \,=\, \int_{\epsilon L\CO_\alpha}\mskip -10mu 
\CD\!U \, \exp{\Big[ i\,{\Rc\Rs}_\alpha\big(U; A|_C\big)\Big]}\,,}
with no errant prefactor.  We will explain the need for the effective
rescaling ${\CO_\alpha \mapsto \epsilon\,\CO_\alpha}$ later, in Section $4.3$.

We are left to establish the relationship between the semi-classical
description of the Wilson loop operator in \NEWWCIIIRES\ and its
conventional description as the trace in $R$ of the holonomy of $A$
around $C$.  As we indicated initially, this task amounts to 
demonstrating the correspondence  
\eqn\ISQMII{\eqalign{
\SH \,&\longleftrightarrow\,R\,,\cr
P\exp{\left(-i \oint_C\!{\bf H}\right)}\,&\longleftrightarrow\,
P\exp{\left(-\oint_C\! A\right)}\,.\cr}}

We have already discussed how the quantization of $U$ leads to the
first identification in \ISQMII\ before we couple to $A$, and the
corresponding story in the gauged sigma model proceeds essentially as
before.  From the gauge-invariant action in \SRUIII, we immediately
deduce that $U$ satisfies the equation of motion ${d_A U^n = 0}$ and
hence must be covariantly constant.\foot{Equivalently, the lift $g$
satisfies ${[\alpha, g^{-1} d_A g] = 0}$ and is covariantly constant
up to the right-action of $G_\alpha$.}  As a result, all classical
trajectories for $U$ are determined uniquely by parallel transport
from an arbitrary initial value $U_0$.  

Very explicitly, if we pass from the curve $C$ with a chosen basepoint
to the universal cover $\BR$, then the classical time-evolution of $U$
is given by 
\eqn\COVDGII{ U(\tau) \,=\, P\exp{\left(-\int^\tau_0\!\!A\right)} \cdot
U_0\,,\qquad\qquad U_0 \in \CO_\alpha\,.}
Here $\tau$ is a time-coordinate along $\BR$, with ${0 \in \BR}$ being
a lift of the basepoint on $C$.  The minus sign in \COVDGII\ arises
from our convention that ${d_A = d + A}$.  We now identify the classical phase
space for $U$ with the coadjoint orbit $\CO_\alpha$, as parametrized
by the initial value $U_0$ in \COVDGII.  Similarly, the Poisson
bracket on the phase space is still determined by the coadjoint
symplectic form $\nu_{\alpha}$, and the Hilbert space $\SH$ for $U$ is
again isomorphic to $R$.

To establish the second identification in \ISQMII, we must consider
the time-evolution in the gauged sigma model.  As \COVDGII\ indicates,
the classical time-evolution in this theory is not completely trivial,
insofar as it depends upon parallel transport using the restriction of
the connection $A$ to $C$.  Hence the classical time-evolution around
$C$ is given by the holonomy of $A$, acting as an element of $G$ on
the phase space $\CO_\alpha$.

Since we identify the Hilbert space for $U$ with the space of
holomorphic sections of the prequantum line bundle $\FL(\alpha)$, the
action of the classical time-evolution operator on $\CO_\alpha$
immediately lifts to a corresponding quantum action on ${\SH \cong
R}$.  Thus as we require in \ISQMII, the quantum time-evolution
operator in the gauged sigma model is given by the holonomy of $A$
around $C$, acting as an element of $G$ on $R$.

The following remark is not essential, but it may resolve a small
puzzle for some readers.  To pass from the cover $\BR$ to $C$ in the
discussion above, we make the periodic identification ${\tau \sim \tau
+ 1}$.  For $U(\tau)$ in \COVDGII\ to be single-valued under this
identification, the holonomy ${P\exp{\!\left(-\oint_C A\right)}}$ must
preserve the initial value ${U_0\in\CO_\alpha}$ under the adjoint
action of $G$.  As a result, the classical holonomy of $A$ around $C$
is not arbitrary but lies in the centralizer of $U_0$, a subgroup of
$G$ conjugate to the stabilizer $G_\alpha$.

At first glance, this restriction on ${P\exp{\!\left(-\oint_C 
A\right)}}$ might seem to conflict with the usual definition of the
Wilson loop operator, for which the holonomy around $C$ is
arbitrary.  However, the stabilizer group $G_\alpha$ always contains
the maximal torus $T$, so as $U_0$ ranges over points in $\CO_\alpha$,
the centralizer of $U_0$ ranges over all of $G$.  Once we integrate over
$U_0$ in the semi-classical description \NEWWCIIIRES\ of $W_R(C)$, the
holonomy of $A$ around $C$ is therefore unrestricted.  For this
reason, the appearance in \NEWWCIIIRES\ of the free, as opposed to
based, loopspace $L\CO_\alpha$ is crucial.

\subsec{A New Formulation of Chern-Simons Theory, Part II}

In obtaining the identifications in \ISQMII, we treat the gauge
field $A$ as a fixed background connection on $M$.  As a result, the 
semi-classical description of the Wilson loop operator in \NEWWCIIIRES\
is completely general, applicable to any gauge theory in any
dimension.  This elegant little idea seems not to have found significant
application in four-dimensional Yang-Mills theory, which may be one
reason that the idea has remained somewhat obscure.  Nonetheless, as a
small bit of cosmic justice, the semi-classical description of
$W_R(C)$ proves to be tailor-made for our study of Wilson loop
operators in Chern-Simons theory.

We return to the basic Wilson loop path integral in \PZCSWL, which
we now formulate using \NEWWCIIIRES\ as a path integral over the product
${\CA \times \epsilon L\CO_\alpha}$,
\eqn\NWWLPZ{ Z(\epsilon; C, R) \,=\, {1 \over {\Vol(\CG)}} \, \left({1
\over {2 \pi \epsilon}}\right)^{\Delta_{\CG}} \,\int_{\CA \times
\epsilon L\CO_\alpha} \mskip
-10mu\CD \! A \; \CD\!U \,\exp{\!\left[{i \over {2 \epsilon}} \,
\RC\RS\big(A\big) \,+\, i\, {\Rc\Rs}_\alpha\big(U; A|_C\big)\right]}\,.}
Our primary goal in the remainder of this section is to cast the
Wilson loop path integral \NWWLPZ\ into a shift-invariant form,
just as we did for the Chern-Simons partition function in Section $3$.
Yet before we perform any path integral manipulations, let us quickly
discuss the moduli space of classical solutions which arise as
critical points of the joint action
\eqn\SJT{ S_0(A,U) \,=\, \RC\RS(A) \,+\, 2\epsilon\,
{\Rc\Rs}_\alpha(U;A|_C)\,.}

\bigskip\noindent{\it On Classical Wilson Loops in Chern-Simons
Theory}\smallskip

In order to discuss the equations of motion for $A$ and ${U = g 
\alpha g^{-1}}$, we first find it convenient to rewrite the
topological sigma model action for $U$ in terms of a bulk integral
over $M$, 
\eqn\SRGAC{
{\Rc\Rs}_\alpha\big(U; A|_C\big) \,=\, \oint_C \Tr\big(\alpha \cdot g^{-1}
d_A g\big) \,=\, \int_M \delta_C \^ \Tr\big(\alpha \cdot g^{-1} d_A
g\big)\,.}
In passing to the second expression in \SRGAC, we have introduced a
two-form $\delta_C$ with delta-function support along $C$ to represent 
the Poincar\'e dual of this curve.\countdef\DeltaC=88\DeltaC=\pageno

Varying $\RC\RS(A)$ and ${\Rc\Rs}_\alpha(U; A|_C)$ with respect to $A$ and
$g$, we find that the classical equations of motion which follow from
the action in \SJT\ are given by 
\eqn\EOMGGIII{ F_A \,+\, \epsilon \, \big(g \alpha g^{-1}\big) \cdot 
\delta_C \,=\, 0\,, \qquad\qquad \left[\alpha\,,\,g^{-1} d_A
g\right] \,=\, 0\,,}
or equivalently in terms of $U$,
\eqn\EOMGGII{ F_A \,+\, \epsilon \, U \cdot \delta_C \,=\, 0\,,
\qquad\qquad d_A U \,=\, 0\,.}
These first equation in \EOMGGII\ implies that $A$ is a connection on
$M$ which is flat away from $C$ and otherwise has delta-function
curvature along $C$ fixed by the value of $U$ in
${\CO_\alpha\subset\Fg}$.  Also, as we applied previously, the second 
equation in \EOMGGII\ asserts that $U$ is covariantly constant along $C$.

By way of notation, we let $\SM(C,\alpha)$ denote the space of pairs
$(A,U)$ solving the system in \EOMGGII, modulo gauge transformations.  
In the special case ${\alpha = 0}$, $\SM(C,\alpha)$ immediately
reduces to the moduli space $\SM$ of flat connections on $M$, and as
well-known, that moduli space admits a concrete, finite-dimensional
presentation as the moduli space of homomorphisms $\varrho$ from the
fundamental group $\pi_1(M)$ to the gauge group $G$,
\eqn\HOMSIG{ \SM \,=\,\Big\{ \varrho:\pi_1(M) \rightarrow G\Big\}\Big/G\,.}
In terms of the gauge theory, $\varrho$ encodes the holonomies of a
flat connection on $M$.\countdef\Flat=83\Flat=\pageno

As we review now, the extended moduli space $\SM(C,\alpha)$
admits a completely analogous presentation, again as a moduli space of
homomorphisms to $G$.  Besides elucidating the classical
interpretation of the Wilson loop operator, this global perspective 
on $\SM(C,\alpha)$ proves to be a critical ingredient for our  
localization computations in Section $7$.

To analyze $\SM(C,\alpha)$, we first provide a local model for
classical configurations of $(A,U)$ which satisfy the equations of
motion \EOMGGII\ on a small tubular neighborhood $N_C$ of the curve
$C$.  Topologically $N_C$ is a solid torus, upon which we introduce
cylindrical coordinates $(r,\varphi,\tau)$.  Here $(r,\varphi)$ are
polar coordinates on a plane transverse to $C$, which passes through
the origin at ${r=0}$, and $\tau$ is an axial coordinate along $C$
with unit length.  We assume that the standard orientation on
$(r,\varphi,\tau)$ agrees with the given orientations on $M$ and $C$.
Explicitly in terms of the cylindrical coordinates $(r,\varphi,\tau)$,
the standard orientation on $N_C$ is given by the three-form 
${dr \^ d\varphi \^ d\tau}$, and $C$ itself is oriented by $d\tau$. 

Up to gauge transformations, local solutions to \EOMGGII\ can then 
be presented in terms of parameters $(U_0,V_0)$ as 
\eqn\LOCAC{ A \,=\, -{{\epsilon\,U_0}\over{2\pi}} \,
d\varphi \,+\, V_0 \, d\tau\,,\qquad\qquad U(\tau) \,=\,
U_0\,\in\,\CO_{\alpha}\,.}
Here $U_0$ is a constant taking values in the orbit ${\CO_{\alpha}
\subset \Fg}$, and $V_0$ is any element of $\Fg$ commuting with $U_0$,
so that 
\eqn\COMMUV{ \big[U_0, V_0\big] \,=\, 0\,.}
Using the identity ${d(d\varphi) = 2\pi\delta_C}$, one can immediately
check that the \LOCAC\ solves the equations of motion in \EOMGGII.

The description of $A$ in \LOCAC\ holds on the neighborhood $N_C$.
Away from $N_C$, classical configurations for $A$ are given by flat
connections on the complement $M^o$,
\eqn\SFRCOMP{ M^o \,=\, M - N_C\,.}
\countdef\KnotComp=85\KnotComp=\pageno Such a flat connection is
determined up to gauge-equivalence by its holonomies, which are now
encoded by a homomorphism $\varrho^o$ from the fundamental group of
the complement $M^o$ to $G$,
\eqn\BIGHMRIII{ \varrho^o:\pi_1(M^o) \longrightarrow G\,.}
In the special case that ${M=S^3}$, the group $\pi_1(M^o)$ is just the
classical knot group of $C$.  We will recall a few interesting 
examples of knot groups later, when we arrive at the computations in
Section $7$. 

Of course, $\varrho^o$ cannot be an arbitrary homomorphism in
\BIGHMRIII, since the holonomies of $A$ around $C$ and its meridian
are already fixed in terms of $\big(U_0, V_0\big)$ in \LOCAC.  As
standard, by the meridian of $C$ we mean the distinguished element 
${\Rm \in \pi_1(M^o)}$ which is represented by the small circle about
$C$ parametrized by $\varphi$ in \LOCAC\ and oriented according to
$d\varphi$.\foot{To characterize $\Rm$ more intrinsically, we note
that the boundary $\partial N_C$ of $N_C$ is a two-torus.  As a curve
on that boundary, $\Rm$ is then determined up to orientation as the
generator of the kernel of the map from ${\pi_1(\partial N_C) \cong
\BZ \times \BZ}$ to ${\pi_1(N_C) \cong \BZ}$.   To fix the orientation
of $\Rm$, we require that the curve in $\partial N_C$ which is homotopic to
$C$ itself to have positive intersection with $\Rm$, where $\partial
N_C$ is oriented according to the outward orientation on ${M^o = M -
N_C}$.}  Hence the homomorphism $\varrho^o$ in \BIGHMRIII\ must be
related to the constant $U_0$ in \LOCAC\ via 
\eqn\JFCOND{ \varrho^o(\Rm) \,=\, P\exp{\!\left(-\oint_\Rm A\right)} \,=\,
\exp{\!\left(\epsilon\,U_0\right)}\,,\qquad\qquad \epsilon \,=\,
{{2\pi}\over k}\,.}
\countdef\Meridian=86\Meridian=\pageno
As $U_0$ varies in $\CO_{\alpha}$, the holonomy around the meridian of
$C$ therefore takes values in the conjugacy class ${\FC_{\alpha/k}
\subset G}$ containing the group element $\exp{\!(2\pi\alpha\!/k)}$,
\eqn\JFCONDA{ \varrho^o(\Rm) \,\in\, \FC_{\alpha/k} \,=\,
\Cl\!\big[\!\exp{\!(2\pi\alpha\!/k)}\big]\,.}
\countdef\ConjCl=84\ConjCl=\pageno
Here $\Cl[\,\cdot\,]$ indicates the conjugacy class in $G$ containing
the given element.  We alert the reader that we have included a
convenient factor of $2\pi$ in the definition of $\FC_{\alpha/k}$.

By the same token, the holonomy of $A$ around $C$ is fixed by $V_0$ in
\LOCAC.  Moreover, since ${[U_0,V_0] = 0}$, the holonomy around $C$
necessarily commutes with the holonomy around $\Rm$, implying
${\left[\varrho^o(C),\varrho^o(\Rm)\right]=1}$.  At first glance, one 
might worry that this condition must be imposed as an additional
constraint on $\varrho^o(C)$, analogous to the constraint in
\JFCONDA.  Thankfully, that worry is misplaced for the following
elementary reason.  Both $C$ and $\Rm$ are represented by curves on the
boundary $\partial N_C$, a two-torus.  Trivially, $C$ and $\Rm$ commute as
elements of $\pi_1(\partial N_C)$, so they also commute as elements of
$\pi_1(M)$.  Hence the images of $C$ and $\Rm$ under any homomorphism
from $\pi_1(M)$ to $G$ automatically commute.  As a result, the condition
${\left[\varrho^o(C),\varrho^o(\Rm)\right]=1}$ is not an independent
constraint on $\varrho^o$, and we immediately eliminate the parameter
$V_0$ in favor of the holonomy $\varrho^o(C)$ from the description of
$\SM(C,\alpha)$.

Combining the local and global descriptions of $A$ on $N_C$ and
$M^o$ respectively, we present the extended moduli space
$\SM(C,\alpha)$ concretely as the space of pairs  $(\varrho^o, U_0)$  
which satisfy the compatibility condition in \JFCOND, modulo the
diagonal action of $G$.  So as a formal quotient,
\eqn\MAONE{ \SM(C,\alpha) \,=\,\Big\{(\varrho^o, U_0) \;\big|\;
\varrho^o(\Rm)=\exp{\!\left(\epsilon\,U_0\right)}\Big\}\Big/G\,.}

The description of $\SM(C,\alpha)$ in \MAONE\ is still somewhat
redundant.  For generic ${k\in\BZ}$ and ${\alpha\in\Gamma_{\rm wt}}$,
the relation in \JFCOND\ can be smoothly inverted to determine $U_0$
as a function of $\varrho^o(\Rm)$, thereby also eliminating $U_0$ from
the description of $\SM(C,\alpha)$.  Hence $\SM(C,\alpha)$ can be
presented more succinctly as the moduli space of homomorphisms from
$\pi_1(M^o)$ to $G$ which satisfy the necessary condition in \JFCONDA,
\eqn\MAONEII{\eqalign{
\SM(C,\alpha) \,&=\, \Big\{\varrho^o:\pi_1(M^o)\rightarrow G \;\big|\;
\varrho^o(\Rm) \in \FC_{\alpha/k}\Big\}\Big/G\,,\cr 
\FC_{\alpha/k} \,&=\,
\Cl\!\big[\!\exp{\!(2\pi\alpha\!/k)}\big]\,.\cr}}
\countdef\ExtFlat=87\ExtFlat=\pageno
Though the technical conditions on $k$ and $\alpha$ under which $U_0$
can be eliminated from \MAONE\ are not so important now, they will be
important later, and we state those conditions precisely in Section
$7.1$.  (See also Section $5.2$ for a warm-up discussion.)

To conclude our discussion of the classical Wilson loop, let us give
two very simple examples of the extended moduli space $\SM(C,\alpha)$.
The first example will really be a non-example, meant only to
illustrate that homomorphisms ${\varrho^o:\pi_1(M^o)\rightarrow G}$
satisfying the necessary condition in \JFCONDA\ may or may not exist,
depending upon the structure of $\pi_1(M^o)$.  So for a case in which
$\SM(C,\alpha)$ is actually empty, let $M$ be the product ${S^2 \times 
S^1}$, and for any point in $S^2$, let $C$ be the corresponding $S^1$
fiber over that point.  Then the meridian $\Rm$ of $C$ is contractible
in $M^o$, and a homomorphism $\varrho^o$ satisfying \JFCONDA\ exists
only if ${\alpha = 0}$.

As a slightly more interesting example, one which we will considerably
generalize in Section $7$, let us consider the case that $M$ is $S^3$
and $C$ is the unknot.  Then ${\pi_1(M^o) \cong \BZ}$ is freely
generated by $\Rm$, so the homomorphism $\varrho^o$ is determined
once its value on $\Rm$ is fixed.  Via \JFCOND, $\varrho^o(\Rm)$ is in
turn determined by ${U_0 \in \CO_{\alpha}}$, and $\SM(C,\alpha)$ is the
quotient $\CO_\alpha/G$.  Since $G$ acts transitively on $\CO_\alpha$,
this quotient is just a point, with a non-trivial stabilizer $G_\alpha$.

\bigskip\noindent{\it The Shift-Invariant Wilson Loop
Operator}\smallskip

We now arrive at our first main result, which is to extend the
path integral manipulations in Section $3$ to provide a
shift-invariant reformulation of the Wilson loop path
integral $Z(\epsilon; C, R)$.  Indeed, once we apply the
semi-classical description of $W_R(C)$ to rewrite $Z(\epsilon; C, R)$
as below, 
\eqn\NWWLPZA{ Z(\epsilon; C, R) \,=\, {1 \over {\Vol(\CG)}} \,
\left({1 \over {2 \pi \epsilon}}\right)^{\Delta_{\CG}} \,
\int_{\CA \times \epsilon L\CO_\alpha} \mskip
-10mu\CD \! A \; \CD\!U \,\exp{\!\left[{i \over {2 \epsilon}} \,
\RC\RS\big(A\big) \,+\, i\, {\Rc\Rs}_\alpha\big(U; A|_C\big)\right]}\,,}
the ideas in Section $3$ extend in an almost embarrassingly
straightforward fashion.

As in \PZCSWLII, we first consider the generalization of \NWWLPZA\
obtained by replacing $A$ with the shift-invariant combination ${A -
\kappa \, \Phi}$,
\eqn\PZCSWLII{ Z(\epsilon; C, R) \,=\, {1 \over
{\Vol(\CG)}} \, {1 \over {\Vol(\CS)}} \, \left({1 \over {2 \pi
\epsilon}}\right)^{\Delta_{\CG}} \, \int \CD\!A \, \CD\!U \, \CD\!\Phi
\, \exp{\!\left[{i \over {2 \epsilon}} \, S\big(A\,, \Phi\,, U\big)\right]}\,,}
where 
\eqn\SAPHIU{ S\big(A\,, \Phi\,, U\big) \,=\, \RC\RS\big(A - \kappa \,
\Phi\big) \,+\, 2 \epsilon \, {\Rc\Rs}_\alpha\big(U; A - \kappa \,
\Phi\big)\,.}
We assume that the shift symmetry $\CS$ acts on $A$ and $\Phi$
just as before, and $\CS$ acts trivially on $U$.  Upon setting
${\Phi = 0}$ with the shift symmetry, we reproduce \NWWLPZA\ 
as before.

On the other hand, to underscore the significance of \PZCSWLII, let us
expand the shift-invariant sigma model action ${\Rc\Rs}_\alpha\big(U; A
- \kappa \, \Phi\big)$ in terms of $\Phi$.  From \SRGAC, we
immediately find 
\eqn\SRGAPHI{ {\Rc\Rs}_\alpha\big(U; A - \kappa \, \Phi\big) \,=\,
{\Rc\Rs}_\alpha\big(U; A\big) \,-\, \int_M \kappa\^\delta_C \,
\Tr\big[(g \alpha g^{-1}) \, \Phi\big]\,.}
The essential observation to make about \SRGAPHI\ is simply that
$\Phi$ appears linearly.  Thus $\Phi$ still enters the total
shift-invariant action $S(A,\Phi, U)$ quadratically.  

To be explicit, we expand $S(A,\Phi,U)$ in terms of $\Phi$ to obtain    
\eqn\SAPHIUII{ S(A\,, \Phi\,, U) \,=\, \RC\RS(A) \,+\, 2 \epsilon \,
{\Rc\Rs}_\alpha(U; A) \,-\, \int_M \Big[ 2 \kappa \^ \Tr\left(\Phi \,
\CF_A\right) \,-\, \kappa \^ d \kappa \, \Tr(\Phi^2)\Big]\,.}
Here as a convenient shorthand, we introduce a `generalized' curvature
$\CF_A$ which includes the delta-function contribution from \SRGAPHI,
so that 
\eqn\NEWF{ \CF_A \,=\, F_A \,+\, \epsilon \left(g \alpha 
g^{-1}\right) \delta_C\,.}
\countdef\CurlyF=89\CurlyF=\pageno

Because the integral over $\Phi$ in \PZCSWLII\ is Gaussian, we
perform it exactly as before.  By virtue of the shift symmetry, the
remaining integral over the affine space $\CA$ then reduces to an
integral over the quotient ${\bar\CA = \CA / \CS}$, and we obtain
the promised shift-invariant reformulation of the general Wilson loop
path integral in Chern-Simons theory.  Thus,
\eqn\PZCSWLIII{ Z(\epsilon; C, R) \,=\, {1 \over
{\Vol(\CG)}} \, \left({{-i} \over {2 \pi
\epsilon}}\right)^{\Delta_{\CG}/2} \, \int_{\bar\CA \times \epsilon
L\CO_\alpha} \mskip -10mu \bar{\CD \! A} \,\CD\!U \,\exp{\!\left[ {i
\over {2 \epsilon}} S\big(A, U\big)\right]}\,,}
where 
\eqn\SAPHIUIII{ S(A, U) \,=\, \RC\RS(A) \,+\, 2 \epsilon \,
{\Rc\Rs}_\alpha(U; A|_C) - \int_M {1 \over {\kappa \^ d \kappa}}
\Tr\Big[ (\kappa \^ \CF_A)^2 \Big]\,.}

By construction, $S(A, U)$ is invariant under the shift ${\delta A =
\sigma \kappa}$, where $\sigma$ is an arbitrary function on $M$
valued in the Lie algebra $\Fg$ of the gauge group $G$.
Alternatively, using the description of ${\Rc\Rs}_\alpha(U; A|_C)$ in
\SRGAC, one can verify the shift-invariance of $S(A, U)$ directly.
Under the shift ${\delta A = \sigma \kappa}$, a new term linear 
in $\sigma$ arises from the defect action ${\Rc\Rs}_\alpha(U;A|_C)$.  
This term is then cancelled by the cross-term proportional to $\sigma$
which appears in the square of $\CF_A$.

\bigskip\noindent{\it A Word About the Wilson Link}\smallskip

Though our discussion throughout will focus for simplicity on the
case of a single Wilson loop operator, the previous path integral
manipulations extend immediately to Wilson links in $M$.

To state the general result, we consider a product of Wilson loop
operators associated to oriented curves $C_\ell$ which are linked in $M$
and decorated by irreducible representations $R_\ell$ with highest 
weights $\alpha_\ell$ for ${\ell=1,\ldots,\RL}$.  On each curve we
introduce a corresponding sigma model field $U_\ell$, and we apply the
semi-classical description of $W_R(C)$ in \NEWWCIIIRES\ to write the
obvious generalization of \NWWLPZ,
\eqn\WLLK{\eqalign{
&Z\big(\epsilon; (C_1, R_1),\ldots,(C_\RL,R_\RL)\big) \,=\, {1 \over
{\Vol(\CG)}} \, \left({1 \over {2 \pi \epsilon}}\right)^{\Delta_{\CG}}
\, \times\cr 
&\qquad\times\,\int_{\CA \times \epsilon L\CO_{\alpha_1} \times \cdots \times
\epsilon L\CO_{\alpha_\RL}} \mskip -45mu \CD \! A \; \CD\!U_1 \cdots
\CD\!U_\RL\, \,\exp{\!\Biggr[{i \over {2 \epsilon}} \, \RC\RS(A) \,+\, i\,
\sum_{\ell=1}^\RL \, {\Rc\Rs}_{\alpha_\ell}\!\big(U_\ell;
A|_C\big)\Biggr]}\,.\cr}}

Through the same manipulations as before, we find that the shift
invariant version of the Wilson link path integral in \WLLK\ is given by 
\eqn\WLLKII{\eqalign{
&Z\big(\epsilon; (C_1, R_1),\ldots,(C_\RL,R_\RL)\big) \,=\, {1 \over
{\Vol(\CG)}} \, \left({{-i} \over {2 \pi
\epsilon}}\right)^{\Delta_{\CG}/2} \, \times\,\cr
&\qquad \times\, \int_{\bar\CA \times \epsilon L\CO_{\alpha_1} 
\times \cdots \times \epsilon L\CO_{\alpha_\RL}} \mskip -45mu \bar{\CD \! A}
\;\CD\!U_1 \cdots \CD\!U_\RL \,\,\exp{\!\left[ {i \over {2 \epsilon}}
S\big(A,\, U_1,\, \cdots,\, U_\RL\big)\right]}\,,\cr}}
where
\eqn\WLSAPHIU{
S\big(A,\, U_1,\, \cdots,\, U_\RL\big) \,=\, \RC\RS\big(A\big) \,+\, 2
\epsilon \, \sum_{\ell=1}^\RL \,  {\Rc\Rs}_{\alpha_\ell}\big(U_\ell;
A|_C\big) - \int_M {1 \over {\kappa \^ d \kappa}} \Tr\Big[ (\kappa \^
\CF_A)^2 \Big]\,,} 
with
\eqn\CFII{
\CF_A \,=\,  F_A \,+\, \epsilon \, \sum_{\ell=1}^\RL \,
\left[\left(g \alpha g^{-1}\right) \delta_C\right]_\ell\,.}
In order to suppress the proliferation of subscripts in \CFII, the
component index `$\ell$' on the bracketed quantity applies to all
elements therein.

Via \WLLKII, \WLSAPHIU, and \CFII, we have succeeded in reformulating
Chern-Simons theory on an arbitrary contact three-manifold $M$ in such
a way that one component of $A$ completely decouples from all
Wilson loop observables.  We now turn to a wonderful geometric
consequence of this fact.

\subsec{The Seifert Loop Path Integral as a Symplectic Integral}

Although the reformulation of the Wilson loop path integral in
\PZCSWLIII\ is completely general, to interpret that path integral
symplectically we must again specialize to the Seifert case.  

As in Section $3.3$, we assume $M$ to be a Seifert manifold, with a
distinguished $U(1)_\RR$ action and an invariant contact form
$\kappa$.  We also assume that the curve $C$ in \PZCSWLIII\ is a
generic orbit of the $U(1)_\RR$ action on $M$ and hence is the Seifert
fiber over a smooth, non-orbifold point on the Riemann surface
$\Sigma$ sitting at the base of $M$.  With this assumption, we do not
need to specify the particular point ${p \in \Sigma}$ over which $C$
sits, since the isotopy class of the embedding ${C \subset 
M}$ does not change under continuous variations of $p$.  In the
terminology of the Introduction, $W_R(C)$ is then a Seifert loop
operator in $M$.  Later, in Section $7.1$, we classify the possible 
Seifert loop operators when $M$ is $S^3$.

So far, an implicit question has been hanging over our quest to
recast the Seifert loop path integral as a symplectic integral of the
canonical form \PZSM.  As we reviewed in Section $3$, the Chern-Simons
partition function on a Seifert manifold is given by such an 
integral, determined by the Hamiltonian action of the group ${\CH =
U(1)_\RR \ltimes \wt\CG_0}$ on the symplectic space $\bar\CA$.  But
since the data of $\bar\CA$ and $\CH$ are intrinsically associated to
Chern-Simons theory, and since the form of the canonical symplectic
integral is, well, canonical, how can we hope to find yet {\sl
another} symplectic integral to describe the Seifert loop path
integral?

As soon as we ask this question, a glance at \PZCSWLIII\ suffices to
answer it.  There we have written the general Wilson loop path
integral as an integral over the product 
\eqn\BARCAA{ \bar\CA_{\alpha} \,=\, \bar\CA \times
\epsilon L\CO_{\alpha}\,.}
\countdef\BarAalpha=90\BarAalpha=\pageno
To summarize the chief miracle in this paper, when $M$ and $C$ are
both Seifert, the space $\bar\CA_\alpha$ is also a symplectic space
upon which the Hamiltonian group $\CH$ acts, and the Seifert loop path
integral in \PZCSWLIII\ is the canonical symplectic integral
associated to the data of $\bar\CA_\alpha$ and $\CH$.

\bigskip\noindent{\it Hamiltonian Symmetries of $L\CO_\alpha$}\smallskip

To explain the statement above, let us first consider the Hamiltonian
action of $\CH$ on $\bar\CA_\alpha$.  The group $\CH$ will act on
$\bar\CA_\alpha$ in a diagonal fashion; since we have already
discussed the Hamiltonian action of $\CH$ on $\bar\CA$, we need only
discuss the Hamiltonian action of $\CH$ on $L\CO_\alpha$.  

Briefly, $\CH$ acts on $L\CO_\alpha$ in the natural way.  Gauge
transformations on $M$ act on $L\CO_\alpha$ by restriction to $C$,
just as in \GT.  Because $C$ as a Seifert fiber of $M$ is preserved
under rotations in $U(1)_\RR$, the loopspace $L\CO_\alpha$ also inherits the
natural action by $U(1)_\RR$.  Finally, the central $U(1)_\RZ$ in the
extension $\wt\CG_0$ acts trivially on $L\CO_\alpha$, just as it does
on $\bar\CA$.  Explicitly, if $(p,\phi,a)$ is a generator in the Lie
algebra of $\CH$ as in Section $3.3$, the infinitesimal action of
$(p,\phi,a)$ on $L\CO_\alpha$ is given by 
\eqn\DLHG{ \delta g \,=\, p \, \lie_\RR g\,+\, \phi|_C \cdot g\,.}
As in Section $4.2$, $g$ is a section of the principal $G$-bundle
$P|_C$ over $C$ which we use to represent a point in $L\CO_\alpha$.

Besides a natural action by $\CH$, the loopspace $L\CO_\alpha$ also
carries a natural symplectic form $\Upsilon_\alpha$, inherited from
the canonical coadjoint symplectic form $\nu_{\alpha}$ on the orbit 
$\CO_\alpha$ itself.  As for any sigma model, the tangent space
to $L\CO_\alpha$ at the point corresponding to a given sigma model map
${U:C\rightarrow \CO_\alpha}$ is the vector space of sections of the
pullback by $U$ of the tangent bundle $T\CO_\alpha$ to $\CO_\alpha$.  To present
$\Upsilon_\alpha$ explicitly, we let $\eta$ and $\xi$ be sections of
$U^*(T\CO_\alpha)$ representing tangent vectors to $L\CO_\alpha$.
Then 
\eqn\BIGCOADJ{ \Upsilon_\alpha\big(\eta\,,\xi\big) \,=\,
\oint_C\!\kappa\,\,\nu_{\alpha}(\eta\,,\xi)\,.}
\countdef\BigUpsilon=91\BigUpsilon=\pageno
We abuse notation slightly in \BIGCOADJ, but hopefully the
meaning of this expression is clear.  The symplectic form $\nu_{\alpha}$
on $\CO_\alpha$ induces a pointwise pairing on $C$ between the
sections $\eta$ and $\xi$ of $U^*(T\CO_\alpha)$, and we integrate 
the resulting function over $C$ using the contact form $\kappa$.

One can easily check that $\Upsilon_\alpha$ is a symplectic form on
$L\CO_\alpha$.  For instance, because $\nu_\alpha$ is closed on
$\CO_\alpha$, the two-form $\Upsilon_\alpha$ is immediately closed on
$L\CO_\alpha$.  Also, the non-degeneracy of $\Upsilon_\alpha$ follows
from the non-degeneracy of $\nu_{\alpha}$, along with the observation
that the pullback of $\kappa$ to $C$ is nowhere
vanishing.\foot{Because $C$ is a Seifert fiber of $M$, the vector
field $\RR$ which generates the action of $U(1)_\RR$ is everywhere
tangent to $C$.  By its construction as an abelian connection on the
Seifert fibration, $\kappa$ satisfies ${\langle\kappa,\RR\rangle = 1}$
and hence provides a nowhere vanishing one-form on $C$.}  Finally,
since $\nu_{\alpha}$ is invariant under the action of $G$ on
$\CO_\alpha$ and since $\kappa$ is invariant under the action of
$U(1)_\RR$ on $C$, the symplectic form $\Upsilon_\alpha$ is manifestly
invariant under $\CH$.

We are left to compute the moment map $\mu$ which describes the
infinitesimal action \DLHG\ of $\CH$ on $L\CO_\alpha$.  This moment
map can actually be determined in two different ways, both of
which are illuminating.  To start, we take the direct approach.

From the expression for $\nu_{\alpha}$ in \COADJ, we see that the
contraction of $\Upsilon_\alpha$ with the vector field $V(p,\phi,a)$
on $L\CO_\alpha$ generated by $(p,\phi,a)$ in \DLHG\ is given by 
\eqn\CNTRUP{ \iota_{V(p,\phi,a)} \Upsilon_\alpha \,=\, \oint_C\!\kappa
\,\, \Tr\Big(\big(p \, g^{-1} \lie_\RR g \,+\, g^{-1}\,\phi\,
g\big)\cdot\big[\alpha\!, g^{-1} \delta g\big]\Big)\,.}
Here $g^{-1} \delta g$ is the left-invariant Cartan form on
$LG$, where as usual, $\delta$ is best thought of as an
infinite-dimensional version of the de Rham operator.  With a bit of
calculation, the result in \CNTRUP\ can be rewritten as 
\eqn\CNTRUPII{ \iota_{V(p,\phi,a)} \Upsilon_\alpha \,=\, -\delta
\oint_C\! \kappa \,\, \Tr\Big[ \alpha\cdot \big(p \, g^{-1} \lie_\RR
g \,+\, g^{-1} \, \phi \, g\big)\Big]\,.}
So via the defining Hamiltonian relation in \MOMMAPEQ, the moment map
$\mu$ for the action of $\CH$ on $L\CO_\alpha$ is given up to a
constant by  
\eqn\MOMCU{ \big\langle\mu\,, (p,\phi,a)\big\rangle \,=\, -\oint_C\!
\kappa \,\, \Tr\Big[ \alpha \cdot \big(p \, g^{-1} \lie_\RR g \,+\,
g^{-1} \, \phi \, g\big)\Big]\,.}

We are left to check that the Hamiltonian condition in \HOMMUII\ is
obeyed for $\mu$ in \MOMCU.  This condition ultimately fixes the
arbitrary constant that could otherwise appear in \MOMCU\ to be zero.
So we must verify that the expression for $\mu$ in \MOMCU\ satisfies 
\eqn\HAMCU{\eqalign{
\Big\{ \big\langle\mu,(p,\phi,a)\big\rangle,
\big\langle\mu,(q,\psi,b)\big\rangle \Big\} \,&=\, 
\Big\langle\mu,\big[(p,\phi,a),(q,\psi,b)\big]\Big\rangle\,,\cr
&=\,\Big\langle\mu,\big(0\,,[\phi,\psi] + p \lie_\RR \psi - q \lie_\RR
\phi\,,c(\phi,\psi)\big)\Big\rangle\,.\cr}}
Here in the second line of \HAMCU, we recall from Section $3.3$ the
explicit form for the bracket on the Lie algebra of $\CH$.

Before we perform any computations, let us make one observation.
A noteworthy feature of the moment map in \MOMCU\ is that $\mu$
vanishes when contracted with the central generator $(0,0,a)$ in the
Lie algebra of $\CH$.  As a result, the cocycle $c(\phi,\psi)$ can
effectively be set to zero when checking \HAMCU.  The Hamiltonian
condition for the Poisson bracket associated to generators of the form
$(0,\phi,0)$ and $(0,\psi,0)$ then follows just as it does for the
moment map \COADJMOM\ on the finite-dimensional orbit $\CO_\alpha$.

Otherwise, the only non-trivial Poisson bracket to check is the one below,
\eqn\HAMCUII{\eqalign{
\Big\{ \big\langle\mu,(p,0,0)\big\rangle,
\big\langle\mu,(0,\psi,0)\big\rangle \Big\} \,&=\, \Upsilon_\alpha\Big(p \,
\lie_\RR g\,, \psi|_C \cdot g\Big)\,,\cr
&=\, p \, \oint_C\!\kappa \,\, \Tr\Big(g^{-1} \lie_\RR g\,, \big[\alpha\!,
g^{-1}\,\psi\,g\big]\Big)\,,\cr
&=\, p \, \oint_C\!\kappa \,\, \Tr\Big(\alpha \cdot
\big[g^{-1}\,\psi\,g\,,g^{-1} \lie_\RR g\big]\Big)\,.\cr}}
To simplify the last line in \HAMCUII, we observe that 
\eqn\LIEP{ \lie_\RR\big(g^{-1}\,\psi\,g\big) \,=\, \big[
g^{-1}\,\psi\,g\,, g^{-1}\lie_\RR g\big] \,+\, g^{-1} \lie_\RR\psi\, g\,.}
Applying the identity in \LIEP\ to \HAMCUII\ and integrating by parts,
we obtain 
\eqn\HAMCUIV{ \Big\{ \big\langle\mu,(p,0,0)\big\rangle,
\big\langle\mu,(0,\psi,0)\big\rangle \Big\} \,=\, -p \, \oint_C\!\kappa\,\,
\Tr\Big[\alpha \cdot \big(g^{-1} \lie_\RR \psi \, g\big)\Big] \,=\,
\Big\langle\mu\,,\big(0, p \lie_\RR\psi,0\big)\Big\rangle\,,}
as required by \HAMCU.  Thus the action of $\CH$ on $L\CO_\alpha$ is
Hamiltonian with moment map $\mu$ in \MOMCU.

\bigskip\noindent{\it $L\CO_\alpha$ as a Coadjoint Orbit of
$\CH$}\smallskip

At this point, an excellent question to ask is why $L\CO_\alpha$
should even admit a Hamiltonian action by $\CH$.  After all,
the fact that $\CH$ acts in a Hamiltonian fashion on $\bar\CA$ is not 
obvious, given that the appearance of $\CH$ itself is rather
unexpected.

A bit more abstractly, if we are given a connected Lie group $H$, we
can ask which symplectic spaces admit a Hamiltonian action by $H$.  As
we have already discussed, the coadjoint orbits of $H$ in the dual
$\Fh^*$ of its Lie algebra furnish canonical examples of such
symplectic spaces.  Thus, the simplest way to explain why $\CH$ should
act in Hamiltonian fashion on $L\CO_\alpha$ is to identify
$L\CO_\alpha$ with a particular coadjoint orbit in the dual
$\Lie(\CH)^*$ of the Lie algebra of $\CH$.

To establish the interpretation of $L\CO_\alpha$ as a coadjoint orbit,
we first observe that central elements in $\CH$ of course act trivially on
the Lie algebra and therefore on its dual $\Lie(\CH)^*$.  So
for the purpose of discussing coadjoint orbits of $\CH$, we need only
consider the orbits in $\Lie(\CH)^*$ of the quotient group 
\eqn\CALK{ \bar\CH \,=\, \CH/U(1)_\RZ \,=\, U(1)_\RR \ltimes \CG_0\,.}

We now claim that $L\CO_\alpha$ can be formally identified as the
orbit of $\bar\CH$ which passes through the element
${\gamma_0\in\Lie(\CH)^*}$ defined by the pairing 
\eqn\DISLH{ \left\langle\gamma_0, (p,\phi,a)\right\rangle\,=\,
-\int_M \kappa\^\delta_C \, \Tr(\alpha\,\phi) \,=\, -\oint_C \kappa \,
\Tr(\alpha\,\phi)\,.}
Equivalently, in coordinates dual to $(p,\phi,a)$, we write 
\eqn\DISLHDL{ \gamma_0 \,=\, \left(0, \alpha \, \kappa\^\delta_C,
0\right).}
Here we regard ${\alpha \, \kappa\^\delta_C}$ as a section of the bundle
${\Omega^3_M\otimes\Fg}$ of adjoint-valued three-forms on $M$ and hence 
as an element in the dual of the Lie algebra of the group $\CG_0$,
which is the identity component of the group $\CG$ of all gauge
transformations on $M$.

Let us consider the action of $\bar\CH$ on $\gamma_0$.  Since both
$\kappa$ and $\delta_C$ are invariant under $U(1)_\RR$, $\gamma_0$ is
fixed under $U(1)_\RR$.  Hence the orbit of the semi-direct product
${\bar\CH = U(1)_\RR \ltimes \CG_0}$ through $\gamma_0$ further
reduces to the orbit of $\CG_0$ through $\gamma_0$.

In general, elements of the group $\CG$ of gauge transformations are
described geometrically by maps from $M$ to $G$, and elements in the
identity component $\CG_0$ correspond to those maps ${f:M\rightarrow
G}$ which are homotopically trivial (and hence can be continuously
connected to the identity).  Because $\gamma_0$ has delta-function
support along $C$, $f$ acts on $\gamma_0$ by restriction to $C$
and so determines a point in $L\CO_\alpha$.  Upon setting ${g =
f|_C}$, the point is simply ${g \alpha g^{-1} \in L\CO_\alpha}$. 
Conversely, if ${g\in LG}$ represents the point ${g \alpha g^{-1} \in
L\CO_\alpha}$, then $g$ can always be extended over $M$ to some 
homotopically trivial $f$.\foot{By standard obstruction theory,
${g\!:C\rightarrow G}$ can always be extended from $C$ to $M$, so we
need only argue that the extension can be chosen to be homotopically
trivial.  With our assumption that $G$ is compact, connected,
simply-connected, and simple, homotopy classes of maps
${f\!:M\rightarrow  G}$ are represented by elements in $H^3(M;\BZ)$, and
homotopy classes of maps ${\wt f\!:M\rightarrow G}$ such that ${\wt
f|_C = 1}$ are represented by elements in the relative group
$H^3(M,C;\BZ)$.  But ${H^3(M, C;\BZ) \cong H^3(M;\BZ) \cong \BZ}$ by the usual
exact sequence.  Hence if $f$ happens to be a homotopically nontrivial 
extension of $g$, we can always find another map ${\wt f\!:M\rightarrow
G}$ such that ${\wt f|_C = 1}$ and the product $\wt f \cdot f$
(taken pointwise on $M$) is a homotopically trivial extension of $g$.}  
Hence the orbit of $\CG_0$ through $\gamma_0$ is the loopspace $L\CO_\alpha$.

Once $L\CO_\alpha$ is identified as the coadjoint orbit of
$\CH$ through the element $\gamma_0$ in \DISLHDL, $\Upsilon_\alpha$
immediately becomes the canonical coadjoint symplectic form.
In precise analogy to \BIGTH\ and \COADJ, $\Upsilon_\alpha$ also
derives from a pre-symplectic one-form $\Xi_\alpha$, 
\eqn\BIGLTH{ \Upsilon_\alpha \,=\, \delta \Xi_\alpha\,,}
where 
\eqn\BIGLTHII{ \Xi_\alpha \,=\, \oint_C \!\kappa \,\, \Tr\big(\alpha
\cdot g^{-1}\delta g\big)\,.}
\countdef\BigXi=92\BigXi=\pageno
Consequently, the moment map $\mu$ in \MOMCU\ is given just as in
\COADJMOM\ by the contraction 
\eqn\BIGCOADJMOM{ \mu = - \iota_{V(p,\phi,a)} \Xi_\alpha\,,} 
where $V(p,\phi,a)$ is the vector field on $L\CO_\alpha$ appearing in
\DLHG.

As a coadjoint orbit of $\CH$, the loopspace $L\CO_\alpha$ also
carries an invariant complex structure which is compatible with the
symplectic form $\Upsilon_\alpha$, in the sense that together these
data determine an invariant K\"ahler metric on $L\CO_\alpha$.
Concretely, the invariant complex structure on $L\CO_\alpha$ is
inherited from the corresponding complex structure $\SJ$ on
$\CO_\alpha$, such that the complex structure on $L\CO_\alpha$ is
given by the pointwise action of $\SJ$ on sections of
$U^*(T\CO_\alpha)$.  

Since the symplectic form $\Upsilon_\alpha$ on  $L\CO_\alpha$ is also
induced pointwise from the coadjoint symplectic form $\nu_\alpha$ on
$\CO_\alpha$, the invariant K\"ahler metric on $L\CO_\alpha$ is then
given by the natural pairing 
\eqn\BIGKAHL{ \big(\eta\,,\xi\big) \,=\, \oint_C\!\kappa\,\,
\nu_{\alpha}(\eta\,,\SJ \cdot \xi)\,.}
Again, $\eta$ and $\xi$ are sections of $U^*(T\CO_\alpha)$ representing
tangent vectors to $L\CO_\alpha$.  Of course, the metric in \BIGKAHL\
is nothing more than the usual sigma model metric derived from the
invariant K\"ahler metric on $\CO_\alpha$ itself.  As a formal
consequence, the Riemannian path integral measure $\CD\!U$ associated 
to the sigma model metric in \BIGKAHL\ can be identified with the
symplectic measure on $L\CO_\alpha$ induced from $\Upsilon_\alpha$,
\eqn\BIGKAHLII{ \CD\!U \,=\, \exp{\!\left(\Upsilon_\alpha\right)}\,.}

By the preceding observations, the shift-invariant path integral
describing $Z(\epsilon; C, R)$ in \PZCSWLIII\ becomes a symplectic
integral over ${\bar\CA_\alpha = \bar\CA \times \epsilon L\CO_\alpha}$,
\eqn\PZCSWLIV{
Z\big(\epsilon; C, R\big) \,=\, {1 \over
{\Vol(\CG)}} \, \left({{-i} \over {2 \pi
\epsilon}}\right)^{\Delta_{\CG}/2} \, \int_{\bar\CA_\alpha} 
\exp{\!\left[\Omega_\alpha \,+\, {i \over {2
\epsilon}} S\big(A, U\big)\right]}\,,}
where we introduce the total symplectic form 
\eqn\OMEPS{ \Omega_\alpha \,=\, \Omega \,+\, \epsilon 
\Upsilon_\alpha\,.}
Here $\Omega$ is the symplectic form on $\bar\CA$ in \BO, and in
passing from \PZCSWLIII\ to \PZCSWLIV, we have been careful to recall
that the K\"ahler metric on $\epsilon L\CO_\alpha$ is scaled by
$\epsilon$ relative to the metric on $L\CO_\alpha$ in \BIGKAHL.  Hence
a crucial factor of $\epsilon$ multiplies $\Upsilon_\alpha$ in \OMEPS.
The need for this factor will become clear momentarily.

\countdef\BigOmalpha=93\BigOmalpha=\pageno

\bigskip\noindent{\it The Action $S(A,U)$ as the Square of the Moment
Map}\smallskip

We are left to show that the shift-invariant Seifert loop action $S(A,
U)$ in \SAPHIUIII\ is precisely the square of the moment map for the
Hamiltonian action of $\CH$ on $\bar\CA_\alpha$.  Given the
corresponding result \CSMU\ for the shift-invariant action $S(A)$, 
this claim is not so unexpected, but it remains (at least to me) a fairly
miraculous statement.

At this stage, we can explain the fundamental reason for the relative
factor of $\epsilon$ appearing in the symplectic form $\Omega_\alpha$
in \OMEPS.  With this factor, the moment map which describes the
Hamiltonian action of $\CH$ on the product ${\bar\CA_\alpha = \bar\CA
\times \epsilon L\CO_\alpha}$ is the sum of the moment map for
$\bar\CA$ in \MUIV\ with $\epsilon$ times the moment map for
$L\CO_\alpha$ in \MOMCU, so that the total moment map on
$\bar\CA_\alpha$ is given by 
\eqn\MUVX{\eqalign{
&\Big\langle\mu,(p,\phi,a)\Big\rangle \,=\, a \,-\, p \int_M
\kappa\^\Tr\!\left[\ha \, \lie_\RR A\^A \,+\, \epsilon \, \alpha 
\big(g^{-1} \lie_\RR g\big) \, \delta_C\right] \,-\, \int_M 
\kappa\^\Tr\big(\phi \, \CF_A\big) \,+\,\cr
&\qquad\qquad\,+\, \int_M d\kappa\^\Tr\big(\phi \, A\big),}}
where 
\eqn\NEWFII{ \CF_A \,=\, F_A \,+\, \epsilon \left(g \, \alpha 
g^{-1}\right) \delta_C\,.}
Again, $\CF_A$ is the generalized curvature \NEWF\ appearing already in the
shift-invariant action $S(A,U)$.  Indeed, we were careful to arrange
for the factor of $\epsilon$ in \OMEPS\ to ensure the appearance of
$\CF_A$ in \MUVX.

As in Section $3$, we proceed by computing directly the square of $\mu$ in
\MUVX.  From the description of the invariant form on the Lie algebra
of $\CH$ in \FRM, we see that 
\eqn\MUSQ{ \big(\mu, \mu\big) \,=\, \int_M
\kappa\^\Tr\Big[\lie_\RR A\^A \,+\, 2 \epsilon \alpha 
\big(g^{-1} \lie_\RR g\big) \, \delta_C\Big] \,-\, \int_M \kappa\^d\kappa \,
\Tr\!\left[\left({{\kappa \^ \CF_A - d\kappa \^ 
A}\over{\kappa\^d\kappa}}\right)^2\right]\,.}
To simplify \MUSQ, let us expand the last term therein as 
\eqn\MUSQI{\eqalign{
&\int_M \kappa\^d\kappa \,
\Tr\!\left[\left({{\kappa \^ \CF_A - d\kappa \^ 
A}\over{\kappa\^d\kappa}}\right)^2\right]\,=\,\cr
&\qquad\qquad\qquad\qquad\int_M {1 \over {\kappa \^ d\kappa}} \,
\Tr\Big[\big(\kappa \^ \CF_A\big)^2 \,-\,
2\big(\kappa\^\CF_A\big)\big(d\kappa\^ A\big) \,+\,
\big(d\kappa\^A\big)^2\Big].\cr}}
The term in \MUSQI\ which is quadratic in $\CF_A$ appears explicitly
in $S(A, U)$, and as for the term linear in $\CF_A$, we need only
extract the new contribution from the Seifert loop operator,
\eqn\MULM{ -2\epsilon\int_M \kappa \^ \delta_C \, \Tr\Biggr[\left(g
\, \alpha g^{-1}\right) \left({{d\kappa \^ A} \over {d\kappa \^
\kappa}}\right)\Biggr] \,=\, -2\epsilon\oint_C\!\kappa \,\,
\Tr\Big[\alpha \cdot \left(g^{-1} \, \iota_\RR A \,
g\right)\Big]\,.}
Here we have applied the identity in \IXA.  

After a little bit of algebra, we thus rewrite $(\mu,\mu)$ using
\MULM\ as 
\eqn\MUSQII{\eqalign{
(\mu, \mu) \,&=\, \int_M \kappa\^\Tr\Big(\lie_\RR
A\^A\Big) \,+\, 2 \int_M \kappa \^ \Tr\Big[\big(\iota_\RR A\big) \,
F_A\Big] \,-\, \int_M \kappa \^d\kappa \, \Tr\Big[\big(\iota_\RR
A\big)^2\Big] \,+\,\cr 
&+\, 2 \epsilon \oint_C\!\kappa \,\, \Tr\Big[\alpha \cdot \big(g^{-1}
\lie_\RR g \,+\, g^{-1} \, \iota_\RR A \, g \big)\Big] \,-\, \int_M
{1 \over {\kappa \^ d\kappa}} \, \Tr\Big[\big(\kappa \^
\CF_A\big)^2\Big]\,.\cr}} 
At this stage, we apply our baroque identity in \CSID\ to recognize
the first line in \MUSQII\ as the Chern-Simons action $\RC\RS(A)$.  We
also have the much more transparent identity 
\eqn\CSIDII{
\Rc\Rs_\alpha\big(U; A|_C\big) \,=\, \oint_C \Tr\Big(\alpha \cdot g^{-1} d_A
g\Big)\,=\, \oint_C\!\kappa\,\,\Tr\Big[\alpha \cdot \big(g^{-1}
\lie_\RR g \,+\, g^{-1} \, \iota_\RR A \, g \big)\Big]\,.}
The identity in \CSIDII\ follows immediately if we recall that the
vector field $\RR$ is tangent to $C$ and satisfies
${\langle\kappa\,,\RR\rangle = 1}$.  

So from \CSID, \MUSQII, and \CSIDII, we finally obtain the beautiful 
result  
\eqn\MUSQIII{\eqalign{ 
\big(\mu, \mu\big) \,&=\, \RC\RS\big(A\big) \,+\, 2 \epsilon \,
{\Rc\Rs}_\alpha\big(U; A|_C\big) - \int_M {1 \over {\kappa \^ d \kappa}}
\Tr\Big[\big(\kappa \^ \CF_A\big)^2\Big]\,,\cr
&=\, S\big(A,U\big)\,.\cr}}
Consequently the Seifert loop path integral in \PZCSWLIV\ assumes the
canonical symplectic form required for non-abelian localization,
\eqn\PZCSWLV{ Z\big(\epsilon; C, R\big) \,=\, {1 \over
{\Vol(\CG)}} \, \left({{-i} \over {2 \pi
\epsilon}}\right)^{\Delta_{\CG}/2} \, \int_{\bar\CA_\alpha} 
\exp{\!\left[\Omega_\alpha \,+\, {i \over {2
\epsilon}} (\mu, \mu)\right]}\,.}

\bigskip\noindent{\it Extension to Multiple Seifert Loop
Operators}\smallskip

Although for simplicity we have focused throughout on the case of a
single Seifert loop operator, the preceding discussion extends
immediately to the case of multiple Seifert loop operators in $M$.

To state the general result, we let $C_\ell$ for ${\ell=1,\ldots,\RL}$ be a
set of disjoint Seifert fibers of $M$, each fiber labelled by an
irreducible representation $R_\ell$ with highest weight $\alpha_\ell$.
We then consider the symplectic space 
\eqn\BARCAAII{ \bar\CA_{\hbox{\boldmath$_\alpha$}} \,=\, \bar\CA
\times \epsilon L\CO_{\alpha_1} \times \cdots \times \epsilon
L\CO_{\alpha_\RL}\,,}
with symplectic form 
\eqn\OMEPSII{ \Omega_{\hbox{\boldmath$_\alpha$}} \,=\, \Omega \,+\,
\epsilon\, \sum_{\ell=1}^\RL \Upsilon_{\alpha_\ell}\,,}
where ${\hbox{\boldmath$\alpha$} = (\alpha_1,\ldots,\alpha_\RL)}$
serves as a multi-index.

The group ${\CH = U(1)_\RR \ltimes \wt\CG_0}$ now acts on
$\bar\CA_{\hbox{\boldmath$_\alpha$}}$ in a Hamiltonian fashion with moment map 
\eqn\MUVXLK{\eqalign{
\Big\langle\mu, (p,\phi,a)\Big\rangle \,&=\, a \,-\, p \int_M
\kappa\^\Tr\Biggr(\ha \, \lie_\RR A\^A \,+\, \epsilon
\,\sum_{\ell=1}^\RL \Big[\alpha \big(g^{-1} \lie_\RR g\big) \,
\delta_C\Big]_\ell\Biggr) \,-\,\cr
&-\, \int_M \kappa\^\Tr\big(\phi \, \CF_A\big) \,+\, \int_M
d\kappa\^\Tr\big(\phi \, A\big),\cr}}
where 
\eqn\CFIILK{
\CF_A \,=\,  F_A \,+\, \epsilon \, \sum_{\ell=1}^\RL \,
\Big[\big(g\, \alpha g^{-1}\big)\,\delta_C\Big]_\ell\,.}
Once more, to suppress the proliferation of subscripts, the index
`$\ell$' applies simultaneously to all quantities in brackets.

By the same calculations leading to \MUSQIII, the shift-invariant
action $S(A,U_1,\cdots,U_\RL)$ in \WLSAPHIU\ is precisely the
square of the moment map \MUVXLK\ for the Hamiltonian action of $\CH$
on $\bar\CA_{\hbox{\boldmath$_\alpha$}}$.  So when applied to multiple Seifert
loop operators, the shift-invariant path integral in \WLLKII\ can also
be rewritten in the canonical symplectic form,
\eqn\WLLKIII{ Z\big(\epsilon; (C_1, R_1),\ldots,(C_\RL, R_\RL)\big)
\,=\, {1 \over {\Vol(\CG)}} \, \left({{-i} \over {2 \pi
\epsilon}}\right)^{\Delta_{\CG}/2} \,
\int_{\bar\CA_{\hbox{\boldmath$_\alpha$}}} 
\exp{\!\left[\Omega_{\hbox{\boldmath$_\alpha$}} \,+\, {i\over{2\epsilon}} 
(\mu, \mu)\right]}.}

\newsec{Monodromy Operators in Two-Dimensional Yang-Mills Theory}

We began in Section $2$ by recalling the well-known symplectic
description for the Yang-Mills partition function, which we then
extended in Section $3$ to the partition function of Chern-Simons
theory on a Seifert manifold.  In this light, given the symplectic
description for the Seifert loop operator in Section $4$, one might
ask which operator, if any, in two-dimensional Yang-Mills theory plays
a symplectic role analogous to that of the Seifert loop operator in
Chern-Simons theory.

The answer to the preceding question is well-known, at least to
aficionados of two-dimensional Yang-Mills theory, and it will be
important when we perform localization computations for Seifert loop
operators in Section $7$.  In a nutshell, the analogue on $\Sigma$ of
the Seifert loop operator in $M$ is a local ``monodromy'' operator
which inserts a singularity into the gauge field $A$ at a marked point
of $\Sigma$.  These monodromy operators were introduced by Witten
\WittenWE\ to model the current algebra vertex operators which describe a Wilson
line puncturing $\Sigma$ in the canonical quantization of Chern-Simons
theory on ${\Sigma \times \BR}$, and in that context they were given a
beautiful symplectic interpretation by Atiyah (see \S $5.2$ of
\AtiyahKN).  Though much of the following material is standard, our
goal at present is thus to review a few essential ideas about
monodromy operators in two-dimensional Yang-Mills theory.
  
The monodromy operator in Yang-Mills theory on $\Sigma$ is perhaps the 
simplest example disorder operator in gauge theory.  By this
statement, we mean that the monodromy operator is defined not in
terms of a classical, gauge-invariant functional of $A$, as for
instance we originally defined the Wilson loop operator in \DEFWVC,
but as a prescription to perform the two-dimensional Yang-Mills path
integral \YMD\ over connections with specified classical
singularities.  Here by a `classical' singularity, we mean a
singularity that can appear in a solution to the Yang-Mills equations
on $\Sigma$.  Only for such singularities does one obtain a sensible
path integral in the presence of the corresponding disorder operator.

The classical singularity which defines the monodromy operator in
two-dimensional Yang-Mills theory will simply be the reduction to
$\Sigma$ of the classical singularity which appears in the
Chern-Simons gauge field in the background of a Seifert loop operator.
According to \LOCAC, the Chern-Simons gauge field behaves classically
near a Wilson loop wrapping ${C\subset M}$ as 
\eqn\LOCACII{ A \,=\, -{\alpha \over k} \, d\varphi \,+\, V_0 \,
d\tau\,,}
at least up to gauge transformations.  In comparison to \LOCAC,
we have set ${\epsilon=2\pi/k}$ and ${U_0 = \alpha}$ without loss;
$V_0$ is then any element in $\Fg$ commuting with $\alpha$.  

To reduce \LOCACII\ to two dimensions, we take ${p\in\Sigma}$ be the
basepoint of $C$, now a Seifert fiber of $M$, around which
$(r,\varphi)$ serve as local polar coordinates.  We also introduce a
parameter ${\lambda\in\Ft}$ to play the role of the ratio ${\alpha /
k}$.  In particular, despite the fact that in three dimensions
${\alpha\in\Gamma_{\rm wt}}$ is quantized as a weight of $G$, in two
dimensions we allow $\lambda$ to vary continuously in $\Ft$.  The
monodromy operator $\RV_\lambda(p)$ is then defined as the disorder
operator in two-dimensional Yang-Mills theory which creates a
singularity in $A$ at the point $p$ of the form 
\eqn\LOCA{ A \,=\, -\lambda \, d\varphi\,,\qquad\qquad \lambda\,\in\,\Ft\,,}
again up to gauge transformations.  Of course, in reducing to two
dimensions, we omit the component of $A$ proportional to $d\tau$ in
\LOCACII.\countdef\MonoOp=94\MonoOp=\pageno

Let us make two elementary comments about \LOCA.  First, because the
connection in \LOCA\ is flat away from $p$, it trivially satisfies the
classical Yang-Mills equation ${d_A\*F_A = 0}$ on a punctured
neighborhood of $p$.  Second, as our terminology for $\RV_\lambda(p)$
suggests, the connection in \LOCA\ has non-trivial monodromy around $p$
given by 
\eqn\BIGL{\Lambda \,=\, \exp{\!\left(2 \pi \lambda\right)}\,.}
The parameter $\lambda$ appears with a positive sign in \BIGL\ since
the holonomy of $A$ is defined with a negative sign, as in \JFCOND.

Before we proceed further, we need to consider how the operator
$\RV_\lambda(p)$ depends upon the parameter $\lambda$.  For instance, 
even after we fix the maximal torus ${T \subset G}$, the Weyl group
$\FW$ of $G$ remains as a residual discrete symmetry acting on
$\lambda$.\countdef\WeylGp=95\WeylGp=\pageno

Somewhat less obviously, as pointed out by Gukov and
Witten \GukovJK\ in relation to surface operators in four-dimensional
gauge theory (for which the same codimension two singularity in $A$
plays an essential role), the monodromy operator $\RV_\lambda(p)$ is
also invariant under any shift of the form ${\lambda \,\mapsto\, \lambda +
y}$, where ${y\in\Ft}$ satisfies the integrality condition
${\exp{\!(2\pi y)}\,=\,1}$.  This shift in $\lambda$ leaves the
monodromy $\Lambda$ in \BIGL\ invariant and is induced by a singular
gauge transformation generated locally at $p$ by the $T$-valued function 
\eqn\SINGP{ (r\,,\varphi) \,\longmapsto\, \exp{\!(\varphi\,y)}\,.}
Intrinsically as in Section $4.1$, $y$ is characterized as an element
of the cocharacter lattice
\eqn\COCHARII{ \Gamma_{\rm cochar} \,=\, \Hom\!\big(U(1),\,T\big),}
which becomes isomorphic to the coroot lattice $\Gamma_{\rm cort}$
when $G$ is simply-connected.  In the present discussion of
two-dimensional Yang-Mills theory, we will not always assume $G$ to be
simply-connected, so we are careful to distinguish $\Gamma_{\rm cort}$ and
$\Gamma_{\rm cochar}$.  Thus if we consider both the action of the
Weyl group $\FW$ on $\lambda$ as well as shifts ${\lambda\mapsto\lambda + y}$
generated by the singular gauge transformations in \SINGP, the
gauge-invariant label for the monodromy operator $\RV_\lambda(p)$ is
not the parameter ${\lambda\in\Ft}$ {\it per se} but rather the image
of $\lambda$ in the quotient $\Ft/\FW_{\rm aff}$,  where $\FW_{\rm
aff}$ is the affine Weyl group of $G$,
\eqn\AFFWEYL{ \FW_{\rm aff} \,=\, \FW \ltimes \Gamma_{\rm
cochar}\,.}
\countdef\AffWeylGp=96\AffWeylGp=\pageno

The fact that the monodromy operator is actually labelled by elements
of the quotient $\Ft/\FW_{\rm aff}$, as opposed to elements of $\Ft$,
has a natural geometric interpretation.  Because the maximal torus $T$
of $G$ can itself be presented as ${T = \Ft/\Gamma_{\rm cochar}}$,
we identify ${\Ft/\FW_{\rm aff} \cong T/\FW}$.  On the other hand,
points in $T/\FW$ correspond to conjugacy classes in $G$.  As a
result, the monodromy operator $\RV_\lambda(p)$ is naturally labelled
in a gauge-invariant fashion by the conjugacy class ${\FC_\lambda =
\Cl[\Lambda]}$ containing the monodromy
${\Lambda=\exp{\!(2\pi\lambda)}}$.

\bigskip\noindent{\it Hamiltonian Interpretation}\smallskip

As an aside, the monodromy operators which create singularities in $A$
of the form \LOCA\ are well-known in two-dimensional Yang-Mills
theory, but they are often described in a slightly different way.
Instead of working on the punctured Riemann surface ${\Sigma^o =
\Sigma - \{p\}}$ as we have done so far, we consider a small disc ${D
\subset \Sigma}$ containing the point $p$, and we replace the local
operator $\RV_\lambda(p)$ by an external state $|\FC_\lambda\rangle$ in the
Hilbert space constructed by quantizing Yang-Mills theory on the
circle bounding $D$.  Tautologically, the 
state $|\FC_\lambda\rangle$ is obtained by performing the path integral for
Yang-Mills theory on $D$ with the monodromy operator inserted at $p$.
But because two-dimensional Yang-Mills theory is such a simple theory,
the state $|\FC_\lambda\rangle$ can be given an absolutely explicit
description, which was applied by Witten in \WittenWE\ to perform exact
computations with the monodromy operator in the Hamiltonian
formulation of two-dimensional Yang-Mills theory.

Very briefly, to describe the state $|\FC_\lambda\rangle$, we recall
that the only gauge-invariant data carried by a connection on $S^1$ is
the conjugacy class of its holonomy $W$ as an element of $G$,
\eqn\HOLWP{ W\,=\,P\exp{\!\left(-\oint_{S^1} A\right)}\,\in\, G\,.}
Hence the Hilbert space $\SH$ of two-dimensional Yang-Mills theory on
the disc $D$ is the space of square-integrable class functions
$\Psi(W)$ on $G$.  By the Peter-Weyl theorem, ${\SH = L^2(G)^{G}}$ 
is spanned by the characters of the irreducible representations $R$ of
$G$, so the Yang-Mills Hilbert space can be presented in a basis of
states $|R\rangle$ corresponding to each irreducible representation of
$G$.

By definition, the monodromy operator $\RV_\lambda(p)$ enforces the
condition that $A$ have holonomy around the boundary of $D$ which lies
in the conjugacy class ${\FC_\lambda = \Cl[\Lambda]}$.  Thus, as a
formal class function on $G$, the corresponding external state
$|\FC_\lambda\rangle$ must be a delta-function supported on $\FC_\lambda$.
Concretely, via the standard orthonormality of characters, the state
$|\FC_\lambda\rangle$ can be expanded in the basis of representations as 
\eqn\STFC{ |\FC_\lambda\rangle \,=\, \sum_R \, |R\rangle\langle
R|\FC_\lambda\rangle \,=\, \sum_R \, \bar{\ch_R(\FC_\lambda)} \; |R\rangle\,.}
Here $\ch_R$ is the character associated to the representation $R$,
and ${\langle\FC_\lambda|R\rangle = \ch_R(\FC_\lambda)}$ by
definition.\foot{As standard, $\langle R|\FC_\lambda\rangle$ is the
complex conjugate of $\langle\FC_\lambda|R\rangle$, accounting for
appearance of $\bar{\ch_R(\FC_\lambda)}$ in \STFC.}  Among other
advantages, the description of $|\FC_\lambda\rangle$ in 
\STFC\ makes manifest the fact that the monodromy operator
$\RV_\lambda(p)$ depends only upon the gauge-invariant data of the
conjugacy class $\FC_\lambda$, or equivalently upon the point in the
quotient $\Ft/\FW_{\rm aff}$, and not upon the particular
representative ${\lambda\in\Ft}$ introduced initially.

Actually, as noted by Witten in \WittenWE, the normalization
of the delta-function associated to $|\FC_\lambda\rangle$ is subtle, since
this delta-function depends upon the choice of a measure on the space of
conjugacy classes of $G$.  We have made a particular choice in \STFC,
but another choice would multiply $|\FC_\lambda\rangle$ by an arbitrary class
function on $G$.  Equivalently, we have not been very careful so far
to fix the absolute normalization of $\RV_\lambda(p)$.  In this paper,
we implicitly normalize $\RV_\lambda(p)$ through its symplectic path
integral, to which we now turn.

\subsec{The Monodromy Operator Path Integral as a Symplectic Integral}

Although the monodromy operator in two-dimensional Yang-Mills theory
is perhaps interesting in its own right as a very simple, almost 
pedagogical, example of a disorder operator in gauge theory, our
interest in the monodromy operator stems from the fact that it admits
a symplectic description precisely analogous to that established for
the Seifert loop operator in Section $4.3$.  The essentials of the
symplectic description for $\RV_\lambda(p)$ were explained long ago by
Atiyah in \S $5.2$ of his beautiful lectures \AtiyahKN\ on
Chern-Simons theory, which we follow shamelessly here.

Let us first introduce the monodromy operator path integral,
\eqn\PZYMRV{ Z(\epsilon; p, \lambda) \,=\,  {1
\over {\Vol(\CG(P))}} \, \left({1 \over {2 \pi
\epsilon}}\right)^{\Delta_{\CG(P)}/2} \, \int_{\CA(P)} \mskip -10mu\CD
\! A \,\; \RV_\lambda(p) \; \exp{\left[ {1 \over {2 \epsilon}}
\int_\Sigma \! \Tr\left(F_A \^ \* F_A\right)\right]}.}
\countdef\Zepsplam=97\Zepsplam=\pageno
By definition, $Z(\epsilon;p,\lambda)$ is now a path integral over
connections on $\Sigma$ with a singularity at $p$ of the form \LOCA,
up to gauge transformations.  Though we suppress analytic 
details, see for instance \refs{\JeffreyLCII,\DaskW} for rigorous 
models of the relevant spaces of singular connections associated to
the monodromy operator.  Nonetheless, one small analytic detail concerning
\PZYMRV\ will be important.  In our previous discussion of the
monodromy operator, we emphasized that $\RV_\lambda(p)$ depends on the
parameter ${\lambda\in\Ft}$ only up to the action of the affine Weyl
group $\FW_{\rm aff}$.  Yet to discuss the path integral in \PZYMRV,
especially in its symplectic incarnation, we find it necessary to fix
at the outset a particular value for $\lambda$ in its orbit under
$\FW_{\rm aff}$.

To explain why fixing the value of $\lambda$ under the action of
$\FW_{\rm aff}$ is necessary, let us reconsider the geometric meaning of
$\lambda$ in Yang-Mills theory.   We began in Section $2$ with a given 
principal $G$-bundle $P$ over $\Sigma$, on which $A$ is a connection.
However, as noted by Gukov and Witten \GukovJK\ in the same situation,
when $A$ has a singularity at a point ${p\in\Sigma}$, the $G$-bundle
$P$ is only naturally defined on the punctured Riemann surface
${\Sigma^o \,=\, \Sigma - \{p\}}$.  We are free to extend $P$ 
over the puncture at $p$, but there is no natural way to do so.
Instead, different extensions of $P$ are labelled by the various ways
to lift the monodromy $\Lambda$ as an element of the torus 
${T\cong\Ft/\Gamma_{\rm cochar}}$ to a corresponding Lie algebra
element ${\lambda\in\Ft}$, and these extensions are all related by the
singular gauge transformations in \SINGP.

To give a well-known example, let us assume that the Yang-Mills gauge
group $G$ takes the adjoint form ${G = \wt G/ \CZ(\wt G)}$, where
$\wt G$ is simply-connected and $\CZ(\wt G)$ is the center of $\wt 
G$.  In this case, the extension of $P$ even as a smooth $G$-bundle
over $p$ is not unique.  Rather, the extension depends upon the choice
of a characteristic class ${\zeta\in\CZ(\wt G)}$, where $\zeta$
represents the possible monodromy in $\wt G$ that obstructs $P$ from
extending smoothly as a $\wt G$-bundle over $p$.  (Since $\zeta$
becomes trivial once we pass from $\wt G$ to $G$, the principal bundle
$P$ does extend smoothly as a ${\wt G}$-bundle.)  If ${y\in\Gamma_{\rm
cochar}}$ is an element of $\Ft$ satisfying ${1\neq\exp{\!(2\pi
y)}\in\CZ(\wt G)}$, the corresponding singular gauge transformation
in \SINGP\ shifts the value of $\zeta$ and therefore changes the
topology of
$P$.\countdef\Gtilde=98\Gtilde=\pageno\countdef\Center=99\Center=\pageno 

As usual in Yang-Mills theory, the need to choose a particular
extension of $P$ over the puncture at $p$ and hence a particular lift
from $\Lambda$ to $\lambda$ can be obviated by summing over all such
choices.  For instance, when $G$ is the adjoint form of a
simply-connected group $\wt G$, one computes the physical Yang-Mills
partition function by summing over all topological types of $P$.
Nevertheless, from the purely semi-classical perspective here, the
Yang-Mills path integral is most naturally defined 
for a fixed $G$-bundle $P$ and hence, in the case of \PZYMRV, for a
particular choice of $\lambda$ in its orbit under $\FW_{\rm aff}$.

Fixing the residual action of $\FW_{\rm aff}$ on $\lambda$ has two
important consequences.  First, once $P$ and $\lambda$ are fixed, the
group $\CG(P)$ of gauge transformations appearing in \PZYMRV\ consists
of those gauge transformations which are strictly non-singular at $p$,
precisely as in Section $2$.  Singular gauge transformations at $p$,
such as those in \SINGP, would otherwise shift $\lambda$.  Second,
once the extension of $P$ over $p$ is chosen, the curvature of the
singular connection in \LOCA\ is well-defined at $p$ and can be
evaluated as 
\eqn\LOCF{ F_A \,=\, -2\pi\lambda \, \delta_p\,.}
\countdef\deltap=100\deltap=\pageno
Here $\delta_p$ is a two-form on $\Sigma$ with delta-function support
which represents the Poincar\'e dual of $p$, and the formula for the
curvature $F_A$ in \LOCF\ follows from the local description \LOCA\ of
$A$ via the naive relation ${d(d\varphi) = 2 \pi \delta_p}$.  Note
that under shifts of $\lambda$ generated by gauge transformations
which are singular at $p$, the curvature $F_A$ in \LOCF\ also shifts,
so that the formula in \LOCF\ is really only sensible once $\lambda$
is fixed.  Not surprisingly, the delta-function curvature of $F_A$ at
$p$ will be an important ingredient in the symplectic description of the
monodromy operator.

Although the particular choice for $\lambda$ will not matter until
Section $5.2$, for concreteness let us make that choice now.  In
specifying a distinguished representative for $\lambda$ under 
the action of $\FW_{\rm aff}$, we assume for convenience that the
Yang-Mills gauge group ${G = \wt G}$ is simply-connected, with an eye
towards the eventual application to Chern-Simons theory.  As explained for
instance in Ch.~$5$ of \PressleySG, the Cartan subalgebra of $G$ then
divides into a countable set of alcoves under the action of $\FW_{\rm
aff}$, and each alcove serves as a fundamental domain for $\FW_{\rm
aff}$.  So to make a particular choice for $\lambda$ in its orbit
under $\FW_{\rm aff}$, we simply pick a distinguished Weyl alcove
$\RD_+$, in which we assume $\lambda$ lies.

\countdef\FundWeyl=102\FundWeyl=\pageno  
Following the discussion in Section $4.1$, for which ${\lambda
\ge 0}$ by convention, we take the distinguished Weyl alcove ${\RD_+
\subset \RC_+}$ to sit in the positive Weyl chamber of $\Ft$.
According to \POSWEYL, the positive Weyl chamber $\RC_+$ is a
simplicial cone, bounded by walls associated to the positive simple
roots of $G$.  At the tip of this cone sits a unique alcove containing
the origin in $\Ft$, and we take $\RD_+$ to be that 
alcove.  Explicitly, $\RD_+$ is the simplex in $\RC_+$ for which
${\lambda\ge 0}$ satisfies the additional bound 
\eqn\INEQII{ \langle\vartheta\,,\lambda\rangle \,\le\, 1\,,}
where $\vartheta$ is the highest root
of $G$.  Stated more geometrically, for each orbit of $\FW_{\rm aff}$
in $\Ft$, we take ${\lambda\in\RD_+}$ to be the positive
representative which lies at minimal distance from the origin.

\countdef\Highestrt=101\Highestrt=\pageno
Since these conventions may seem a little bit abstract, let us give a
concrete example, corresponding to the case ${G=SU(r+1)}$.  The Cartan
subalgebra is then represented by diagonal matrices of the form ${i
\diag(\lambda_1,\ldots\,\lambda_{r+1})}$ such that ${\lambda_1 +
\cdots + \lambda_{r+1} = 0}$.  A standard set of positive simple roots for
$SU(r+1)$ is given by the successive differences ${\lambda_j -
\lambda_{j+1}}$ for ${j=1,\ldots,r}$, so the positive Weyl chamber is
described by the inequalities ${\lambda_1 \ge \lambda_2 \ge \cdots \ge
\lambda_{r+1}}$.  Finally, the highest root $\vartheta$ of $SU(r+1)$
is given by the difference ${\lambda_1 - \lambda_{r+1}}$, so the
fundamental Weyl alcove $\RD_+$ is described by the additional
constraint ${\lambda_1 - \lambda_{r+1} \le 1}$.  For instance, if
${G=SU(2)}$, then ${\lambda_1}$ lies in the interval
${\RD_+=\left[0,\ha\right]}$.

\bigskip\noindent{\it Monodromy Operators and Coadjoint
Orbits}\smallskip

One satisfying aspect of our work in Section $4.3$ is that the Seifert
loop path integral in its symplectic form appears as an elegant
extension of the path integral which describes the Chern-Simons 
partition function on $M$.  To describe the Seifert loop operator
symplectically, we merely replace the symplectic space $\bar\CA$ with
the product ${\bar\CA_\alpha = \bar\CA \times \epsilon L\CO_\alpha}$,
where the loopspace $L\CO_\alpha$ can be considered as a coadjoint
orbit for the Hamiltonian group $\CH$ that acts on $\bar\CA$.

To recast the monodromy operator path integral \PZYMRV\ as a
symplectic path integral of the canonical form, we proceed in
complete analogy to the case of the Seifert loop operator.  As we
reviewed in Section $2$, the canonical symplectic integral which
describes the basic Yang-Mills partition function on $\Sigma$ is
determined by the Hamiltonian action of the group $\CG(P)$ on the
affine space $\CA(P)$.  To describe the monodromy operator by analogy
to the Seifert loop operator, we just consider the product of $\CA(P)$
with an appropriate coadjoint orbit of $\CG(P)$.

In fact, the correct coadjoint orbit is easy to guess.  Under 
reduction from ${C \subset M}$ to ${p \in \Sigma}$, the analogue of
the loopspace $\epsilon L\CO_\alpha$ is the finite-dimensional orbit
$\CO_\lambda$ itself, and $\CO_\lambda$ can be immediately embedded  
as a coadjoint orbit of $\CG(P)$.  Briefly, we recall that the dual of
the Lie algebra of $\CG(P)$ is formally the space of sections of the
bundle ${\Omega^2_\Sigma \otimes \ad(P)}$.  We then regard
$\CO_\lambda$ as the orbit of $\CG(P)$ passing through the singular
section of ${\Omega^2_\Sigma \otimes \ad(P)}$ which is given by 
\eqn\DISLHII{ \gamma_0 \,=\, \lambda \, \delta_p\,,}
in complete analogy to \DISLHDL.

Because $\gamma_0$ has delta-function support at $p$, an
element of $\CG(P)$ acts on $\CO_\lambda$ by restriction to the point
$p$.  Hence if we endow $\CO_\lambda$ with the coadjoint symplectic
form $\nu_\lambda$, the action of $\CG(P)$ on $\CO_\lambda$ is
Hamiltonian with moment map 
\eqn\COADJMOMII{ \langle\mu,\phi\rangle \,=\, -\int_\Sigma
\delta_p \; \Tr\Big[\big(g\,\lambda\,g^{-1}\big)\cdot \phi\Big] \,=\,
-\Tr\Big[\big(g\,\lambda\,g^{-1}\big) \cdot \phi\big|_p\Big]\,.}
Here $\phi$ transforms as a section of the bundle $\ad(P)$
on $\Sigma$, and $g$ is an element of $G$ which we use to specify the
point ${g\,\lambda\,g^{-1}}$ on $\CO_\lambda$.  We obtain the final
expression in \COADJMOMII\ by performing the integral over $\Sigma$
using the delta-function, thereby restricting $\phi$ to $p$. 
Clearly \COADJMOMII\ then agrees with the moment map \COADJMOM\ for
the action of $G$ on $\CO_\lambda$.

To recast the monodromy operator path integral as a symplectic path
integral of the canonical form, we consider the product 
\eqn\DEFAPLM{ \CA(P)_\lambda \,=\, \CA(P) \times 2\pi\CO_{\lambda}\,,}
\countdef\CurlyAPlam=103\CurlyAPlam=\pageno
with symplectic form 
\eqn\OMLM{ \Omega_\lambda \,=\, \Omega \,+\, 2\pi\nu_\lambda\,.}
\countdef\Omlam=104\Omlam=\pageno
Here $\Omega$ is the symplectic form \OMYM\ on the affine space
$\CA(P)$ of connections, and the factors of $2\pi$ in \DEFAPLM\ and
\OMLM\ are ultimately necessary to agree with our conventions for
$\RV_\lambda(p)$ in \LOCA.

As in Section $4.3$, we consider the diagonal action of $\CG(P)$
on $\CA(P)_\lambda$.  The moment map for this action is of course the
sum of the moment maps which describe the action of $\CG(P)$ on each
factor in \DEFAPLM.  Accounting for the coefficient of $2\pi$ in
\OMLM, we thus write the total moment map for the action of $\CG(P)$
on $\CA(P)_\lambda$ as 
\eqn\APMOM{ \langle\mu,\phi\rangle \,=\, -\int_\Sigma
\Tr\big(\CF_A \, \phi\big)\,.}
Here $\CF_A$ is a generalized curvature on $\Sigma$ which includes
the effective delta-function contribution at $p$ from \COADJMOMII,
\eqn\BIGFS{ \CF_A \,=\, F_A \,+\, 2\pi \, \big(g \,\lambda\,g^{-1}\big) \,
\delta_p\,.}
Based upon \APMOM\ and \BIGFS, the action ${S = \ha(\mu,\mu)}$ for the
canonical symplectic integral over $\CA(P)_\lambda$ then takes
precisely the same form as the Yang-Mills action, but expressed in
terms of $\CF_A$,
\eqn\SLAM{ S \,=\, \ha \big(\mu, \mu\big) \,=\, -\ha
\int_\Sigma \Tr\big(\CF_A \^ \*\CF_A\big)\,.}

With \SLAM\ in hand, we now claim that the monodromy operator path
integral can be interpreted as the canonical symplectic integral
determined by the Hamiltonian action of $\CG(P)$ on $\CA(P)_\lambda$,
\eqn\PZYMIII{ Z\big(\epsilon; p, \lambda\big) \,=\, {1
\over{\Vol(\CG(P))}} \, \left({1 \over {2 \pi
\epsilon}}\right)^{\Delta_{\CG(P)}/2} \, \int_{\CA(P)_\lambda} \!
\exp{\left[ \Omega_\lambda - {1 \over {2 \epsilon}}
\left(\mu,\mu\right)\right]}\,.}

To relate \PZYMIII\ to the starting path integral in \PZYMRV, let us
consider carefully the generalized Yang-Mills action in \SLAM.  Unlike
the usual Yang-Mills action for $F_A$, which is a perfectly
well-behaved functional of smooth connections on $\Sigma$, the
Yang-Mills action for $\CF_A$ is badly divergent for smooth 
configurations of $A$, due to the fact that it implicitly involves the 
square of a delta-function.  Since the Yang-Mills action is also 
positive-definite, the only connections on $\Sigma$ for which the 
action in \SLAM\ is finite are those for which $A$ has the appropriate
singularity to cancel the explicit delta-function curvature in
$\CF_A$.  Of course, because of the exponential suppression in the
integrand of \PZYMIII, the path integral only receives contributions
from connections with finite action.  Thus all non-zero contributions to
the symplectic integral over $\CA(P)_\lambda$ arise from singular
connections on $\Sigma$ whose curvatures near $p$ take the form  
\eqn\MONF{ F_A \,=\, -2\pi \, \big(g \,\lambda\,g^{-1}\big) \,
\delta_p \,+\, \cdots\,,}
where the `$\cdots$' indicate terms in $F_A$ which are regular at
$p$.  Comparing \LOCF\ to \MONF, we see that the singular connections
which define the monodromy operator $\RV_\lambda(p)$ are precisely
those which contribute to the symplectic path integral in \PZYMIII.

To place \PZYMIII\ into its proper physical context, we note that
disorder operators in quantum field theory can often be described in
terms of an auxiliary defect theory living on the submanifold in
spacetime where the operator is inserted.  In this language, the path
integral in \PZYMIII\ describes the monodromy operator inserted at $p$
via a very simple defect theory, one which incorporates only finitely-many
degrees of freedom valued in $\CO_\lambda$.

In Section $7.3$, at the very end of the paper, we will revisit the
symplectic interpretation of the monodromy operator $\RV_\lambda(p)$
in Yang-Mills theory on $\Sigma$.

\subsec{More About the Classical Monodromy Operator}

So far, we have described the monodromy operator $\RV_\lambda(p)$ as a
quantum operator in two-dimensional Yang-Mills theory.  In preparation
for the localization computations in Section $7$, we now wish to
consider $\RV_\lambda(p)$ in somewhat more detail from a purely
classical perspective, as reflected in the structure of the moduli
space of Yang-Mills solutions with monodromy on $\Sigma$.  

Actually, we restrict attention throughout to only the most basic
solutions of Yang-Mills theory in the presence of the monodromy
operator.  Namely, we consider connections which are flat on the
punctured Riemann surface ${\Sigma^o = \Sigma - \{p\}}$ and otherwise have
a singularity at $p$ of the form \LOCA, up to gauge transformations.
For such connections, ${\CF_A = 0}$ everywhere on $\Sigma$, so these
solutions make the dominant contribution to the monodromy operator
path integral.

By analogy to the Wilson loop moduli space $\SM(C,\alpha)$
introduced in Section $4.2$, we let $\SN(P;p,\lambda)$ denote the
moduli space of flat connections on $\Sigma$ with a singularity at $p$
of the form \LOCA, up to gauge transformations.  Here we are careful
to keep track of the data for both the principal $G$-bundle $P$ over
$\Sigma$ and the point $p$ at which the operator $\RV_\lambda$ is inserted.
However, because the topology of ${\SN(P;p,\lambda) \equiv \SN(P;\lambda)}$
does not vary with the continuous choice of $p$, we frequently omit 
$p$ from the notation.
\countdef\CurlyNP=105\CurlyNP=\pageno
\countdef\CurlyNPlam=106\CurlyNPlam=\pageno

In the special case ${\lambda = 0}$, the extended moduli space
$\SN(P;\lambda)$ reduces to the moduli space $\SN(P)$ of non-singular
flat connections on $\Sigma$.  Yet even when ${\lambda>0}$ is non-zero,
the extended moduli space $\SN(P;\lambda)$ is still related to
$\SN(P)$ in two essential ways.  First, $\SN(P;\lambda)$
is the total space of a natural symplectic fibration over
$\SN(P)$.  Second, $\SN(P;\lambda)$ is the splitting manifold for the
universal bundle on $\SN(P)$ (when that universal bundle exists).
These statements are entirely standard, but since they both feature
heavily in the cohomological computations in Section $7.3$, we take
some time to review them now.

\bigskip\noindent{\it $\SN(P;\lambda)$ as a Moduli Space of
Homomorphisms}\smallskip

To start, let us describe $\SN(P;\lambda)$ very concretely as a moduli
space of homomorphisms.  We proceed in complete analogy to our
discussion of the classical Wilson loop moduli space $\SM(C,\alpha)$
in Section $4.2$.

As well-known, if the Yang-Mills gauge group ${G = \wt G}$ is simply-connected
and $P$ is topologically trivial, the moduli space $\SN(P)$ of
(non-singular) flat connections appears directly as the
moduli space of homomorphisms ${\varrho:\pi_1(\Sigma)\rightarrow G}$,
where the fundamental group $\pi_1(\Sigma)$ is generated by elements
$\Ra_\ell$ and $\Rb_\ell$ for ${\ell=1,\ldots,h}$, subject to the
single relation  
\eqn\PIONEH{ \prod_{\ell=1}^h \left[\Ra_\ell, \Rb_\ell\right] \,=\,
1\,, \qquad\qquad \left[ \Ra_\ell , \Rb_\ell \right] \,\equiv\,
\Ra^{}_\ell \, \Rb^{}_\ell \, \Ra^{-1}_\ell \, \Rb^{-1}_\ell\,.}
Here $h$ is the genus of $\Sigma$.

By essentially the same observations as in Section $4.2$, the extended
moduli space $\SN(P;\lambda)$ can then be presented as the moduli space of
pairs $(\varrho^o,U_0)$, where $\varrho^o$ is now a homomorphism from the
fundamental group of the punctured Riemann surface ${\Sigma^o = \Sigma
- \{p\}}$ to
$G$,\countdef\Sigmao=107\Sigmao=\pageno\countdef\Rhoo=108\Rhoo=\pageno
\eqn\BIGHMRIV{ \varrho^o:\pi_1(\Sigma^o) \longrightarrow G\,,}
and ${U_0 = g\,\lambda\,g^{-1}}$ is an element of $\CO_\lambda$ related
to $\varrho^o$ by 
\eqn\YMCOND{ \varrho^o(\Rc) \,=\, \exp{\!\left(2 \pi U_0\right)}\,.}
Here ${\Rc\in\pi_1(\Sigma^o)}$ is the distinguished element which 
represents the small one-cycle about $p$ parametrized by $\varphi$ in
\LOCA, so that 
\eqn\PIONEHC{ \prod_{\ell=1}^h \left[\Ra_\ell, \Rb_\ell\right] \,=\,
\Rc\,,}
and the role of $U_0$ is to encode the monodromy of the connection at $p$.

In discussing Yang-Mills theory on $\Sigma$, we do not necessarily
wish to assume that the gauge group $G$ is simply-connected nor that
the bundle $P$ is trivial.  So more generally, we take $G$ to be 
the quotient of its simply-connected cover $\wt G$ by a subgroup of
the center $\CZ(\wt G)$.  The topology of $P$ as a principal
$G$-bundle over $\Sigma$ is then encoded by an element ${\zeta \in
\CZ(\wt G)}$.  The central element $\zeta$ represents the possible
monodromy in $\wt G$ at a generic point ${q\in\Sigma}$ which
otherwise obstructs $P$ from extending smoothly as a $\wt G$-bundle 
over $q$.

To incorporate the topology of $P$ into the general description of
$\SN(P;\lambda)$, we introduce the doubly-punctured Riemann surface
$\Sigma^{oo}$, 
\eqn\PSIG{ \Sigma^{oo} \,=\, \Sigma - \{p\} - \{q\}\,.}
\countdef\Sigmaoo=109\Sigmaoo=\pageno
As usual, the fundamental group $\pi_1(\Sigma^{oo})$ is generated by
elements $\Ra_\ell$, $\Rb_\ell$, $\Rc$, and $\Rd$ for
${\ell=1,\ldots,h}$, subject to the doubly-extended relation 
\eqn\PIMSII{ \prod_{\ell=1}^h \, \left[ \Ra^{}_\ell , \Rb^{}_\ell
\right] \,=\, \Rc\,\Rd\,.}
Here $\Rc$ and $\Rd$ represent small, suitably-oriented one-cycles
about the respective punctures at $p$ and $q$.  

To describe flat connections on $P$ with monodromy, we consider
homomorphisms $\varrho{}^{oo}$ from the fundamental group
$\pi_1(\Sigma^{oo})$ to the simply-connected group $\wt G$,
\eqn\BIGHMRV{ \varrho^{oo}:\pi_1(\Sigma^{oo}) \longrightarrow
\wt G\,,}
\countdef\Rhooo=110\Rhooo=\pageno
such that $\varrho^{oo}$ satisfies 
\eqn\YMCONDII{ \varrho^{oo}(\Rc) \,=\, \exp{\!\left(2 \pi
U_0\right)}\,,\qquad\qquad \varrho^{oo}(\Rd) \,=\, \zeta\,.}
As before, $U_0$ is an element of $\CO_\lambda$ encoding the monodromy at
$p$, and in the special case that ${\lambda = U_0 = 0}$, the
homomorphism $\varrho^{oo}$ in \YMCONDII\ describes a non-singular
flat connection on the $G$-bundle $P$.

By way of notation, we let $\wt\SN(P;\lambda)$ be the space of pairs 
$(\varrho^{oo},U_0)$ satisfying \YMCONDII, modulo the diagonal action
of $\wt G$,
\eqn\MNONE{ \wt\SN(P;\lambda)\,=\,\Big\{(\varrho^{oo}, U_0)
\;\big|\; \varrho^{oo}(\Rc)=\exp{\!\left(2 \pi
U_0\right)},\;\varrho^{oo}(\Rd)=\zeta\Big\}\Big/{\wt G}\,.}
Though $U_0$ will be useful to have around in a moment, if $\lambda$
is generic we can solve for $U_0$ in terms of $\varrho^{oo}(\Rc)$ via
\YMCONDII, so that in terms of $\varrho^{oo}$ alone,
\eqn\MNONEII{\eqalign{
\wt\SN(P;\lambda)\,&=\,\Big\{ \varrho^{oo}\;\big|\;\varrho^{oo}(\Rc)\,
\in\,\FC_\lambda,\;\varrho^{oo}(\Rd)=\zeta\Big\}\Big/{\wt G}\,,\cr
\FC_\lambda\,&=\, \Cl\!\left[\exp{\!(2\pi\lambda)}\right]\,.}}
\countdef\tildeCurlyNP=113\tildeCurlyNP=\pageno
The description of $\wt\SN(P;\lambda)$ here is entirely analogous to the
description of $\SM(C,\alpha)$ in \MAONEII.  Moreover, when we
specialize in Section $7.1$ to the case that $M$ is a Seifert
manifold, we will make the relationship between $\wt\SN(P;\lambda)$
and $\SM(C,\alpha)$ even more precise.

The moduli space $\wt\SN(P;\lambda)$ of homomorphisms in \MNONE\ is
almost, but not quite, $\SN(P;\lambda)$.  Rather,
$\wt\SN(P;\lambda)$ is an unramified cover of $\SN(P;\lambda)$ of degree
${|\wt G\!:\!G|^{2h}}$, where $|\wt G\!:\!G|$ is the order of the basic
covering ${\wt G\rightarrow G}$.  For instance, if ${G = \wt G/\CZ(\wt
G)}$ is the adjoint form of $\wt G$, then ${|\wt G\!:\!G| = |\CZ(\wt
G)|}$.  We obtain a covering of $\SN(P;\lambda)$ in \MNONE\ because
the holonomies associated to the cycles $\Ra_\ell$ and $\Rb_\ell$ are
specified as elements of $\wt G$, not $G$, in \BIGHMRV.  This caveat
aside, the distinction between $\wt\SN(P;\lambda)$ and
$\SN(P;\lambda)$ will at most affect an overall numerical factor in the
cohomological formulae in Section $7.3$ and is otherwise inessential.

\bigskip\noindent{\it $\SN(P;\lambda)$ as a Symplectic Fibration over
$\SN(P)$}\smallskip

From the perspective of the present paper, we now arrive at the
first important geometric fact about $\SN(P;\lambda)$.  As we have already
mentioned, in the trivial case that ${\lambda = 0}$, $\SN(P;\lambda)$
immediately reduces to the moduli space $\SN(P)$ of non-singular flat
connections on $\Sigma$.  More generally if $\lambda$ is non-zero but
small, in a sense to be made precise, then $\SN(P;\lambda)$ is still
related to $\SN(P)$ in a simple way, and this relationship underlies
the cohomological interpretation for both the monodromy and the
Seifert loop operators.

Specifically, when ${\lambda \ge 0}$ obeys the strict bound
${\langle\vartheta,\lambda\rangle< 1}$ in \INEQII\ and $\SN(P)$ itself 
is non-singular, the extended moduli space $\SN(P;\lambda)$ fibers 
\refs{\JeffreyLCII,\DaskW,\DonaldsonII} smoothly over $\SN(P)$,
\eqn\FBSMOLII{\matrix{
&2\pi\CO_{-\lambda}\,\longrightarrow\,\SN(P;\lambda)\cr
&\mskip 115mu\Big\downarrow\lower 0.5ex\hbox{$^{\Rq}$}\cr
&\mskip 100mu\SN(P)\cr}\,,\qquad\qquad \langle\vartheta,\lambda\rangle
< 1\,.}
Here the fiber of $\SN(P;\lambda)$ is the coadjoint orbit of $G$
through ${-2\pi\lambda}$, as indicated by the prefactor and the sign in
\FBSMOLII.  Of course, at the level of topology, $2\pi\CO_{-\lambda}$
is indistinguishable from the basic orbit $\CO_\lambda$, but
$2\pi\CO_{-\lambda}$ carries the symplectic form ${-2\pi\nu_\lambda}$.

The symplectic structure on the fiber of \FBSMOLII\ is relevant, because
$\SN(P)$ and $\SN(P;\lambda)$ also carry natural symplectic forms.
These moduli spaces are determined by the vanishing of the respective   
moment maps ${\mu = F_A}$ and ${\mu = \CF_A}$ for the Hamiltonian
action of $\CG(P)$ on the symplectic spaces $\CA(P)$ and
$\CA(P)_\lambda$, from which $\SN(P)$ and $\SN(P;\lambda)$ inherit
symplectic forms under the symplectic quotient construction.
Abusing notation somewhat, we let $\Omega$ be the symplectic form on
$\SN(P)$ inherited from the form \OMYM\ on $\CA(P)$, and we let
$\Omega_\lambda$ be the symplectic form on $\SN(P;\lambda)$ inherited
from the corresponding form \OMLM\ on $\CA(P)_\lambda$.

The smooth fibration in \FBSMOLII\ is now compatible with the
symplectic data on $\CO_{\lambda}$, $\SN(P)$, and $\SN(P;\lambda)$ in
the following sense.  Again provided that ${\lambda\ge 0}$ satisfies
the strict inequality ${\langle\vartheta,\lambda\rangle < 1}$ in \INEQII,
the symplectic form $\Omega_\lambda$ on $\SN(P;\lambda)$ decomposes as
a sum 
\eqn\OMLAM{ \Omega_\lambda \,=\, {\Rq}^*\Omega \,-\,
2\pi \Re_\lambda\,.}
\countdef\Relam=112\Relam=\pageno\countdef\OmCurlyNP=114\OmCurlyNP=\pageno
Here ${\Rq}^*\Omega$ is the pullback of the symplectic form
$\Omega$ on $\SN(P)$ to $\SN(P;\lambda)$, and $\Re_\lambda$ is a
closed two-form on $\SN(P;\lambda)$ which restricts fiberwise to the
coadjoint symplectic form $\nu_\lambda$.  The relative factor of
$-2\pi$ in \OMLAM\ is responsible for the appearance of
the same factor in \FBSMOLII.

Following the exposition in \S $3.5$ of \GukovJK, let us quickly
sketch how the fibration of $\SN(P;\lambda)$ arises.  First, the fiber
of $\SN(P;\lambda)$ is parametrized by the element
${U_0\in\CO_\lambda}$ appearing in \MNONE, which thereby determines
the monodromy $\varrho^{oo}(\Rc)$ around the puncture at $p$.  Otherwise,
the base of $\SN(P;\lambda)$ is parametrized by the holonomies
associated to the remaining generators $\Ra_\ell$ and $\Rb_\ell$ of
$\pi_1(\Sigma^{oo})$.  As $\varrho^{oo}$ is a group homomorphism,
those holonomies necessarily satisfy 
\eqn\BSSNP{ \prod_{\ell=1}^h \, \Big[\varrho^{oo}(\Ra^{}_\ell)
\,,\, \varrho^{oo}(\Rb^{}_\ell)\Big] \,=\, \zeta \cdot
\exp{\!\left(2 \pi U_0\right)}\,.}
For ${\lambda = U_0 = 0}$, a homomorphism $\varrho^{oo}$ satisfying
\BSSNP\ determines a non-singular flat connection on the bundle $P$.
So for ${U_0 \neq 0}$ fixed and sufficiently small (we discuss the
precise bound in a moment), the relation \BSSNP\ is just a deformation
of constraint defining $\SN(P)$.  Provided that the moduli space
$\SN(P)$ is smooth, its topology is unchanged under continuous
deformations, so $\varrho^{oo}(\Ra_\ell)$ and $\varrho^{oo}(\Rb_\ell)$
effectively parametrize a copy of $\SN(P)$.

As one instance when the smoothness condition holds, we take $\Sigma$
to be a Riemann surface of genus ${h \ge 1}$.  Then $\SN(P)$ is smooth 
if ${G =SU(r+1)/\BZ_{r+1}}$ is the adjoint form of $SU(r+1)$, and $P$ is a
topologically non-trivial $G$-bundle over $\Sigma$ characterized by a
generator $\zeta$ of $\BZ_{r+1}$.  For example, if ${G=SO(3)}$, $P$
is the $SO(3)$-bundle over $\Sigma$ with non-vanishing Stiefel-Whitney
class ${w_2 \neq 0}$.

At the level of topology, we have sketched why $\SN(P;\lambda)$ fibers
over $\SN(P)$, but the symplectic nature of the fibration in
\FBSMOLII\ is also extremely important.  Indeed, the symplectic
decomposition \OMLAM\ of $\Omega_\lambda$ turns out to be a basic
ingredient in our cohomological analysis of the Seifert loop 
operator in Section $7.3$.

The formula \OMLAM\ for $\Omega_\lambda$ can be understood in at
least two ways.  As exploited by Jeffrey \JeffreyLCII, one way to
understand \OMLAM\ is as a general consequence of symplectic reduction
at a non-zero value of the moment map.  We have already noted that
${\SN(P;\lambda)}$ can be constructed as the quotient under 
$\CG(P)$ of the vanishing locus for the moment map ${\mu = \CF_A}$ in
the product ${\CA(P)_\lambda \,=\, \CA(P) \times 2\pi\CO_\lambda}$.
But since ${\CF_A = F_A \,+\, 2\pi\,(g\lambda g^{-1})\,\delta_p}$, the
extended moduli space $\SN(P;\lambda)$ can be equivalently constructed
as the quotient under $\CG(P)$ of the locus in the original affine
space $\CA(P)$ where the moment map ${\mu = F_A}$ takes values in the
coadjoint orbit through ${-2\pi\lambda\,\delta_p}$.  For $\lambda$
sufficiently small, the formula for $\Omega_\lambda$ in \OMLAM\ then
follows from general facts about symplectic reduction at a non-zero
value of the moment map.
 
Alternatively, we can check the formula for $\Omega_\lambda$ in
\OMLAM\ directly, by evaluating $\Omega_\lambda$ on two tangent
vectors to $\SN(P;\lambda)$.  Concretely, if $\SX$ is a tangent vector to
$\SN(P;\lambda)$ at a point corresponding to a given flat connection
$A$ with singularity at $p$ of the form \LOCA, then $\SX$ can be
represented as a sum 
\eqn\TANFBS{ \SX \,=\, \eta \,+\, d_A \phi\,.}
Here $\eta$ is a smooth, non-singular section of ${\Omega^1_\Sigma
\otimes \ad(P)}$ which satisfies ${d_A^{} \eta \,=\, d_A^\dagger \eta \,=\,
0}$, where ${d_A^\dagger = -\*\, d_A \*}$ is the adjoint of $d_A$.  By
the usual Hodge theory, $\eta$ thereby represents a tangent vector to
the base $\SN(P)$ in a local trivialization of the fibration in \FBSMOLII.  

Of course, we must allow $\SX$ to have a component along the
fiber of $\SN(P;\lambda)$.  For this reason, we have introduced
besides $\eta$ a gauge-trivial term $d_A\phi$ in $\SX$, where $\phi$
is an arbitrary smooth section of $\ad(P)$.  To explain the role of
$\phi$, we recall that the fiber of $\SN(P;\lambda)$ parametrizes the
possible monodromies of $A$ at $p$.  Since non-trivial gauge
transformations at $p$ act transitively upon the possible monodromies,
$d_A \phi$ thus represents a tangent vector to the fiber in \FBSMOLII.

If $\SX_1$ and $\SX_2$ are two such tangent vectors to
$\SN(P;\lambda)$, associated to pairs $(\eta,\phi)$ and
$(\xi,\psi)$ as in \TANFBS, then the symplectic pairing on
$\SN(P;\lambda)$ is given explicitly by 
\eqn\TANPRI{\eqalign{
\Omega_\lambda(\SX_1, \SX_2) \,&=\, -\int_\Sigma\!\Tr\big(\SX_1 \^ 
\SX_2\big)\,,\cr
&=\, -\int_\Sigma\!\Tr\big[(\eta + d_A \phi)\^(\xi + d_A
\psi)\big]\,,\cr
&=\, -\int_\Sigma\!\Tr\big(\eta \^ \xi\big) \,-\,
\int_\Sigma\!\Tr\big(d_A\phi\^d_A\psi\big)\,.}}
In passing to the third line of \TANPRI, we have noted that the
cross-terms otherwise appearing in the second line of \TANPRI\ vanish
upon integration by parts, since ${d_A \eta = d_A \xi = 0}$.  We then
immediately recognize the pairing between $\eta$ and $\xi$ in 
\TANPRI\ as the natural symplectic pairing on $\SN(P)$, 
\eqn\TANPRII{ \Omega(\eta,\xi) \,=\, -\int_\Sigma
\Tr\big(\eta\^\xi)\,.}

We are left to consider the pairing between $\phi$ and $\psi$.
By assumption, the background connection $A$ which represents a point in
$\SN(P;\lambda)$ satisfies ${d_A^2 = F_A = -2\pi\lambda\,\delta_p}$.
Integrating by parts, we find  
\eqn\TANPRIII{ 
-\int_\Sigma\!\Tr\big(d_A\phi\^d_A\psi\big) \,=\,
\int_\Sigma\!\Tr\big(\phi\,[F_A, \psi]\big) \,=\, -2 \pi
\Tr\big(\phi\,[\lambda, \psi]\big)\big|_p\,.}
Comparing \TANPRIII\ to \COADJ\ and being careful about signs, we thus
see that the restriction of $\Omega_\lambda$ to the fiber of
$\SN(P;\lambda)$ is given by $-2\pi\nu_\lambda$, as claimed in
\OMLAM.

To finish up our discussion of the symplectic fibration in \FBSMOLII,
let us quickly consider the regime in which the fibration is 
valid.  As we have been careful to emphasize, the formula \OMLAM\ for 
$\Omega_\lambda$ holds only when ${\lambda \ge 0}$ satisfies the
strict bound ${\langle\vartheta,\lambda\rangle < 1}$, where
$\vartheta$ is the highest root of $G$.  According to \INEQII, this
bound is saturated on the far wall of the Weyl alcove $\RD_+$, and the
reader may wonder what happens to $\SN(P;\lambda)$ when $\lambda$ hits
that wall.  In brief, when $\lambda$ satisfies
${\langle\vartheta,\lambda\rangle=1}$, a cycle (generically, a
two-cycle) collapses in $\SN(P;\lambda)$, and the symplectic form
$\Omega_\lambda$ degenerates.\foot{Of course, along the walls of
$\RD_+$ which coincide with the walls of the full Weyl chamber
$\RC_+$, cycles also collapse in $\SN(P;\lambda)$, for the basic
reason that $\lambda$ ceases to be regular on those walls. 
Nonetheless, on the walls of $\RC_+$, the formula \OMLAM\ for
$\Omega_\lambda$ itself remains valid.}

The collapsing cycle in $\SN(P;\lambda)$ is not hard to see 
if we simply compare the two descriptions for $\SN(P;\lambda)$ in
\MNONE\ and \MNONEII.  In passing from \MNONE\ to \MNONEII, we used
the relation ${\varrho^{oo}(\Rc) = \exp{\!(2\pi U_0)}}$ to eliminate
the parameter ${U_0 \in \CO_\lambda}$ in favor of the monodromy
${\varrho^{oo}(\Rc) \in \FC_\lambda}$.  For generic values of 
$\lambda$, the element $U_0$ is determined uniquely by
$\varrho^{oo}(\Rc)$, and the conjugacy class $\FC_\lambda$ which
parametrizes the monodromy is diffeomorphic to the coadjoint orbit
$2\pi\CO_\lambda$, from which $\FC_\lambda$ inherits the natural
symplectic form (up to sign).  However, at exceptional values of
$\lambda$, the relation between $\varrho^{oo}(\Rc)$ and $U_0$ ceases
to be invertible, and a non-trivial cycle in $2\pi\CO_\lambda$
collapses under the exponential map to $\FC_\lambda$.  Since
$\FC_\lambda$ is identified with the fiber of $\SN(P;\lambda)$, as we
implicitly used in the gauge theory computation \TANPRI\ of
$\Omega_\lambda$, the same cycle collapses in $\SN(P;\lambda)$.

To present the collapsing cycle explicitly, we observe
that both $\CO_\lambda$ and $\FC_\lambda$ can be described as quotients
\eqn\RNQUOT{ \CO_\lambda \,=\, G/G_\lambda\,,\qquad\qquad \FC_\lambda \,=\,
G/\CZ_\Lambda\,,}
\countdef\Centralizerlam=111\Centralizerlam=\pageno
where $G_\lambda$ is the stabilizer of $\lambda$ under the adjoint
action of $G$, and $\CZ_\Lambda$ is the centralizer of ${\Lambda =
\exp{\!(2\pi\lambda)}}$ in $G$.  The stabilizer $G_\lambda$ is
always a subgroup of the centralizer $\CZ_\Lambda$, so the exponential
map from $2\pi\CO_\lambda$ to $\FC_\lambda$ fits generally into a
sequence 
\eqn\EXPSEQ{\matrix{
&\CZ_\Lambda/G_\lambda\,\longrightarrow\,2\pi\CO_\lambda\cr
&\mskip 125mu\Big\downarrow\lower 0.5ex\hbox{$^{\exp}$}\cr
&\mskip 110mu\FC_\lambda\cr}\,.}
As apparent from \EXPSEQ, whenever $\CZ_\Lambda$ is enhanced beyond
$G_\lambda$, the exponential map from $2\pi\CO_\lambda$ to
$\FC_\lambda$ is not smoothly invertible, and the cycle
${\CZ_\Lambda/G_\lambda}$ collapses in $2\pi\CO_\lambda$.

The relevant criterion for the centralizer $\CZ_\Lambda$ to enhance is
well-known, so we will merely state it.  See \S ${\rm VII}.4$ of 
\Helgason\ for a textbook discussion.  For convenience, we take $G$ to
be simply-connected.  The fiber ${\CZ_\Lambda/G_\lambda}$ in \EXPSEQ\
is then non-trivial precisely when $\lambda$ satisfies the integrality
condition 
\eqn\CONDL{ \langle\beta,\lambda\rangle \,\in\, \BZ -
\{0\}\,,}
for some root $\beta$ of $G$.  In fact, because the roots of $G$
appear in positive and negative pairs $\pm\beta$, the condition
\CONDL\ is always satisfied by $\lambda$ for an even number (possibly
zero) of roots.

As an elementary example of the integrality condition \CONDL, we
consider the case ${G=SU(2)}$, for which we parametrize ${\lambda\in\Ft}$ in
terms of a real variable $x$ as ${\lambda = i \diag(x,-x)}$.  If
${x=0}$, both $2\pi\CO_\lambda$ and $\FC_\lambda$ are points, and the
exponential map is trivial.  Similarly, if $x$ is generic,
$2\pi\CO_\lambda$ and $\FC_\lambda$ are mutually diffeomorphic to the
quotient ${SU(2)/U(1) = S^2}$.  Finally, in the special case that
${x=j/2}$ for some non-zero integer $j$, corresponding precisely to
the integrality condition in \CONDL, $2\pi\CO_\lambda$ is again a
two-sphere, but ${\exp{\!(2\pi\lambda)} = \pm 1}$ is central in
$SU(2)$, so $\FC_\lambda$ is a point.

For general $G$, the same pattern holds.  Generically, if $\lambda$
satisfies the integrality condition in \CONDL, it does so for a single
pair of roots $\pm\beta$, and a two-cycle collapses under the
exponential map to $\FC_\lambda$.

Returning to our analysis of $\SN(P;\lambda)$, we take $\lambda$ to
sit in the fundamental Weyl alcove $\RD_+$.  Then as $\lambda$ moves
away from the origin in $\RD_+$, the integrality condition \CONDL\ is
first satisfied when $\lambda$ hits the wall defined by the highest
root $\vartheta$ of $G$, so that ${\langle\vartheta,\lambda\rangle=1}$.
Along that wall, the gauge symmetry preserved by $\RV_\lambda(p)$
enhances, and the cycle $\CZ_\Lambda/G_\lambda$ collapses in the fiber
of $\SN(P;\lambda)$.

\bigskip\noindent{\it $\SN(P;\lambda)$ as a Splitting
Manifold for the Universal Bundle}\smallskip

Beyond its mere existence, the fibration \FBSMOLII\ of $\SN(P;\lambda)$ over
$\SN(P)$ plays an important theoretical role, since it presents
$\SN(P;\lambda)$ as the splitting manifold for the universal bundle on
$\SN(P)$.  This observation appears for instance as Proposition $3.5$
in \JeffreyLCII, where it is applied to a cohomological computation
very similar to the one in Section $7.3$, and it is what we wish to
explain now.

We start by introducing the universal bundle $\SV$ associated to the
basic moduli space $\SN(P)$.  So far, we have interpreted $\SN(P)$ as
the moduli space of flat connections on the principal $G$-bundle $P$
over $\Sigma$.  On the other hand, the classic theorem of
Narasimhan and Seshadri \Narasimhan\ establishes a one-to-one
correspondence between flat connections on $P$ and suitable   
holomorphic bundles over $\Sigma$, so that $\SN(P)$ admits an
algebraic description as well.  For the eventual application in
Section $7.3$, we are most interested in the standard case that ${G =
SU(r+1)/\BZ_{r+1}}$ is the adjoint form of ${\wt G = SU(r+1)}$, and $P$ is a
topologically non-trivial $G$-bundle over $\Sigma$ characterized by a
generator ${\zeta \in \BZ_{r+1}}$.  If $\Sigma$ is a Riemann surface
of genus ${h \ge 1}$, these assumptions ensure that $\SN(P)$ is smooth.
Then in its algebraic incarnation, $\SN(P)$ is the moduli space of
stable holomorphic vector bundles of rank ${r+1}$ on $\Sigma$, with
first Chern class equal to $\zeta$ mod ${r+1}$ and with fixed
determinant.

In the situation above,  the universal bundle $\SV$ exists as a
holomorphic vector bundle of rank ${r+1}$ over the product ${\Sigma
\times \SN(P)}$,
\eqn\UNIVB{\matrix{
&\BC^{r+1}\,\longrightarrow\,\SV\cr
&\mskip 85mu\big\downarrow\cr
&\mskip 115mu {\Sigma \times \SN(P)}\cr}\,,}
\countdef\CurlyV=115\CurlyV=\pageno
such that $\SV$ possesses the following universal property.  By definition,
$\SN(P)$ parametrizes holomorphic vector bundles of rank ${r+1}$ on
$\Sigma$, so each point ${y\in\SN(P)}$ determines a corresponding 
vector bundle $V_y$ on $\Sigma$.  On the other hand, the restriction
of $\SV$ to ${\Sigma \times \{y\}}$ also determines a rank ${r+1}$
holomorphic vector bundle on $\Sigma$.  According to the universal
property of $\SV$, these bundles are isomorphic,
\eqn\UNIVBII{ V_y \,\cong\, \SV\big|_{\Sigma \times
\{y\}}\,,\qquad\qquad y\,\in\,\SN(P)\,.}

In general, the universal property does not determine $\SV$ uniquely.
For if a bundle $\SV$ satisfies \UNIVBII, then so does the tensor product
of $\SV$ with any holomorphic line bundle on ${\Sigma \times \SN(P)}$
which is itself the pullback from a holomorphic line bundle on
$\SN(P)$.  Nevertheless, this ambiguity in $\SV$ can be fixed (see \S
$9$ of \AtiyahYM\ for details), and a `normalized' version of the
universal bundle exists which is unique up to isomorphism.  Though the
details will not matter here, for concreteness we take $\SV$ to be
that normalized universal bundle.

Just as the restriction of $\SV$ to ${\Sigma \times \{y\}}$ determines
a holomorphic vector bundle on $\Sigma$ for each point ${y\in\SN(P)}$,
the restriction of $\SV$ to ${\{p\}\times \SN(P)}$ determines a
holomorphic vector bundle on $\SN(P)$ for each point ${p\in\Sigma}$.
By way of abbreviation, we set 
\eqn\DEFSVP{ \SV_p \,\equiv\, \SV|_{\{p\}\times \SN(P)}\,.}
\countdef\CurlyVp=116\CurlyVp=\pageno  
Because $p$ varies continuously in $\Sigma$, the topology of $\SV_p$
is independent of $p$.  In particular, the Chern classes of $\SV_p$ do
not depend on the point $p$ and therefore provide distinguished 
elements in the integral cohomology ring of $\SN(P)$.

Throughout the following discussion, we impose the crucial condition
that ${\lambda > 0}$ be regular.  Given that condition, we now claim
that ${\SN(P;p,\lambda) \equiv \SN(P;\lambda)}$ is tautologically the 
splitting manifold for the vector bundle $\SV_p$ over $\SN(P)$.  That
is, under the map ${\Rq:\SN(P;\lambda)\rightarrow \SN(P)}$ in the
symplectic fibration \FBSMOLII, the pullback $\Rq^*\SV_p$ splits
smoothly into a direct sum of complex line bundles $\SL_j$ for
${j=1,\ldots,r+1}$ over $\SN(P;\lambda)$, 
\eqn\SPLITVP{ \Rq^*\SV_p \,\cong\, \bigoplus_{j=1}^{r+1} \,
\SL_j\,.}
\countdef\CurlyLj=119\CurlyLj=\pageno
Hence the Chern classes of the line bundles $\SL_j$ realize the
Chern roots $\Ru_j$ of $\SV_p$,
\eqn\CHRNRTS{ \Rq^*c(\SV_p) \,=\, \prod_{j=1}^{r+1} \left(1 \,+\,
\Ru_j\right),\qquad\qquad  \Ru_j \,\equiv\, c_1(\SL_j) \in
H^2\!\big(\SN(P;\lambda);\BZ\big)\,,}
\countdef\Ruj=120\Ruj=\pageno
where $c(\SV_p)$ is the total Chern class,
\eqn\TOTCH{ c(\SV_p) \,=\, 1 \,+\, c_1(\SV_p) \,+\, \cdots \,+\,
c_{r+1}(\SV_p)\,.}
As apparent from \CHRNRTS\ and essential later, any symmetric function
of the Chern roots $\Ru_j$ is the pullback from $\SN(P)$ of an
associated function of the Chern classes $c_j(\SV_p)$.

The splitting in \SPLITVP\ can be understood from various
perspectives.  One approach is to interpret $\SN(P;\lambda)$
algebraically, via the classic theorem of Mehta and Seshadri
\MehtaSH, as a moduli space of holomorphic bundles on $\Sigma$ with
parabolic structure at $p$.  See \S $3.4$ of \GukovJK\ for an
excellent review of parabolic bundles as they occur in gauge theory.
From the parabolic perspective, $\SN(P;\lambda)$ is then precisely the 
splitting manifold constructed on general grounds in Chapter $21$ of \BottT.

Rather than follow the algebraic route here, we will take an
equivalent but slightly more hands-on approach, which allows us to
relate the discussion to some ideas already appearing in Section
$4.1$.  We begin by unraveling a few definitions.

Let us consider arbitrary points ${y\in\SN(P)}$ and
${z\in\SN(P;\lambda)}$ satisfying ${y=\Rq(z)}$, so that $z$ is a 
point in the fiber of $\SN(P;\lambda)$ over $y$.  Relative to a local
trivialization of the fibration \FBSMOLII, we can always write ${z =
\left(U_0,y\right)}$, where ${U_0 \in \CO_\lambda}$ parametrizes the
coadjoint fiber over $y$.  

Now by definition, the fiber of the pullback $\Rq^*\SV_p$ over $z$
is the fiber of $\SV_p$ itself over $y$.  But according to the
universal property \UNIVBII, the fiber of $\SV_p$ over $y$ is
isomorphic to the fiber of $V_y$ over $p$,
\eqn\UNIVBIII{ \SV_p\big|_y \,=\, \SV\big|_{\{p\}\times\{y\}} \,\cong\,
V_y\big|_p\,,\qquad\qquad \{p\}\times\{y\} \in \Sigma \times \SN(P)\,,}
where we recall that $V_y$ is the holomorphic vector bundle over
$\Sigma$ determined by ${y\in\SN(P)}$.  The splitting in \SPLITVP\ is thus
equivalent to a smooth splitting of $V_y\big|_p$ for each
${z=\left(U_0,y\right)}$, such that the splitting respects the action
of $G$ on $U_0$ induced by gauge transformations at ${p\in\Sigma}$.

To obtain the requisite splitting, we consider the infinitesimal
action of $U_0$ on the fiber of $V_y$ at $p$.  Here we use the embedding
${\CO_\lambda \subset \Fg}$, where the Lie algebra ${\Fg =
\Fs\Fu(r+1)}$ acts via the standard, fundamental representation on the
fiber ${V_y\big|_p \cong \BC^{r+1}}$.  If $\lambda$ is regular, then
$U_0$ acts on $V_y\big|_p$ with distinct eigenvalues, and each
eigenvalue is associated to a one-dimensional complex eigenspace
$L_j$ for ${j=1,\ldots,r+1}$.  Thus for each ${U_0 \in \CO_\lambda}$,
the fiber of $V_y$ at $p$ decomposes into a sum of eigenspaces 
\eqn\SPLTVPL{ V_y\big|_p \,=\, \bigoplus_{j=1}^{r+1} \, L_j\,.}
Moreover, the eigenspace decomposition \SPLTVPL\ naturally respects
the simulaneous action of $G$ (or more precisely, the universal cover $\wt
G$) on both $U_0$ and $V_y\big|_p$, so the decomposition does not
depend on the choice of the local trivialization we used to write
${z=(U_0,y)}$.

As $z$ varies, the eigenspaces $L_j$ themselves vary smoothly as the
fibers of corresponding complex line bundles $\SL_j$ over
$\SN(P;\lambda)$.  Via \UNIVBIII\ and \SPLTVPL, these line bundles
then provide the tautological splitting \SPLITVP\ of $\Rq^*\SV_p$.

Actually, we can go a bit further in identifying the line bundles
$\SL_j$ over $\SN(P;\lambda)$.  By construction, each $\SL_j$ respects
the action of $G$ on the coadjoint fibers of $\SN(P;\lambda)$, so
$\SL_j$ must restrict fiberwise to a homogeneous line bundle.  That
is, 
\eqn\SLJA{ \SL_j\big|_{\CO_\lambda} \,=\, \FL(-\alpha_j)\,,\qquad\qquad
\alpha_j \in \Gamma_{\rm wt}\,,}
where $\FL(-\alpha_j)$ is the homogeneous line bundle on $\CO_\lambda$
determined by an appropriate weight $\alpha_j$ of ${SU(r+1)}$.  The
slightly perverse sign in \SLJA\ becomes useful in a moment.  Implicit in
\SLJA, as the fiber of $\SN(P;\lambda)$ varies continuously over
$\SN(P)$, the particular weight $\alpha_j$ associated to each line
bundle $\SL_j$ does not jump.  Thus the tautological splitting \SPLITVP\
is characterized by a finite set of weights 
$\big\{\alpha_1,\ldots,\alpha_{r+1}\big\}$, which we would like to
determine.

To do so, we once again unravel definitions.  As in Section $4.1$, to
discuss homogeneous line bundles on $\CO_\lambda$, we take ${G =
SU(r+1)}$ to be simply-connected for convenience.  According to the
convention in \LRHO, the homogeneous line bundle $\FL(\alpha)$ is then 
given by a quotient of ${G  \times \BC}$ under the action of ${T =
U(1)^r}$ as 
\eqn\LRHOII{ t\cdot\left(g\,, v\right)
\,=\,\left(g\,t^{-1},\,\varrho_\alpha(t) \cdot v\right)\,,\qquad g \in
G\,,\qquad v \in \BC\,,\qquad t \in T\,,}
where 
\eqn\HOMRHOII{ \varrho_\alpha(t) \,=\,
\exp{\!\big[i\,\langle\alpha\,,\xi\rangle\big]}, 
\qquad\qquad t=\exp(\xi)\,,\qquad\qquad \xi\in\Ft\,.}
In these terms we identify ${\CO_\lambda = G/T}$, with ${U_0 = g
\,\lambda\, g^{-1}}$.  Without loss we set 
\eqn\COMPLAM{ \lambda \,=\, i
\diag(\lambda_1,\cdots,\lambda_{r+1})\,,\qquad\qquad \lambda_1 \,+\, 
\cdots \,+\, \lambda_{r+1} \,=\, 0\,,}
where regularity of ${\lambda > 0}$ implies 
\eqn\REGLAM{ \lambda_1 > \lambda_2 > \cdots > \lambda_{r+1}\,.}

At the identity in $G$, the eigenspaces $\big\{L_1,\ldots,L_{r+1}\big\}$
appearing in the decomposition \SPLTVPL\ of ${V_y\big|_p \cong
\BC^{r+1}}$ are spanned by the standard basis vectors $\big\{e_1,
\ldots, e_{r+1}\big\}$, upon which ${U_0 = \lambda}$ acts as  
${U_0 \cdot e_j \,=\, \lambda_j \, e_j}$ for ${j=1,\ldots,r+1}$.
Explicitly, ${e_1 = (1,0,\ldots,0)}$,  ${e_2=(0,1,\ldots,0)}$, and so
on.  More generally, if ${U_0 = g\,\lambda\, g^{-1}}$, then each eigenspace
$L_j$ is the subspace of $\BC^{r+1}$ spanned by ${g \cdot e_j}$, where
${g\in SU(r+1)}$ acts on $e_j$ by ordinary matrix multiplication.
Therefore, under the action ${g \mapsto g \, t^{-1}}$ as in \LRHOII, 
elements in $L_j$ are multiplied by the phase
\eqn\PHLJ{ \varrho_j(t) \,=\, \exp{\!\big(-i\,\xi_j\big)},
\qquad\qquad t \,=\,
\diag\!\big(\e{i\xi_1},\cdots,\e{i\xi_{r+1}}\big).} 

To specify the associated line bundle on $\CO_\lambda$, let us
introduce a standard set of generators for the weight lattice
$\Gamma_{\rm wt}$ of $SU(r+1)$.  If ${\xi =
i\diag(\xi_1,\ldots,\xi_{r+1})}$ is any element in the Cartan
subalgebra of $SU(r+1)$, we define weights 
$\big\{\hat\omega_1,\ldots,\hat\omega_{r+1}\big\}$ such that 
\eqn\SUWTII{\eqalign{
&\left\langle\hat\omega_j,\xi\right\rangle \,=\, \xi_j\,,
\qquad\qquad\qquad\qquad j=1,\ldots,r+1\,,\cr
&\xi \,=\, i\diag\!\big(\xi_1,\ldots,\xi_{r+1}\big),\qquad\qquad \xi_1
\,+\, \cdots \,+\, \xi_{r+1}\,=\,0\,.}}
\countdef\Hatomj=118\Hatomj=\pageno
This generating set of weights is slightly redundant, since 
\eqn\SUWT{ \hat\omega_1 \,+\, \cdots \,+\, \hat\omega_{r+1} \,=\,
0\,,}
but these generators also have the virtue of being simply
permuted by the Weyl group of $SU(r+1)$.  As usual, in terms of
$\big\{\hat\omega_1,\ldots,\hat\omega_{r+1}\big\}$, the positive
roots of $SU(r+1)$ are given by the successive differences
${\hat\omega_j - \hat\omega_\ell}$ for all pairs ${j < \ell}$.

Clearly from \PHLJ\ and \SUWTII,
\eqn\PHLJII{ \varrho_j(t) \,=\,
\exp{\!\left[-i\,\langle\hat\omega_j,\xi\rangle\right]}\,.}
So according to the definition in \LRHOII, the eigenspace $L_j$
transforms as the fiber of the homogeneous line bundle
$\FL(-\hat\omega_j)$ over $\CO_\lambda$, and the weights $\alpha_j$
appearing in \SLJA\ are precisely the fundamental weights of $SU(r+1)$,
\eqn\PHAJ{ \SL_j\big|_{\CO_\lambda} \,=\, \FL(-\hat\omega_j)\,,\qquad\qquad
j = 1,\ldots,r+1\,.}
As a special case, for ${G = SU(2)}$ and ${\CO_\lambda \cong \BC\BP^1}$,
the minus sign in \PHAJ\ is consistent with the standard
identification of the tautological line bundle on $\BC\BP^1$ as the
line bundle of degree $-1$.

Since $\SL_j$ restricts fiberwise on $\SN(P;\lambda)$ to the the
homogeneous line bundle $\FL(-\hat\omega_j)$, the Chern root $\Ru_j$ in
\CHRNRTS\ likewise restricts fiberwise to the Chern class of
$\FL(-\hat\omega_j)$.  But as we reviewed in the context of geometric
quantization in Section $4.1$, the Chern class of $\FL(-\hat\omega_j)$ is
represented by the coadjoint form ${-\nu_{{\hat\omega}_j}/2\pi}$.  The Chern
root $\Ru_j$ is therefore represented fiberwise on $\SN(P;\lambda)$ by 
\eqn\CHRNRTSII{ \Ru_j\big|_{\CO_\lambda} \,=\,
c_1\!\left(\FL(-\hat\omega_j)\right) \,=\,
-{\nu_{{\hat\omega}_j}\over{2\pi}}\,,\qquad\qquad j\,=\,1,\ldots,r+1\,.}
In \CHRNRTSII, we divide the curvature of the invariant connection on
$\FL(-\hat\omega_j)$ by $2\pi$ to ensure that $\Ru_j$ has integral
periods.

The formula in \CHRNRTSII\ has an important consequence.  As before,
the weights $\big\{\hat\omega_1,\ldots,\hat\omega_r\big\}$ provide a
basis for the weight lattice $\Gamma_{\rm wt}$ of $SU(r+1)$.  Under
the isomorphism ${\Gamma_{\rm wt} \cong H^2(G/T;\BZ)}$ in \ISOI, the
corresponding forms
$\{\nu_{\hat\omega_1}/2\pi,\ldots,\nu_{\hat\omega_r}/2\pi\}$ then span
$H^2\!\left(G/T;\BZ\right)$.  So via \CHRNRTSII, the fiberwise 
component of any class in ${H^2\!\left(\SN(P;\lambda);\BZ\right)}$ can
be expanded in terms of the Chern roots $\Ru_j$ of the universal
bundle $\SV_p$.

Let us apply this observation to the two-form $\Re_\lambda$ which
is the fiberwise component of the symplectic form $\Omega_\lambda$ on
$\SN(P;\lambda)$.  With ${\lambda = i
\diag(\lambda_1,\cdots,\lambda_{r+1})}$ as before, $\lambda$ can 
be expanded in terms of the weights $\hat\omega_j$ as\foot{Again,
we use the invariant metric given by $-\Tr$ to identify
${\Ft \cong \Ft^*}$.}
\eqn\LAMWT{ \lambda \,=\, \lambda_1 \, \hat\omega_1 \,+\, \cdots \,+\,
\lambda_{r+1} \, \hat\omega_{r+1}\,.}
Being careful about the prefactor in \CHRNRTSII, we thence obtain 
\eqn\CHRNRTSIII{ \Re_\lambda \,=\, - \sum_{j=1}^{r+1} 2\pi\lambda_j \,
\Ru_j\,.}

Equivalently, we regard ${\Ru = i
\diag\!\left(\Ru_1,\ldots,\Ru_{r+1}\right)}$ as a two-form on 
$\SN(P;\lambda)$ taking values in the Cartan subalgebra $\Ft$ of 
$SU(r+1)$, 
\eqn\FUNNYU{ \Ru \,=\, i \diag\!\big(\Ru_1,\cdots,
\Ru_{r+1}\big) \,\in\, H^2\!\big(\SN(P;\lambda);\BZ\big) \otimes
\Ft\,,}
\countdef\Rut=117\Rut=\pageno
such that 
\eqn\FUNNYUII{ \Ru_j \,=\,
\big\langle\hat\omega_j, \Ru\big\rangle\,,\qquad\qquad
j\,=\,1,\ldots,r+1\,.} 
So more invariantly, the description of $\Re_\lambda$ in
\CHRNRTSIII\ becomes 
\eqn\CHRNRTSIII{ \Re_\lambda \,=\, -2\pi \big(\lambda, \Ru\big).}

Together, \OMLAM\ and \CHRNRTSIII\ finally imply that the symplectic form
$\Omega_\lambda$ on $\SN(P;\lambda)$ is given by 
\eqn\OMLAMII{ \Omega_\lambda \,=\, \Rq^*\Omega \,+\,
4\pi^2\!\left(\lambda,\Ru\right).}
In Section $7.3$, we will make essential use of this expression
relating $\Omega_\lambda$ to the symplectic form $\Omega$ on $\SN(P)$
and the Chern roots $\Ru$ of $\SV_p$.

\newsec{General Aspects of Non-Abelian Localization}

In preparation for the computations in Section $7$, we quickly review
in this section some general aspects of non-abelian localization.  Our
goals are to recall the philosophy which underlies computations
via non-abelian localization and to state the basic localization
formula derived in \BeasleyVF.  Since the latter result may be useful
elsewhere, I have decided to present it in somewhat greater generality 
than we will actually require for our study of the Seifert loop
operator.

Very broadly, non-abelian localization provides a general means to
study a symplectic integral of the canonical form 
\eqn\ZE{ Z(\epsilon) \,=\, {1 \over {\Vol(H)}} \,
\left({1 \over {2 \pi \epsilon}}\right)^{\Delta_H/2} \,
\int_X \! \exp{\left[ \Omega - {1 \over {2 \epsilon}}
\left(\mu,\mu\right) \right]}\,,\qquad \Delta_H = \dim H\,.}
Here $X$ is a symplectic manifold with symplectic form
$\Omega$, and $H$ is a Lie group which acts on $X$ in a Hamiltonian
fashion with moment map $\mu$.  Finally, $(\,\cdot\,,\,\cdot\,)$ is an
invariant, positive-definite\foot{In the case of Chern-Simons theory,
the corresponding quadratic form \FRM\ on $\Fh$ has indefinite
signature, due to the hyperbolic summand associated to the two extra
$U(1)$ generators of ${\CH = U(1)_\RR \ltimes \wt\CG_0}$ relative to
the group of gauge transformations $\CG_0$.  As a related fact,
invariance under large gauge transformations requires the Chern-Simons
symplectic integral \PZCSV\ to be oscillatory, instead of
exponentially damped.  These features do not essentially change our
general discussion of localization.} quadratic form on the Lie algebra
$\Fh$ of $H$ and dually on $\Fh^*$ which we use to define the
``action'' $S = \ha (\mu,\mu)$ and the volume $\Vol(H)$ of $H$ that
appear in \ZE.

To apply non-abelian localization to an integral of the form \ZE,
we first observe that $Z(\epsilon)$ can be rewritten as 
\eqn\ZEII{ Z(\epsilon) \,=\, {1 \over {\Vol(H)}} \, \int_{\Fh \times X}
\left[{{d\phi} \over {2\pi}}\right] \exp{\left[\Omega - i \,
\langle\mu,\phi\rangle - {\epsilon \over 2}
(\phi,\phi)\right]}\,.} 
Here $\phi$ is an element of the Lie algebra $\Fh$ of $H$, and
$\left[d\phi\right]$ is the Euclidean measure on $\Fh$ that is
determined by the same invariant form $\left(\,\cdot\,,\,\cdot\,\right)$ 
which we use to define the volume $\Vol(H)$ of $H$.  To avoid
confusion, the measure $[d\phi/2\pi]$ includes a factor of $1/2\pi$
for each real component of $\phi$.  The Gaussian integral over $\phi$
in \ZEII\ then leads immediately to the expression for $Z$ in \ZE.

\countdef\BigX=121\BigX=\pageno
\countdef\BigOmX=122\BigOmX=\pageno
\countdef\BigH=123\BigH=\pageno
\countdef\Gothh=124\Gothh=\pageno
\countdef\Gothhdual=125\Gothhdual=\pageno
\countdef\MomentX=126\MomentX=\pageno
\countdef\EpsilonR=127\EpsilonR=\pageno
\countdef\ActionS=128\ActionS=\pageno
\countdef\BigDelH=133\BigDelH=\pageno

\subsec{BRST Symmetry}

The great advantage of writing $Z$ in the form \ZEII\ is that, once we
introduce $\phi$, the integrand of $Z$ becomes invariant under a BRST
symmetry which leads directly to a localization result for \ZE.
Specifically, as first demonstrated by Witten in \WittenXU, the
canonical symplectic integral in \ZE\ can be evaluated as a sum of
local contributions from the neighborhoods of the critical points of
the function ${S = \ha (\mu,\mu)}$.

To describe the BRST symmetry of \ZEII, we recall that the moment map
satisfies 
\eqn\MOMMAPEQII{d\langle\mu,\phi\rangle \,=\, \iota_{V(\phi)}
\Omega\,,}
where $V(\phi)$ is the vector field on $X$ associated to
the infinitesimal action of $\phi$.  Because of the relation
\MOMMAPEQII, the argument of the exponential in \ZEII\ is immediately
annihilated by the BRST operator $D$ defined by
\eqn\CRTD{ D \,=\, d + i \, \iota_{V(\phi)}\,.}

\countdef\Vphi=129\Vphi=\pageno
\countdef\CartanD=130\CartanD=\pageno

To exhibit the action of $D$ locally, we choose a basis $\phi^a$ for
$\Fh$, and we introduce local coordinates $x^m$ on $X$.  We also
introduce the fermionic notation $\chi^m \equiv dx^m$ for the
corresponding basis of local one-forms on $X$, and we expand the
vector field $V(\phi)$ into components as ${V(\phi) = \phi^a \, V_a^m
\, \partial / \partial x^m}$.  Then the action of $D$ in \CRTD\ is
described in terms of these local coordinates by
\eqn\LCRTD{\eqalign{
&D x^m \,=\, \chi^m\,,\cr
&D \chi^m \,=\, i \, \phi^a \, V^m_a\,,\cr
&D \phi^a \,=\, 0\,.\cr}}
From this local description \LCRTD, we see that the action of $D$
preserves a ghost number, or grading, under which $x$ carries charge
$0$, $\chi$ carries charge $+1$, $\phi$ carries charge $+2$, and $D$
itself carries charge $+1$.

The most important property of a BRST operator is that it squares to
zero.  In this case, either from \CRTD\ or from \LCRTD, we see that
$D$ squares to the Lie derivative along the vector field $V(\phi)$,
\eqn\DSQR{ D^2 = i \, \{d, \iota_{V(\phi)}\} = i \, \lie_{V(\phi)}\,.}
Thus, $D^2 = 0$ exactly when $D$ acts on the subspace of $H$-invariant
functions $\CO(x, \chi, \phi)$ of $x$, $\chi$, and $\phi$.  

For simplicity, we restrict attention to functions $\CO(x,\chi,\phi)$
which are polynomial in $\phi$.  Then an arbitrary function of this
form can be expanded as a sum of terms
\eqn\PWRO{ \CO(x)_{m_1 \ldots m_p \, a_1 \ldots a_q} \; \chi^{m_1} \cdots
\chi^{m_p} \, \phi^{a_1} \cdots \phi^{a_q}\,,}
for some $0 \le p \le \dim X$ and $q \ge 0$.  (The restriction on $p$
arises from the fact that $\chi$ satisfies Fermi statistics, whereas
$\phi$ satisfies Bose statistics.)  

Globally, each term of the form \PWRO\ is specified by a section of
the bundle ${\Omega^p_X \otimes \Sym^q(\Fh^*)}$ of $p$-forms on $X$
taking values in the $q$-th symmetric tensor product of the dual
$\Fh^*$ of the Lie algebra of $H$.  Thus, if we consider the complex
${\big(\Omega_X^* \otimes \Sym^*(\Fh^*)\big){}^H}$ of all $H$-invariant
differential forms on $X$ which take values in the ring of polynomial
functions on $\Fh$, then we see that $D$ defines a cohomology theory
associated to the action of $H$ on $X$.  This cohomology theory is
known as the Cartan model for the $H$-equivariant cohomology
$H^*_H(X)$ of $X$.  See \refs{\AtiyahRB, \GuilleminSII} for nice 
references on equivariant cohomology in the spirit of the present
paper.

\countdef\HequivX=131\HequivX=\pageno

For later discussion we require a few elementary properties of
equivariant cohomology; let us record them now.  First, in the special 
case that $X$ is a point, the $H$-equivariant cohomology ring of $X$
is simply the ring of invariant functions on $\Fh$.  That is,
\eqn\HEQI{ H^*_H(pt) \,=\, {\Sym}^*(\Fh^*)^H\,.}
Similarly, if $X$ is not necessarily a point but $H$ still acts
trivially on $X$, the $H$-equivariant cohomology ring of $X$ is the
tensor product of the ordinary cohomology ring of $X$ and the ring of
invariant functions on $\Fh$, so that 
\eqn\HEQII{ H^*_H(X) \,=\, H^*(X) \otimes H^*_H(pt)\,.}
Both of these statements follow immediately from our definition of the
Cartan model.  Finally, at the opposite extreme that $H$ acts freely
on $X$, the $H$-equivariant cohomology of $X$ is isomorphic to the ordinary
cohomology of the quotient $X/H$,
\eqn\HEQIII{ H^*_H(X) \,=\, H^*(X/H)\,.}
This exceedingly natural statement is less obvious in the Cartan
model, but it is often taken as the starting point in other
topological models for equivariant cohomology, as explained in Chapter~$1$
of \GuilleminSII.

\bigskip\noindent{\it Localization for $Z$}\smallskip

Because the argument of the exponential in \ZEII\ is annihilated by
$D$ and because this argument is manifestly invariant under $H$, the
integrand of the canonical symplectic integral determines an equivariant
cohomology class on $X$.  Furthermore, by the usual arguments, $Z$ in
\ZEII\ is formally unchanged under the addition of any $D$-exact
invariant form to its integrand.  This formal statement can fail if
$X$ is not compact and $Z$ suffers from divergences.  To avoid such
complications, we assume that $X$ is compact --- but see Appendix A of
\BeasleyVF\ for an example of the sort of analysis required when this
assumption is relaxed.  Thus $Z$ depends only on the equivariant
cohomology class of its integrand.

We now explain how this observation leads immediately to a localization
result for $Z$.  We first note that we can add to the argument of
the exponential in \ZEII\ an arbitrary term of the form $s \, D
\Psi$, where $\Psi$ is any $H$-invariant one-form on $X$ and $s$
is a real parameter.  As a result,
\eqn\ZEIII{ Z(\epsilon) = {1 \over {\Vol(H)}} \, \int_{\Fh \times X}
\left[{{d\phi} \over {2\pi}}\right] \exp{\left[\Omega - i \,
\langle\mu,\phi\rangle - {\epsilon \over 2} (\phi,\phi) + s \, D
\Psi\right]}\,.}
This deformation of the integrand of \ZEII\ is $D$-exact and does not
change $Z$.  In particular, $Z$ does not depend on $s$.

By definition, since $\Psi$ is a one-form, $D \Psi$ is given explicitly by
\eqn\DLAM{ D \Psi \,=\, d \Psi + i \, \langle\Psi,
V(\phi)\rangle\,.}
As before, $\langle\,\cdot\,,\,\cdot\,\rangle$ denotes the canonical
dual pairing, so that in components the last term of \DLAM\ is given
by ${i \, \Psi_m V^m_a \phi^a}$.  Thus, apart from a polynomial in $s$ that
arises from expanding the exponential term $\exp{\!(s \, d\Psi)}$ in
\ZEIII, all of the dependence on $s$ in the integrand of $Z$ arises
from the factor ${\exp{\!\left[i \, s \, \langle\Psi,
V(\phi)\rangle\right]}}$ that now appears in \ZEIII.  So if we
consider the limit ${s\rightarrow \infty}$, then the stationary phase
approximation to the integral is valid, and all contributions to $Z$
localize around the critical points of the function $\langle\Psi,
V(\phi)\rangle$.

To determine the critical points of $\langle\Psi,
V(\phi)\rangle$, we expand this function in the basis $\phi^a$ for
$\Fh$ which we introduced previously,
\eqn\DLAMII{ \langle\Psi, V(\phi)\rangle \,=\, \phi^a \,
\langle\Psi, V_a\rangle\,.}
Thus, the critical points of $\langle\Psi, V(\phi)\rangle$ arise
from the simultaneous solutions in $\Fh \times X$ of the equations
\eqn\DLAMIII{\eqalign{
\langle\Psi, V_a\rangle \,&=\, 0\,,\cr
\phi^a \, d \langle\Psi, V_a\rangle \,&=\, 0\,.\cr}}
The first equation in \DLAMIII\ implies that $Z$ necessarily localizes
on points in $\Fh \times X$ for which $\langle\Psi, V_a\rangle$
vanishes.  As for the second equation in \DLAMIII, we see that it is
invariant under an overall scaling of $\phi$ in the vector space $\Fh$.
Consequently, upon integrating over $\phi$ in \ZEIII, we see that the
critical locus of the function $\langle\Psi, V(\phi)\rangle$ in
$\Fh \times X$ projects onto the vanishing locus of $\langle\Psi,
V_a\rangle$ in $X$.  So $Z$ localizes on the subset of $X$ where
${\langle\Psi, V_a\rangle = 0}$.

By making a specific choice of the one-form $\Psi$, we can describe
the localization of $Z$ more precisely.  In particular, we now show
that $Z$ localizes on the set of critical points of the invariant
function ${S = \ha (\mu,\mu)}$ on $X$.

We begin by choosing an almost complex structure ${\bf J}$ on $X$.  That is,
${\bf J}: TX \rightarrow TX$ is a linear map from $TX$ to itself such that
${\bf J}^2 = -1$.  We assume that ${\bf J}$ is compatible with the
symplectic form $\Omega$ in the sense that $\Omega$ is of type $(1,1)$
with respect to ${\bf J}$ and the associated metric
${(\,\cdot\,,\,\cdot\,) \equiv \Omega(\,\cdot\,,\,{\bf J}\, \cdot\,)}$
on $X$ is positive-definite.  Such an almost complex structure always
exists.

\countdef\AcJX=132\AcJX=\pageno

Using ${\bf J}$ and $S$, we now introduce the invariant one-form
\eqn\LCLM{ \Psi \,=\, {\bf J} \, dS \,=\, (\mu, {\bf J} \, d\mu)\,.}
In components, ${\Psi \,=\, dx^m {\bf J}_m^n \partial_n S \,=\, dx^m
\mu^a {\bf J}^n_m \partial_n \mu_a}$.

The integral $Z$ now localizes on the subset of $X$ where
${\langle\Psi, V_a\rangle= 0}$.  Comparing to \LCLM, we see that
this subset certainly includes all critical points of $S$, since by
definition $dS = 0$ at these points.  Conversely, we now show that if
${\langle\Psi, V_a\rangle = 0}$ at some point on $X$, then this point
is a critical point of $S$.  To prove this assertion, we use the
inverse $\Omega^{-1}$ to $\Omega$, which arises by considering the
symplectic form as an isomorphism ${\Omega: TM \rightarrow T^*M}$ with
inverse ${\Omega^{-1}:T^*M \rightarrow TM}$.  In components,
$\Omega^{-1}$ is defined by ${(\Omega^{-1})^{l m} \, \Omega_{m n}
\,=\, \delta^l_n}$.

In terms of $\Omega^{-1}$, the moment map equation \MOMMAPEQII\ is
equivalent to the relation
\eqn\INVMOMEQ{ V \,=\, \Omega^{-1} \, d\mu\,,}
or ${V_a^m \,=\, (\Omega^{-1})^{m n} \, \partial_n \mu_a}$.  Thus,
\eqn\SII{ \Omega^{-1} \, dS \,=\, \left(\mu, \, \Omega^{-1} d\mu\right)
\,=\, \left( \mu, V\right)\,,}
or ${(\Omega^{-1})^{m n} \, \partial_n S \,=\, \mu^a V^m_a}$.

In particular, the condition that ${\langle\Psi, V_a\rangle = 0}$
implies that
\eqn\SIII{ 0 \,=\, \left(\mu,\,\langle\Psi, V\rangle\right) \,=\,
\langle\Psi,\,\Omega^{-1} dS\rangle \,=\, \langle {\bf J}\, dS,\,
\Omega^{-1} dS\rangle\,,}
or more explicitly, ${0 \,=\, \mu^a \Psi_m V^m_a \,=\, \Psi_m \,
(\Omega^{-1})^{m n} \, \partial_n S \,=\, (\Omega^{-1})^{m n} {\bf J}^l_m \,
\partial_l S \,\partial_n S}$.  We recognize the last expression in
\SIII\ as the norm of the one-form $dS$ with respect to the metric on
$X$.  Because this metric is positive-definite, we conclude that the
condition ${\langle\Psi, V_a\rangle = 0}$ implies the vanishing of
$dS$.  Thus the symplectic integral $Z$ localizes exactly on the
critical set of ${S = \ha(\mu,\mu)}$.

\subsec{A Non-Abelian Localization Formula}

In order to pass from the preceding general statement to a precise
localization formula for $Z$, we must introduce a model for the local
symplectic geometry near a critical point of $S$.  In this context,
one of the important technical results of \BeasleyVF\ was to identify
a ``canonical'' symplectic model appropriate to describe the local
geometry in $\CA(P)$ near a critical point of the two-dimensional
Yang-Mills action, or alternatively, to describe the local geometry in
$\bar\CA$ near a critical point of the shift-invariant Chern-Simons
action.  Happily, the canonical symplectic model deserves its name,
insofar as it suffices also to describe the local symplectic geometry
in the loopspace $L\CO_\alpha$ relevant for our computations of the
Seifert loop path integral.

To keep the paper to a reasonable length, we will present the
canonical symplectic model and state the attendant localization
formula for $Z$ more or less as a {\it fait accompli}.  See \S $4.3$
and \S $5.3$ of \BeasleyVF\ for a detailed explanation of how this
model arises from two-dimensional Yang-Mills theory, along with a 
careful derivation of the non-abelian localization formula.  We refer
the reader to the work \ParadanPE\ of Paradan for a related general 
analysis of non-abelian localization.  Specific applications of
non-abelian localization to gauge theory have also been discussed by
Teleman and Woodward in \refs{\TelemanC,\TelemanCW,\WoodwardCT} and by
Blau and Thompson in \refs{\BlauGH,\BlauRS}.

\bigskip\noindent{\it The Canonical Symplectic Model in Brief}\smallskip

Before we can even state the non-abelian localization formula, we need
to establish a model for the local symplectic geometry to which it
applies.  Somewhat abstractly, let us consider a connected component
$\CC$ in the critical locus of the function ${S = \ha(\mu,\mu)}$ on
$X$.  Since $S$ is invariant under the connected Hamiltonian group
$H$, the action of $H$ on $X$ automatically preserves $\CC$.  The
following local model for a symplectic neighborhood of $\CC$ in 
$X$ is certainly not the most general, and in particular it relies on
two simplifying assumptions about the way $H$ acts on $\CC$.  Roughly
speaking, both of these assumptions mean that $H$ acts uniformly on
$\CC$.

\countdef\CurlyC=134\CurlyC=\pageno
\countdef\Hnought=135\Hnought=\pageno

First, we assume that the stabilizer of each point in $\CC$ is
isomorphic to a fixed subgroup ${H_0 \subseteq H}$.  Of course, we
allow the possibility that $H$ acts freely near $\CC$, so that $H_0$
is trivial.  Because all points of $\CC$ have the same stabilizer,
which is certainly a restrictive condition in general, the quotient
${\SM = \CC / H}$ is smooth.  (But see work \JeffreyKY\ of Jeffrey and
collaborators for a very interesting example of non-abelian
localization in a case where $\SM$ is singular.)

Second, we assume that the moment map $\mu$ on $\CC$ takes values in a
single, fixed coadjoint orbit in $\Fh^*$.  Somewhat perversely, we
again dually parametrize the value of the moment map on $\CC$ by an
element $\gamma_0$ in $\Fh$, the Lie algebra of $H$.  

By the assumptions so far, $\gamma_0$ cannot be an arbitrary element
of $\Fh$ but must be a central element in the Lie algebra $\Fh_0$ of
the stabilizer group $H_0$.  In identifying $\gamma_0$ as 
an element of $\Fh_0$, we note that at a point $x_0$ in $\CC$ for
which ${d S = (\mu, d\mu) = 0}$, the moment map $\mu(x_0)$ satisfies
${\big(\mu(x_0), V\big) = \mu(x_0)^a\, V_a^m = 0}$ via \INVMOMEQ\ and
\SII.  Since $\gamma_0$ is the dual of $\mu(x_0)$, the vector field
generated by the action of $\gamma_0$ vanishes at $x_0$, and
$\gamma_0$ therefore lies in $\Fh_0$.  Because $H_0$ fixes $x_0$, the
Hamiltonian condition on the moment map $\mu$ at $x_0$ further implies  
\eqn\INVGAM{ H_0^{} \, \gamma_0 \, H_0^{-1} \,=\, \gamma_0\,.}
Thus $\gamma_0$ is central in $\Fh_0$.  As a small consistency
check, we note that the restriction ${S|_\CC = \ha (\gamma_0,\gamma_0)}$ is
automatically constant on $\CC$ by the second assumption above.  Of
course, if $\CC$ is a component of the critical locus of $S$, then $S$
is necessarily constant on $\CC$.

\countdef\GothHnought=136\GothHnought=\pageno
\countdef\Littlegamnought=138\Littlegamnought=\pageno

We now describe an $H$-invariant neighborhood $N$ of ${\CC \subset X}$
as the total space of an $H$-equivariant bundle with fiber $F$ over
the quotient ${\SM = \CC / H}$,
\eqn\PRESII{\matrix{
&F \,\longrightarrow\, N\cr
&\mskip 70mu\big\downarrow\lower 0.5ex\hbox{$^{\rm pr}$}\cr
&\mskip 55mu \SM\cr}\;,\qquad\qquad \SM \,=\, \CC/H\,.}
In these terms, our canonical symplectic model amounts to a
description of the fiber $F$ in \PRESII\ as a symplectic manifold
equipped with a Hamiltonian action of $H$.  Since we assume $F$ to be
symplectic (and the fibration over $\SM$ to respect that symplectic
structure), the symplectic form on $N$ immediately induces a
corresponding symplectic form on the base $\SM$ of \PRESII.

\countdef\CurlyM=137\CurlyM=\pageno

Let us first recall the topological model for $F$.  Because $H$ acts
on each point in $\CC$ with stabilizer $H_0$, $F$ contains
a copy of the homogeneous space $H/H_0$.  To incorporate the normal
directions to $\CC$ in $F$ as well, we take $F$ to be a
homogeneous vector bundle over $H/H_0$ of the form   
\eqn\COTFBII{ F \,=\, H \times_{H_0} \!\left(\Fh^\perp \oplus
E_1\right)\,,\qquad \Fh^\perp \,\equiv\, \Fh \ominus \Fh_0 \ominus 
E_0\,.}

\countdef\SympF=139\SympF=\pageno
\countdef\Enought=140\Enought=\pageno
\countdef\Gothhperp=141\Gothhperp=\pageno

The description of $F$ in \COTFBII\ requires a bit of explanation.  
Besides the data of $H_0$ and $\gamma_0$, the canonical symplectic
model for $F$ is parametrized by the choice of vector spaces $E_0$ and
$E_1$.  More precisely, $E_0$ is a subspace of $\Fh$ which has trivial
intersection with $\Fh_0$ and is preserved under the adjoint action of
$H_0$, so that infinitesimally ${\big[\Fh_0, E_0\big]\subseteq E_0}$.
Given the invariant metric $(\,\cdot\,,\,\cdot\,)$ on $\Fh$, we then
define ${\Fh^\perp = \Fh \ominus \Fh_0 \ominus E_0}$ as the
orthocomplement to the direct sum ${\Fh_0 \oplus E_0}$ in $\Fh$.
Similarly, $E_1$ is a vector space on which $H_0$ acts in some
representation (not necessarily the adjoint), and we assume that $E_1$
also carries a Euclidean metric invariant under the action of $H_0$.  

\countdef\Eone=142\Eone=\pageno

Throughout the following, we use $h$ to denote an element of $H$,
$\gamma$ to denote an element of  $\Fh^\perp$, and $v$ to denote an
element of $E_1$.  With this notation, we define $F$ in 
\COTFBII\ to be the quotient of the product ${H \times (\Fh^\perp
\oplus E_1)}$ under the action\foot{Our 
convention that $H_0$ acts on $H$ from the right is opposite to the
convention in \BeasleyVF\ but agrees with the convention for coadjoint
orbits in Section $4$.  As a result, some of the expressions in this
section differ by signs from those in \BeasleyVF.} of $H_0$ as 
\eqn\KACTII{ h_{0} \cdot \big(h, \gamma, v \big) \,=\,
\big(h\,h_{0}^{-1},\, h_{0}\,\gamma\,h_{0}^{-1},\, h_{0} \cdot
v\big)\,,\qquad h_{0} \in H_0\,.}

To specify completely the local model, we must also discuss the
symplectic structure and the Hamiltonian $H$-action on $F$.  I will be 
fairly succinct here, since this material is covered in detail in \S
$4.3$ of \BeasleyVF\ and is based upon standard techniques for
constructing symplectic bundles, as explained for instance in
Ch.~$35$--$41$ of \GuilleminS.

In order to place a symplectic structure on $F$, we must make an
additional assumption about the representations $E_0$ and $E_1$ of
$H_0$.  Namely, we assume that ${\gamma_0\in\Fh_0}$ acts
non-degenerately on both $E_0$ and $E_1$, with non-zero eigenvalues.
Because the action of $\gamma_0$ preserves the invariant metrics on
both $E_0$ and $E_1$, the action of $\gamma_0$ on each of these vector
spaces is represented by a real, anti-symmetric matrix.   The positive
and negative eigenspaces of the hermitian operator $-i\gamma_0$
thereby determine complex structures on $E_0$ and $E_1$.  Each of 
these complex structures is automatically preserved by $H_0$, and the
invariant metrics on $E_0$ and $E_1$ are hermitian.

The complex structures on $E_0$ and $E_1$ will be essential when we
present the non-abelian localization formula, so let us describe them
more precisely.  By convention, we take elements of
holomorphic/anti-holomorphic type $(1,0)$ in $E_0$ to lie in the {\sl
positive} eigenspace of ${-i\gamma_0}$, and we take corresponding
elements of type $(1,0)$ in $E_1$ to lie in the {\sl negative}
eigenspace of ${-i\gamma_0}$.  That is, when acting on a standard
holomorphic basis, 
\eqn\HOLCONV{ -i\gamma_0>0 \,\hbox{ on }\, E_0\,,\qquad\qquad
-i\gamma_0<0 \,\hbox{ on }\, E_1\,.}
With this convention for the respective complex structures on $E_0$
and $E_1$, extraneous signs are suppressed in the
non-abelian localization formula.  Later, in Section $6.3$, we
illustrate the convention \HOLCONV\ in several elementary examples.

Without further ado, the symplectic form $\Omega$ on $F$ is given by 
\eqn\OMVW{ \Omega \,=\, -d\big(\gamma + \gamma_0\,,\theta\big) \,-\,
d\big\langle\mu_{E_1}, \theta\big\rangle + \Omega_{E_1}\,,\qquad
\theta = h^{-1} dh\,.}
Here $\Omega_{E_1}$ is the invariant symplectic form on $E_1$
determined by the aforementioned metric and complex structure.  Also,
$\mu_{E_1}$ is the moment map for the Hamiltonian action of $H_0$ on
$E_1$.  As we shall use in a moment, because the action of $H_0$ on
$E_1$ is linear, the moment map $\mu_{E_1}$ is a homogeneous quadratic
function of the vector ${v \in E_1}$.

The two-form $\Omega$ in \OMVW\ is manifestly closed and invariant
under $H_0$.  Further, if $V(\psi)$ denotes the vector field on
${H\times(\Fh^\perp\oplus E_1)}$ generated by an element ${\psi \in
\Fh_0}$, then $\Omega$ vanishes upon contraction with $V(\psi)$.
This statement can be checked directly from \OMVW, once we note that
$V(\psi)$ is given explicitly by 
\eqn\VECTH{ \delta h \,=\, -h \cdot \psi\,,\qquad \delta \gamma \,=\,
\left[\psi,\gamma\right]\,,\qquad \delta v \,=\, \psi \cdot v.}
Under \VECTH, the terms involving $\gamma$ and $\gamma_0$ in
$\Omega$ are separately annihilated when contracted with $V(\psi)$,
and we have carefully arranged the terms involving $\mu_{E_1}$ and
$\Omega_{E_1}$ to cancel when contracted with $V(\psi)$.  Thus
$\Omega$ descends to a closed two-form on $F$.

We are left to check that $\Omega$ is non-degenerate as a symplectic
form on a neighborhood of $H/H_0$ in $F$, where we embed $H/H_0$ as
the base of the homogeneous bundle \COTFBII\ at ${\gamma = v = 0}$.
Restricting $\Omega$ to $H/H_0$, thereby setting ${\gamma = v = 0}$ in
\OMVW, we immediately obtain 
\eqn\OMVWRES{ \Omega\big|_{H/H_0} \,=\, -\big(d\gamma\,,\theta\big)
\,-\, \big(\gamma_0\,,d\theta\big) \,+\, \Omega_{E_1}\,.}
Here we use that $\mu_{E_1}$, being quadratic in $v$, satisfies
${\mu_{E_1} = d\mu_{E_1} = 0}$ at ${v=0}$, so that no terms involving
$\mu_{E_1}$ appear above.  The term involving $d\gamma$ in \OMVWRES\
then provides a non-degenerate, cotangent pairing between the tangent
directions to $H/H_0$ which lie in ${\Fh^\perp = \Fh  \ominus \Fh_0
\ominus E_0}$ and the corresponding directions in the fiber over
$H/H_0$.  The term involving $\gamma_0$ similarly pairs the remaining
tangent directions to $H/H_0$ in $E_0$ via the coadjoint symplectic
form introduced in \COADJ.  Finally, $\Omega_{E_1}$ is non-degenerate on
$E_1$ by definition.  So $\Omega$ in \OMVW\ defines a symplectic form
on a neighborhood of $H/H_0$ in $F$.  

As for the Hamiltonian symmetry, we assume that $H$ acts from the
left on $H/H_0$, so that 
\eqn\GACT{ h' \cdot \big(h,\,\gamma,\, v\big) \,=\, \big(h' \, h,\,
\gamma,\, v\big)\,,\qquad h' \in H\,.}
The corresponding element $\phi$ in $\Fh$ generates the vector field 
\eqn\VPACT{ \delta h \,=\, \phi \cdot h\,,\qquad \delta \gamma
\,=\,0\,,\qquad \delta v\,=\, 0\,.}
Since the one-form $\theta$ appearing in $\Omega$ is left-invariant, the
symplectic form on $F$ is manifestly invariant under $H$.  Finally,
using \OMVW\ and \VPACT, one can easily check that the action 
of $H$ on $F$ is Hamiltonian with moment map $\mu$ given by
\eqn\MOMTWGV{ \big\langle\mu,\phi\big\rangle \,=\, \big(\gamma +
\gamma_0,\, h^{-1} \phi \, h\big) \,+\,
\big\langle\mu_{E_1},\,h^{-1} \phi \, h\big\rangle\,.}
As a small check, we note that the value of $\mu$ in
\MOMTWGV\ on ${H/H_0 \subset F}$ is given by $\gamma_0$, as we
initially assumed, and $H/H_0$ sits on the critical locus of ${S = \ha
(\mu,\mu)}$.

\bigskip\noindent{\it An Equivariant Euler Class From $F$}\smallskip

Having fixed a symplectic model for $F$ in \PRESII, we can now present
the basic localization formula for the canonical symplectic integral
in \ZE.  To start, we define the local contribution to $Z$ from the
component ${\CC \subset X}$ by the following symplectic integral
over $N$,
\eqn\ZEIV{ Z(\epsilon)\big|_\SM = {1 \over {\Vol(H)}}
\, \int_{\Fh \times N} \left[{{d\phi} \over {2\pi}}\right]
\exp{\left[\Omega - i \, \langle\mu,\phi\rangle - {\epsilon \over 2}
(\phi,\phi) + s \, D \Psi\right]}\,.}
So long as $s$ is non-zero and $\Psi$ is given by \LCLM, the integral
\ZEIV\ over the non-compact space $N$ is both convergent and
independent of $s$, so that $Z(\epsilon)|_\SM$ is well-defined.

In principle, to obtain a localization formula, we simply integrate
over ${(\Fh\ominus\Fh_0)\times F}$ in \ZEIV, thereby reducing
$Z(\epsilon)|_\SM$ to an integral over the reduced space ${\Fh_0
\times \SM}$.  In practice, because the integral in \ZEIV\ does not
become Gaussian even in the limit ${s\rightarrow\infty}$, the analysis
of \ZEIV\ is somewhat involved and occupies a significant fraction of
\BeasleyVF.  Rather than repeat that analysis here, I will simply
summarize its result.  

In a nutshell, the non-abelian localization formula of \BeasleyVF\
states that $Z(\epsilon)|_\SM$ in \ZEIV\ is given by the following
integral over ${\Fh_0 \times \SM}$,
\eqn\ZEV{ Z(\epsilon)\big|_\SM = {1 \over {\Vol(H_0)}}
\, \int_{\Fh_0 \times \SM} \left[{{d\psi} \over {2\pi}}\right] \,
{{e_{H_0}\!\left(\SM, E_0\right)}\over{e_{H_0}\!\left(\SM,
E_1\right)}} \; \exp{\!\left[\Omega \,+\, \epsilon \, \Theta \,-\, i \,
(\gamma_0,\psi) \,-\, {\epsilon \over 2}
(\psi,\psi)\right]}\,.}
This formula deserves a number of
comments.\countdef\ZepsSM=143\ZepsSM=\pageno

Let us first explain our notation.  The integration variable $\psi$ in
\ZEV\ is an element in $\Fh_0$, and $\Omega$ and $\Theta$ are classes 
in the ordinary de Rham cohomology ring $H^*(\SM)$ of $\SM$.  As the
notation suggests, $\Omega$ is the symplectic form on $\SM$ inherited
from $X$, and $\Theta$ is a degree-four characteristic class on $\SM$
that we specify more precisely below.  Finally,
${e_{H_0}\!\left(\SM,E_0\right)}$ and
${e_{H_0}\!\left(\SM,E_1\right)}$ are the $H_0$-equivariant Euler
classes of complex vector bundles over $\SM$ associated to $E_0$ and
$E_1$.  Again, we provide a more explicit description of
${e_{H_0}\!\left(\SM,E_0\right)}$ and
${e_{H_0}\!\left(\SM,E_1\right)}$ below.  For now, we simply note  
that these classes are defined as elements in the $H_0$-equivariant 
cohomology ring $H^*_{H_0}(\SM)$ of $\SM$.

In fact, all terms in the integrand of \ZEV\ can be understood as
classes in $H^*_{H_0}(\SM)$.  Because $H_0$ acts trivially on $\SM$,
the $H_0$-equivariant cohomology ring of $\SM$ is given by the tensor
product of the ordinary cohomology ring of $\SM$ with the ring of
invariant functions on $\Fh_0$.  Identifying the ring of invariant
functions on $\Fh_0$ with the $H_0$-equivariant cohomology of a point,
we thus write 
\eqn\ISOHOM{ H_{H_0}^*(\SM) \,=\, H^*(\SM) \otimes H^*_{H_0}(pt)\,.}
Perhaps a bit more concretely, elements in $H^*_{H_0}(\SM)$ can be
written as sums of terms having the form $x \cdot f(\psi)$, where $x$
is an ordinary cohomology class on $\SM$ and $f(\psi)$ is an invariant
function of $\psi$.  Recalling that ${\gamma_0 \in \Fh_0}$ is
invariant under $H_0$, we thus see that the terms appearing in the
argument of the exponential in \ZEV\ can also be considered as classes
in $H^*_{H_0}(\SM)$.

The appearance of $H^*_{H_0}(\SM)$ in the non-abelian localization
formula is no accident.  As we observed in Appendix C of \BeasleyVF,
the pullback ${\rm pr}^*$ from $\SM$ to $N$ in \PRESII\ induces an
isomorphism, at least rationally, between the $H_0$-equivariant
cohomology ring of $\SM$ and the $H$-equivariant cohomology ring of
$N$.  Under this isomorphism, the $H$-equivariant classes on $N$
represented by $[\Omega - i \, \langle\mu,\phi\rangle]$ and $[-\ha
(\phi,\phi)]$ in \ZEIV\ are the pullbacks from corresponding
$H_0$-equivariant classes on $\SM$.

To identify the relevant classes in $H^*_{H_0}(\SM)$, let us quickly
recall how the identification between $H^*_H(N)$ and $H^*_{H_0}(\SM)$
formally arises.  Given the symplectic model \COTFBII\ for $F$ as an
equivariant vector bundle over $H/H_0$, the neighborhood $N$
equivariantly retracts onto a bundle $\bar N$ with fibers $H/H_0$ over
$\SM$,
\eqn\PRESIII{\matrix{
&H/H_0 \,\longrightarrow\, \bar N\cr
&\mskip 105mu\big\downarrow\lower 0.5ex\hbox{$^{\rm pr}$}\cr
&\mskip 90mu \SM\cr}\;.}
Under this retraction, we identify ${H^*_H(N) = H^*_H(\bar N)}$.  

For the moment, let us suppose that $\bar N$ can be constructed as a quotient
of a principal $H$-bundle $P_H$ over $\SM$.  Thus $P_H$ is a 
bundle with fiber $H$ over $\SM$.  Besides the given action of $H$,
we assume that $P_H$ admits a free action of $H_0$ commuting with
the action of $H$, so that ${\bar N = P_H/H_0}$.  On the other hand,
if we take the quotient of $P_H$ by the free action of $H$, we identify 
${\SM = P_H/H}$.  Considering these quotients in successive orders,
we obtain the required isomorphism 
\eqn\ISOMEN{ H^*_H(\bar N) \,=\, H^*_{H \times H_0}(P_H) \,=\,
H^*_{H_0}(\SM)\,.}
Finally, though the requisite $P_H$ in \ISOMEN\ may or may not exist
on the nose, $P_H$ always exists rationally, which suffices for our
application to the Cartan model.

We can now identify the elements in $H^*_{H_0}(\SM)$ which pull back
to the $H$-equivariant classes on $N$ represented by $[\Omega - i \,
\langle\mu,\phi\rangle]$ and $[-\ha (\phi,\phi)]$ in \ZEIV.  Namely,
we have the correspondences 
\eqn\HHZM{\eqalign{
\Omega - i \, \langle\mu,\phi\rangle
&\;\buildrel{{\rm pr}^*}\over\longleftrightarrow\; \Omega 
\,-\, i \, (\gamma_0,\psi)\,,\cr
-\ha(\phi,\phi) &\;\buildrel{{\rm pr}^*}\over\longleftrightarrow\; \Theta
\,-\, \ha (\psi,\psi)\,.\cr}}
We abuse notation slightly in the first line of \HHZM, since on the
left $\Omega$ is the symplectic form on $X$, whereas on the right
$\Omega$ is the symplectic form on $\SM$.  The identification between
these equivariant cohomology classes of degree-two is manifest once we
recall that the value of the moment map $\mu$ on $\SM$ is given
dually by $\gamma_0$.

\countdef\BigTheta=145\BigTheta=\pageno

The identification between the equivariant classes of degree-four in
the second line of \HHZM\ will be similarly transparent as soon as we
define $\Theta$ as a characteristic class in $H^*(\SM)$.
Specifically, we once again consider the fiber bundle $\bar N$ in
\PRESIII.  By assumption, $H$ acts on $\bar N$ with stabilizer $H_0$,
so $\bar N$ is not the total space of a principal $H$-bundle over
$\SM$.  However, if we let $K$ be the maximal subgroup of $H$
commuting with $H_0$, so that ${K \times H_0 \subseteq H}$, then $K$
does act freely on $\bar N$, and we regard $\bar N$ as a principal
$K$-bundle over $\SM$.  We then define $\Theta$ to be the degree-four
characteristic class of $\bar N$ which is determined via the
Chern-Weil homomorphism by the restriction of ${-\ha(\phi,\phi)}$ to
the Lie algebra of $K$.  The correspondence in the second line of
\HHZM\ then follows directly by definition of $\Theta$.

With the identifications in \HHZM, we see how the classes ${[\Omega 
\,-\, i \, (\gamma_0,\psi)]}$ and ${[\Theta \,-\, \ha (\psi,\psi)]}$ in
\ZEV\ appear under the reduction from \ZEIV.  From a physical
perspective, these terms in the localization formula arise by
classical reduction.

Conversely, the ratio of equivariant Euler classes in \ZEV\ arises by
integrating the localization form $\exp{\!(s\,D\Psi)}$ over the
directions ${(\Fh\ominus\Fh_0) \times F}$ in \ZEIV.  As in the
Duistermaat-Heckman formula \refs{\AtiyahRB,\Duistermaat}, where an
equivariant Euler class makes its appearance for much the same reason,
this prefactor can be described physically as a ratio of one-loop
determinants associated to the normal directions to $\CC$ in $X$.

Since we will need to compute the analogues of $e_{H_0}\!\left(\SM,
E_0\right)$ and $e_{H_0}\!\left(\SM, E_1\right)$ for the Seifert
loop operator in Section $7$, let us recall a more concrete
description of the equivariant Euler class.  We let $E$ be any complex
representation of $H_0$ which is fibered over $\SM$ to determine an
$H_0$-equivariant bundle (with $H_0$ acting trivially on $\SM$ as
throughout).  The $H_0$-equivariant  Euler class $e_{H_0}(\SM, E)$
then incorporates both the algebraic data associated to the action of
$H_0$ on $E$ as well as the topological data that describes the
twisting of $E$ over $\SM$.

To encode the algebraic data related to the action of $H_0$ on $E$, we
assume without loss that $H_0$ is abelian.  (Otherwise, we simply
restrict to a maximal torus in $H_0$.)  We then decompose $E$ under the
action of $H_0$ into a sum of one-dimensional complex eigenspaces
\eqn\DECE{ E \,=\, \bigoplus_{j=1}^{\dim E} \, E_{\alpha_j}\,,}
where each $\alpha_j$ is a weight in $\Fh_0^*$ which describes the
action of $H_0$ on the eigenspace $E_{\alpha_j}$.

To encode similarly the topological data associated to the vector bundle
determined by $E$ over $\SM$, we apply the splitting principle in
topology, as explained for instance in Chapter $21$ of \BottT\ and as
illustrated at the conclusion of Section $5.2$.  By this
principle, we can assume that the vector bundle determined by $E$ over
$\SM$ splits equivariantly into a sum of line-bundles associated to
each of the eigenspaces $E_{\alpha_j}$ for the action of $H_0$.  Under
this assumption, we let $e_j = c_1(E_{\alpha_j})$ be the first Chern class
of the corresponding line-bundle.  These virtual Chern roots $e_j$
determine the total Chern class of $E$ as
\eqn\CHRRTS{ c(E) \,=\, \prod_{j=1}^{\dim E} \, (1 + e_j)\,.}
In particular, the ordinary Euler class of $E$ over $\SM$ is given by
\eqn\CHRRTSII{ e(\SM, E) \,=\, \prod_{j=1}^{\dim E} e_j\,.}

The equivariant Euler class $e_{H_0}(\SM, E)$ is now determined in
terms of the weights $\alpha_j$ and the Chern roots $e_j$.  We
recall that $e_{H_0}(\SM, E)$ is defined as an element of
${H^*_{H_0}(\SM, E) = H^*(\SM) \otimes H_{H_0}^*(pt)}$. Thus 
$e_{H_0}(\SM, E)$ will be a function of ${\psi\in \Fh_0}$ with values in
the ordinary cohomology ring of $\SM$.  Very briefly, the necessary
function is given by the product 
\eqn\HEQE{ e_{H_0}(\SM, E) \,=\, \prod_{j=1}^{\dim E} \left(
{{i\,\langle\alpha_j, \psi\rangle} \over {2 \pi}} + e_j\right)\,.} 
\countdef\EquivEuler=144\EquivEuler=\pageno
When $H_0$ acts trivially on $E$ so that all $\alpha_j$ vanish, we see
that this expression for $e_{H_0}(\SM,E)$ reduces to the ordinary
Euler class in \CHRRTSII.  At the opposite extreme, when $\SM$ is only
a point, the Chern roots $e_j$ do not appear in \HEQE\ for dimensional
reasons, and the product over the weights $\alpha_j$ in \HEQE\ reduces
to the determinant of $\left[\psi/2 \pi\right]$ acting on $E$.

\bigskip\noindent{\it Two Special Cases}\smallskip

To conclude our description of the non-abelian localization formula in 
\ZEV, let us mention two particularly simple special cases.

At one extreme, we suppose that $H$ acts freely on a neighborhood of
the vanishing locus ${\CC=\mu^{-1}(0) \subset X}$ of the moment map
$\mu$.  Thus $H_0$ is trivial, and ${\gamma_0 = E_0 = E_1 = 0}$.  As
first shown by Witten in \WittenXU, the non-abelian localization
formula in this case reduces to the following integral over ${\SM =
\mu^{-1}(0)/H}$,
\eqn\ZEVI{ Z(\epsilon)\big|_\SM \,=\, \int_{\SM} \exp{\left[\Omega
\,+\, \epsilon \, \Theta\right]}\,.}
Here $\Theta$ is now the degree-four characteristic class associated
to $\mu^{-1}(0)$, regarded as a principal $H$-bundle over $\SM$, and
determined under the Chern-Weil homomorphism by ${-\ha(\phi,\phi)}$.

At the opposite extreme, we allow the stabilizer ${H_0 \subset H}$ to be
non-trivial, but we assume that $\SM$ is simply a point.  The
non-abelian localization formula for $Z|_\SM$ in \ZEV\ then reduces to an
integral over the Lie algebra $\Fh_0$,
\eqn\ZVWHKIV{ Z(\epsilon)\big|_\SM = {1 \over {\Vol(H_0)}} \, \int_{\Fh_0}
\left[{{d\psi} \over {2\pi}}\right] \, \det\!\left({{\psi} \over {2
\pi}}\Big|_{E_0}\right) \det\!\left({{\psi} \over {2
\pi}}\Big|_{E_1}\right)^{-1} \exp{\left[-i \left(\gamma_0,
\psi\right) - {\epsilon \over 2} \left(\psi, \psi\right)\right]}\,.}
Here we have written the $H_0$-equivariant Euler classes in \ZEV\ more
explicitly as determinants of ${\psi\in\Fh_0}$ acting on the
respective vector spaces $E_0$ and $E_1$.

The expression for $Z|_\SM$ in \ZVWHKIV\ makes clear a necessary caveat in
our general localization formula.  If $E_0$ and $E_1$ are
finite-dimensional representations of $H_0$, the ratio of determinants 
in \ZVWHKIV\ is just a ratio of invariant polynomials on $\Fh_0$.
Thus the integral over $\Fh_0$ in \ZVWHKIV\ is clearly convergent at
large $\psi$, due to the exponential suppression of the integrand when
$\epsilon$ is non-zero.  However, depending upon $E_0$ and $E_1$, the
integrand in \ZVWHKIV\ might also have singularities at points in
$\Fh_0$ for which the determinant of $\psi$ acting on $E_1$ vanishes.
For instance, the determinant of $\psi$ acting on $E_1$ always
vanishes at ${\psi = 0}$, and without a compensating zero from the
determinant of $\psi$ acting on $E_0$, the integral over $\Fh_0$ could
fail to converge near the origin.  So strictly speaking, the integral
over $\Fh_0$ in the non-abelian localization formula \ZEV\ is not
convergent in general.

We discussed this issue at some length in \S $4.3$ and Appendix A of
\BeasleyVF.  To summarize the results of that discussion, the
non-abelian localization formula is valid whenever the integral over
$\Fh_0$ turns out to be convergent.  In all our applications to
Chern-Simons theory, this condition will hold.

\subsec{Elementary Examples}

To provide some additional intuition for the localization formula in
\ZEV, let us discuss a couple of elementary examples in which the
canonical symplectic integral can be evaluated directly and the
result compared to \ZEV.  In the first example, we will illustrate the
role of $E_0$ in the local model for $F$.  In the second example, we
will illustrate the role of $E_1$. 

\bigskip\noindent{\it The Symplectic Volume of a Coadjoint Orbit}\smallskip

In our first example, we consider the canonical symplectic integral
defined for ${X = \CO_\lambda}$, the adjoint orbit of $G$ through
${\lambda\in\Ft}$ introduced in Section $4.1$.  Here, the role of
the Hamiltonian group $H$ acting on $X$ is played by $G$ itself.  We
recall that the coadjoint symplectic form $\nu_\lambda$ on
$\CO_\lambda$ is given by 
\eqn\COADJII{ \nu_\lambda \,=\, -\ha\big(\theta\,,[\lambda\,,
\theta]\big)\,,\qquad \theta \,=\, g^{-1} \, dg\,,}
and the moment map $\mu$ which describes the action of $G$ on
$\CO_\lambda$ is given by 
\eqn\COADJMOMIII{ \langle\mu,\phi\rangle \,=\,
\big(g\,\lambda\,g^{-1}, \phi\big) \,=\,
-\Tr\big[(g\,\lambda\,g^{-1}) \cdot \phi\big]\,,\qquad\qquad
\phi\in\Fg\,.}

Trivially, the norm-square ${(\mu,\mu) = (\lambda,\lambda)}$ is constant on
$\CO_\lambda$.  As a result, the canonical symplectic integral
$Z(\epsilon)$ in \ZE\ is proportional to the symplectic volume
of $\CO_\lambda$, 
\eqn\LOCEXI{\eqalign{
&Z(\epsilon) \,=\, {1 \over {\Vol(G)}} \, \left({1 \over
{2\pi\epsilon}}\right)^{\Delta_G/2}\,
\exp{\!\left[-{{(\lambda,\,\lambda)}\over{2\epsilon}}\right]} \cdot 
\int_{\CO_\lambda} \e{\!\nu_\lambda}\,,\cr
&\Delta_G \,=\, \dim G\,.}}
By convention, we take ${\lambda > 0}$ to lie in the positive Weyl
chamber, and we orient $\CO_\lambda$ so that the symplectic volume is
positive.  Then from the description of $\nu_\lambda$ in \COADJII,
\eqn\SYMPV{ \int_{\CO_\lambda} \e{\!\nu_\lambda} \,=\,
{{\Vol(G)}\over{\Vol(G_\lambda)}} \, \prod_{\langle\beta,\lambda\rangle>0}
\langle\beta,\lambda\rangle\,,\qquad\qquad \beta\in\FR\,.}

\countdef\DeltaG=146\DeltaG=\pageno

Here ${G_\lambda \subseteq G}$ is the stabilizer of $\lambda$ under
the adjoint action, and the product in \SYMPV\ runs over roots $\beta$
of $G$ whose associated rootspaces lie in the holomorphic tangent
space $\Fg^{(1,0)}$ to $\CO_\lambda$ defined in \HOLTN.  As usual,
this product arises after diagonalizing the adjoint action of
$\lambda$ on $\Fg^{(1,0)}$, such that $\lambda$ acts on each rootspace
${\Fe_\beta \subset \Fg^{(1,0)}}$ by ${[\lambda,\Fe_\beta] =
i\,\langle\beta,\lambda\rangle\,\Fe_\beta}$.  Finally, we recall that
the Riemannian volumes $\Vol(G)$ and $\Vol(G_\lambda)$ are both
defined with respect to the invariant form $(\,\cdot\,,\,\cdot\,)$
determined by `$-\Tr$' on the Lie algebra $\Fg$.

Thus according to \LOCEXI\ and \SYMPV, the canonical symplectic
integral in this example is given directly by 
\eqn\LOCEXIA{ Z(\epsilon) \,=\, {1 \over {\Vol(G_\lambda)}} \, \left({1 \over
{2\pi\epsilon}}\right)^{\Delta_G/2} \,
\exp{\!\left[-{{(\lambda,\,\lambda)}\over{2\epsilon}}\right]} \cdot
\prod_{\langle\beta,\lambda\rangle>0} \langle\beta,\lambda\rangle\,.}

We want to compare \LOCEXIA\ to what we obtain when we apply the
non-abelian localization formula.  To describe $\CO_\lambda$ in terms
of the local symplectic model for $F$, we set 
\eqn\MODEXI{ H \,=\, G\,,\qquad H_0 \,=\, G_\lambda\,,\qquad \gamma_0
\,=\, \lambda\,,\qquad E_0 \,=\, \Fg^{(1,0)}\,,\qquad E_1 \,=\,
\{0\}\,.}
With these assignments, $F$ in \COTFBII\ reduces to
$\CO_\lambda$ itself, and the space $\SM$ sitting beneath $F$ in 
the symplectic fibration \PRESII\ is just a point.

Since $\SM$ is only a point, the localization formula in \ZVWHKIV\ 
presents $Z(\epsilon)$ as an integral over the Lie algebra
$\Fg_\lambda$ of $G_\lambda$,
\eqn\LOCEXIB{ Z(\epsilon) \,=\, {1 \over {\Vol(G_\lambda)}} \,
\int_{\Fg_\lambda} \left[{{d\psi} \over {2\pi}}\right] \,
\det\!\left({{\psi} \over {2 \pi}}\Big|_{\Fg^{(1,0)}}\right) \,
\exp{\left[-i \left(\lambda, \psi\right) - {\epsilon \over 2}
\left(\psi, \psi\right)\right]}\,.} 
Just as in \SYMPV, the determinant of ${\psi\in\Fg_\lambda}$ acting
via the adjoint action on $\Fg^{(1,0)}$ is given by a product over
roots $\beta$ of $G$,
\eqn\LOCDETP{\eqalign{
&\det\!\left({{\psi} \over {2 
\pi}}\Big|_{\Fg^{(1,0)}}\right) \,=\,
\left({i\over{2\pi}}\right)^{(\Delta_G - \Delta_{G_\lambda})/2} \cdot  
\prod_{\langle\beta,\lambda\rangle>0}
\langle\beta,\psi\rangle\,,\cr
&\Delta_G \,=\, \dim G\,,\qquad\qquad \Delta_{G_\lambda} \,=\, \dim
G_{\lambda}\,.}}
As a result,
\eqn\LOCEXIC{ Z(\epsilon) \,=\, {1 \over {\Vol(G_\lambda)}} \,
\left({i\over{2\pi}}\right)^{(\Delta_G - \Delta_{G_\lambda})/2} \, 
\int_{\Fg_\lambda} \left[{{d\psi} \over {2\pi}}\right] 
\prod_{\langle\beta,\lambda\rangle>0} \! \langle\beta,\psi\rangle\,
\exp{\left[-i \left(\lambda, \psi\right) - {\epsilon \over 2}
\left(\psi, \psi\right)\right]}\,.}

\countdef\DeltaGlam=147\DeltaGlam=\pageno

To evaluate the integral over $\Fg_\lambda$ in \LOCEXIC, we employ the
standard trick to rewrite $Z(\epsilon)$ as the derivative of a
Gaussian function,
\eqn\LOCEXID{\eqalign{
Z(\epsilon) &= {1 \over {\Vol(G_\lambda)}} 
\left({i\over{2\pi}}\right)^{(\Delta_G - \Delta_{G_\lambda})/2} 
\int_{\Fg_\lambda} \left[{{d\psi} \over {2\pi}}\right] 
\prod_{\langle\beta,\lambda\rangle>0} \! \left(\beta, i\, 
{\partial\over{\partial\lambda}}\right) \, 
\exp{\left[-i \left(\lambda, \psi\right) - {\epsilon \over 2}
\left(\psi, \psi\right)\right]}\,,\cr
&= {1 \over {\Vol(G_\lambda)}} 
\left({{i}\over{2\pi}}\right)^{(\Delta_G -
\Delta_{G_\lambda})/2} \cdot \left[\prod_{\langle\beta,\lambda\rangle>0} \!
\left(\beta,\, i\,{\partial\over{\partial\lambda}}\right)\right]
{\rm I}(\lambda,\epsilon)\,,}}
where ${\rm I}(\lambda,\epsilon)$ is the Gaussian function obtained
from the first line of \LOCEXID\ once the derivatives are pulled
outside the integral,
\eqn\BIGI{ {\rm I}(\lambda,\epsilon) \,=\,
\left({1\over{2\pi\epsilon}}\right)^{\Delta_{G_\lambda}/2} \,
\exp{\!\left[-{{(\lambda,\lambda)}\over{2\epsilon}}\right]}\,.}

When the differential operator ${\prod_{\langle\beta,\lambda\rangle>0} 
\left(\beta,\, i\,{\partial/{\partial\lambda}}\right)}$
acts on ${\rm I}(\lambda,\epsilon)$, we obtain two sorts of terms.
First, each derivative with respect to $\lambda$ can act separately on
the argument of the exponential in ${\rm I}(\lambda,\epsilon)$ to
bring down a factor ${-i\,\langle\beta,\lambda\rangle/\epsilon}$.
Alternatively, successive derivatives in the operator
${\prod_{\langle\beta,\lambda\rangle>0}  \left(\beta,\,
i\,{\partial/{\partial\lambda}}\right)}$ can act on the monomial 
prefactor in $\lambda$ which is generated by preceding derivatives.
So schematically,
\eqn\BIGITWO{  \left[\prod_{\langle\beta,\lambda\rangle>0} \!
\left(\beta,\, i\,{\partial\over{\partial\lambda}}\right)\right]
{\rm I}(\lambda,\epsilon) \,=\,
\left({1\over{i\epsilon}}\right)^{(\Delta_G - \Delta_{G_\lambda})/2}
 \left[\prod_{\langle\beta,\lambda\rangle>0}
\langle\beta,\lambda\rangle \,+\, \CO(\epsilon)\right] \cdot {\rm
I}(\lambda,\epsilon)\,,}
where $\CO(\epsilon)$ indicates polynomial terms of higher order
in $\epsilon$ which arise {\it a priori} in the second way.  In
the case ${G = SU(3)}$, we present the $\CO(\epsilon)$ terms
explicitly below.

Via \BIGITWO, the expression for $Z(\epsilon)$ in \LOCEXID\ becomes 
\eqn\LOCEXIE{ Z(\epsilon) \,=\,  {1 \over {\Vol(G_\lambda)}} \,
\left({1 \over {2\pi\epsilon}}\right)^{\Delta_G/2} \,
\exp{\!\left[-{{(\lambda,\,\lambda)}\over{2\epsilon}}\right]} \cdot \left[
\prod_{\langle\beta,\lambda\rangle>0} \langle\beta,\lambda\rangle
\,+\, \CO(\epsilon)\right].}
Comparing \LOCEXIE\ to \LOCEXIA, we see that the product over $\beta$
in the righthand bracket produces precisely the same expression for
$Z(\epsilon)$ that we found previously by direct integration.  

That said, consistency of \LOCEXIE\ with \LOCEXIA\ also requires
the $\CO(\epsilon)$ terms in \BIGITWO\ to vanish identically.  Rather
than attempt a general proof of this statement, we 
will content ourselves to check it in the simplest 
non-trivial case, for which ${G = SU(3)}$ and ${\lambda = 
i\diag(\lambda_1,\lambda_2,\lambda_3)}$ is regular.  The
$\CO(\epsilon)$ terms in \LOCEXIE\ are then proportional to the 
following cyclic sum over the three positive roots $\beta_1$,
$\beta_2$, and ${\beta_3=\beta_1+\beta_2}$ of $SU(3)$, 
\eqn\SUMSU{ \sum_{(i,j,k) \,\,\hbox{cyclic}}\!\!\! (\beta_i, \beta_j) \, 
\langle\beta_k,\lambda\rangle \,=\, (\beta_1,\beta_2) \,
\langle\beta_3,\lambda\rangle 
\,+\, (\beta_3,\beta_1) \,  
\langle\beta_2,\lambda\rangle \,+\, (\beta_2,\beta_3) \, 
\langle\beta_1,\lambda\rangle.}
As standard, 
\eqn\SUMSUII{ \langle\beta_1,\lambda\rangle \,=\, \lambda_1 -
\lambda_2\,,\qquad \langle\beta_2,\lambda\rangle \,=\, \lambda_2 -
\lambda_3\,,\qquad \langle\beta_3,\lambda\rangle \,=\, \lambda_1 -
\lambda_3.}
Also, the pairwise inner products of the roots in the invariant metric
`$-\Tr(\,\cdot\,)$' on $\Fs\Fu(3)$ are given by 
\eqn\SUMSUIII{ (\beta_1,\beta_2) \,=\, -1\,,\qquad (\beta_2, \beta_3)
\,=\, 1\,,\qquad (\beta_1,\beta_3) \,=\, 1\,.}
The cyclic sum in \SUMSU\ then vanishes trivially according to \SUMSUII\ and
\SUMSUIII, as required by the non-abelian localization formula.\foot{I
have also checked directly that the $\CO(\epsilon)$ terms in \BIGITWO\ vanish 
when $\lambda$ is regular and the group is $G_2$, a computation which
is considerably more involved.}

\bigskip\noindent{\it Non-Abelian Localization on $S^2$}\smallskip

For our second example, we take ${X = S^2}$.  The discussion here
will be a slight elaboration on the example provided by Witten in the
appendix of \WittenXU.  We parametrize $S^2$ globally in
terms of angular coordinates $(\theta,\varphi)$, where
${\theta\in[0,\pi]}$ is the latitude and ${\varphi\in[0,2\pi)}$ is the
longitude.  As standard, we take the symplectic form $\Omega$ to be 
\eqn\LOCEXIIA{ \Omega \,=\, \sin\theta \, d\theta\^d\varphi\,.}

As a special case of the previous example, we considered the
transitive action of $SU(2)$ on $S^2$, intepreted as a coadjoint
orbit of $SU(2)$.  Here we consider a different group action on $S^2$,
which will not be transitive.  So we take ${H=U(1)}$ to act by
rotations on $S^2$, with fixed points at the poles ${\theta=\{0,\pi\}}$. 
This $U(1)$ action is generated by the vector field ${V =
{\partial/{\partial\varphi}}}$, with moment map 
\eqn\LOCEXIIB{ \mu \,=\, \cos\theta \,+\, \mu_0\,,}
where $\mu_0$ is an arbitrary constant.  As hopefully clear, we regard
the moment map as taking values in ${\Fu(1)^* \cong \Fu(1) = \BR}$.  
For convenience, we set ${\mu_0 = 0}$ in the following.

By convention, we orient $S^2$ so that $\Omega$ is
positive, and we normalize the metric on $U(1)$ so that
${\Vol(U(1))=2\pi}$.  The canonical symplectic integral $Z(\epsilon)$
then becomes 
\eqn\LOCEXIIC{\eqalign{
Z(\epsilon) \,&=\, {1\over{\Vol(U(1))}} \, {1\over{\sqrt{2\pi\epsilon}}} \,
\int_{S^2} \exp{\!\left[\Omega \,-\, {1\over{2\epsilon}}
(\cos\theta)^2\right]}\,,\cr
&= {1\over{\sqrt{2\pi\epsilon}}} \, \int_{-1}^{1} dx \,
\exp{\!\left[-{{x^2}\over{2\epsilon}}\right]}\,.}}
In passing to the second line of \LOCEXIIC, we integrate trivially 
over $\varphi$ to obtain a factor of $2\pi$, which cancels the volume of
$U(1)$, and we make the substitution ${x = \cos\theta}$.  Because the
integration variable $x$ ranges only over the finite interval $[-1,1]$,
$Z(\epsilon)$ is given by the Gaussian error-function and has no simple
analytic expression in terms of more elementary functions.

Once again, we want to compare the result for $Z(\epsilon)$ in
\LOCEXIIC\ with what we obtain from the non-abelian localization
formula.

The local contributions to $Z(\epsilon)$ arise from critical points of 
the function
\eqn\LOCEXS{ S = \ha(\mu,\mu)=\ha(\cos\theta)^2\,.} 
These critical points are located along the equator ${\theta = \pi/2}$, where
the moment map vanishes, as well as the poles ${\theta = \{0,\pi\}}$,
where the moment map takes the values $\pm 1$.  So under localization,
$Z(\epsilon)$ can be expressed as a sum of three terms,
\eqn\LOCEXIID{ Z(\epsilon) \,=\, Z(\epsilon)\big|_{\theta=\pi/2} \,+\,
Z(\epsilon)\big|_{\theta=0} \,+\, Z(\epsilon)\big|_{\theta=\pi}\,,}
each of which we now compute.

We first compute the contribution from the equator, where the moment
map vanishes.  In a neighborhood of the equator, $U(1)$ acts freely,
so the special case of the localization formula in \ZEVI\ is
applicable.  Specifically, in terms of the local symplectic fibration
\PRESII, the symplectic fiber $F$ is the cotangent bundle $T^*U(1)$,
and the base $\SM$ is a point.  So trivially,
\eqn\LOCEXZI{ Z(\epsilon)\big|_{\theta=\pi/2} \,=\, \int_{\SM =
\{pt\}} \mskip -10mu \exp{\left[\Omega \,+\, \epsilon \,
\Theta\right]} \,=\, 1.}
Equivalently, as will be useful in comparing to \LOCEXIIC, we rewrite
the constant `$1$' as 
\eqn\LOCEXZII{  Z(\epsilon)\big|_{\theta=\pi/2} \,=\,
{1\over{\sqrt{2\pi\epsilon}}} \, \int_{-\infty}^{+\infty}
dx\,\exp{\!\left[-{{x^2}\over{2\epsilon}}\right]}\,.}

More interesting are the local contributions from the higher critical
points at ${\theta = \{0,\pi\}}$.  Both critical points make an
identical contribution to $Z(\epsilon)$, so we need only compute the
contribution from ${\theta = 0}$.  In this case, the point ${\theta=0}$
is fixed under the $U(1)$ action, and the symplectic fiber ${F = \BC}$
in \PRESII\ is the tangent space to $S^2$ that point.  To describe $F$
in terms of the canonical local model, we set\foot{By convention, the
generator ${\gamma_0 \in \Fu(1)}$ is anti-hermitian and related to
$\mu$ by ${\gamma_0 = i\,\mu}$.} 
\eqn\MODEXII{ H \,=\, H_0 \,=\, U(1)\,,\qquad \gamma_0\,=\,i\,,\qquad
E_0\,=\,\{0\}\,,\qquad E_1\,=\,\BC\,.}

Since $\SM$ is now a point with a non-trivial stabilizer, we apply the
special case of the localization formula in \ZVWHKIV\ to obtain 
\eqn\LOCEXZIII{  Z(\epsilon)\big|_{\theta=0} \,=\, {1 \over
{\Vol(U(1))}} \, \int_\BR \left[{{d\psi} \over {2\pi}}\right]
\,\det\!\left({{\psi} \over {2 \pi}}\Big|_{\BC}\right)^{-1}
\exp{\left[-i\,\psi - {\epsilon \over 2}\,\psi^2\right]}\,.}
Here we normalize ${\psi\in\Fu(1)}$ to act with unit weight on ${E_1 =
\BC}$.  Thus  
\eqn\LOCDETPSI{ \det\!\left({{\psi} \over {2 \pi}}\Big|_{\BC}\right)
\,=\, -{{i\,\psi}\over{2\pi}}\,.}
The minus sign on the right in \LOCDETPSI\ is a consequence of
the convention for complex structures on $E_0$ and $E_1$ in \HOLCONV.
Indeed, this example provides a check of the relative sign for
$\gamma_0$ in \HOLCONV.  

Recalling that ${\Vol(U(1))=2\pi}$ in \LOCEXZIII, we are left to
analyze the integral 
\eqn\LOCEXZIV{ Z(\epsilon)\big|_{\theta=0} \,=\, 
\int_\BR \left[{{d\psi} \over {2\pi}}\right] \, i\,\psi^{-1} 
\exp{\left[-i\,\psi - {\epsilon \over 2}\,\psi^2\right]}\,.}
The existence of the integral in \LOCEXZIV\ is rather delicate, due to
the singularity in the integrand at ${\psi=0}$.  To make sense of
\LOCEXZIV, we follow the discussion surrounding $(4.90)$ in
\BeasleyVF\ and consider instead the derivative of $Z(\epsilon)$  
with respect to $\epsilon$, for which the same non-abelian
localization formula implies 
\eqn\LOCEXDZ{\eqalign{
{{\partial Z(\epsilon)}\over{\partial\epsilon}}\Big|_{\theta=0} \,&=\,
-{i\over 2} \int_\BR \left[{{d\psi} \over {2\pi}}\right] \, \psi \,  
\exp{\!\left[-i\,\psi - {\epsilon \over 2}\,\psi^2\right]},\cr
&=\, -{1\over{2\sqrt{2\pi}}} \, \epsilon^{-3/2} \,
\exp{\!\left[-{1\over{2\epsilon}}\right]}\,.}}
In passing to the second line of \LOCEXDZ, we evaluate the integral
over $\psi$ after completing the square in the argument of the
exponential.

The result in \LOCEXDZ\ suffices to determine
$Z(\epsilon)\big|_{\theta=0}$ as a function of $\epsilon$ up to an
arbitrary integration constant $c_0$,
\eqn\LOCEXZV{\eqalign{
 Z(\epsilon)\big|_{\theta=0} \,&=\, -{1\over{2\sqrt{2\pi}}}
 \int_0^\epsilon dy \, y^{-3/2} \,
 \exp{\!\left[-{1\over{2y}}\right]} \,+\, c_0\,,\cr
&=\, -{1\over{\sqrt{2\pi\epsilon}}} \int_1^{\infty} dx \,
\exp{\!\left[-{{x^2}\over{2\epsilon}}\right]} \,+\, c_0\,,}}
where in the second line of \LOCEXZV, we substitute ${x =
\sqrt{\epsilon/y}}$.  The same argument shows that
$Z(\epsilon)\big|_{\theta=\pi}$ is given by an identical expression,
which we write as 
\eqn\LOCEXZVII{ Z(\epsilon)\big|_{\theta=\pi} \,=\,
-{1\over{\sqrt{2\pi\epsilon}}} \int_{-\infty}^{-1} dx \, 
\exp{\!\left[-{{x^2}\over{2\epsilon}}\right]} \,+\, c_0\,.}
For ${c_0 = 0}$, the sum over the local contributions from the
critical points in \LOCEXZII, \LOCEXZV, and \LOCEXZVII\ then exactly
reproduces the error-function in \LOCEXIIC.

\newsec{Localization for Seifert Loops in Chern-Simons Theory}

We are finally prepared to apply non-abelian localization to the
Seifert loop path integral, which as in Section $4$ takes the
canonical form 
\eqn\PZCSWLVI{ Z\big(\epsilon; C, R\big) \,=\, {1 \over
{\Vol(\CG)}} \, \left({{-i} \over {2 \pi
\epsilon}}\right)^{\Delta_{\CG}/2} \, \int_{\bar\CA_\alpha} 
\exp{\!\left[\Omega_\alpha \,+\, {i \over {2
\epsilon}} (\mu, \mu)\right]}\,.}
By the general properties of the canonical symplectic integral,
$Z(\epsilon; C, R)$ localizes onto the critical points in ${\bar\CA_\alpha
= \bar\CA \times \epsilon L\CO_\alpha}$ of the shift-invariant action 
\eqn\MUSQIV{ S(A, U) \,=\, \RC\RS\big(A\big) \,+\, 2 \epsilon \,
\oint_C \Tr\big(\alpha \cdot g^{-1} d_A g\big) \,-\, \int_M {1 \over
{\kappa \^ d \kappa}} \Tr\Big[\big(\kappa \^ \CF_A\big)^2\Big]\,,}
where 
\eqn\NEWFII{ \CF_A \,=\, F_A \,+\, \epsilon \, \big(g \alpha
g^{-1}\big) \, \delta_C\,.}
Thus our first task in analyzing the path integral in \PZCSWLVI\ is to
characterize the critical points of $S(A, U)$.

Varying $S(A,U)$ in \MUSQIV\ with respect to $A$, we immediately find one
classical equation of motion,
\eqn\EOMA{ F_A \,+\, \epsilon \, \big(g \alpha g^{-1}\big) \,
\delta_C \,-\, \left({{\kappa \^ \CF_A}\over{\kappa \^
d\kappa}}\right) d\kappa \,-\, \kappa \^ d_A\!\!\left({{\kappa \^
\CF_A}\over{\kappa \^ d\kappa}}\right)\,=\, 0\,.}
Varying with respect to $g$, we find the other equation of motion,  
\eqn\EOMU{ \left[\alpha,\,g^{-1} d_A g \,-\, \kappa 
\; g^{-1}\!\left({{\kappa \^ \CF_A}\over{\kappa \^
d\kappa}}\right) g\right] \,=\, 0\,.}
To avoid cluttering the notation, we have suppressed an overall factor
of $\delta_C$ in \EOMU, so the equation of motion derived by varying
$g$ must be understood to hold upon restriction to $C$.

Because we have yet to fix a gauge for the shift symmetry $\CS$, the
equations of motion in \EOMA\ and \EOMU\ are invariant under the
transformation ${\delta A = \sigma\,\kappa}$, with $\sigma$ being 
an arbitrary function on $M$ valued in the Lie algebra $\Fg$ of the
gauge group $G$.  To fix a gauge for $\CS$, we observe that the quantity
$\kappa\^\CF_A$ transforms very simply under the shift of $A$,
\eqn\SHFTFB{ \kappa \^ \CF_A \,\longmapsto\, \kappa \^ \CF_A \,+\,
\sigma \, \kappa\^d\kappa\,.}
Since $\kappa\^d\kappa$ is a non-vanishing three-form on $M$, we can
then unambiguously fix the shift symmetry by the condition  
\eqn\SHFTGG{ \kappa\^\CF_A \,=\, 0\,.}  
In this gauge, the shift-invariant equations of motion in \EOMA\ and
\EOMU\ reduce to 
\eqn\EOMGGIV{ \CF_A \,=\, F_A \,+\, \epsilon \, \big(g \alpha
g^{-1}\big) \cdot \delta_C \,=\, 0\,, \qquad\qquad
\left[\alpha,\,g^{-1} d_A g\right] \,=\, 0\,,}
or in terms of ${U = g \alpha g^{-1}}$,
\eqn\EOMGGV{ F_A \,+\, \epsilon \, U \cdot \delta_C \,=\, 0\,,
\qquad\qquad d_A U \,=\, 0\,.}

We have seen these equations before, precisely as \EOMGGIII\ and
\EOMGGII\ in Section $4.2$.  There we noted that solutions to
\EOMGGV\ correspond to classical configurations for the fields $A$ and
$U$ in the background of a Wilson loop operator in the standard
(non-shift-invariant) formulation of Chern-Simons theory.  Hence the
Seifert loop path integral localizes onto the moduli space
$\SM(C,\alpha)$ of connections which are flat on the complement to $C$
in $M$ and otherwise have delta-function curvature along $C$
determined by the covariantly constant value of $U$.

\subsec{More About the Classical Seifert Loop Operator}

For generic ${k\in\BZ}$ and ${\alpha\in\Gamma_{\rm wt}}$, we recall the
explicit description of $\SM(C,\alpha)$ as a moduli space of
homomorphisms from the fundamental group of the knot complement 
${M^o = M - C}$ to $G$,
\eqn\MCALPHA{\eqalign{
\SM(C,\alpha) \,&=\, \Big\{\varrho^o:\pi_1(M^o)\rightarrow G \;\big|\;
\varrho^o(\Rm) \in \FC_{\alpha/k}\Big\}\Big/G\,,\cr 
\FC_{\alpha/k} \,&=\,
\Cl\!\big[\!\exp{\!(2\pi\alpha/k)}\big]\,,\qquad\qquad \Rm =
\hbox{meridian of }\, C\,.\cr}}
The description of $\SM(C,\alpha)$ in \MCALPHA\ holds for general $M$
and $C$, but for the localization computations to be performed in
Sections $7.2$ and $7.3$, we must specialize to the Seifert case.  
Not surprisingly, when ${C\subset M}$ is Seifert, the classical moduli
space $\SM(C,\alpha)$ acquires additional geometric structure.  This
structure will be essential for the cohomological interpretation of
the Seifert loop operator, and it is what we wish to explore now.

According to \MCALPHA, the form of $\SM(C,\alpha)$ is largely  
determined by the fundamental group $\pi_1(M^o)$, so let us
quickly present $\pi_1(M^o)$ when ${C \subset M}$ is Seifert. 
As in Section $3.2$, we suppose that the Seifert manifold $M$ is
characterized by the Seifert invariants 
\eqn\SFRTII{\Big[h;n;(a_1,b_1), \ldots, (a_N,b_N)\Big]\,,\qquad\qquad
\gcd(a_j,b_j) = 1\,.}
Here $h$ is the genus of the Riemann surface $\Sigma$ which sits at
the base of $M$, $n$ is the degree of the Seifert fibration over
$\Sigma$, and the relatively-prime pairs $(a_j,b_j)$ for
${j=1,\ldots,N}$ specify the topology of $M$ over each orbifold point
of $\Sigma$.

As we briefly discussed in Section $5.1$ of \BeasleyVF, the
fundamental group $\pi_1(M)$ of the closed Seifert manifold $M$ with
Seifert invariants \SFRTII\ is generated by elements 
\eqn\GENPM{\matrix{
\eqalign{
&\Ra_\ell^{}\,,\, \Rb_\ell^{}\,,\cr 
&\Rc_j\,,\cr
&\Rf\,,}&\qquad
\eqalign{
&\ell=1,\ldots,h\,,\cr
&j=1,\ldots,N\,,\cr
&\cr}}}
\countdef\Orbgen=151\Orbgen=\pageno
\countdef\Fibgen=152\Fibgen=\pageno
subject to the following relations,
\eqn\PIM{\eqalign{
&\left[\Ra_\ell^{}, \Rf\right] \,=\, \left[\Rb_\ell^{}, \Rf\right] \,=\,
\left[\Rc_j, \Rf\right] \,=\, 1\,,\cr
&\Rc_j^{a_j} \, \Rf_{}^{b_j}\,=\, 1\,,\cr
&\prod_{\ell=1}^h \, \left[ \Ra_\ell^{} , \Rb_\ell^{} \right] \,
\prod_{j=1}^N \, \Rc_j \,=\, \Rf^n\,,\qquad\qquad \left[ \Ra_\ell^{} ,
\Rb_\ell^{} \right] \,\equiv\,
\Ra_\ell^{} \, \Rb_\ell^{} \, \Ra_\ell^{-1} \, \Rb_\ell^{-1}\,.\cr}}
The generator $\Rf$, which is a central element in $\pi_1(M)$ by the
first line of \PIM, arises geometrically from the $S^1$ fiber
over $\Sigma$.  Similarly, the generators $\Ra_\ell$ and $\Rb_\ell$ arise
from the $2h$ non-contractible cycles present when $\Sigma$ is a
smooth Riemann surface of genus $h$.  Finally, for ${N > 0}$ the
surface $\Sigma$ carries a non-trivial orbifold structure, and the
generators $\Rc_j$ for ${j=1,\ldots,N}$ correspond to small one-cycles
around each of the orbifold points of $\Sigma$.

To present the Seifert loop moduli space $\SM(C,\alpha)$, we actually
want the fundamental group of the complement ${M^o = M - C}$, as
opposed to $M$ itself.  However, when $C$ is a generic Seifert fiber
of $M$, the presentation of $\pi_1(M)$ in \GENPM\ and \PIM\ can be
immediately extended to a presentation of $\pi_1(M^o)$.  We simply
adjoin a single generator $\Rm$ to the set in \GENPM, where,
consistent with the previous notation, $\Rm$ represents the meridian
of $C$.  As for relations,  $\Rm$ commutes with the fiber $\Rf$ (which
is represented geometrically by $C$ itself),
\eqn\COMM{ \left[\Rm\,, \Rf\right] \,=\, 1\,,}
and the last relation in \PIM\ must be modified to include $\Rm$ as   
\eqn\PIMII{ \prod_{\ell=1}^h \, \left[ \Ra_\ell^{} , \Rb_\ell^{} \right] \,
\prod_{j=1}^N \, \Rc_j \,=\, \Rm \, \Rf^n\,.}
To explain \PIMII\ briefly, we assume that $C$ sits as the fiber over
a smooth, non-orbifold point ${p \in \Sigma}$.  Then $\Rm$ is
represented geometrically by a small, suitably-oriented 
one-cycle around $p$ in $\Sigma$.  Hence $\Rm$ appears in 
\PIMII\ in the same way that the generators $\Rc_j$ enter the
corresponding relation in \PIM.

\bigskip\noindent{\it A Two-Dimensional Interpretation for 
$\SM(C,\alpha)$}\smallskip

The preceding presentation for $\pi_1(M^o)$ is clearly reminiscent of
the presentation for $\pi_1(\Sigma^o)$ in Section $5.2$, where
${\Sigma^o = \Sigma - \{p\}}$ is the punctured Riemann surface at the
base of $M^o$.  Indeed, the presentation for $\pi_1(M^o)$ directly
reduces to the presentation for  $\pi_1(\Sigma^o)$ once we eliminate the
generator ${\Rf\in\pi_1(M^o)}$, setting ${\Rf = 1}$ in the relations
\PIM, \COMM, and \PIMII.  Here when $\Sigma$ has orbifold points, we take
$\pi_1(\Sigma^o)$ to be the orbifold fundamental group.  Hence as
groups, $\pi_1(M^o)$ is a central extension of $\pi_1(\Sigma^o)$,
\eqn\ZPIONE{ 1 \longrightarrow \BZ \longrightarrow \pi_1(M^o)
\longrightarrow \pi_1(\Sigma^o) \longrightarrow 1\,,}
where $\Rf$ generates $\BZ$ above.

The relationship between the groups $\pi_1(M^o)$ and $\pi_1(\Sigma^o)$ in
\ZPIONE\ is naturally reflected in a corresponding relationship
between the moduli spaces $\SM(C,\alpha)$ and $\SN(P;\lambda)$, where
we recall from Section $5.2$ that $\SN(P;\lambda)$ describes flat
connections on $\Sigma$ with monodromy at $p$ specified by
${\lambda\in\Ft}$.  In the special case ${\alpha = \lambda = 0}$, that
relationship reduces to a correspondence between the respective moduli
spaces $\SM$ and $\SN(P)$ of (non-singular) flat connections on $M$
and $\Sigma$, as we discussed in \S $5.1$ of \BeasleyVF\ following
much earlier observations by Furuta and Steer \Furuta.  Because the
two-dimensional interpretation of $\SM(C,\alpha)$ fundamentally
underlies our work on non-abelian localization, let us briefly review
the observations in \BeasleyVF\ as applied to $\SM(C,\alpha)$.

To explain the relationship between $\SM(C,\alpha)$ and
$\SN(P;\lambda)$, we consider a fixed homomorphism
${\varrho^o\!:\pi_1(M^o)\rightarrow G}$, where we assume $G$ to be 
simply-connected.  Since $\Rf$ is central in $\pi_1(M^o)$, the image
of $\varrho^o$ must lie in the centralizer ${\CZ_{\varrho^o(\Rf)}
\subseteq G}$ of the element ${\varrho^o(\Rf)\in G}$.  To simplify the
following discussion, we suppose that $\varrho^o(\Rf)$ actually lies
in the center $\CZ(G)$ of $G$, implying ${\CZ_{\varrho^o(\Rf)} = G}$.
This condition is necessary whenever the connection described by
$\varrho^o$ is irreducible, which is one of the two main cases we
consider when we perform computations for the Seifert loop operator.
In the irreducible case, we will take $\Sigma$ to have genus ${h \ge
1}$.  For the other case we consider, $\Sigma$ has genus zero, and we 
do not require $\varrho^o(\Rf)$ to be central nor $\varrho^o$
irreducible.  As often happens, the genus zero case is somewhat
special, and we postpone consideration of it until the end of this 
section.

Clearly if ${\varrho^o(\Rf)=1}$, so that the corresponding flat
connection on $M^o$ has trivial holonomy around the Seifert fiber,
then $\varrho^o$ factors through the extension \ZPIONE\ to determine
a homomorphism ${\varrho^o\!:\pi_1(\Sigma^o)\rightarrow G}$.
As in \LOCACII\ and \LOCA, the parameters $\alpha$ and $\lambda$ which
specify the respective monodromies for connections in $\SM(C,\alpha)$
and $\SN(P;\lambda)$ are related by 
\eqn\LAMALK{ \lambda \,=\, {\alpha\over k}\,.}
Under the identification in \LAMALK, $\varrho^o$ specifies a flat
connection with monodromy on $M$ which is merely the pullback from a
corresponding flat connection with monodromy on $\Sigma$.

Otherwise, if $\varrho^o(\Rf)$ is a non-trivial element in the center 
$\CZ(G)$, then the flat connection (with monodromy) on $M$ has
non-trivial holonomy around the Seifert fiber and cannot be the
pullback of a flat $G$-connection (with monodromy) on $\Sigma$.
However, if we pass from the simply-connected group $G$ to the
quotient ${G_{\rm ad} = G/\CZ(G)}$, \countdef\Gad=150\Gad=\pageno so
that we consider the connection on $M$ as a connection on the trivial
$G_{\rm ad}$-bundle, then the holonomy of this connection around the
Seifert fiber becomes trivial.  As a result, the homomorphism
$\varrho^o$ can now be interpreted as describing a flat connection
with monodromy on $M$ that arises as the pullback from a corresponding
flat connection on a generally non-trivial $G_{\rm ad}$-bundle $P$
over $\Sigma$.  Specializing for convenience to the case that $\Sigma$
is smooth, with no orbifold points, we see by comparing the relation
\PIMII\ in $\pi_1(M^o)$ to the relation \PIMSII\ in
$\pi_1(\Sigma^{oo})$ that the holonomy $\varrho^o(\Rf)$ is related to
the central element ${\zeta\in\CZ(G)}$ which characterizes the
topology of the principal $G_{\rm ad}$-bundle $P$ on $\Sigma$ via   
\eqn\LEUIII{ \zeta \,=\, \varrho^o(\Rf)^n\,.}

For instance, if we consider the case that the gauge group $G$ is
$SU(2)$ and the degree $n$ of $M$ as a principal $U(1)$-bundle over
$\Sigma$ is odd, then flat connections on $M$ whose holonomies satisfy
${\varrho^o(\Rf) = \varrho^o(\Rf)^n = -1}$ correspond bijectively to 
flat $SU(2)$ connections on $\Sigma$ which have monodromy $-1$ around
a specified puncture.  Such flat $SU(2)$ connections can then be
identified with flat connections on the topologically non-trivial
principal $SO(3)$-bundle over $\Sigma$.  On the other hand, if the
degree $n$ is even, then ${\varrho^o(\Rf)^n = 1}$ for both
${\varrho^o(\Rf) = \pm 1}$, so points in both of these components of
the moduli space on $M$ are identified with flat $SU(2)$ connections
on $\Sigma$.

If $\Sigma$ is not smooth but is an orbifold, the prior discussion 
extends immediately when additional monodromies on $\Sigma$ associated
to the elements $\Rf^{\beta_\ell}$ in \PIM\ are similarly taken into
account.

Let us make a concluding parenthetical remark.  In obtaining a
two-dimensional interpretation for the irreducible homomorphism
$\varrho^o$ on $M^o$, we naturally pass from the simply-connected
group $G$ to the quotient ${G_{\rm ad} = G/\CZ(G)}$.  If we correctly
account for this quotient, the corresponding smooth component
$\SM_0(C,\alpha)$ in the full moduli space $\SM(C,\alpha)$ is
identified most literally with the unramified cover
${\wt\SN(P;\lambda) \to \SN(P;\lambda)}$ appearing in \MNONEII,
\eqn\IDMZERO{ \SM_0(C,\alpha) \,\cong\,
\wt\SN(P;\lambda)\,,\qquad\qquad \lambda \,=\, \alpha/k\,,}
as opposed to the moduli space $\SN(P;\lambda)$ itself.  When ${\alpha =
\lambda = 0}$, a similar statement holds for smooth components $\SM_0$
in the moduli space $\SM$ of flat connections on $M$,
\eqn\IDMZEROII{ \SM_0 \,\cong\, \wt\SN(P)\,.}
Nonetheless, as we noted following \MNONEII, the distinction
between $\SN(P;\lambda)$ and the finite cover
$\wt\SN(P;\lambda)$ in \IDMZERO\ will not be essential for our work.

\bigskip\noindent{\it Consequences for the Symplectic Geometry of
$\SM(C,\alpha)$}\smallskip

According to discussion in Section $5.2$, when $P$ and $\lambda$ are
appropriately chosen, the extended moduli space $\SN(P;\lambda)$ is a
smooth manifold which fibers symplectically over the more basic moduli 
space $\SN(P)$ of flat connections on $P$,
\eqn\FBSMOLIII{\matrix{
&2\pi\CO_{-\lambda}\,\longrightarrow\,\SN(P;\lambda)\cr
&\mskip 115mu\Big\downarrow\lower 0.5ex\hbox{$^{\Rq}$}\cr
&\mskip 100mu\SN(P)\cr}\,,\qquad\qquad \langle\vartheta,\lambda\rangle
< 1\,.}
However, as we reviewed just a moment ago, $\SN(P;\lambda)$ also bears a
natural relation to $\SM(C,\alpha)$ when ${C \subset M}$ is Seifert.
Via the identification in \IDMZERO, the fibration in \FBSMOLIII\ implies a
corresponding fibration for suitable components of $\SM(C,\alpha)$
over the moduli space $\SM$ of flat connections on $M$,
\eqn\FBMCA{\matrix{
&\epsilon\CO_{-\alpha}\,\longrightarrow\,\SM_0(C,\alpha)\cr
&\mskip 100mu\Big\downarrow\lower 0.5ex\hbox{$^{\Rq}$}\cr
&\mskip 97mu\SM_0\cr}\,,\qquad\qquad \langle\vartheta,\alpha\rangle <
k\,.}
\countdef\SMnought=148\SMnought=\pageno
\countdef\SMalpha=153\SMalpha=\pageno
\countdef\Romanq=154\Romanq=\pageno

In obtaining \FBMCA, we recall the identification ${\lambda =
\alpha/k}$ and the definition ${\epsilon = 2\pi/k}$.  Also, both
$\SM_0$ and $\SM_0(C,\alpha)$ denote connected components of 
the typically disconnected moduli spaces $\SM$ and $\SM(C,\alpha)$.  
In general, the fibration in \FBMCA\ holds for only some
components of $\SM(C,\alpha)$, since only some components of 
$\SM(C,\alpha)$ are smooth.  For the application in Section $7.3$, we
will be most interested in the case that $\Sigma$ is a smooth Riemann
surface of genus ${h \ge 1}$ (with no orbifold points), and the central
holonomy ${\varrho^o(\Rf)\in\CZ(G)}$ around the Seifert fiber of $M$
is such that the associated component $\SM_0(C,\alpha)$ is smooth.  In
that situation, the fibration in \FBMCA\ does hold.
 
Just as for \FBSMOLIII, the symplectic form \OMEPS\ on
$\bar\CA_\alpha$ induces a symplectic form $\Omega_\alpha$ on
$\SM_0(C,\alpha)$ which is compatible with the fibration in \FBMCA, so
that $\Omega_\alpha$ admits a decomposition 
\eqn\OMALPH{ \Omega_\alpha \,=\, {\Rq}^*\Omega \,-\,
\epsilon\,\Re_\alpha\,,\qquad\qquad \epsilon\,=\,2\pi/k\,.}
Here $\Omega$ is the symplectic form on $\SM_0$ induced from the
corresponding form \BO\ on $\bar\CA$, and $\Re_\alpha$ is a closed
two-form on $\SM_0(C,\alpha)$ which restricts fiberwise to the
coadjoint symplectic form on $\CO_\alpha$.  Of course, the description
of $\Omega_\alpha$ in \OMALPH\ is nothing more than the transcription
of \OMLAM\ to the Seifert setting.

In Section $5.2$, we were careful to consider the dependence of
$\SN(P;\lambda)$ on the parameter ${\lambda\in\Ft}$ and to delineate
the regime of validity for the fiberwise symplectic decomposition in
\OMLAM.  For instance, due to singular gauge transformations taking
the form \SINGP, we saw that the affine Weyl group $\FW_{\rm aff}$ of
$G$ acts as a group of discrete gauge symmetries on $\lambda$.  We
then fixed the residual action of $\FW_{\rm aff}$ on $\lambda$ by
requiring $\lambda$ to lie in the fundamental Weyl alcove ${\RD_+
\subset \Ft}$.  Briefly, we recall that $\RD_+$ is the bounded region
in $\Ft$ specified by the inequalities ${\lambda \ge 0}$ and
${\langle\vartheta,\lambda\rangle \le 1}$, where $\vartheta$ is the
highest root of $G$.  Finally, we observed that the symplectic
fibration in \FBSMOLIII\ is valid away from the wall of $\RD_+$
defined by ${\langle\vartheta,\lambda\rangle =1}$, where the
symplectic form on $\SN(P;\lambda)$ degenerates.

Under the identification ${\lambda = \alpha/k}$, precisely the same
conclusions apply to $\SM_0(C,\alpha)$.  Gauge transformations on $M$
which are singular along $C$, again of the form \SINGP, alter the
local behavior \LOCACII\ of $A$ near $C$ by shifting 
\eqn\DISCA{ \alpha \longmapsto \alpha \,+\, k \, y\,,\qquad\qquad
y\in\Gamma_{\rm rt}\,,}
for $y$ an arbitrary element in the root lattice $\Gamma_{\rm
rt}$ of the simply-connected group $G$.  This discrete gauge symmetry
can be fixed by taking the weight $\alpha$ to lie in ${k \cdot
\RD_+}$, implying that $\alpha$ satisfies the bounds ${\alpha \ge 0}$ and
${(\vartheta,\alpha) \le k}$.\foot{For once, we regard both the root
$\vartheta$ and the weight $\alpha$ as elements of $\Ft^*$.  The
pairing between $\vartheta$ and $\alpha$ is then defined using the
invariant metric $(\,\cdot\,,\cdot\,)$ on $\Ft^*$ induced by
`$-\Tr$'.}  Finally, the symplectic fibration for $\SM_0(C,\alpha)$ in
\FBMCA\ is valid for positive weights $\alpha$ which satisfy the strict
inequality ${(\vartheta,\alpha) < k}$.

In the context of Chern-Simons theory, the discrete gauge symmetry
\DISCA\ is especially interesting, since it is the manifestation of a 
well-known quantum mechanical equivalence among the set of all Wilson
loop operators in the theory.  As well-known, in the classical limit
${k\rightarrow\infty}$, the Wilson loop operators $W_R(C)$ defined for 
a fixed curve $C$ are distinct operators in Chern-Simons theory as
$R$ ranges over the set of irreducible representations of $G$.  Yet if ${k <
\infty}$ is finite, a quantum mechanical basis \ElitzurNR\ for the 
set of Wilson loops wrapping $C$ is given by only the {\sl finite} set of
operators associated to those representations which are integrable in
the affine Lie algebra $\hat\Fg$ at level $k$.  Explicitly, the integrable
representations of $\hat\Fg$ at level $k$ are characterized by the
bound (see for instance \S $14.3.1$ of \DiFrancesco) 
\eqn\INTCOND{ (\vartheta\,,\alpha) \,\le\, k\,,}
where ${\alpha\ge 0}$ is the highest weight of $R$.  In terms of
${\Ft^* \!\cong \Ft}$, the integrability condition in \INTCOND\ simply
says that $\alpha$ lies in the bounded region ${k \cdot \RD_+}$, as we
already observed.

In the literature, one sometimes finds the integrability condition
\INTCOND\ alternatively expressed by means of the strict inequality 
\eqn\INTCONDII{ (\vartheta\,,\alpha + \rho) \,<\, k +
\check{c}_\Fg\,,}
where $\rho$ is the distinguished weight which is half the sum of the
positive roots of $G$, 
\eqn\WEYLRH{ \rho \,=\, \ha \, \sum_{\beta>0}\,\beta\,,\qquad\qquad
\beta\in\FR\,,}
and $\check{c}_\Fg$ is the dual Coxeter number of $G$.  For instance, if 
${G = SU(r+1)}$, then ${\check{c}_\Fg = r+1}$.  The strict form
\INTCONDII\ of the integrability condition on the weight ${\alpha\ge 0}$ is
entirely equivalent to the condition in \INTCOND, due to the standard identity
${(\vartheta,\rho) =\, \check{c}_\Fg - 1}$.

\countdef\CoxG=157\CoxG=\pageno
\countdef\Weylv=155\Weylv=\pageno

For the semi-classical computations in Section $7.3$, we will need to
assume that ${\alpha \ge 0}$ satisfies the strict bound
\eqn\SRCTBND{ (\vartheta,\alpha) \,<\, k\,,}
so that the symplectic fibration in \FBMCA\ is valid.  This bound is
slightly stronger than the integrability condition in \INTCOND, which
is the appropriate bound on $\alpha$ in the quantum theory.  

Nonetheless, the bound \INTCOND\ on $\alpha$ derived
from two-dimensional conformal field theory and the bound \SRCTBND\ on
$\alpha$ derived from symplectic geometry are not unrelated.  In its
strict form \INTCONDII, the integrability condition can be very
naturally interpreted as the quantum analogue of the classical
constraint in \SRCTBND.  As often happens for Chern-Simons theory, in passing
from the classical to the quantum bound, the level $k$ is replaced
by ${k + \check{c}_\Fg}$, corresponding for ${G=SU(2)}$ to the infamous
shift ${k \mapsto k+2}$, and the weight $\alpha$ itself is replaced by
${\alpha + \rho}$.  Eventually in the course of the localization
computations in Sections $7.2$ and $7.3$, we will see how both of
these rather delicate quantum corrections arise.

\bigskip\noindent{\it Analogue for Seifert Homology Spheres}\smallskip

The symplectic fibration \FBMCA\ for $\SM(C,\alpha)$ is most relevant
when $\Sigma$ has genus ${h \ge 1}$, in which case smooth components
$\SM_0$ and $\SM_0(C,\alpha)$ of positive dimension generally exist.
Yet before we proceed to computations, we also wish to consider the
structure of $\SM(C,\alpha)$ in the opposite topological regime, for
which $\Sigma$ has genus zero and $M$ is a Seifert homology sphere.

To start, let us recall what it means for $M$ to be a Seifert homology
sphere.  In general, a compact three-manifold $M$ is a
homology sphere if and only if ${H_1(M) = 0}$.  Furthermore, $H_1(M)$
is always the abelianization of $\pi_1(M)$.  Given the explicit
description of $\pi_1(M)$ in \GENPM\ and \PIM, we can easily
characterize the conditions on the Seifert invariants \SFRTII\ of $M$
such that $H_1(M)$ vanishes.

First, the base $\Sigma$ of the Seifert fibration must have genus zero
if $M$ is to be a homology sphere, since otherwise the homology of
$\Sigma$ would contribute to the homology of $M$.  Specializing to
this case in \GENPM\ and \PIM, the fundamental group $M$ is then
generated by the elements $\Rc_j$ for ${j=1,\ldots,N}$ and $\Rf$, subject to
the relations 
\eqn\PIMII{\eqalign{
\left[\Rc_j, \Rf\right] \,&=\, 1\,,\qquad j=1,\ldots,N,\cr
\Rc_j^{a_j} \, \Rf^{b_j} \,&=\, 1\,,\cr
\prod_{j=1}^N \, \Rc_j \,&=\, \Rf^n\,.\cr}}

The vanishing of $H_1(M)$ implies that the relations in
\PIMII\ must be non-degenerate in the following sense.  To make
the notation more uniform, let us set ${\Rf \equiv \Rc_{N+1}}$.  Then
$H_1(M)$ is the abelian group generated by the $\Rc_j$ for
${j=1,\ldots,N+1}$, subject to $N+1$ relations of the form
${\prod_{j=1}^{N+1} \Rc_j^{{\bf K}_{j,l}} = 1}$, where 
${\bf K}_{j,l}$ is an ${(N+1)\times(N+1)}$ integer-valued matrix derived 
from the last two lines of \PIMII.  An arbitrary element
${\prod_{j=1}^{N+1} \Rc_j^{v_j}}$ in $H_1(M)$ is trivial if and
only if the vector $v_j$ can be written as ${v_j = \sum_l {\bf K}_{j, l}
\, w_l}$ for some other vector $w_l$ of integers.  In terms of ${\bf
K}_{j, l}$, the vanishing of $H_1(M)$ is then equivalent to the condition
${\det {\bf K} = \pm 1}$.  As one can easily verify, the latter
requirement amounts to the arithmetic condition that the Seifert
invariants \SFRTII\ of $M$ satisfy 
\eqn\SFRTHOM{ n + \sum_{j=1}^N {{b_j}\over{a_j}} \,=\, \pm
\prod_{j=1}^N {1\over{a_j}}\,.}
Hence if ${h=0}$ and the arithmetic condition in \SFRTHOM\ holds, $M$
is a Seifert homology sphere.

As in \S $5.2$ of \BeasleyVF, the present discussion will actually
apply to a slightly more general class of three-manifolds than Seifert
homology spheres.  To indicate the broader class of Seifert 
manifolds we consider, we note that the quantity appearing on the
left of \SFRTHOM\ is the Chern class of the line $V$-bundle $\CL$
over $\Sigma$ associated to $M$, 
\eqn\CHRNCL{ c_1(\CL) \,=\, n + \sum_{j=1}^N {{b_j} \over
{a_j}}\,>\,0\,,}
where we assume without loss that $M$ is oriented so that $c_1(\CL)$ is
positive.  If the arithmetic condition on $c_1(\CL)$ in \SFRTHOM\
holds, then the orders $a_j$ of the orbifold points on $\Sigma$ are
necessarily pairwise coprime,
\eqn\COPRIMA{ \gcd(a_j, a_{j'})=1\,,\qquad j\neq j'\,,}
and as explained for instance in \S $1$ of \Furuta, $\CL$ generates
the Picard group of line $V$-bundles over $\Sigma$.  That is, every
line $V$-bundle on $\Sigma$ is an integral multiple of $\CL$.  

Besides the Seifert homology sphere associated to the fundamental
line $V$-bundle $\CL$, we also consider the Seifert manifolds which arise
from any multiple $\CL^d$ with ${d\ge 1}$.  The Seifert manifold $M$
derived from $\CL^d$ is a quotient by the cyclic group $\BZ_d$ of the
integral homology sphere associated to $\CL$, and in this case, ${H_1(M)
= \BZ_d}$.  Thus the integer $d$ can be characterized topologically as
the order of $H_1(M)$,
\eqn\INTD{ d = |H_1(M)|\,.}\countdef\HoneM=159\HoneM=\pageno
Such Seifert manifolds are rational homology spheres, with
${H_1(M;\BR) = 0}$.  As a simple example, if $\CL$ is the smooth
line-bundle of degree one over $\BC\BP^1$ which describes $S^3$, then
the rational homology spheres associated to multiples $\CL^d$ are lens
spaces.

Regardless of whether $M$ is a Seifert homology sphere or a cyclic
quotient thereof, the vanishing of $H_1(M;\BR)$ implies that the trivial
connection is isolated as a flat connection on $M$.  Hence the trivial
connection determines a distinguished point ${\{0\} \in \SM}$, whose
contribution to the Chern-Simons path integral is natural to consider
under localization.

One can ask whether an analogous statement holds for the extended
moduli space $\SM(C,\alpha)$.  That is, if $M$ is a Seifert homology
sphere or a cyclic quotient thereof, does $\SM(C,\alpha)$ contain an
distinguished, isolated point ${\{\varrho_*\} \in \SM(C,\alpha)}$, whose
local contribution to the Seifert loop path integral is equally
natural to consider?  As our notation indicates,
${\varrho_*:\pi_1(M^o)\rightarrow G}$ must provide a generally
non-trivial representation of the knot group $\pi_1(M^o)$, since
$\varrho_*(\Rm)$ lies by assumption in the generally non-trivial
conjugacy class ${\FC_{\alpha/k} = \Cl[\exp{\!(2\pi\alpha/k)}]}$.

To answer this question, we will directly exhibit $\varrho_*$.  We do
so by the time-tested means of following our nose.

Specializing to the case of Seifert homology spheres, we first observe
that the knot group $\pi_1(M^o)$ is generated by elements $\Rc_j$, $\Rf$,
and $\Rm$, subject to the relations 
\eqn\PIMIII{\eqalign{
 \left[\Rm, \Rf\right] \,&=\, \left[\Rc_j, \Rf\right] \,=\, 1\,,\qquad
 j=1,\ldots,N,\cr
\Rc_j^{a_j} \, \Rf^{b_j} \,&=\, 1\,,\cr
\prod_{j=1}^N \, \Rc_j \,&=\, \Rm \, \Rf^n\,.\cr}}
According to \MCALPHA, if $\varrho_*$ determines a point in
$\SM(C,\alpha)$, then ${\varrho_*(\Rm) = \exp{\!(2\pi U_0/k)}}$ for some
${U_0 \in \CO_\alpha}$.  Without loss, we similarly parametrize the
values of $\varrho_*$ on the generators ${\Rc_1,\ldots,\Rc_N}$ and
${\Rf \equiv c_{N+1}}$ in terms of elements ${\delta_j \in \Fg}$, such
that ${\varrho_*(\Rc_j) = \exp{\!(\delta_j)}}$ for ${j=1,\ldots,N+1}$.  

So far, we have made no assumptions about $\varrho_*$.  However, as the
analogue of ${\{0\} \in \SM}$, the homomorphism $\varrho_*$ should be 
suitably `close' to the trivial homomorphism, for which ${\delta_j =
0}$ for all ${j=1,\ldots,N+1}$.  Motivated by this observation, we initially
work to first order in the parameters $(\delta_1,\ldots,\delta_{N+1})$
in our attempt to construct $\varrho_*$.

At first order, the relations among the generators in \PIMIII\ imply
corresponding linear relations among the parameters
$(\delta_1,\ldots,\delta_{N+1})$ which specify $\varrho_*$.
Specifically, if ${\bf K}_{j, l}$ is the ${(N+1)\times(N+1)}$
integer-valued matrix that appeared following \PIMII\ in our
discussion of $H_1(M)$, then ${\sum_j {\bf K}_{j, l}\,\delta_j \,=\,
0}$ for ${l=1,\ldots,N}$, and ${\sum_j {\bf K}_{j, l}\,\delta_j \,=\,
2\pi U_0/k}$ for ${l=N+1}$.  Since $M$ is a Seifert homology sphere or
a cyclic quotient thereof, the matrix ${\bf K}_{j, l}$ is
non-degenerate over $\BR$.  Therefore, if ${U_0 \in \CO_\alpha}$ is
held fixed, the inhomogeneous linear system for the parameters
$(\delta_1,\ldots,\delta_{N+1})$ has a unique solution.  Moreover, by
linearity, each $\delta_j$ for ${j=1,\ldots,N+1}$ is proportional to
$U_0$ as an element of $\Fg$.  Explicitly, in terms of the inverse
matrix $({\bf K}^{-1})_{l, j}$, we have ${\delta_j = ({\bf
K}^{-1})_{N+1, j} \cdot (2\pi U_0/k)}$.  In particular, the elements 
$(\delta_1,\ldots,\delta_{N+1})$ all commute, and we obtain 
an honest homomorphism ${\varrho_*:\pi_1(M^o)\rightarrow G}$, even
though we began by working only to first order in
${(\delta_1,\ldots,\delta_{N+1})}$.

Necessarily $\varrho_*$ is a reducible representation of $\pi_1(M^o)$,
since $\varrho_*$ is invariant under constant gauge transformations on
$M$ which preserve $U_0$.  Such gauge transformations form a subgroup
of $G$ conjugate to the stabilizer $G_\alpha$.  Also, any
infinitesimal deformation of $\varrho_*$ with fixed $U_0$ must lie in
the kernel of ${\bf K}_{j, l}$.  But since ${\bf K}_{j, l}$ is
non-degenerate over $\BR$, the kernel of ${\bf K}_{j, l}$
trivial. Therefore, like the trivial flat connection on $M$, the
reducible connection described by $\varrho_*$ is isolated as a flat
connection on $M^o$ with monodromy $\varrho_*(\Rm)$ fixed by
${U_0\in\CO_\alpha}$.  Since $\varrho_*$ derives from a
maximally-reducible, abelian representation of $\pi_1(M^o)$, we
henceforth set ${\varrho_* \equiv \varrho_{\rm ab}}$ to indicate this
fact.

We are not quite done, though, since the parameter $U_0$ still varies
in the orbit $\CO_\alpha$.  As $U_0$ varies, the homomorphism
$\varrho_{\rm ab}$ also varies, so we actually have a family of
homomorphisms parameterized by $\CO_\alpha$.  But by the same token,
constant gauge transformations on $M$ induce the adjoint action of $G$
on $\CO_\alpha$.  Thus in total, $\{\varrho_{\rm ab}\}$ represents a
distinguished point in $\SM(C,\alpha)$ isomorphic to the quotient
$\CO_\alpha/G$, 
\eqn\VARRHOST{ \{\varrho_{\rm ab}\} \cong \CO_\alpha/G \in \SM(C,\alpha)\,.}
Despite the fact that ${\{\varrho_{\rm ab}\}\cong\CO_\alpha/G}$ is
naively just a point, keeping track of both the orbit $\CO_\alpha$ and
the $G$-action on $\CO_\alpha$ will be extremely important for the
localization computation in Section $7.2$.  To illustrate why such an
equivariant perspective is useful, let us consider the analogue for Seifert
homology spheres of the smooth fibration in \FBMCA.

\countdef\Rhoab=156\Rhoab=\pageno

If ${\alpha=0}$, the orbit $\CO_\alpha$ reduces to a point, and
$\varrho_{\rm ab}$ describes the trivial connection on $M^o$.  So if 
${\alpha/k}$ is small, the reducible connection on $M^o$ described by 
$\varrho_{\rm ab}$ is only a small deformation of the trivial connection.
Via the same reasoning as in our discussion of the symplectic
fibration for $\SN(P;\lambda)$ in Section $5.2$, we then obtain the 
$G$-equivariant fibration 
\eqn\FBMCAHS{\matrix{
&\CO_\alpha/G\,\longrightarrow\,\CO_\alpha/G\cr
&\mskip 110mu\Big\downarrow\lower 0.5ex\hbox{$^\Rq$}\cr
&\mskip 95mu {pt}/G\cr}\,.}
Here ${\{0\} \cong {pt}/G}$ is the point in $\SM$ corresponding to
the trivial connection on $M$.  Since constant gauge transformations
on $M$ fix the trivial connection, we write ${\{0\}\cong {pt}/G}$
to indicate that $\{0\}$ is equivariantly a point equipped with the
trivial action of $G$.

Because the base of \FBMCAHS\ is the equivariant point ${pt/G}$, the 
total space is just the equivariant fiber ${\CO_\alpha/G}$.  So
admittedly, the equivariant fibration in \FBMCAHS\ may appear rather
formal at first glance.  Nonetheless, the analogy between \FBMCA\ and
\FBMCAHS\ will eventually be helpful to explain certain similarities
between localization on the smooth moduli space $\SM_0(C,\alpha)$ and
localization at the maximally-reducible, abelian point
${\{\varrho_{\rm ab}\} \cong \CO_\alpha/G}$.

\bigskip\noindent{\it Example: Torus Knots in $S^3$}\smallskip

Most intentionally, our discussion of the classical Seifert loop
moduli space $\SM(C,\alpha)$ has been couched in general terms.
Oftentimes, though, concrete examples can be even more illuminating.
With this thought in mind, we now turn our attention to the simplest,
and also the most interesting, Seifert loop operators.

The simplest Seifert manifold is $S^3$, so the simplest
Seifert loop operators will wrap curves embedded as knots ${\SK
\subset S^3}$.  Here we meet a basic topological question.
Namely, which knots can be realized as Seifert fibers of $S^3$?
Clearly, for each locally-free $U(1)_\RR$ action on $S^3$, we 
obtain a corresponding knot as the generic orbit.  So to
approach our question in a systematic fashion, let us consider the
possible locally-free $U(1)_\RR$ actions on $S^3$.

We regard $S^3$ as the unit sphere in $\BC^2$, which carries
complex coordinates $(X, Y)$.  For each pair of integers $({\bf
p},{\bf q})$, we obtain a natural $U(1)$ action on $S^3$ under which
$X$ and $Y$ transform with respective charges ${\bf p}$ and ${\bf q}$,
\eqn\SFRTUO{\big(X, Y\big)\,{\buildrel\e{\!i\theta}\over\longmapsto}\,
\big(\e{i {\bf p} \theta}\cdot X,\, \e{i {\bf q} \theta}\cdot Y\big)\,,
\qquad\qquad \gcd({\bf p},{\bf q}) \,=\, 1\,.}
Here $\theta$ is an angular parameter on $U(1)$, and without loss we
assume in \SFRTUO\ that ${\bf p}$ and ${\bf q}$ are relatively-prime.
Otherwise, if ${\bf p}$ and ${\bf q}$ share a common factor 
${f = \gcd({\bf p},{\bf q})}$, then a cyclic subgroup $\BZ_f$ of
$U(1)$ acts trivially on $S^3$ and can be factored out.  

The classification of locally-free $U(1)_\RR$ actions on $S^3$ is now
very easy to state.  So long as both ${\bf p}$ and ${\bf q}$ are
non-zero, the vector field on $S^3$ which generates the $U(1)$ action
in \SFRTUO\ is nowhere vanishing, and the action is locally-free.
Conversely, as for instance in \S $1.5$ of \OrlikPK, every
locally-free $U(1)_\RR$ action on $S^3$ appears as an action of
the form \SFRTUO\ with some ${{\bf p},{\bf q} \neq 0}$.

One interesting corollary of the classification of locally-free $U(1)_\RR$
actions on $S^3$ is that $S^3$ admits many distinct presentations as a
Seifert manifold.  For instance, in the case ${{\bf p} = {\bf
q} = 1}$, the $U(1)_\RR$ action in \SFRTUO\ is actually free, and we
obtain the standard Hopf presentation of $S^3$ as a smooth
$S^1$-bundle over $\BC\BP^1$, on which ${[X:Y]}$ serve as homogeneous
coordinates.  As we mentioned in the Introduction, here the Hopf fiber
is just the unknot.

More generally, dividing $S^3$ by the $U(1)_\RR$ action in \SFRTUO, we
obtain a fibration of $S^3$ over the weighted projective space
$\BW\BC\BP^1_{{\bf p},{\bf q}}$,
\eqn\PQHOPF{\matrix{
&S^1\,\longrightarrow\,S^3\cr
&\mskip 57mu \big\downarrow\cr
&\mskip 72mu \BW\BC\BP^1_{{\bf p},{\bf q}}\cr}\,.}
Topologically, $\BW\BC\BP^1_{{\bf p},{\bf q}}$ is a genus-zero Riemann
surface with two orbifold points, of orders ${\bf p}$ and ${\bf q}$.
The orbifold points sit under the two exceptional orbits in $S^3$
where either $X$ or $Y$ vanishes, and hence a corresponding cyclic
subgroup $\BZ_{\bf q}$ or $\BZ_{\bf p}$ in $U(1)_\RR$ acts trivially.

For each Seifert presentation of $S^3$ in \PQHOPF, let us quickly compute the
associated Seifert invariants $[h; n, (a_1, b_1), (a_2, b_2)]$, in the
notation of \SFRTII.  Because the base ${\Sigma = \BW\BC\BP^1_{{\bf p},{\bf
q}}}$ is a genus-zero Riemann surface with orbifold points of orders
${\bf p}$ and ${\bf q}$, we obtain immediately 
\eqn\AONE{ h=0\,,\qquad\qquad a_1 = {\bf p}\,,\qquad\qquad a_2 = {\bf q}\,.}
Otherwise, the remaining Seifert invariants $(n, b_1, b_2)$ are
determined by the arithmetic condition in \SFRTHOM\ which
characterizes Seifert homology spheres.  Explicitly, the condition in
\SFRTHOM\ becomes 
\eqn\SFRTSTHR{ n \,+\, {{b_1}\over{a_1}} \,+\, {{b_2}\over{a_2}} \,=\,
{1 \over {a_1 a_2}}\,, \qquad\qquad 0 \le b_1 < a_1\,, \qquad 0 \le
b_2 < a_2\,.}
To solve \SFRTSTHR, we note that because ${a_1 = \bf p}$ and ${a_2 =
\bf q}$ are relatively-prime, there exist unique integers ${\bf r}$
and ${\bf s}$ such that 
\eqn\RLPRIME{ {\bf p}{\bf s} \,-\, {\bf q}{\bf r} \,=\,
1\,,\qquad 0 < {\bf r} < {\bf p}\,,\qquad 0 < {\bf s} < {\bf q}\,.}
In terms of ${\bf r}$ and ${\bf s}$, the unique integers $(n, b_1, b_2)$
satisfying \SFRTSTHR\ are then  
\eqn\NBONEBTWO{ n \,=\,-1\,,\qquad\qquad b_1\,=\, {\bf p} - {\bf
r}\,,\qquad\qquad b_2\,=\, {\bf s}\,.}

At this stage, from the explicit description of the $U(1)_\RR$ action
in \SFRTUO, we see immediately that the knots which can be realized
as Seifert orbits in $S^3$ are exactly \Moser\ the torus knots.  As
the name suggests, the torus knot $\SK_{{\bf p},{\bf q}}$ is 
described by a trigonometric embedding of $S^1$ into a two-torus
${S^1\times S^1}$, itself embedded in $S^3$, under which 
\eqn\TORK{ \e{\! i \theta} \longmapsto \left(\e{\!i {\bf p} \theta},
\e{i {\bf q} \theta}\right)\,,\qquad\qquad \gcd({\bf p},{\bf
q})\,=\,1\,.}\countdef\SKpq=158\SKpq=\pageno
Here $\theta$ is an angular parameter along $S^1$, and ${\bf p}$ and
${\bf q}$ are again non-zero, relatively-prime\foot{If ${\bf p}$ and
${\bf q}$ share a common factor ${f>1}$, the map in \TORK\ is a degree
$f$ covering of the embedding determined by the reduced pair 
$({\bf p}/f,\, {\bf q}/f)$.} integers which determine the embedding
${S^1 \hookrightarrow S^1 \times S^1}$.  Our reuse of $\theta$
as an angular parameter on $S^1$ in \TORK\ is no accident.  Comparing
\SFRTUO\ to \TORK, we see that so long as both ${X,Y \neq 0}$,
the torus knot $\SK_{{\bf p},{\bf q}}$ is embedded in $S^3$ as a
$U(1)_\RR$ orbit in the two-torus parametrized by the phases of the
complex coordinates $X$ and $Y$.

For later use, let us quickly recall a few basic facts about the
classification of torus knots.   If either ${\bf p}$ or ${\bf q}$ is
equal to $1$, then $\SK_{{\bf p},{\bf q}}$ is equivalent to the
unknot, as we already observed for the Hopf fiber ${\SK_{{\bf 1},{\bf 1}}}$.
Also, for fixed $({\bf p},{\bf q})$, the knots $\SK_{{\bf p},{\bf q}}$
and $\SK_{{\bf q},{\bf p}}$ are equivalent under isotopy in $S^3$.
By definition, $\SK_{-{\bf p},-{\bf q}}$ agrees with $\SK_{{\bf
p},{\bf q}}$ but carries the opposite orientation.  Finally,
$\SK_{{\bf p},-{\bf q}}$ is the mirror image of $\SK_{{\bf p},{\bf
q}}$.  All torus knots (other than the unknot) are chiral, meaning
that $\SK_{{\bf p},{\bf q}}$ is not isotopic to its mirror.  Indeed,
one way to prove the latter statement is to compute the Jones
polynomial for $\SK_{{\bf p},{\bf q}}$, as we shall do using
localization in the next section.

\topinsert
$$\matrix{
&\hbox{\epsfxsize=1in\epsfbox{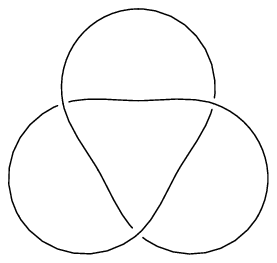}}\qquad
&\hbox{\epsfxsize=1in\epsfbox{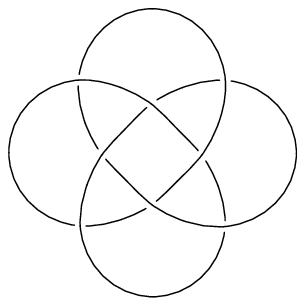}}\qquad
&\hbox{\epsfxsize=1in\epsfbox{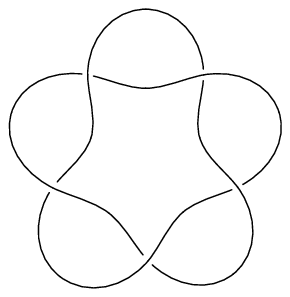}}\cr}$$
\smallskip\centerline{\it Figure 1. The Torus Knots $\SK_{{\bf 3},{\bf
2}}$, $\SK_{{\bf 4},{\bf 3}}$, and $\SK_{{\bf 5},{\bf 2}}$}\smallskip
\endinsert

Up to orientation and reflection, we can thus restrict attention to
the torus knots associated to distinct pairs\foot{We restrict
attention to positive ${\bf q}$ merely to suppress the proliferation of
signs in the computations in Section $7.2$.} ${{\bf p} > {\bf q} >
1}$.  These knots are all non-trivial and topologically distinct.  By
way of illustration, we present in Figure $1$ three torus knots which
can be drawn on the plane with few crossings, including the trefoil,
which is $\SK_{{\bf 3},{\bf 2}}$.  As hopefully clear, the set of Seifert loop
operators is a restricted but nonetheless quite interesting subset of
the possible Wilson loop operators in $S^3$.

\bigskip\noindent{\it Irreducible Points in $\SM(\SK_{{\bf
p},{\bf q}}, \alpha)$}\smallskip

With the torus knot $\SK_{{\bf p},{\bf q}}$ in hand, we finally want to
make some remarks about the structure of the associated Seifert loop
moduli space $\SM(\SK_{{\bf p},{\bf q}},\alpha)$.  At least for gauge
group ${G = SU(2)}$, this moduli space has been extensively studied, 
and our goal here is just to sketch a few of its salient features.  See for
instance \refs{\Klassen,\Boden} for a much more thorough analysis.

To discuss the structure of $\SM(\SK_{{\bf p},{\bf q}},\alpha)$, we
first need to discuss the structure of the knot group ${\pi_1(S^3 -
\SK_{{\bf p},{\bf q}})}$.  Most famously, $\pi_1(S^3 - \SK_{{\bf
p},{\bf q}})$ is the group generated by two elements $\Rx$ and $\Ry$
subject to the single relation ${\Rx^{\bf p} = \Ry^{\bf q}}$.  That is, 
\eqn\TORUSKN{ \pi_1(S^3 - \SK_{{\bf p},{\bf q}}) \,\cong\, \big\langle \Rx,
\Ry \,|\, \Rx^{\bf p} = \Ry^{\bf q}\big\rangle\,.}
When ${{\bf p} = {\bf q} = 1}$ so that $\SK_{{\bf 1},{\bf 1}}$ is the unknot,
${\pi_1(S^3 - \SK_{{\bf 1},{\bf 1}}) \cong \BZ}$ is freely-generated
by ${\Rx=\Ry}$.  Otherwise, for ${{\bf p} > {\bf q} > 1}$, the knot
group is a non-abelian group of infinite order.

Of course, a finitely-generated group can be presented in many
different ways.  Via the general description of $\pi_1(M^o)$ in
\PIMIII, we have already provided a distinct, Seifert set of
generators and relations for the knot group in \TORUSKN.  According to
our computation of the Seifert invariants in \AONE\ and \NBONEBTWO, the
Seifert presentation of $\pi_1(S^3 - \SK_{{\bf p},{\bf q}})$ is given
by generators $\{\Rc_1, \Rc_2, \Rm, \Rf\}$ subject to relations 
\eqn\TORUSKNSF{\eqalign{
 &\left[\Rm, \Rf\right] \,=\, \left[\Rc_j, \Rf\right] \,=\, 1\,,\qquad
 j=1,2,\cr
&\Rc_1^{{\bf p}} \, \Rf^{{\bf p} - {\bf r}} \,=\, 1\,,\cr
&\Rc_2^{{\bf q}} \, \Rf^{{\bf s}} \,=\, 1\,,\cr
&\Rc_1 \Rc_2 \,=\, \Rm \, \Rf^{-1}\,.\cr}}
We recall from \RLPRIME\ that the coprime integers $({\bf r},{\bf s})$
satisfy ${{\bf p}{\bf s} - {\bf q}{\bf r} = 1}$.

The presentation of $\pi_1(S^3 - \SK_{{\bf p},{\bf q}})$ in \TORUSKN\
is certainly simpler than the Seifert presentation in \TORUSKNSF,
but the Seifert presentation has the virtue of making manifest the
meridian $\Rm$ which enters the definition of $\SM(\SK_{{\bf p},{\bf
q}},\alpha)$.  With a little bit of work, one can check that the
Seifert presentation of $\pi_1(S^3 - \SK_{{\bf p},{\bf q}})$ reduces
to the presentation in \TORUSKN\ under the identifications 
\eqn\TORUSKNII{ \Rx \,=\, \Rc_1^{\bf q} \Rc_2^{\bf q} \Rf^{\bf
q}\,,\qquad\qquad \Ry = \Rc_1^{\bf p} \Rc_2^{\bf p} \Rf^{\bf p}\,.}
The meridian $\Rm$ of the knot $\SK_{{\bf p},{\bf q}}$ is then
specified in terms of the generators $\Rx$ and $\Ry$ 
by 
\eqn\TORUSKNM{ \Rm \,=\, \Rc_1 \Rc_2 \Rf \,=\, \Rx^{-{\bf r}} \Ry^{\bf
s}\,.}
For the special case of the unknot $\SK_{{\bf 1},{\bf 1}}$ (with ${{\bf
r}=0}$, ${{\bf s}=1}$), the meridian ${\Rm = \Ry}$ is the generator of
${\pi_1(S^3 - \SK_{{\bf 1},{\bf 1}})\cong\BZ}$.

We are now in an excellent position to analyze the structure of the
classical moduli space $\SM(\SK_{{\bf p},{\bf q}},\alpha)$, which
parametrizes homomorphisms ${\varrho^o\!:\pi_1(S^3 - \SK_{{\bf p},{\bf
q}})\rightarrow G}$ with monodromy ${\varrho^o(\Rm)\in\FC_{\alpha/k}}$.
According to the presentation of ${\pi_1(S^3 - \SK_{{\bf p},{\bf
q}})}$ in \TORUSKN, any homomorphism $\varrho^o$ is uniquely
determined by its values on the generators $\Rx$ and $\Ry$ of the knot
group.  To parametrize $\varrho^o$, let us therefore set 
\eqn\PRMVRHO{ \varrho^o(\Rx) \,=\, \RX\,,\qquad\qquad\varrho^o(\Ry)
\,=\, \RY\,,} 
for elements ${\RX\,, \RY \in G}$.  In terms of $\RX$ and $\RY$, the
moduli space $\SM(\SK_{{\bf p},{\bf q}},\alpha)$ is given very concretely by 
\eqn\MCATORUS{\eqalign{
\SM\big(\SK_{{\bf p},{\bf q}},\alpha\big) \,&=\,
\Big\{\left(\RX,\RY\right) \in G \times G \;\big|\; \RX^{\bf
p}\,=\,\RY^{\bf q}\,,\,\, \RX^{-{\bf r}} \cdot \RY^{\bf s} \in
\FC_{\alpha/k}\Big\}\Big/G\,,\cr  
\FC_{\alpha/k} \,&=\, \Cl\!\big[\!\exp{\!(2\pi\alpha\!/k)}\big]\,.\cr}}

As we have already observed for the general Seifert homology sphere,
$\SM(\SK_{{\bf p},{\bf q}},\alpha)$ always contains a distinguished, 
reducible point ${\{\varrho_{\rm ab}\} \cong \CO_\alpha/G}$, which fibers
equivariantly via \FBMCAHS\ over the trivial connection on $S^3$.  To
exhibit $\varrho_{\rm ab}$ in the notation of \MCATORUS, we set 
\eqn\PRMVRHOII{ \RX \,=\,
\exp{\!\left(\delta\RX\right)}\,,\qquad\qquad \RY \,=\,
\exp{\!\left(\delta\RY\right)}\,,}
where $\delta\RX$ and $\delta\RY$ are elements in the Lie algebra $\Fg$.
According to \MCATORUS, to first order the logarithms $\delta\RX$ and
$\delta\RY$ satisfy the linear relations 
\eqn\LINTORUS{\eqalign{
{\bf p} \, \delta\RX \,-\, {\bf q} \, 
\delta\RY \,&=\, 0\,,\cr
-{\bf r} \, \delta\RX \,+\, {\bf s} \, 
\delta\RY \,&=\, {{2 \pi} \over k} U_0\,,\qquad\qquad U_0 \,\in\,
\CO_\alpha\,.}}
If $U_0$ is fixed, the system in \LINTORUS\ then admits the unique
solution 
\eqn\LINTORUSII{ \delta\RX \,=\, {{2 \pi {\bf q}}\over
k}\,U_0\,,\qquad\qquad \delta\RY \,=\, {{2 \pi {\bf p}\over
k}}\,U_0\,.}
Since $\delta\RX$ and $\delta\RY$ are both proportional to $U_0$, the
group elements $\RX$ and $\RY$ automatically commute, and we obtain a
family of abelian homomorphisms $\varrho_{\rm ab}$ labelled by $U_0$.
These homomorphisms correspond geometrically to the reducible point
${\{\varrho_{\rm ab}\}\cong\CO_\alpha/G}$ in $\SM(\SK_{{\bf p},{\bf q}},\alpha)$.

If ${{\bf p} = {\bf q} = 1}$ so that $\SK_{{\bf 1},{\bf 1}}$ is the
unknot, then $\Rm$ freely-generates ${\pi_1(S^3 - \SK_{{\bf 1},{\bf
1}})\cong\BZ}$, and any homomorphism $\varrho^o$ is entirely
determined by its value on $\Rm$.  Consequently, the reducible point
$\{\varrho_{\rm ab}\}$ is indeed the only point in $\SM(\SK_{{\bf 1},{\bf
1}},\alpha)$.  

When $\SK_{{\bf p},{\bf q}}$ is a non-trivial torus knot, the
structure of the moduli space $\SM(\SK_{{\bf p},{\bf q}},\alpha)$ is
much more interesting.  Beyond the distinguished reducible point,
$\SM(\SK_{{\bf p},{\bf q}},\alpha)$ may contain other 
points associated to {\sl irreducible}\foot{When $G$ 
has higher rank than $SU(2)$, the moduli space $\SM(\SK_{{\bf p},{\bf
q}},\alpha)$ also generally contains points associated to
partially-reducible homomorphisms $\varrho^o$, whose image lies in a
subgroup $G'$ such that ${T \subsetneq G' \subsetneq G}$.  For the purpose
of the following discussion, these partially-reducible points can be
lumped together with the irreducible points.} homomorphisms 
${\varrho^o\!:\pi_1(S^3 - \SK_{{\bf p},{\bf q}})\rightarrow G}$.  In
the case ${G = SU(2)}$, the existence of such points was essentially
noted long ago by Klassen \Klassen, and their presence depends upon
the precise value of the parameter $\alpha/k$.

To exhibit the irreducible points in $\SM(\SK_{{\bf p},{\bf
q}},\alpha)$, we first observe that ${\Rx^{\bf p} = \Ry^{\bf
q}}$ is automatically a central element in the knot group ${\pi_1(S^3 -
\SK_{{\bf p},{\bf q}})}$.  Hence if $\varrho^o$ is irreducible, the
element ${\RX^{\bf p} = \RY^{\bf q}}$ must lie in the center of
$G$.  For simplicitly, we focus on the standard case ${G = SU(2)}$,
for which ${\RX^{\bf p} = \RY^{\bf q} = \pm 1}$.  Thus $\RX$ and $\RY$
are conjugate to elements ${\RX_0, \RY_0 \in SU(2)}$ which take the
diagonal forms   
\eqn\SUPQ{\matrix{
\RX_0 = \pmatrix{\e{\!{{i\pi}\over{\bf p}} u}&0\cr
0&\e{\!-{{i\pi}\over{\bf p}} u}\cr}\,,
& \qquad 0 \le u < {\bf p}\,, &\cr\cr
\RY_0 = \pmatrix{\e{\!{{i\pi}\over{\bf q}} v}&0\cr
0&\e{\!-{{i\pi}\over{\bf q}} v}\cr}\,,
& \qquad 0 \le v < {\bf q}\,, & \qquad v \,\equiv\, u \,\mod
2\,,}}
for some integers $u$ and $v$ which are equal modulo $2$ as above.

We assume without loss that ${\RX = \RX_0}$ takes precisely the diagonal
form in the first line of \SUPQ.  Up to the action of $SU(2)$, $\RY$
can then be parametrized in terms of an angle $\phi$ as 
\eqn\NEWRY{ \RY \,=\, g^{}_\phi \cdot \RY_0 \cdot
g^{-1}_\phi\,,\qquad\qquad g^{}_\phi = 
\pmatrix{\cos\!\left({\phi\over 2}\right)&
i\sin\!\left({\phi\over2}\right)\cr 
i\sin\!\left({\phi\over2}\right)&
\cos\!\left({\phi\over 2}\right)\cr}.} 
The angle $\phi$ is not arbitrary but is fixed by the requirement in
\MCATORUS\ that $\RX^{-{\bf r}} \cdot \RY^{\bf s}$ be conjugate to the
group element $\exp{\!\left(2\pi\alpha\!/k\right)}$, where the weight
$\alpha$ is given explicitly by  
\eqn\WTSUTWO{ \alpha =\, {m\over
2} \pmatrix{i&0\cr0&-i\cr}\,,\qquad\qquad m \in \BZ_{\ge 0}\,.}
Tracing over the entries of the conjugate elements
$\exp{\!\left(2\pi\alpha\!/k\right)}$ and $\RX^{-{\bf r}} \cdot
\RY^{\bf s}$ with the ansatz for $\RY$ in \NEWRY, we obtain a
trigonometric relation which determines $\phi$ in terms of the
integers $(m, k, {\bf p}, {\bf q}, u, v)$,
\eqn\CONDTH{ \cos\!\left({\pi m}\over k\right) \,=\, \cos\!\left({\pi
u {\bf r}}\over {\bf p}\right) \cos\!\left({\pi v {\bf 
s}}\over {\bf q}\right) \,+\, \sin\!\left({\pi u {\bf r}}\over {\bf
p}\right) \sin\!\left({\pi v {\bf s}}\over {\bf q}\right)
\cos(\phi)\,.}
So long as $\phi$ is not equal to $0$ or $\pi$, any solution of
\CONDTH\ describes an irreducible homomorphism ${\varrho^o\!:\pi_1(S^3
- \SK_{{\bf p},{\bf q}})\rightarrow SU(2)}$ with
${\varrho^o(\Rm)\in\FC_{\alpha/k}}$.  Geometrically, $\{\varrho^o\}$
corresponds to an isolated, smooth point in $\SM(\SK_{{\bf p},{\bf
q}},\alpha)$, distinct from the reducible point ${\{\varrho_{\rm ab}\}
\cong \CO_\alpha/G}$.

As one example, for the trefoil knot $\SK_{{\bf 3},{\bf 2}}$ (with
${{\bf r} = {\bf s} = 1}$), we take ${u = v = 1}$.  Then \CONDTH\
reduces to the condition 
\eqn\TRFLCOND{ \cos\!\left({\pi m}\over k\right) \,=\,
\sin\!\left({\pi \over 3}\right) \cos(\phi)\,,}
which can be satisfied whenever $m/k$ lies in the interval
$\left[{1\over 6}, {5\over 6}\right]$.  

This example illustrates an important, well-known general phenomenon.
Namely, the irreducible points in $\SM\big(\SK_{{\bf p},{\bf q}}, \alpha\big)$
exist only when ${\alpha / k > 0}$ is a finite, non-zero distance from
the origin.  In contrast, the maximally-reducible, abelian point
$\{\varrho_{\rm ab}\}$ exists even when $\alpha/k$ is arbitrarily
small, as in the classical limit ${k\rightarrow\infty}$ with $\alpha$
held fixed.\foot{More broadly, both statements are true if $\SK$ is
any knot in $S^3$, not necessarily a torus knot.}

To summarize the discussion so far, the moduli space 
$\SM(\SK_{{\bf p},{\bf q}},\alpha)$ always contains the distinguished,
reducible point ${\{\varrho_{\rm ab}\} \cong \CO_\alpha/G}$.  Depending on
the ratio $\alpha/k$ as well as the pair $({\bf p},{\bf q})$, the
moduli space $\SM(\SK_{{\bf p},{\bf q}},\alpha)$ may contain
additional points which arise from irreducible, non-abelian 
representations of the knot group $\pi_1(S^3 - \SK_{{\bf p},{\bf
q}})$.  

Under localization, the Seifert loop path integral for the torus knot
$\SK_{{\bf p}, {\bf q}}$ is generally evaluated as a sum over
contributions from the connected components of $\SM\big(\SK_{{\bf p},{\bf
q}},\alpha\big)$.  Although evaluating the individual contribution
from each component of $\SM\big(\SK_{{\bf p},{\bf q}}\big)$ is relatively
straightforward (such computations are the essential content of
Sections $7.2$ and $7.3$), summing the results over all components
of $\SM\big(\SK_{{\bf p},{\bf q}}\big)$ would appear to be a 
much harder task, requiring a detailed analysis of $\SM\big(\SK_{{\bf
p},{\bf q}}\big)$ for all values of ${\bf p}$, ${\bf q}$, and
${\alpha/k}$.

In fact, the situation turns out to be a good deal simpler than at
first glance, and we are able to evaluate the Seifert loop path
integral for the torus knot $\SK_{{\bf p},{\bf q}}$ entirely by
localization at the distinguished reducible point ${\{\varrho_{\rm
ab}\} \cong \CO_\alpha/G}$.\foot{I thank E.~Witten for remarks which
substantially clarified the observations here.}  As
follows from the exact expressions obtained via conformal field
theory, the Seifert loop path integral $Z\big(\epsilon;\SK_{{\bf p},{\bf
q}},R\big)$ is an analytic function of the coupling ${\epsilon =
2\pi/k}$.  Hence the behavior of $Z\big(\epsilon;\SK_{{\bf p},{\bf
q}},R\big)$ for all $\epsilon$ is determined by the behavior of
$Z\big(\epsilon;\SK_{{\bf p},{\bf q}},R\big)$ for $\epsilon$ in any
neighborhood of zero.  But in the limit ${\epsilon \to 0}$ 
with $\alpha$ held fixed, so that $\alpha/k$ is arbitrarily small, the
only point in the moduli space $\SM\big(\SK_{{\bf p},{\bf q}}\big)$ is the
reducible point ${\{\varrho_{\rm ab}\}}$.  Hence by analyticity,
$Z\big(\epsilon;\SK_{{\bf p},{\bf q}},R\big)$ is completely determined
under localization by the contribution from the single point ${\{\varrho_{\rm
ab}\} \cong \CO_\alpha/G}$.  

We now turn to evaluating that contribution.

\subsec{Localization at the Trivial Connection on a Seifert Homology 
Sphere}

In this section, we perform our first localization computation for the
Seifert loop operator.  Throughout, we assume that $M$ is a Seifert
homology sphere, or a cyclic $\BZ_d$ quotient thereof.  Our goal is to
compute the local contribution to the Seifert loop path integral from
the reducible point ${\{\varrho_{\rm ab}\} \cong \CO_\alpha/G}$ in the
Seifert loop moduli space $\SM(C,\alpha)$.

From the perspective of \FBMCAHS, the point $\{\varrho_{\rm ab}\}$ can be
considered to fiber equivariantly over the trivial connection on $M$.
Essentially for this reason, the localization computation at
$\{\varrho_{\rm ab}\}$ turns out to be a very natural extension of the
computation in Section $5.2$ of \BeasleyVF, where we evaluated the
contribution from the trivial connection to the Chern-Simons partition
function.  Among the highlights of the present work, we will extract
the renowned Weyl character formula from the Seifert loop path
integral.  Moreover, since we effectively treat the Seifert loop
operator as a disorder operator, this computation provides a nice
example of how, at least in principle, the expectation value of a
disorder operator can be evaluated beyond the leading classical
approximation.

\bigskip\noindent{\it The Results of Lawrence and Rozansky
Revisited}\smallskip

To prepare for our localization computation of the Seifert loop path
integral, let us first discuss what results to expect based upon the
exact expression for $Z(\epsilon;C,R)$ as computed from conformal
field theory.  Once again, we rely on the work of Lawrence and
Rozansky in \LawrenceRZ, where the authors provide an exceedingly
simple formula for $Z(\epsilon;C,R)$ in the case that $C$ is the
generic fiber of a Seifert homology sphere (or a cyclic $\BZ_d$
quotient thereof) and the gauge group $G$ is $SU(2)$.

To express $Z(\epsilon;C,R)$ in a manner which makes the
semi-classical interpretation of the Seifert loop operator 
manifest, we find it useful to introduce the quantities 
\eqn\QRZE{\eqalign{
&\epsilon_r \,=\, {{2 \pi} \over {k + 2}}\,,\cr
&\RP \,=\, \prod_{j=1}^N \, a_j \quad\hbox{if } N \ge 1\,,\qquad \RP
= 1 \quad\hbox{otherwise}\,,\cr
&\theta_0 \,=\, 3 - {d \over \RP} + 12 \, \sum_{j=1}^N \, s(b_j,
a_j)\,.\cr}}\countdef\EpsR=160\EpsR=\pageno\countdef\RomanP=161\RomanP=\pageno 
Here $\epsilon_r$ is the renormalized coupling incorporating the
famous shift ${k \to k + 2}$ in the Chern-Simons level in the case
$G=SU(2)$, and $s(b,a)$ is the Dedekind sum,
\eqn\DEDEK{ s(b, a) \,=\, {1 \over {4 a}} \, \sum_{l =
1}^{a - 1} \, \cot\!\left({{\pi l} \over a}\right)
\cot\!\left({{\pi l b} \over
a}\right)\,.}\countdef\Dedekind=162\Dedekind=\pageno
For sake of brevity, we also introduce the analytic functions
\eqn\FFSS{\eqalign{
F(z) \,&=\, \left(2 \sinh{\!\left({z \over 2}\right)}\right)^{2 - N}
\cdot \prod_{j=1}^N\, \left(2 \sinh{\!\left({z \over {2
a_j}}\right)}\right)\,,\cr
G^{(l)}(z) \,&=\, {i \over {4 \epsilon_r}} \left({d \over \RP}\right) z^2
- {{2 \pi \, l} \over {\epsilon_r}} \, z\,.\cr}}
Last but not least, we introduce the character $\ch_\CMj(z)$ for the
irreducible representation $\CMj$ of $SU(2)$ with dimension $j$,
\eqn\CHNII{ \ch_\CMj(z) \,=\, {\sinh(j \, z) \over \sinh(z)} \,=\, \e{(j-1)
z} \,+\, \e{(j-3) z} \,+\, \cdots \,+\, \e{-(j-3) z} \,+\, \e{-(j-1)
z}\,.}\countdef\SUcharj=163\SUcharj=\pageno\countdef\SUrep=164\SUrep=\pageno

According to the results of \LawrenceRZ, the Seifert loop path
integral on $M$ can then be written exactly as 
\eqn\ZLWRZ{\eqalign{ 
&Z(\epsilon;C,\CMj) \,=\, (-1) \, {{\exp{\!\left[{{3 \pi i} \over 4} - {i
\over 4}\big(\theta_0 + (j^2-1) \, \RP\big)\,\epsilon_r\right]}} \over
{4 \sqrt{\RP}}}\,\times\,\cr
&\qquad\times\,\Bigg\{ \sum_{l=0}^{d-1} \, {1 \over {2 \pi i}} 
\, \int_{\CC^{(l)}} \! dz \; \ch_\CMj\!\left({z\over 2}\right) \, F(z) \,
\exp{\!\left[G^{(l)}(z)\right]}\,-\,\cr 
&\qquad\qquad\,-\,\sum_{t=1}^{2 \RP - 1}
\Res\!\left({{\ch_\CMj\!\left({z\over 2}\right) \, F(z) \,
\exp{\!\left[G^{(0)}(z)\right]}} \over {1 - \exp{\!\left(- {{2 \pi}
\over {\epsilon_r}} \, z\right)}}}\right)\Bigg|_{z= 2 \pi i \, t}\,-\,\cr
&\qquad\qquad\qquad\,-\,\sum_{l=1}^{d-1} \sum_{t=1}^{[{{2 \RP l} \over d}]}
\Res\!\left(\ch_\CMj\!\left({z\over 2}\right) \, F(z) \,
\exp{\!\left[G^{(l)}(z)\right]}\right)\Bigg|_{z=- 2 \pi i \, t} 
\,\Bigg\}\,.\cr}}
Here $\CC^{(l)}$ for $l=0,\ldots,d-1$ denote a set of contours in the
complex plane over which we evalute the integrals in the first line of
\ZLWRZ. In particular, $\CC^{(0)}$ is the diagonal line contour
through the origin, 
\eqn\CCRC{ \CC^{(0)} \,=\, \e{{i \pi} \over 4} \times \BR\,,} 
and the other contours $\CC^{(l)}$ for $l > 0$ are diagonal line
contours parallel to $\CC^{(0)}$ running through the stationary phase
point of the integrand, given by $z = - 4 \pi i \, l \, (\RP / d)$.
Also, ``$\Res$'' denotes the residue of the given analytic function
evaluated at the given point.\countdef\Czero=165\Czero=\pageno
\countdef\Cell=166\Cell=\pageno\countdef\Resf=167\Resf=\pageno

Our formula for $Z(\epsilon;C,\CMj)$ in \ZLWRZ\ is a marginal
extension of the corresponding formula in \LawrenceRZ, where the
authors focus attention on torus knots in $S^3$.  However, \ZLWRZ\
follows very easily from general results in \LawrenceRZ.  A bit later,
we will indeed specialize \ZLWRZ\ to the particular case of torus knots.

We now wish to point out a few interesting features of \ZLWRZ\ from
the perspective of non-abelian localization.

First, if ${\CMj = {\bf 1}}$ is the trivial representation,
$\ch_{\CMj}(z)$ becomes the identity, and the formula for
$Z(\epsilon;C,\CMj)$ in \ZLWRZ\ immediately reduces to a corresponding 
formula for the Chern-Simons partition function $Z(\epsilon)$
exhibited in \S $5.2$ of \BeasleyVF.  In the case of the partition
function, each summand in \ZLWRZ\ is naturally identified with the
contribution to $Z(\epsilon)$ from an associated component in the
moduli space $\SM$ of flat connections on $M$.  Briefly, the $d$
integrals over the contours $\CC^{(l)}$ appearing in 
the second line of \ZLWRZ\ represent the local contributions to
$Z(\epsilon)$ from the $d$ reducible flat connections on $M$,
and the remaining residues in the final two lines of \ZLWRZ\
can be identified with the contributions to $Z(\epsilon)$ from
irreducible flat connections on $M$.

Second and perhaps more remarkably, we see from the formula in \ZLWRZ\
that $Z(\epsilon;C,\CMj)$ has exactly the same structure as
$Z(\epsilon)$ even when $\CMj$ is non-trivial.  Again,
$Z(\epsilon;C,\CMj)$ appears as a sum of terms associated to each
component in the moduli space $\SM$, and the Seifert loop operator is
{\sl universally} described on each component by the character $\ch_\CMj$.
Of the terms in \ZLWRZ, the contour integral for ${l=0}$ represents the
contribution from the trivial connection, which is given explicitly by 
\eqn\ZLWRZII{\eqalign{ 
&Z(\epsilon;C,\CMj)\big|_{\{0\}} \,=\, (-1) \,
{{\exp{\left[{{3 \pi i} \over 4}  - {i \over 4}\big(\theta_0 + (j^2-1) \, 
\RP\big)\,\epsilon_r\right]}} \over {4 \sqrt{\RP}}} \,\times\,\cr 
&\times\, {1 \over {2 \pi i}} \, \int_{\CC^{(0)}} \! dz \;
\ch_\CMj\!\left({z\over 2}\right) \, \exp{\left[{i \over{4 \epsilon_r}}
\left({d \over \RP}\right) z^2\right]} \, \left(2 \sinh{\left({z \over
2}\right)}\right)^{2 - N} \cdot \prod_{j=1}^N\, \left(2 \sinh{\left({z
\over {2 a_j}}\right)}\right)\,.\cr}}

Of course, the Seifert loop path integral does not localize on the
moduli space $\SM$ but on its extended cousin $\SM(C,\alpha)$, so the
apparent structure in \ZLWRZ\ deserves further explanation.  Although
the formula \ZLWRZ\ for $Z(\epsilon;C,\CMj)$ is valid for arbitrary
values of $\epsilon$ and $\CMj$, its semi-classical interpretation
holds in the particular regime ${\epsilon\to 0}$ with $\CMj$ held
fixed.  In that regime, the parameter ${\alpha/k}$ is small, and each
component of $\SM(C,\alpha)$ fibers over a corresponding component of
$\SM$, with fiber $\CO_\alpha$.  Pushing down over $\CO_\alpha$, we
then identify the contributions from such fibered components of
$\SM(C,\alpha)$ with contributions from $\SM$, just as in \ZLWRZ.
Indeed, the major result in this section is to reproduce the
particular term in \ZLWRZII\ by localization at the
distinguished reducible point ${\{\varrho_{\rm ab}\} \cong
\CO_\alpha/G}$, which fibers over the trivial connection $\{0\}$ on
the Seifert homology sphere $M$.

The preceding limit for the parameters $\epsilon$ and $\CMj$ should be
contrasted with the alternative semi-classical limit in which $\epsilon$ is
taken to zero with the product ${\epsilon\CMj}$ held fixed.
Equivalently, the Chern-Simons level $k$ goes to infinity with
${\alpha/k}$ non-zero and finite.  In the latter limit, relevant for
the generalized volume conjecture \Murakami, the moduli space
$\SM(C,\alpha)$ can have components which do not fiber over components
of $\SM$, as we exhibited explicitly for torus knots ${\SK_{{\bf p},{\bf
q}} \subset S^3}$ at the end of Section $7.1$.  The contributions to
$Z(\epsilon;C,\CMj)$ from the additional, non-fibered components in
$\SM(C,\alpha)$ then become visible only when the exact expression in
\ZLWRZ\ is rewritten in an asymptotic form compatible with the
stationary-phase approximation at finite (non-zero) values for
${\alpha/k}$.  See for example the ``Main Theorem'' in \Dubois,
where such an asymptotic expansion is carried out for torus knots
in $S^3$.

Finally, the phase of $Z(\epsilon;C,\CMj)$ in \ZLWRZ\ is quite subtle.
This phase depends upon both a two-framing of the manifold $M$ as well
as a framing of the curve $C$.  By definition, a two-framing of $M$ is
a trivialization of the direct sum $TM\oplus TM$ of two copies of the
tangent bundle $TM$.  As explained in \AtiyahFM, a canonical choice 
(up to homotopy) of two-framing exists for each three-manifold.

Similarly, a framing of the curve $C$ is specified by non-vanishing
normal vector field on $C$.  Such a vector field determines a small
displacement $C'$ of $C$ inside $M$, as would be used for instance in
a point-splitting regularization of the Wilson loop operator.  If $M$
is a homology sphere, Alexander duality (see for instance
Theorem $3.44$ in \Hatcher) implies that ${H_1(M^o;\BZ) 
\cong \BZ}$ is freely-generated by the meridian $\Rm$ of $C$.  In this
case, $C$ also carries a canonical framing, determined by the 
condition that $C'$ and $C$ have zero linking number inside $M$, where
the linking number is defined by ${\lk(C',C) = [C']}$ in $H_1(M^o;\BZ)$.

Because $M$ carries a canonical two-framing and $C$ carries a
canonical framing, $Z(\epsilon;C,\CMj)$ can be presented with a
definite phase, as given in \ZLWRZ.  The phase of $Z(\epsilon;C,\CMj)$
that arises naturally when we define Chern-Simons theory via
localization actually differs from the canonical phase, and we discuss
this fact at the end of the section.

\bigskip\noindent{\it Special Case: The Jones Polynomial of a Torus
Knot}\smallskip

To gain a bit more insight into the empirical formula \ZLWRZ\ for
$Z(\epsilon;C,\CMj)$, let us again specialize to the case of torus
knots $\SK_{{\bf p},{\bf q}}$ in $S^3$.  

With the Seifert invariants given in \AONE\ and \NBONEBTWO, the
formula for $Z(\epsilon;C,\CMj)$ becomes 
\eqn\ZLWRZTN{\eqalign{ 
&Z\big(\epsilon;\SK_{{\bf p},{\bf q}}, \CMj\big) \,=\, (-1) \,
{{\exp{\left[{{3 \pi i} \over 4}  - {i \over 4}\big({{\bf p}\over{\bf q}} +
{{\bf q}\over{\bf p}} + {\bf p} {\bf q} \,
(j^2-1)\big)\,\epsilon_r\right]}} \over 
{\sqrt{{\bf p} {\bf q}}}} \,\times\,\cr  
&\times\,\Bigg\{{1 \over {2 \pi i}} \, \int_{\CC^{(0)}} \! dz \;
\ch_\CMj\!\left({z\over 2}\right) \sinh{\!\left({z
\over {2 {\bf p}}}\right)} \sinh{\!\left({z
\over {2 {\bf q}}}\right)} \exp{\!\left[{i \over{4 \epsilon_r}}
\left({1 \over {\bf p}{\bf q}}\right) z^2\right]}\,+\,\cr
&\qquad\,+\,\left({j \over{k+2}}\right) \sum_{t=1}^{2{\bf p}{\bf q} -
1} \, (-1)^{t (j + 1)}\,\sin{\!\left({\pi t}\over{\bf
p}\right)}\,\sin{\!\left({\pi t}\over{\bf
q}\right)}\,\exp{\!\left({{-i \pi (k+2)}\over{2 {\bf p} {\bf q}}} 
t^2\right)}\Bigg\}\,.\cr}}
\countdef\Ztorus=168\Ztorus=\pageno

In passing from \ZLWRZ\ to \ZLWRZTN, we have explicitly evaluated the
phase $\theta_0$ in \QRZE\ for the Seifert presentation of $S^3$ with fiber 
$\SK_{{\bf p},{\bf q}}$.  Here we use two arithmetic properties of the
Dedekind sum $s(\,\cdot\,,\,\cdot\,)$ that enters $\theta_0$.  First,
as follows more or less directly from the definition in \DEDEK,
\eqn\DEDSYM{ s({\bf p}-{\bf r},{\bf p}) \,=\, s({\bf q},{\bf
p})\,,\qquad\qquad s({\bf s},{\bf q}) = s({\bf p},{\bf
q})\,, \qquad\qquad {\bf p} {\bf s} - {\bf q} {\bf r} \,=\, 1\,.}  
Much more non-trivially, we also use Dedekind reciprocity, which
states that 
\eqn\DEDREC{ 12 \, {\bf p}{\bf q} \, \Big[s({\bf p},{\bf q}) \,+\,
s({\bf q},{\bf p})\Big] \,=\, {\bf p}^2 \,+\, {\bf q}^2 \,-\, 
3 {\bf p}{\bf q} \,+\, 1\,,\qquad \gcd({\bf p},{\bf q})\,=\, 1\,.}
See \S $3.8$ of \Apostol\ for an elementary (but by no means obvious)
proof of Dedekind reciprocity.  Together, we apply \DEDSYM\ and \DEDREC\
to compute $\theta_0$ as 
\eqn\QRZEKII{ \theta_0 \,=\, 3 \,-\, {1 \over {{\bf p}{\bf q}}} \,+\, 
12 \big[s({\bf p} - {\bf r},{\bf p}) \,+\, s({\bf s},{\bf q})\big] \,=\, 
{{\bf p}\over{\bf q}} \,+\, {{\bf q}\over{\bf p}}\,.}

We have also evaluated the residues appearing in the
empirical formula for $Z(\epsilon;C,R)$.  These residues appear in the 
sum over $t$ in \ZLWRZTN, in terms of which we decompose 
$Z(\epsilon;\SK_{{\bf p},{\bf q}},\CMj)$ as 
\eqn\ZLWRZTNII{ Z\big(\epsilon;\SK_{{\bf p},{\bf q}},\CMj\big) \,=\,
Z\big(\epsilon;\SK_{{\bf p},{\bf q}},\CMj\big)\big|_{\{0\}} \,+\,
Z\big(\epsilon;\SK_{{\bf p},{\bf q}},\CMj\big)_{\rm res}\,,}
where 
\eqn\JPT{\eqalign{
Z\big(\epsilon;\SK_{{\bf p},{\bf q}},\CMj\big)\big|_{\{0\}} \,&=\, (-1) \,
{{\exp{\!\left[{{3 \pi i} \over 4}  - {i \over 4}\big({{\bf p}\over{\bf q}} +
{{\bf q}\over{\bf p}} + {\bf p} {\bf q} \,
(j^2-1)\big)\,\epsilon_r\right]}} \over 
{\sqrt{{\bf p} {\bf q}}}} \,\times\,\cr  
&\times\,{1 \over {2 \pi i}} \, \int_{\CC^{(0)}} \! dz \;
\ch_\CMj\!\left({z\over 2}\right) \sinh{\!\left({z
\over {2 {\bf p}}}\right)} \sinh{\!\left({z
\over {2 {\bf q}}}\right)} \exp{\!\left[{i \over{4 \epsilon_r}}
\left({1 \over {\bf p}{\bf q}}\right)
z^2\right]}\,,\cr}}\countdef\Ztriv=169\Ztriv=\pageno
and 
\eqn\JPSUM{\eqalign{
Z\big(\epsilon;\SK_{{\bf p},{\bf q}},\CMj\big)_{\rm res} \,&=\,  (-1) \,
{{\exp{\!\left[{{3 \pi i} \over 4}  - {i \over 4}\big({{\bf p}\over{\bf q}} +
{{\bf q}\over{\bf p}} + {\bf p} {\bf q} \,
(j^2-1)\big)\,\epsilon_r\right]}} \over 
{\sqrt{{\bf p} {\bf q}}}} \,\times\,\cr
&\times\,\left({j \over{k+2}}\right) \sum_{t=1}^{2{\bf p}{\bf q} -
1} \, (-1)^{t (j + 1)}\,\sin{\!\left({\pi t}\over{\bf
p}\right)}\,\sin{\!\left({\pi t}\over{\bf
q}\right)}\,\exp{\!\left({{-i \pi (k+2)}\over{2 {\bf p} {\bf q}}} 
t^2\right)}\,.\cr}}

\countdef\Zres=170\Zres=\pageno

As we have already mentioned, $Z(\epsilon;\SK_{{\bf p},{\bf
q}},\CMj)\big|_{\{0\}}$ can be naturally interpreted as the
contribution to the Seifert loop path integral from the reducible
point $\{\varrho_{\rm ab}\}$ in the extended moduli space $\SM(\SK_{{\bf
p},{\bf q}}, \CMj)$.  Equivalently, via equivariant pushdown in the fibration
\FBMCAHS,  $Z(\epsilon;\SK_{{\bf p},{\bf q}},\CMj)\big|_{\{0\}}$ is
the contribution from the trivial connection $\{0\}$ on $S^3$.  But in
the relevant semi-classical limit, for which ${\epsilon\to 0}$ with
${\CMj}$ fixed, $\{\varrho_{\rm ab}\}$ is indeed the only point in $\SM(\SK_{{\bf
p},{\bf q}}, \CMj)$.  Hence our localization result for the Seifert
loop path integral implies that the additional, oscillatory Gaussian
sum in \JPSUM\ must actually vanish,
\eqn\JPSUMII{ Z\big(\epsilon;\SK_{{\bf p},{\bf q}},\CMj\big)_{\rm res}
\,=\, 0\,,\qquad\qquad \gcd({\bf p},{\bf q}) = 1\,.}
As a small check, we verify this statement directly in Appendix B.

After applying the vanishing result \JPSUMII\ from Appendix B, we obtain a
wonderfully compact formula \foot{A related formula appears in Lemma 1
of \KashaevT.} for the expectation value of an arbitrary
Wilson loop operator wrapping the torus knot $\SK_{{\bf p},{\bf q}}$
in $S^3$ and decorated with the irreducible $SU(2)$ representation $\CMj$,
\eqn\JPTII{\eqalign{
Z\big(\epsilon;\SK_{{\bf p},{\bf q}},\CMj\big) \,&=\,
Z\big(\epsilon;\SK_{{\bf p},{\bf q}},\CMj\big)\big|_{\{0\}}\,,\cr
&=\, {1 \over {2 \pi i}} {1 \over {\sqrt{{\bf p}{\bf q}}}} \, \exp{\!\left[-{{i
\pi}\over{2 (k+2)}}\left({{\bf p}\over{\bf q}} +
{{\bf q}\over{\bf p}} + {\bf p} {\bf q} \,
(j^2-1)\right)\right]} \,\times\,\cr  
&\times \, \int_\BR \! dx \;
\ch_\CMj\!\left(\e{{i\pi}\over 4}\,{x\over 2}\right)
\sinh{\!\left(\e{{i\pi}\over 4}\,{x\over {2 {\bf p}}}\right)}
\sinh{\!\left(\e{{i\pi}\over 4}\,{x\over {2 {\bf q}}}\right)} 
\exp{\!\left[-{{(k+2)}\over{8\pi}}
\left({{x^2} \over {\bf p}{\bf q}}\right)\right]}\,.\cr}}
In writing \JPTII, we have rotated the contour ${\CC^{(0)} = \e{{i \pi}
\over 4} \times \BR}$ to the real axis and substituted ${\epsilon_r =
2\pi/(k+2)}$, so that \JPTII\ appears as a simple generalization of the
corresponding formula \UNKNOTWII\ for the unknot ${\bigcirc =
\SK_{{\bf 1},{\bf 1}}}$.

The expression for $Z(\epsilon;\SK_{{\bf p},{\bf q}},\CMj)$ in \JPTII\
has several remarkable properties, which also provide an independent
check that \JPTII\ is correct.  First, if ${\CMj = {\bf 1}}$, then
all dependence on ${\bf p}$ and ${\bf q}$ cancels between the
prefactor and the integral over $x$ in \JPTII, and our formula for
$Z(\epsilon;\SK_{{\bf p},{\bf q}},\CMj)$ reproduces the standard
result for the $SU(2)$ Chern-Simons partition function on
$S^3$,
\eqn\PARTZ{ Z(k) \,=\, \sqrt{2 \over {k+2}} \sin{\left(\pi \over
{k+2}\right)}\,.}

Second, if either ${{\bf p}, {\bf q} = 1}$ and the other is an 
arbitrary non-zero integer, then $\SK_{{\bf p},{\bf q}}$ is equivalent
to the unknot.  As one can verify, \JPTII\ still reproduces the
expected result for $Z(k;\bigcirc,\CMj)$ in that more general case,
\eqn\UNKNOTZ{  Z\big(k; \bigcirc, \CMj\big) \,=\, \sqrt{2 \over
{k+2}} \sin{\left({{\pi \, j} \over {k+2}}\right)}\,.}

Finally, the integral over $x$ in \JPTII\ is just a sum
of Gaussian integrals, which can be evaluated explicitly as 
\eqn\JPTIII{\eqalign{
Z\big(\epsilon;\SK_{{\bf p},{\bf q}},\CMj\big) \,&=\, {1 \over {2 i}}
\sqrt{2\over{k+2}} \, \Rt^{{1 \over 4} {\bf p}{\bf q} (j^2-1)} \,
\sum_{l=0}^{j-1} \, \Rt^{-{1\over 4} l_*^2 {\bf p}{\bf q} \,-\, \ha
l_* {\bf p}} \left[\Rt^{-\ha (l_* {\bf q} + 1)} \,-\, \Rt^{\ha
(l_* {\bf q} + 1)}\right]\,,\cr
l_* \,&=\, 2l - j + 1\,,\qquad\qquad \Rt \,=\, \exp{\!\left(-{{2 \pi
i}\over{k+2}}\right)}\,.\cr}}
The sum over $l$ in \JPTIII\ arises from the sum \CHNII\ over
exponentials in the character $\ch_\CMj$, and we have introduced the
standard variable $\Rt$ that appears in the Jones polynomial.
\countdef\tVariable=172\tVariable=\pageno

When ${\CMj = {\bf 2}}$, the sum in \JPTIII\ must reproduce the
Jones polynomial of the torus knot $\SK_{{\bf p}, {\bf q}}$.
Concretely,
\eqn\JPTIV{ Z\big(\epsilon;\SK_{{\bf p},{\bf q}},{\bf 2}\big) \,=\, {1
\over {2 i}} \sqrt{2\over{k+2}} \, \Rt^{\ha ({\bf p}{\bf q} - {\bf p}
- {\bf q} - 1)} \Big[1 \,+\, \Rt^{{\bf p}+{\bf q}} \,-\, \Rt^{{\bf
p}+1} \,-\, \Rt^{{\bf q}+1}\Big]\,.}
The Jones polynomial $V_\SK(\Rt)$ for a knot $\SK$ is generally  
proportional to $Z\big(\epsilon;\SK,{\bf 2}\big)$.  However,
$V_\SK(\Rt)$ is conventionally normalized so that for the
unknot, ${V_\bigcirc(\Rt) = 1}$.  Rescaling \JPTIV\ by
$Z\big(\epsilon;\SK_{{\bf 1},{\bf 1}},{\bf 2}\big)$, we find 
\eqn\JPTORUS{ V_{\SK_{{\bf p},{\bf q}}}(\Rt) \,=\,
{{Z\big(\epsilon;\SK_{{\bf p},{\bf q}},{\bf
2}\big)}\over{Z\big(\epsilon;\SK_{{\bf 1},{\bf 1}},{\bf 2}\big)}}
\,=\, {{\Rt^{\ha({\bf p}-1)({\bf q}-1)}}\over{1 \,-\,
\Rt^2}} \> \Big[ 1 \,+\, \Rt^{{\bf p}+{\bf
q}} \,-\, \Rt^{{\bf p}+1} \,-\, \Rt^{{\bf q}+1}\Big]\,.}
\countdef\Jones=171\Jones=\pageno
This formula for the Jones polynomial of a torus knot is a classic
result that goes back to Jones himself \JonesVFR, and early 
after Witten's foundational work \WittenHF, the same result was
obtained in \refs{\Labastida,\IsidroFZ} using the Hamiltonian
formulation of Chern-Simons theory.  Among the many motivations
for this paper, one is to explain how $V_{\SK_{{\bf p},{\bf q}}}(\Rt)$
can be alternatively computed using non-abelian localization, as we
now discuss.

\bigskip\noindent{\it A Symplectic Model for ${\CO_\alpha \subset
\bar\CA_\alpha}$}\smallskip

In the remainder of this section, we wish to apply the non-abelian
localization formula in \ZEV\ to compute directly the contribution
from the reducible point ${\{\varrho_{\rm ab}\}\cong\CO_\alpha/G}$ in
$\SM(C,\alpha)$ to the Seifert loop path integral.  Though we
presented the corresponding empirical result \ZLWRZII\ of Lawrence and
Rozansky for the special case ${G=SU(2)}$, no extra effort is required
to treat the case of an arbitrary compact, connected,
simply-connected, and simple Lie group $G$, as we shall do here.  See
also \MarinoFK\ for work extending the results of \LawrenceRZ\ to
arbitrary $G$.

Of course, in order to apply the non-abelian localization formula to
our problem, we first need to determine the local symplectic 
geometry in an equivariant neighborhood of the orbit $\CO_\alpha$ 
embedded as a finite-dimensional submanifold of the
infinite-dimensional product ${\bar\CA_\alpha = \bar\CA \times
\epsilon L\CO_\alpha}$,
\eqn\EMBDOA{ \CO_\alpha \,\subset\, \bar\CA_\alpha = \bar\CA
\times \epsilon L\CO_\alpha\,.}
Following the ansatz in Section $6.2$, we model an equivariant
neighborhood $N$ of ${\CO_\alpha \subset \bar\CA_\alpha}$ on a
symplectic fibration of the general form    
\eqn\NLOOP{\matrix{
&\SF^\alpha \longrightarrow\, N\cr\noalign{\vskip 2 pt}
&\mskip 80mu\Big\downarrow\lower 0.5ex\hbox{$^{\rm pr}$}\cr
&\mskip 65mu \CO_\alpha/G\cr}\,,}
where the fiber $\SF^\alpha$ is itself the total space of a homogeneous
vector bundle over a quotient $\CH/H_0^\alpha$.  Specifically, in
complete correspondence to \COTFBII,
\eqn\FLOOP{ \SF^\alpha \,=\, \CH \times_{H_0^{\alpha}}
\!\left(\Fh^\perp \oplus \CE_1^{\alpha}\right)\,,\qquad\qquad
\Fh^\perp \,\equiv\, \Fh \ominus \Fh_0^{\alpha} \ominus
\CE_0^{\alpha}\,.}\countdef\SFalpha=173\SFalpha=\pageno
\countdef\BigHalpha=174\BigHalpha=\pageno
\countdef\Littlehalpha=175\Littlehalpha=\pageno
\countdef\CurlyEalpha=176\CurlyEalpha=\pageno
\countdef\hyperpal=178\hyperpal=\pageno
Here $H_0^\alpha$ is the subgroup of the Hamiltonian group $\CH$ which
preserves points in $\CO_\alpha$, and we recall from Section $4.3$
that we have already identified $\CH$ to be 
\eqn\HAMCH{ \CH \,=\, U(1)_\RR \ltimes \wt\CG_0\,.} 
So to specify the geometry of $\SF^\alpha$ in \FLOOP, we are left to 
determine  $H_0^\alpha$ as well as the vector spaces
$(\CE_0^\alpha,\CE_1^\alpha)$.

The local model for $\SF^\alpha$ which describes a neighborhood of
$\CO_\alpha$ in $\bar\CA_\alpha$ turns out to be a natural
generalization of the model we applied in \S $5.2$ of \BeasleyVF\ 
to describe a neighborhood of the trivial connection
$\{0\}$ in $\bar\CA$.  Both to compare with our previous work and to
economize the present discussion, let us therefore quickly recall the
analogous data $(H_0, \CE_0, \CE_1)$ which specify the local model
describing a neighborhood of ${\{0\}\in\bar\CA}$.  With these data in
hand, the extension to $(H_0^\alpha,\CE_1^\alpha, \CE_a^\alpha)$ will
be more or less straightforward.

First, under the action of the Hamiltonian group $\CH$, the trivial
connection on $M$ is fixed by the subgroup 
\eqn\HNOUGHT{ H_0 \,=\, U(1)_\RR \times G \times U(1)_\RZ\,,}
where $G$ is identified with the group of constant gauge transformations
on $M$.   To play the roles of $E_0$ and $E_1$ in the abstract
symplectic model of Section $6.2$, we also introduce vector spaces
$(\CE_0, \CE_1)$.  Both $\CE_0$ and $\CE_1$ turn out to be quite
delicate to determine, and we refer the reader to \S $5.1$ of
\BeasleyVF\ for a complete discussion.  Suffice it to say, $\CE_0$ and
$\CE_1$ are given by certain direct sums of Dolbeault cohomology groups,
\eqn\EONE{\eqalign{
\CE_0 \,&=\, \bigoplus_{t \ge 1} H^0_{\bar\partial}\big(\Sigma, \Fg
\otimes (\CL^t \oplus \CL^{-t})\big)\,,\cr
\CE_1 \,&=\, \bigoplus_{t \ge 1} H^1_{\bar\partial}\big(\Sigma, \Fg
\otimes (\CL^t \oplus \CL^{-t})\big)\,.\cr}}
\countdef\CurlyEs=177\CurlyEs=\pageno
Here $\Sigma$ is the Riemann surface (or more generally orbifold)
sitting at the base of $M$, and $\CL$ is the line bundle over $\Sigma$
associated to the Seifert presentation of $M$.  In general, both
${H^0_{\bar\partial}\big(\Sigma, \,\Fg \otimes (\CL^t \oplus \CL^{-t})\big)}$
and ${H^1_{\bar\partial}\big(\Sigma, \,\Fg \otimes (\CL^t \oplus
\CL^{-t})\big)}$ are non-vanishing for arbitrarily large $t$, so
$\CE_0$ and $\CE_1$ have infinite dimension.

To describe a neighborhood of ${\CO_\alpha\subset \bar\CA_\alpha}$, we
must modify both $H_0$ in \HNOUGHT\ and $(\CE_0,\CE_1)$ in \EONE.
Clearly, the stabilizer $H_0^\alpha$ of a point in $\CO_\alpha$ will
only be a subgroup of the stabilizer $H_0$ for the trivial connection.
As we noted following \PIMIII, points in $\CO_\alpha$ are represented
by reducible connections which are preserved by constant gauge
transformations in ${G_\alpha \subset G}$.  These connections are also
invariant under the $U(1)_\RR$ action on $M$, so immediately 
\eqn\HNOUGHTII{ H_0^\alpha \,=\, U(1)_\RR \times G_\alpha \times
U(1)_\RZ\,.}
By way of illustration, we recall that if $\alpha$ is regular, then
${G_\alpha = T}$ is a maximal torus of $G$. At the other extreme, if
${\alpha = 0}$, then ${G_\alpha = G}$ and $H_0^\alpha$ reduces to $H_0$.

We are left to specify the vector spaces $(\CE_0^\alpha,
\CE_1^\alpha)$ in \FLOOP.  By construction, the vector spaces $(\CE_0,
\CE_1)$ in \EONE\ already provide a local model for the normal
directions to ${\CO_\alpha\subset\bar\CA_\alpha = \bar\CA \times
\epsilon L\CO_\alpha}$ which lie along $\bar\CA$.  To account further for the
symplectic geometry of the loopspace $L\CO_\alpha$, we need
only enlarge $\CE_0$ and $\CE_1$ slightly.  We discuss the 
required modifications in turn.

As we explained in \S $4.3$ of \BeasleyVF\ and briefly indicated
in Section $6.2$, the role of $E_0$ in the abstract model for $F$ is to
control the symplectic structure on a neighborhood of ${H/H_0
\subseteq F}$.  In essence, the possible symplectic models for the
embedding of $H/H_0$ inside $F$ interpolate from a cotangent model,
for which $H/H_0$ is embedded as the zero section of the cotangent bundle
$T^*(H/H_0)$, to a coadjoint model, for which $H/H_0$ carries the
coadjoint symplectic form and no cotangent fibers are present in $F$.
In the cotangent model $E_0$ is trivial, and in the coadjoint model,
$E_0$ coincides with the holomorphic tangent space to $H/H_0$ at the
identity, as we illustrated in \MODEXI.

In the case at hand, since $\CH$ contains a copy of $H_0$, the
homogeneous space $\CH/H_0^\alpha$ sitting at  the base of
$\SF^\alpha$ contains a copy of  
\eqn\HNOUGHTIII{ H_0^{}/H_0^\alpha \,=\, G/G_\alpha \,=\, \CO_\alpha\,.}
So we see that ${\CH/H_0^\alpha}$ already contains a copy of the orbit
$\CO_\alpha$ on which we localize.  But we also know that $\CO_\alpha$
must carry the canonical coadjoint symplectic form.  According to
the abstract symplectic model sketched above, if ${\CO_\alpha \subset
\CH/H_0^\alpha}$ is to carry the coadjoint symplectic form, the vector
space $\CE_0^\alpha$ which enters \FLOOP\ must include the holomorphic
tangent space $\Fg^{(1,0)}$ to $\CO_\alpha$ as a summand.  

Otherwise, the original vector space $\CE_0$ in \EONE\ already encodes  
the correct symplectic form on the remaining directions normal to 
$\CO_\alpha$ inside  $\CH/H_0^\alpha$.  Hence the new
vector space $\CE_0^\alpha$ is given merely by the direct sum   
\eqn\ENOUGHTA{ \CE_0^\alpha \,=\, \CE_0^{} \oplus 
\Fg^{(1,0)}\,.}
For future reference, we recall from \HOLTN\ that $\Fg^{(1,0)}$ is
given concretely by the following sum of rootspaces
${\Fe_{\beta}\subset\Fg_\BC}$,
\eqn\POSROOTSPD{
\Fg^{(1,0)} =\, \bigoplus_{(\beta,\alpha) > 0} \Fe_{\beta}\,,\qquad\qquad
\beta\in\FR\,.}
When $\alpha$ is regular, ${\Fg^{(1,0)}\!= \Fg_{+}}$ is the entire 
positive rootspace.

Having identified the analogue of $E_0$ in the abstract symplectic
model, we now consider $E_1$.  The role of $E_1$ in the symplectic
model for $F$ is to describe whatever directions are normal to $H/H_0$
beyond those already captured in the interpolating fiber ${\Fh^\perp =
\Fh \ominus \Fh_0 \ominus E_0}$.  In the case at
hand, the vector space $\CE_1$ in \EONE\ already encodes such
directions normal to ${\CO_\alpha \subset \bar\CA \times \epsilon L\CO_\alpha}$
inside $\bar\CA$.  So we are left to consider the other directions
normal to $\CO_\alpha$ inside the loopspace $L\CO_\alpha$.

In general, the tangent space at a point ${[U] \in L\CO_\alpha}$
corresponding to a given map ${U\!:C\rightarrow \CO_\alpha}$ is the
space of sections of the pullback $U^*(T\CO_\alpha)$.  Localizing on
$\CO_\alpha$, we are interested in the special case that 
${U(\tau) = U_0}$ is constant.  In that case, $U^*(T\CO_\alpha)$ can
be identified up to the action of $G$ as the trivial bundle on $C$
with fiber ${\Fg\ominus\Fg_\alpha}$.  The tangent space to
$L\CO_\alpha$ at a point in $\CO_\alpha$ is then isomorphic to the
space of maps ${\delta U\!:C\rightarrow \Fg\ominus\Fg_\alpha}$.  

By assumption, $C$ is a Seifert fiber of $M$ and hence is preserved 
under the action of $U(1)_\RR$.  We can thus decompose the map $\delta
U$ into eigenmodes of the Lie derivative $\lie_\RR$ along the vector
field $\RR$,
\eqn\FDU{ \delta U \,=\, \sum_{t=-\infty}^{+\infty} \, \delta
U_t\,,}
such that each eigenmode $\delta U_t$ satisfies
\eqn\FDUII{ \lie_\RR \, \delta U_t \,=\, -2\pi i\, t \cdot \delta U_t\,.}
Of course, \FDU\ is nothing more than an infinitesimal version of the
decomposition of $U$ into Fourier modes along $C$.  

The constant eigenmode $\delta U_0$ describes tangent directions to
$\CO_\alpha$ itself, and we have already accounted for these
directions with the summand ${\Fg^{(1,0)} \! \subset \CE_0^\alpha}$.
In contrast, the eigenmodes $\delta U_t$ with ${t \neq 0}$ describe
non-trivial normal directions to $\CO_\alpha$ inside $L\CO_\alpha$.
Thus, the normal fiber $\CN_\alpha$ to $\CO_\alpha$ embedded in
$L\CO_\alpha$ consists of a countable sum of copies of ${\Fg \ominus
\Fg_\alpha}$, graded by the non-zero integer $t$.

Of course, ${\Fg \ominus \Fg_\alpha}$ is a real vector space.  As we
discussed in Section $4.3$, the loopspace $L\CO_\alpha$ carries  
a complex structure induced pointwise from the complex structure on
$\CO_\alpha$.  So as a complex vector space,
\eqn\NLOC{ \CN_\alpha \,=\, \bigoplus_{t \ge 1} \Big[\Fg^{(1,0)}_{t}
\oplus \Fg^{(1,0)}_{-t}\Big]\,.}
Here the subscript on each copy of $\Fg^{(1,0)}$ indicates the Fourier
eigenvalue in \FDUII.\countdef\CurlyN=179\CurlyN=\pageno

Since the role of $\CE_1^\alpha$ is to account for directions normal
to $\CO_\alpha$ in both $\bar\CA$ and $L\CO_\alpha$, one might guess
that $\CE_1^\alpha$ is given by the direct sum ${\CE_1 \oplus
\CN_\alpha}$.  To the first approximation this guess is correct, but
we must be careful about the choice of complex structure on
$\CE_1^\alpha$.

According to the convention in \HOLCONV, the complex structure on the
abstract vector space $E_1$ in the non-abelian localization formula
\ZEV\ is defined so that ${-i \gamma_0}$ acts with strictly negative
eigenvalues on a holomorphic basis of $E_1$.  This convention is
opposite to the convention in \HOLTN\ which defines the complex structure
on $\CO_\alpha$, since $\Fg^{(1,0)}$ is defined to be the positive
eigenspace of ${-i\,\alpha}$ under the adjoint action.  To account for
the relative sign in our two conventions, we define $\CE_1^\alpha$
using not $\CN_\alpha$ in \NLOC\ but rather the conjugate vector space
$\bar\CN_\alpha$,
\eqn\CCNLOC{ \bar\CN_\alpha \,=\, \bigoplus_{t \ge 1} \Big[\Fg^{(0,1)}_{t}
\oplus \Fg^{(0,1)}_{-t}\Big]\,.}\countdef\CurlyNbar=180\CurlyNbar=\pageno
Here $\Fg^{(0,1)}$ is the anti-holomorphic tangent space to
$\CO_\alpha$, which is spanned by the rootspaces $\Fe_{-\beta}$
associated to negative roots ${-\beta < 0}$ of $G$,
\eqn\NEGROOTSPD{
\Fg^{(0,1)} =\, \bigoplus_{(\beta,\alpha) > 0} \Fe_{-\beta}\,,\qquad\qquad
\beta\in\FR\,.}
So if we are careful about the complex structure,
\eqn\EONEA{ \CE_1^\alpha \,=\, \CE_1 \oplus \bar\CN_\alpha\,.}
As will be clear in our later computations, the appearance of
$\bar\CN_\alpha$ as opposed to $\CN_\alpha$ in \EONEA\ is actually
crucial for the interpretation of the Seifert loop operator as the
character associated to the representation $R$.

Altogether, \HNOUGHTII, \ENOUGHTA, and \EONEA\ specify the local
symplectic model for the embedding ${\CO_\alpha \subset
\bar\CA_\alpha}$.

\bigskip\noindent{\it Non-Abelian Localization on $\CO_\alpha/G$}\smallskip

With the symplectic model for $\SF^\alpha$ in hand, we now apply the
non-abelian localization formula in \ZEV\ to the Seifert loop
path integral.  Because ${\{\varrho_{\rm ab}\} \cong \CO_\alpha/G}$ is a
point, the path integral immediately reduces via \ZVWHKIV\ to an
integral over the finite-dimensional Lie algebra ${\Fh_0^\alpha = \BR
\oplus \Fg_\alpha \oplus \BR}$ of the stabilizer $H_0^\alpha$,
\eqn\SFTWYI{\eqalign{
Z\big(\epsilon;C,R\big)\Big|_{\CO_\alpha/G} \,&=\, {{(2\pi\epsilon)}
\over {\Vol(G_\alpha)}} \, \int_{\Fh_0^\alpha} 
\left[{{d\psi} \over {2\pi}}\right] \, \det\!\left({{\psi} \over {2
\pi}}\Big|_{\CE_0^\alpha}\right) \, \det\!\left({{\psi} \over {2
\pi}}\Big|_{\CE_1^\alpha}\right)^{-1}\, \,\times\cr
&\qquad\qquad\qquad\times\,\exp{\left[-i \left(\gamma_0,
\psi\right) - {{i \epsilon} \over 2} \left(\psi,
\psi\right)\right]}\,.}}\countdef\ZOalphaG=181\ZOalphaG=\pageno
Here $\psi$ is an element in the algebra $\Fh_0^\alpha$.  Because the
group ${H_0^\alpha = U(1)_\RR \times G_\alpha \times U(1)_\RZ}$
decomposes as a product, we frequently write $\psi$ in terms of components 
\eqn\LIEPSI{ \psi \,=\, (p,\phi,a) \,\in\, \BR \oplus \Fg_\alpha
\oplus \BR\,,} 
where $p$ and $a$ generate $U(1)_\RR$ and $U(1)_\RZ$ respectively, and
$\phi$ is an element of $\Fg_\alpha$.  

In arriving at the expression for
$Z(\epsilon;C,R)\big|_{\CO_\alpha/G}$ in \SFTWYI, we have multiplied
the result obtained directly from \ZVWHKIV\ by 
\eqn\VOLMUL{ \Vol[U(1)_\RR] \cdot \Vol[U(1)_\RZ] \cdot
2\pi\epsilon\,,} 
which accounts for the prefactor involving $\epsilon$ in \SFTWYI.  The same
multiplicative factor appears in \S $5$ of \BeasleyVF, for precisely
the same reason.  For sake of brevity, we refer the interested reader
to the discussion surrounding $(5.10)$ of \BeasleyVF\ for a simple
explanation of how this normalization factor arises.

In the remainder of this section, most of our effort will be devoted
to evaluating the determinants of ${\psi \in \Fh_0^\alpha}$ acting on
the infinite-dimensional vector spaces $\CE_0^\alpha$ and
$\CE_1^\alpha$.  However, let us first make the argument of the
exponential in \SFTWYI\ a bit more explicit.

By definition, ${\gamma_0\in\Fh_0^\alpha}$ is the dual of the value of
the moment map $\mu$ evaluated at the point ${\alpha\!\in\CO_\alpha}$.
According to \MUVX, $\mu$ is generally given on $\bar\CA_\alpha$ by 
\eqn\MUVXWY{\eqalign{
&\Big\langle\mu,(p,\phi,a)\Big\rangle \,=\, a \,-\, p \int_M
\kappa\^\Tr\!\left[\ha \, \lie_\RR A\^A \,+\, \epsilon \, \alpha 
\big(g^{-1} \lie_\RR g\big) \, \delta_C\right] \,-\, \int_M 
\kappa\^\Tr\big(\phi \, \CF_A\big) \,+\,\cr
&\qquad\qquad\,+\, \int_M d\kappa\^\Tr\big(\phi \, A\big).}}
Points in $\CO_\alpha$ correspond to classical configurations of $(A,
U)$ which are annihilated by $\lie_\RR$ and satisfy ${\CF_A \,=\,
0}$, so only the first and last terms in \MUVXWY\ contribute when
$\mu$ is evaluated at points in $\CO_\alpha$.  

We compute directly  the last term in \MUVXWY\ to be 
\eqn\LSTTRM{\eqalign{
\int_M d\kappa\^\Tr\big(\phi \, A\big) \,&=\, \int_M
\kappa\^\Tr\big(\phi \, F_A\big)\,,\cr
&=\,-\epsilon \int_M \kappa\^\delta_C \, \Tr(\alpha\,\phi)\,,\cr
&=\,-\epsilon\,\Tr(\alpha\,\phi)\,=\,\epsilon\,
\big\langle\alpha,\phi\big\rangle\,.}}
In deducing the first equality of \LSTTRM, we integrate by parts with
${\phi\in\Fg_\alpha}$ constant, and in the second equality, we use that 
${\CF_A = F_A \,+\, \epsilon \alpha \cdot \delta_C = 0}$ at the point 
${\alpha \in \CO_\alpha}$.  Of course, up to the factor of $\epsilon$,
we recognize the last expression in \LSTTRM\ as nothing more than the
moment map \COADJMOM\ for the action of $G$ on the coadjoint orbit
$\CO_\alpha$.  

From \MUVXWY\ and \LSTTRM\ we thereby obtain 
\eqn\MUVXWYII{ \big(\gamma_0,\psi\big) \,=\,
\big\langle\mu,(p,\phi,a)\big\rangle\Big|_{\alpha\in\CO_\alpha} =\, a 
\,+\, \epsilon\,\big\langle\alpha, \phi\big\rangle\,.}
We also recall from \FRM\ that the norm of $\psi$ is given by 
\eqn\FRMSFTW{\eqalign{ 
(\psi,\psi) \,&=\, -\int_M \kappa\^d\kappa \, \Tr\big(\phi^2\big)
\,-\, 2 p a\,,\cr
&= -{d \over \RP} \, \Tr\big(\phi^2\big) \,-\, 2 p a\,.}}
In passing to the second line of \FRMSFTW, we use the description of
$d\kappa$ in \CHRNCLII\ along with the identity in \SFRTHOM\ to compute the
integral ${\int_M \kappa\^d\kappa = d / \RP}$, where $d$ is defined in
\INTD\ and $\RP$ is defined in \QRZE.  Via \MUVXWYII\ and \FRMSFTW,
the integral over $\Fh^\alpha_0$ then takes the more explicit form 
\eqn\SFTWYIITP{\eqalign{ 
Z\big(\epsilon;C,R\big)\Big|_{\CO_\alpha/G} &= {{(2\pi\epsilon)}
\over {\Vol(G_\alpha)}} \int_{\BR \times \Fg_\alpha \times \BR} \left[{{dp}
\over {2 \pi}}\right] \left[{{d\phi} \over {2 \pi}}\right] \left[{{da}
\over {2 \pi}}\right] \det\!\left({{\psi} \over {2
\pi}}\Big|_{\CE_0^\alpha}\right) \det\!\left({{\psi} \over {2
\pi}}\Big|_{\CE_1^\alpha}\right)^{-1}\times\cr
&\qquad\qquad\qquad\times\,\exp{\!\left[-i a \,-\, i \epsilon 
\langle\alpha,\phi\rangle \,+\, {{i \epsilon}\over 2} \left({d \over
\RP}\right) \Tr(\phi^2) \,+\, i \epsilon p a\right]}\,.}}

Let us make one final simplification of \SFTWYIITP\ before we proceed to
honest calculations.  As we have seen, the vector bundles
$\CE_0^\alpha$ and $\CE_1^\alpha$ both decompose into summands
associated to the respective factors in the product ${\bar\CA_\alpha =
\bar\CA \times \epsilon L\CO_\alpha}$, so that 
\eqn\DECES{ \CE_0^\alpha = \CE_0 \oplus \Fg^{(1,0)}\,,\qquad\quad
\CE_1^\alpha = \CE_1 \oplus \bar\CN_\alpha\,.}
Consequently, in any natural regularization, we can factorize the
ratio of determinants appearing in \SFTWYIITP\ as 
\eqn\DETSFAC{ \det\!\left({{\psi} \over {2
\pi}}\Big|_{\CE_0^\alpha}\right) \cdot \det\!\left({{\psi} \over {2
\pi}}\Big|_{\CE_1^\alpha}\right)^{-1} \,=\, e\big(\bar\CA\big)  \cdot
e\big(L\CO_\alpha\big)\,,}
where we introduce the separate ratios 
\eqn\DETSPH{\eqalign{
e\big(\bar\CA\big) \,&=\, \det\!\left({{\psi} \over {2
\pi}}\Big|_{\CE_0}\right) \det\!\left({{\psi} \over {2
\pi}}\Big|_{\CE_1}\right)^{-1}\,,\cr
e\big(L\CO_\alpha\big) \,&=\, \det\!\left({{\psi} \over {2
\pi}}\Big|_{\Fg^{(1,0)}}\right) \det\!\left({{\psi} \over {2
\pi}}\Big|_{\bar\CN_\alpha}\right)^{-1}\,.}}
\countdef\eqEulers=182\eqEulers=\pageno
The integral in \SFTWYIITP\ immediately becomes 
\eqn\SFTWYII{\eqalign{ 
Z\big(\epsilon;C,R\big)\Big|_{\CO_\alpha/G} \,&=\, {{(2\pi\epsilon)}
\over {\Vol(G_\alpha)}} \int_{\BR \times \Fg_\alpha \times \BR} \left[{{dp}
\over {2 \pi}}\right] \left[{{d\phi} \over {2 \pi}}\right] \left[{{da}
\over {2 \pi}}\right] e\big(\bar\CA\big) \cdot e\big(L\CO_\alpha\big)
\; \times\cr
&\qquad\qquad\qquad\times\,\exp{\!\left[-i a \,-\, i \epsilon 
\langle\alpha,\phi\rangle \,+\, {{i \epsilon}\over 2} \left({d \over
\RP}\right) \Tr(\phi^2) \,+\, i \epsilon p a\right]}.}}
The essence of localization on $\CO_\alpha/G$ now lies
in evaluating $e\big(\bar\CA\big)$ and $e\big(L\CO_\alpha\big)$.

\bigskip\noindent{\it Evaluating $e\big(\bar\CA\big)$}\smallskip

Of the determinants in \DETSPH, $e(\bar\CA)$ is by far the more
delicate to compute.  Thankfully, we have already evaluated
$e(\bar\CA)$ in $(5.90)$ of \BeasleyVF, where we used the standard,
but slightly {\it ad hoc}, technique of zeta/eta-function
regularization to define the infinite products of eigenvalues of
$\psi$ acting on $\CE_0$ and $\CE_1$.  Because the central generator
$a$ of $U(1)_\RZ$ acts trivially, $e(\bar\CA)$ depends only on the
generators $(p,\phi)$ of ${U(1)_\RR \times G_\alpha}$ and is given
by\foot{Strictly speaking, we computed $e(\bar\CA)$ in \BeasleyVF\ for
the case that $\phi$ lies in the Lie algebra of $G$, as opposed to the
subgroup ${G_\alpha \subset G}$, but the result in \BeasleyVF\
immediately specializes.}
\eqn\DLETAIV{\eqalign{
&e\big(\bar\CA\big) \,=\, \exp{\left(-{{i \pi} \over 2} \, \eta_0(0)\right)}
\cdot {{\left(2 \pi\right)^{\Delta_G}} \over {(p \, \sqrt{\RP})^{\Delta_T}}}
\,\times\,\cr
&\times\, \exp{\!\left[{{i \, \check{c}_\Fg} \over {4 \pi p^2}} 
\left({d \over \RP}\right) \Tr(\phi^2)\right]} \, \prod_{\beta > 0}
\, \langle\beta, \phi\rangle^{-2} \, \left[2 \sin\!\left({{\langle\beta,
\phi\rangle} \over {2 p}}\right)\right]^{2 - N} \, \prod_{j=1}^N \,
\left[2 \sin\!\left({{\langle\beta, \phi\rangle} \over {2 a_j 
p}}\right)\right]\,,\cr
&\Delta_G \,=\, \dim G\,,\qquad \Delta_T \,=\,
\dim T\,.\cr}}
Here we recall that ${T\subset G}$ is a maximal torus, and in writing
this formula for $e(\bar\CA)$, we assume without loss that $\phi$ lies
in the associated Cartan subalgebra $\Ft$.  Each ${\beta > 0}$ is then
a positive root of $G$, and $\langle\,\cdot\,,\,\cdot\,\rangle$ is the
canonical dual pairing.

One very interesting feature of $e(\bar\CA)$ is the appearance of a 
phase proportional to $\Tr(\phi^2)$ in the second line of \DLETAIV.
This phase also involves the dual Coxeter number $\check{c}_\Fg$ of
the Lie algebra $\Fg$, which for convenience we take to be
simply-laced.\foot{See $(5.88)$ of \BeasleyVF\ for a description of
the $\phi$-dependent phase which does not require $\Fg$ to be
simply-laced.}  As we discussed in detail in \BeasleyVF, the
$\phi$-dependent phase in $e(\bar\CA)$ ultimately leads to the famous
quantum shift ${k \rightarrow k + \check{c}_\Fg}$ in the Chern-Simons
level.  We will shortly encounter a closely related quantum effect
when we compute $e(L\CO_\alpha)$.

Finally, $\eta_0(0)$ in \DLETAIV\ is a rather subtle constant that
arises when we introduce an eta-function to define the phase of
$e(\bar\CA)$.  In the simplest case that $M$ is a circle
bundle of degree $n$ over a smooth Riemann surface $\Sigma$, we
computed $\eta_0(0)$ in $(5.83)$ of \BeasleyVF\ to be 
\eqn\ETAZ{ \eta_0(0) \,=\, -{{n \, \Delta_G}\over 6}\,,\qquad\qquad
\Sigma\;\;\hbox{smooth}\,,}
independent of the genus of $\Sigma$.

On the other hand, to describe a non-trivial torus knot ${\SK_{{\bf
p},{\bf q}} \subset S^3}$, we consider a Seifert structure on
$S^3$ for which the base ${\Sigma = \BW\BC\BP^1_{{\bf p},{\bf q}}}$ is a
non-trivial orbifold.  Because we ultimately wish to compare the
empirical formula for $Z(\epsilon;\SK_{{\bf p},{\bf q}},\CMj)$ in
\JPTII\ with precise results of localization, we need to determine the
value of $\eta_0(0)$ in the orbifold case as well.

Here we are in luck.  By construction, $\eta_0(0)$ is the
eta-invariant associated to a certain ``adiabatic'' Dirac  
operator considered by Nicolaescu \Nicolaescu\ in the context of
Seiberg-Witten theory on a four-manifold bounding the Seifert manifold
$M$.  According to Proposition $1.4$ of \Nicolaescu, the value of
$\eta_0(0)$ for a general Seifert manifold $M$ is given by 
\eqn\ETAZII{ \eta_0(0) \,=\, {{\Delta_G} \over 6} \left[-c_1(\CL)
\,+\, 12 \, \sum_{j=1}^N \, s(b_j,
a_j)\right].}\countdef\Etainv=183\Etainv=\pageno 
In \ETAZII, $\CL$ is the line $V$-bundle associated to the
Seifert presentation of $M$.  Explicitly in terms of Seifert
invariants, 
\eqn\ETAZET{ c_1(\CL) \,=\, n \,+\, \sum_{j=1}^N {{b_j}\over {a_j}}
\,=\, {d\over \RP}\,,}
where the final equality in \ETAZET\ holds when $M$ is a Seifert
homology sphere or a cyclic $\BZ_d$ quotient thereof. 

Of particular note, the general formula \ETAZII\ for $\eta_0(0)$ now
involves the Dedekind sum $s(b_j, a_j)$.  The appearance of
such a complicated arithmetic object in a one-loop determinant may
seem rather mysterious, but it is crucial if we are to make contact
with empirical formulae such as \ZLWRZII, in which the Dedekind sum
enters through the phase $\theta_0$.  That said, at the moment we do
not wish to divert the exposition to review the complete derivation of 
\ETAZII.  Instead, we refer the interested reader to Appendix C for
a self-contained computation of the adiabatic eta-invariant $\eta_0(0)$.

\bigskip\noindent{\it Evaluating $e(L\CO_\alpha)$}\smallskip

We are left to evaluate the product of determinants associated to the
free loopspace $L\CO_\alpha$,
\eqn\DETELO{ e\big(L\CO_\alpha\big) \,=\, \det\!\left({{\psi} \over {2
\pi}}\Big|_{\Fg^{(1,0)}}\right) \, \det\!\left({{\psi} \over {2
\pi}}\Big|_{\bar\CN_\alpha}\right)^{-1}\,.}
Eventually these determinants, along with the moment map on
$\CO_\alpha$ which enters the argument of the exponential in \SFTWYII, 
will determine the invariant function of $\phi$ which represents the
Seifert loop operator under localization at the trivial connection on $M$.

According to \DLHG, the vector field generated by an arbitrary element
${\psi \equiv (p,\phi,a)}$ in the Lie algebra of the Hamiltonian group
$\CH$ is given on $L\CO_\alpha$ by    
\eqn\DLHGII{ \delta g \,=\, p \, \lie_\RR g\,+\, \phi|_C \cdot g\,.}
Upon restriction to the stabilizer ${\Fh_0^\alpha = \BR \oplus \Fg_\alpha
\oplus \BR}$, the generator $\psi$ consequently acts on both the
finite-dimensional vector space $\Fg^{(1,0)}$ and the infinite-dimensional
vector space $\bar\CN_\alpha$ as the first-order differential operator 
\eqn\OPPSI{ \CD_{(p,\phi)} \,=\, p \, \lie_\RR \,+\,
[\phi,\,\cdot\,]\,.}
In complete analogy to the evaluation of $e(\bar\CA)$ in \S $5.2$ of
\BeasleyVF\ (see also Appendix C), we now compute the respective
determinants of $\CD_{(p,\phi)}$ acting on $\Fg^{(1,0)}$ and $\bar\CN_\alpha$.

We begin by evaluating the numerator of \DETELO, which is merely a
finite-dimensional determinant.  Elements in $\Fg^{(1,0)}$ represent
holomorphic tangent vectors to $\CO_\alpha$ embedded as the space of 
constant maps in $L\CO_\alpha$.  Hence the Lie derivative $\lie_\RR$
annihilates $\Fg^{(1,0)}$, so trivially 
\eqn\DETZ{\eqalign{
\det\!\left({\psi \over {2\pi}}\Big|_{\Fg^{(1,0)}}\right) \,&=\,
\det\!\left({\CD_{(p,\phi)} \over {2\pi}}\Big|_{\Fg^{(1,0)}}\right),\cr
&=\,\det\!\left({{[\phi\,,\,\cdot\,]} \over
{2\pi}}\Big|_{\Fg^{(1,0)}}\right), \qquad\qquad \phi \in 
\Fg_\alpha\,.}}

To compute the latter determinant in \DETZ, we assume without loss that $\phi$
lies in the Cartan subalgebra ${\Ft\subseteq \Fg_\alpha}$ associated
to a maximal torus ${T \subseteq G_\alpha}$.\foot{If the weight
$\alpha$ happens to be regular, then ${\Fg_\alpha=\Ft}$ is
automatically such a Cartan subalgebra.}  With this assumption, the
adjoint action of $\phi$ is diagonalized in terms of the roots $\beta$
of $\Fg$ as ${[\phi,x_\beta] \,=\, i \, \langle \beta, \phi\rangle 
\, x_\beta}$, where $x_\beta$ is an element of the rootspace
$\Fe_\beta$.  Thus,
\eqn\DETEO{\eqalign{
&\det\!\left({{[\phi\,,\,\cdot\,]} \over
{2\pi}}\Big|_{\Fg^{(1,0)}}\right) \,=\,
\prod_{(\beta_{+},\,\alpha)>0} \left({i \over {2\pi}}\,
\langle\beta_{+},\phi\rangle\right) \,=\, \left({i \over 
{2\pi}}\right)^{(\Delta_G - \Delta_{G_\alpha})/2} \cdot
\prod_{(\beta_{+},\,\alpha)>0} \, \langle\beta_{+},\phi\rangle\,,\cr
&\qquad\qquad\Delta_G \,=\, \dim G\,,\qquad\qquad \Delta_{G_\alpha}
\,=\, \dim G_\alpha\,.}}
As indicated, the products in \DETEO\ run over those roots $\beta_{+}$ whose
associated rootspaces lie in the holomorphic tangent
space $\Fg^{(1,0)}$ to $\CO_\alpha$.  Equivalently, each root
$\beta_{+}$ satisfies ${(\beta_{+},\,\alpha)>0}$.  Of course, if
$\alpha$ is regular, the latter inequality just says that ${\beta_{+}
> 0}$ is a positive root.  At the opposite extreme, for ${\alpha = 0}$
and $\Fg^{(1,0)}$ empty, the product over $\beta_{+}$ is trivial.

Because the dimension of the normal bundle to $\CO_\alpha$ inside
$L\CO_\alpha$ is infinite, we must work somewhat harder to 
evaluate the denominator in \DETELO. We recall that $\bar\CN_\alpha$
decomposes as a direct sum of anti-holomorphic tangent spaces
$\Fg^{(0,1)}_{t}$ upon which $\lie_\RR$ acts with eigenvalue $-2\pi i\,
t$,
\eqn\NLOCII{ \bar\CN_\alpha \,=\, \bigoplus_{t \ge 1} \Big[\Fg^{(0,1)}_{t}
\oplus \Fg^{(0,1)}_{-t}\Big]\,.}
Diagonalizing the action of $\phi$ on each summand $\Fg^{(0,1)}_{t}$,
we write the determinant of $\CD_{(p,\phi)}$ acting on
$\bar\CN_\alpha$ formally as the product 
\eqn\DETEI{\eqalign{
\det\!\left({\psi \over {2\pi}}\Big|_{\bar\CN_\alpha}\right) \,&=\,
\det\!\left({\CD_{(p,\phi)}\over {2\pi}}\Big|_{\bar\CN_\alpha}\right)\,,\cr
&= \prod_{t \neq 0} \, \prod_{(\beta_{+},\alpha) > 0} 
\left[\left( -i t p \,-\, {i \over {2\pi}}
\langle\beta_{+},\phi\rangle\right)\right]\,,\cr
&=\, \exp{\!\left(-{{i \pi} \over 2} \, \delta(p,\phi) \right)} \times
\prod_{t \ge 1} \left| \left(t p\right)^{(\Delta_G - \Delta_{G_\alpha})} 
\prod_{(\beta_{+},\,\alpha) > 0} \left(1 \,-\,
\left({{\langle\beta_{+}, \phi\rangle} \over {2 \pi t
p}}\right)^2\right)\right|\,.\cr}}
In the second line of \DETEI, we express the determinant of $\phi$
acting on $\Fg^{(0,1)}_t$ as a product over positive roots
${\beta_+>0}$, with a crucial sign relative to the analogous
determinant in \DETEO\ to account for the exchange ${\Fg^{(1,0)}_t
\!\leftrightarrow\, \Fg^{(0,1)}_t}$.  In the final line
of \DETEI, we then encode the phase of the determinant through a
function $\delta(p,\phi)$ generally depending upon both $p$ and
$\phi$.  Otherwise, the terms which appear explicitly in the product
over the Fourier mode $t$ represent the norm.

As often the case for functional determinants, the norm in \DETEI\ is
much easier to evaluate than the phase $\delta(p,\phi)$, so we will
compute the norm first.

To evaluate the product over $t$, we recall the well-known
identity 
\eqn\PRODSIN{ {\sin(x)\over x} \,=\, \prod_{t \ge 1} \, \left(1 \,-\,
{{x^2}\over{\pi^2 t^2}}\right)\,,}
implying 
\eqn\PRODSINII{ \prod_{t\ge 1} \, \left(1 \,-\,
\left({{\langle\beta_{+},\phi\rangle} \over {2 \pi t
p}}\right)^2\right) \,=\, {{2 \, p}\over {\langle\beta_{+},\phi\rangle}} \,
\sin\!\left({{\langle\beta_{+},\phi\rangle}\over{2 p}}\right)\,.}
Just as for the computation in $(5.65)$ of \BeasleyVF, we then use the
Riemann zeta-function $\zeta(s)$ to define the trivial 
but infinite products 
\eqn\PRODZ{\eqalign{
\prod_{t\ge 1} \, p^{(\Delta_G - \Delta_{G_\alpha})} \,&=\,
\exp{\!\left[(\Delta_G - \Delta_{G_\alpha}) \ln p \cdot \zeta(0)\right]}
\,=\, p^{-(\Delta_G - \Delta_{G_\alpha})/2}\,,\cr
\prod_{t \ge 1} \, t^{(\Delta_G - \Delta_{G_\alpha})} \,&=\, 
\exp{\!\left[-(\Delta_G - \Delta_{G_\alpha}) \cdot \zeta'(0)\right]} \,=\,
(2\pi)^{(\Delta_G - \Delta_{G_\alpha})/2}\,.}}
Cancelling overall factors of $p$ from \PRODSINII\ and \PRODZ, we obtain 
\eqn\DETEII{ \det\!\left({\psi \over {2\pi}}\Big|_{\bar\CN_\alpha}\right)
\,=\, (2 \pi)^{(\Delta_G - \Delta_{G_\alpha})/2} \, \exp{\!\left(-{{i \pi}
\over 2} \, \delta(p,\phi) \right)} 
\prod_{(\beta_{+},\,\alpha)>0} \, \left|{2 \over 
{\langle\beta_{+}, \phi\rangle}} \,
\sin\!\left({{\langle\beta_{+}, \phi\rangle}\over{2 p}}\right)\right|.}

Our remaining task is to compute the phase $\delta(p,\phi)$ in \DETEI.
Naively, $\delta(p,\phi)$ is given by the sum 
\eqn\DIFFDEL{ \delta(p,\phi) \,\approx\, \sum_{t\neq 0} \,
\sum_{(\beta_{+},\,\alpha)>0} \,
\sgn\!\big(\lambda(t,\beta_{+})\big)\,,\qquad\qquad \lambda(t,\beta_{+}) \,=\, 
t \,+\, {{\langle\beta_{+}, \phi\rangle}\over{2\pi p}}\,.}
We have not written the expression in \DIFFDEL\ with an equality because
the sum over eigenvalues $\lambda(t,\beta)$ is ill-defined without a
regulator.  To make sense of \DIFFDEL, we follow the philosophy of
\APS\ and introduce an eta-function associated to the spectrum of
$\CD_{(p,\phi)}$.  We thus set 
\eqn\DIFFDELII{ \delta_{(p,\phi)}(s) \,=\, \sum_{t \neq 0} \,
\sum_{(\beta_{+},\,\alpha) > 0} \,
\sgn\!\big(\lambda(t,\beta_{+})\big) \, \big|\lambda(t,\beta_{+})\big|^{-s}\,.}
Here $s$ is a complex parameter.  When the real part of $s$ is
sufficiently large, the sum in \DIFFDELII\ is absolutely convergent,
so that $\delta_{(p,\phi)}(s)$ is defined in that case.  Otherwise,
$\delta_{(p,\phi)}(s)$ is defined by analytic continuation in the
$s$-plane.  Assuming that $\delta_{(p,\phi)}(s)$ remains finite as
${s \to 0}$, we then take 
\eqn\DIFFDELIII{ \delta(p,\phi) \,=\, \delta_{(p,\phi)}(0)\,.}

To evaluate $\delta_{(p,\phi)}(s)$ near ${s=0}$, we first expand the
sum in \DIFFDELII\ as 
\eqn\EVDLS{ \delta_{(p,\phi)}(s) \,=\, \sum_{t\ge 1} \,
\sum_{(\beta_{+},\,\alpha)>0} \, {1 \over {\left(t \,+\,
{{\langle\beta_{+},\phi\rangle}\over{2\pi p}}\right)^s}} \,-\,
\sum_{t\ge 1} \, \sum_{(\beta_{+},\,\alpha)>0} \, {1 \over
{\left(t \,-\,{{\langle\beta_{+},\phi\rangle}\over{2\pi p}}\right)^s}}\,.}
Here we assume that $p$ and $\phi$ satisfy 
\eqn\ASBET{ 0 \,<\, {{\langle\beta_{+},\phi\rangle}\over{2 \pi p}} \,<\,
1\,,}
for each root $\beta_{+}$ appearing in the sum.  Otherwise, when the
quantity in \ASBET\ shifts by an integer, the phase $\exp(-i
\pi \, \delta(p,\phi) / 2)$ is multiplied by a sign $\pm 1$,
depending upon the parity of the shift.  This sign effectively removes
the absolute value bars $|\cdot|$ appearing in \DETEII, so that the
determinant of $\CD_{(p,\phi)}$ depends analytically on $p$ and
$\phi$ as one might naively expect.

We now apply the binomial expansion to the denominators in \EVDLS\ to
obtain 
\eqn\EVDLSII{ \delta_{(p,\phi)}(s) \,=\, \sum_{t \ge 1} \,
\sum_{(\beta_{+},\,\alpha)>0} \, -{{2 s}
\over{t^{s+1}}} \cdot\!\left({{\langle\beta_{+},\phi\rangle}\over{2 \pi
p}}\right) \,+\, \sum_{t \ge 1} \, \sum_{(\beta_{+},\,\alpha)>0}
\, s \cdot \CO\!\left({1 \over {t^{s+2}}}\right)\,.}
In the process of expanding \EVDLS, we collect into $\CO(1/t^{s+2})$
all terms for which the sum over $t$ is absolutely convergent near
${s=0}$.  As a result, when we evaluate $\delta_{(p,\phi)}(s)$ at
${s=0}$, the last term in \EVDLSII\ vanishes.

On the other hand, we note that 
\eqn\EVDLSIII{  \sum_{t \ge 1} \, \sum_{(\beta_{+},\alpha)>0}
\, -{{2 s} \over{t^{s+1}}}
\cdot\!\left({{\langle\beta_{+},\phi\rangle}\over{2 \pi p}}\right)
\,=\, -s \, \zeta(1+s) \cdot \sum_{(\beta_{+},\alpha)>0} \! 
{{\langle\beta_{+},\phi\rangle}\over{\pi p}}\,,}
where $\zeta$ is again the Riemann zeta-function.  Because
$\zeta(1+s)$ has a simple pole with unit residue at ${s=0}$, we see 
that \EVDLSIII\ makes a non-zero contribution to
$\delta_{(p,\phi)}(0)$.  Specifically, 
\eqn\DLPPH{ \delta_{(p,\phi)}(0) \,=\,
-{{2 \, \big\langle\rho^{[\alpha]},\phi\big\rangle}\over{\pi p}}\,\mod
2\,,}
where we introduce the generalized Weyl vector 
\eqn\RHOAL{ \rho^{[\alpha]} \,=\, \ha \, \sum_{(\beta_{+},\,\alpha)>0}
\beta_{+}\,.}\countdef\Rhoalpha=184\Rhoalpha=\pageno

If $\alpha$ is regular, the sum which defines $\rho^{[\alpha]}$ runs
over all positive roots of $\Fg$, and ${\rho^{[\alpha]} = \rho}$ is
the distinguished weight which appeared previously in \WEYLRH.  This
observation motivates the otherwise extraneous factor of $1/2$ in the
definition of $\rho^{[\alpha]}$.  Otherwise, if $\alpha$ is not
regular, the sum which defines $\rho^{[\alpha]}$ runs over only
the subset of positive roots whose corresponding rootspaces lie in
$\Fg^{(1,0)}$.  In this situation, $\rho^{[\alpha]}$ need not be a
weight of $G$,  as for instance when ${G=SU(3)}$ and $\alpha$ is the
weight of the fundamental representation ${\bf 3}$.  Trivially, in the
extreme case ${\alpha = 0}$, then ${\rho^{[\alpha]} = 0}$ as well.

Finally, we emphasize that our computation of $\delta_{(p,\phi)}(0)$
is strictly valid modulo $2$, due to the assumption in \ASBET.  As we
mentioned before, a complete formula for $\delta_{(p,\phi)}(0)$ also
includes locally-constant terms which vanish mod $2$ and which
compensate for the absolute value bars $|\,\cdot\,|$ appearing in
\DETEII.

Physically, the non-trivial value for $\delta_{(p,\phi)}(0)$ in
\DLPPH\ can be understood as a finite renormalization effect, due to
the divergence in the naive sum over eigenvalues in \DIFFDEL.  In
complete analogy to the $\phi$-dependent phase appearing in \DLETAIV,
the determinant of ${\psi \equiv \CD_{(p,\phi)}}$ acting on
$\CN_\alpha$ acquires a similar $\phi$-dependent quantum phase,
\eqn\DETEIII{ \det\!\left({\psi \over {2\pi}}\Big|_{\bar\CN_\alpha}\right)
\,=\, (2 \pi)^{(\Delta_G - \Delta_{G_\alpha})/2} \, \exp{\!\left(i
\,{{\langle\rho^{[\alpha]},\phi\rangle}\over p}\right)}
\prod_{(\beta_{+},\,\alpha)>0} \, {2 \over {\langle\beta_{+},\phi\rangle}} \,
\sin\!\left({{\langle\beta_{+},\phi\rangle}\over{2 p}}\right).}

Taking the ratio between the determinants in \DETEO\ and \DETEIII,
we see that $e(L\CO_\alpha)$ is given succinctly by 
\eqn\FDETELO{\eqalign{
&e\big(L\CO_\alpha\big) \,=\,\left({1 \over
{2\pi}}\right)^{(\Delta_G \,-\, \Delta_{G_\alpha})} \, 
\exp{\!\left[{{i\pi}\over 4} (\Delta_G - \Delta_{G_\alpha}) \,-\, i \,
{{\langle\rho^{[\alpha]},\phi\rangle}\over p}\right]}\,\times\,\cr
&\qquad\qquad\,\times\,
\prod_{(\beta_{+},\,\alpha)>0}
{\langle\beta_{+},\phi\rangle^2}\cdot\left[2\,
\sin\!\left({{\langle\beta_{+},\phi\rangle}\over{2
p}}\right)\right]^{-1},\cr
&\Delta_G \,=\, \dim G\,,\qquad\qquad \Delta_{G_\alpha} \,=\, \dim
G_\alpha\,.}}

\bigskip\noindent{\it Simplifying the Integral over $\Fh_0^\alpha$}\smallskip

As manifest in \DLETAIV\ and \FDETELO, neither $e(\bar\CA)$ nor
$e(L\CO_\alpha)$ depends upon the variable $a$ which parametrizes the
Lie algebra of $U(1)_\RZ$.  Because $U(1)_\RZ$ acts in a completely
trivial fashion on $\bar\CA_\alpha$, the result could hardly have
been otherwise.  Yet this observation does have an important
consequence.

We recall from \SFTWYII\ that the local contribution from
${\{\varrho_{\rm ab}\} \cong \CO_\alpha/G}$ to the Seifert loop path
integral is given by  
\eqn\SFTWYIII{\eqalign{ 
Z\big(\epsilon;C,R\big)\Big|_{\CO_\alpha/G} \,&=\, {{(2\pi\epsilon)}
\over {\Vol(G_\alpha)}} \, \int_{\BR \times \Fg_\alpha \times \BR} \left[{{dp}
\over {2 \pi}}\right] \left[{{d\phi} \over {2 \pi}}\right] \left[{{da}
\over {2 \pi}}\right] \, e\big(\bar\CA\big) \cdot 
e\big(L\CO_\alpha\big) \; \times\cr
&\qquad\qquad\qquad\times\,\exp{\!\left[-i a \,-\, i \epsilon
\langle\alpha,\phi\rangle \,+\, {{i \epsilon}\over 2} \left({d \over 
\RP}\right) \Tr(\phi^2) \,+\, i \epsilon p a\right]}\,.}}
Since $a$ enters the integrand of \SFTWYIII\ only linearly in the
argument of the exponential, we can immediately integrate over $a$
using the elementary identity 
\eqn\DELTAFCN{ \int_{-\infty}^{+\infty} dy \, \exp{\!(-i x y)} \,=\, 2
\pi \, \delta(x)\,.}
Hence the integral over $a$ yields a delta-function ${2\pi \delta
(1 - \epsilon p)}$.  

Next, we use the delta-function to perform the integral over $p$,
thereby setting ${p = 1/\epsilon}$.  In the process, the prefactor of
$2\pi\epsilon$ which appears in the normalization of \SFTWYIII\ is
cancelled, and the integral over ${\BR \oplus \Fg_\alpha \oplus \BR}$
reduces to an integral over $\Fg_\alpha$ alone,
\eqn\SFTWYIV{\eqalign{ 
Z\big(\epsilon;C,R\big)\Big|_{\CO_\alpha/G} \,&=\,
\exp{\!\left[-{{i\pi}\over 2}\left(\eta_0(0) \,-\, \ha 
(\Delta_G - \Delta_{G_\alpha})\right)\right]} \, {1
\over{\Vol(G_\alpha)}} \left({\epsilon \over
\sqrt{\RP}}\right)^{\Delta_T}\,\times\,\cr
&\times \int_{\Fg_\alpha}\!\left[d\phi\right] \, \exp{\!\left[ -i
\epsilon\,\big\langle\alpha + \rho^{[\alpha]},\phi\big\rangle \,+\,
{{i \epsilon}\over 2} \left({d \over \RP}\right) \left(1 \,+\,
{{\epsilon\,\check{c}_\Fg}\over{2 \pi}}\right) \Tr(\phi^2)
\right]}\,\times\,\cr
&\times\,\quad\prod_{\beta > 0}
\, \langle\beta, \phi\rangle^{-2} \, \left[2
\sin\!\left({{\epsilon\,\langle\beta,\phi\rangle} \over
{2}}\right)\right]^{2 - N} \, \prod_{j=1}^N \, 
\left[2 \sin\!\left({{\epsilon\, \langle\beta,\phi\rangle} \over {2
a_j}}\right)\right]\,\times\,\cr
&\times\quad\prod_{(\beta_{+},\,\alpha)>0} 
\langle\beta_{+},\phi\rangle^2 \, \left[2 \sin\!\left({{\epsilon \,
\langle\beta_{+},\phi\rangle}\over 2}\right)\right]^{-1}\,.}}  
Here we have substituted the expressions for $e(\bar\CA)$ and
$e(L\CO_\alpha)$ in \DLETAIV\ and \FDETELO.  Also, as a word of caution, we
emphasize that the products over $\beta$ and $\beta_+$ in \SFTWYIV\ run
over distinct sets of roots whenever $\alpha$ is not regular.

Let us simplify \SFTWYIV\ a bit further.

If the weight $\alpha$ is regular, then ${\Fg_\alpha\cong\Ft}$, and the
integral in \SFTWYIV\ automatically runs over the Cartan subalgebra of
$G$.  More generally, to reduce \SFTWYIV\ to an integral over $\Ft$
even when $\alpha$ is not regular, we note that the integrand is a
function on $\Fg_\alpha$ which is invariant under the adjoint action
of $G_\alpha$.  This invariance ultimately follows from the trivial
invariance of $\alpha$ under the Weyl group $\FW_\alpha$ of
$G_\alpha$.  Here we think of $\FW_\alpha$ as a subgroup of the Weyl
group $\FW$ of $G$, and we note that $\FW_\alpha$ preserves the set of
roots $\beta_{+}$ satisfying ${(\beta_{+},\alpha) > 0}$.  Hence both
$\rho^{[\alpha]}$ and the product over roots $\beta_{+}$ in the last line of
\SFTWYIV\ are invariant under $\FW_\alpha$.  The remaining terms in the
integrand of \SFTWYIV\ are manifestly invariant under $\FW_\alpha$,
from which we deduce invariance under the group $G_\alpha$.

Because the integrand of \SFTWYIV\ is invariant under the adjoint
action of $G_\alpha$, we can apply the Weyl integral formula to reduce
the integral from $\Fg_\alpha$ to $\Ft$.  In its infinitesimal
version, the Weyl integral formula generally states that if $f$ is a
function on a Lie algebra $\Fg$ invariant under the adjoint action of
a group $G$, then 
\eqn\WEYLINT{ \int_\Fg \left[d\phi\right] \, f(\phi) \,=\, {1 \over
|\FW|} \, {\Vol(G) \over \Vol(T)} \, \int_\Ft
\left[d\phi\right] \, \prod_{\beta > 0} \langle\beta,\phi\rangle^2 \,
f(\phi)\,.}
Here $|\FW|$ is the order of the Weyl group of $G$, and the product
over positive roots $\beta$ of $G$ appearing on the right in \WEYLINT\
is a Jacobian factor generalizing the classical van der Monde
determinant.  

In the case at hand, we want to apply the Weyl integral formula
\WEYLINT\ not for $G$ but for $G_\alpha$.  The roots of $G_\alpha$ are
precisely those roots $\beta_\perp$ of $G$ orthogonal to $\alpha$ in
the invariant metric on $\Ft^*$, such that  
\eqn\ROOTSGA{ (\beta_\perp\,, \alpha) \,=\, 0\,.}
Consequently, when we apply the Weyl integral formula to reduce
the integral in \SFTWYIV\ from $\Fg_\alpha$ to $\Ft$, the Weyl
Jacobian for $G_\alpha$ conspires to cancel against the following
product of factors in \SFTWYIV, 
\eqn\WEYLINTII{ \prod_{\beta > 0} \, \langle\beta,\phi\rangle^{-2} \,\cdot
\prod_{(\beta_{+},\,\alpha)>0} \, \langle\beta_{+},\phi\rangle^2 \,=\,
\prod_{\beta_\perp > 0} \, \langle\beta_\perp,\phi\rangle^{-2}\,,}
implying 
\eqn\SFTWYV{\eqalign{ 
Z\big(\epsilon;C,R\big)\Big|_{\CO_\alpha/G} \,&=\, 
\exp{\!\left[-{{i\pi}\over 2}\left(\eta_0(0) \,-\, \ha
(\Delta_G - \Delta_{G_\alpha})\right)\right]} {1 \over
{|\FW_\alpha|}} {1 \over{\Vol(T)}} \left({1 \over
\sqrt{\RP}}\right)^{\Delta_T}\times\cr
&\times \int_{\Ft}\!\left[d\phi\right] \, \exp{\!\left[ -i
\,\big\langle\alpha + \rho^{[\alpha]},\phi\big\rangle \,+\,
{i \over {2\epsilon_{\rm r}}} \left({d \over \RP}\right) \Tr(\phi^2)
\right]}\,\times\,\cr
&\times\, \prod_{\beta > 0} \left[2
\sin\!\left({{\langle\beta,\phi\rangle} \over 
{2}}\right)\right]^{2 - N} \, \prod_{j=1}^N \, \left[2
\sin\!\left({{\langle\beta,\phi\rangle} \over {2
a_j}}\right)\right]\,\times\,\cr 
&\times\, \prod_{(\beta_{+},\,\alpha)>0} 
\left[2 \sin\!\left({{\langle\beta_{+},\phi\rangle}\over
2}\right)\right]^{-1}\,.}}
In passing to \SFTWYV, we have performed a change of variables
${\phi\mapsto\epsilon\,\phi}$ to remove extraneous factors of $\epsilon$.
In the process, we introduce the renormalized coupling $\epsilon_{\rm r}$,
\eqn\EPSILONR{ \epsilon_{\rm r} \,=\, {{2\pi}\over{k +
\check{c}_\Fg}}\,,}
to absorb the explicit shift in the coefficient of $\Tr(\phi^2)$ that
arises from $e(\bar\CA)$ in \DLETAIV.  Also, as hopefully clear,
$|\FW_\alpha|$ denotes the order of the Weyl group of $G_\alpha$.  If
${G_\alpha = T}$ is abelian, then $\FW_\alpha$ is trivial and
${|\FW_\alpha|=1}$.

As a small check, if ${\alpha = 0}$, so that no Wilson loop operator
is actually present in the path integral, the expression for 
$Z\big(\epsilon;C,R\big)\big|_{\CO_\alpha/G}$ in \SFTWYV\ directly
reduces to the expression in $(5.97)$ of \BeasleyVF\ for the local
contribution $Z(\epsilon)\big|_{\{0\}}$ from the trivial connection to
the Chern-Simons partition function.  

Following \BeasleyVF, we now make two further substitutions to relate
the formula in \SFTWYV\ to the empirical result in \ZLWRZII.  First,
we rotate the contour of integration from ${\Ft \equiv \Ft \times
\BR}$ to ${\Ft \times \e{-{{i\pi} \over 4}}}$.  Second, we make
a change of variables ${\phi\mapsto i\,\phi}$.  Hence, 
\eqn\SFTWYV{\eqalign{ 
Z\big(\epsilon;C,R\big)\Big|_{\CO_\alpha/G} \,&=\,
\exp{\!\left(-{{i\pi}\over 2} \, \eta_0(0)\right)} \, {1 \over
{|\FW_\alpha|}} \, {{(-1)^{(\Delta_{G_\alpha} - \Delta_T)/2}}\over{\Vol(T)}}
\left({1 \over {i\sqrt{\RP}}}\right)^{\Delta_T}\times\cr
&\times \int_{\Ft \times \CC^{(0)}}\!\left[d\phi\right] \,
\exp{\!\left[-\big\langle\alpha + \rho^{[\alpha]}, \phi\big\rangle \,-\, 
{i \over {2\epsilon_{\rm r}}} \left({d \over \RP}\right) \Tr(\phi^2)
\right]}\,\times\,\cr
&\times\qquad \prod_{\beta > 0} \left[2
\sinh\!\left({{\langle\beta,\phi\rangle} \over 
{2}}\right)\right]^{2 - N} \, \prod_{j=1}^N \, \left[2
\sinh\!\left({{\langle\beta,\phi\rangle} \over {2
a_j}}\right)\right]\,\times\,\cr 
&\times\qquad \prod_{(\beta_{+},\alpha)>0} \left[2
\sinh\!\left({{\langle\beta_{+},\phi\rangle}\over
2}\right)\right]^{-1}\,.}}
This contour integral should be compared to $(5.99)$ in \BeasleyVF, to
which \SFTWYV\ reduces when ${\alpha = \rho^{[\alpha]} = 0}$.

\bigskip\noindent{\it The Seifert Loop Operator as a Character}\smallskip

We now arrive at the second main result of this paper --- the
interpretation of the Seifert loop operator as a character of
$G$, via the Weyl character formula.  The appearance of the character
formula in the context of localization on the loopspace $L\CO_\alpha$
is not so surprising, given the classic Atiyah-Bott derivation 
\AtiyahRBII\ of the character formula from index theory on
$\CO_\alpha$ and the general relation between index theory on a
manifold and localization on its loopspace \refs{\AtiyahCS,\DeligneII}.
Nonetheless, the emergence of the character formula from the Seifert
loop path integral is quite satisfying.  See also
\refs{\StoneFU,\AlvarezZV} for earlier and somewhat more direct
quantum mechanical derivations of the character formula.

To start, let us quickly recall the essentials of the character
formula.  See for instance \S $24$ of \Fulton\ for a more thorough
introduction to the Weyl character formula and a few of its mathematical
applications.

We first introduce some notation.  For each weight $\alpha$ of $G$, we
define an alternating function $\A_\alpha$ on the Cartan subalgebra 
${\Ft\subset\Fg}$ by  
\eqn\BIGA{ \A_\alpha(\phi) \,=\, \sum_{w \in \FW} \, (-1)^w \,
\e{\!\langle w\cdot\alpha,\,\phi\rangle}\,,\qquad \phi\,\in\,\Ft\,.}
\countdef\BigA=185\BigA=\pageno
Here the sum runs over all elements $w$ in the Weyl group $\FW$ of $G$, with
a sign $(-1)^w$ defined as follows.  Because $\Ft$ carries a metric
invariant under $\FW$, each element ${w\in\FW}$ acts as an orthogonal
transformation with respect to any orthonormal basis of $\Ft$.  Given
such a basis, we then set ${(-1)^w \equiv \det w = \pm 1}$, depending
upon whether $w$ preserves or reverses the orientation of $\Ft$.  As
usual, ${w\cdot\alpha}$ indicates the image of $\alpha$ 
under the dual action of $\FW$ on $\Ft^*$.  Last but not least,
because of the sign $(-1)^w$ appearing in \BIGA, $\A_\alpha$ is
alternating under the Weyl action on $\Ft$, 
\eqn\BIGAA{ \A_\alpha(w\cdot \phi) \,=\, (-1)^w \, \A_\alpha(\phi)\,,
\qquad w\,\in\,\FW\,.}

In terms of $\A_\alpha$, the character $\ch_R$ associated to each 
irreducible representation $R$ of $G$ has a remarkably simple 
description.  As a function on the maximal torus ${T \subset G}$,
the character is given by the ratio 
\eqn\WCF{ \ch_R\big(\e{\phi}\big) \,=\,
{{\A_{\alpha+\rho}(\phi)}\over{\A_{\rho}(\phi)}}\,,\qquad\qquad 
\e{\phi}\in T\,.}\countdef\ChR=186\ChR=\pageno
As throughout, $\alpha$ is the highest weight of $R$, and $\rho$ is
the Weyl vector \WEYLRH\ given by half the sum of the positive roots
of $G$.

Though we will not prove the character formula here, let us make a few  
basic comments about it.  First, because $\A_\alpha$ is alternating under
$\FW$, the ratio appearing on the right of \WCF\ is invariant under
$\FW$, as required of any character when restricted to $T$.  Second,
because $\rho$ is a weight of $G$ (a slightly non-trivial fact), the
expression for $\ch_R$ in \WCF\ is well-defined as a function on $T$,
just as for the discussion surrounding \HOMRHO.  Finally, the
character formula in \WCF\ is manifestly correct in the special case
${\alpha = 0}$, for which $R$ is trivial and ${\ch_R = 1}$.  Also,
when ${G=SU(2)}$ and ${\FW = \BZ_2}$ acting by reflection, one can
readily check that \WCF\ reproduces the formula for $\ch_\CMj$ in
\CHNII\ with a suitable choice of coordinate on ${\Ft\cong\BR}$.

Let us now interpret our result \SFTWYV\ for
${Z\big(\epsilon;C,R\big)\big|_{\CO_\alpha/G}}$ in light of the
character formula.  This interpretation is slightly more
straightforward when $\alpha$ is a regular weight of $G$, so we 
specialize to the regular case first.

When $\alpha$ is regular, ${G_\alpha = T}$, ${|\FW_\alpha|=1}$, and
$\rho^{[\alpha]}$ reduces to the Weyl vector $\rho$ 
itself.  Also, the product over roots $\beta_{+}$ satisfying
${(\beta_{+},\alpha) > 0}$ in \SFTWYV\ is simply the product over all
positive roots ${\beta > 0}$ of $G$.  As a result, the final factor in the
integrand of \SFTWYV\ reduces to the Weyl denominator $\A_\rho$, 
\eqn\ARHO{ \A_\rho(\phi) \,=\, \prod_{\beta > 0} \, 2
\sinh\!\left({{\langle\beta,\phi\rangle}\over 2}\right).}
See Lemma $24.3$ of \Fulton\ for a proof of this well-known 
product formula for the Weyl denominator.  Thus for regular weights,
\eqn\SFTWYVI{\eqalign{ 
&Z\big(\epsilon;C,R\big)\Big|_{\CO_\alpha/G} \,=\,
\exp{\!\left[-{{i\pi}\over 2} \eta_0(0)\right]} \,
{1\over{\Vol(T)}} \left({1
\over {i \sqrt{\RP}}}\right)^{\Delta_T}\times\cr 
&\qquad\qquad\times\,\int_{\Ft \times \CC^{(0)}}\!\left[d\phi\right] \,
{1 \over {\A_\rho(\phi)}}\,\exp{\!\left[-\big\langle\alpha +
\rho,\phi\big\rangle \,-\, {i \over {2\epsilon_{\rm r}}} \left({d
\over \RP}\right) \Tr(\phi^2) \right]}\,\times\,\cr
&\qquad\qquad\times\,\prod_{\beta > 0} \left[2
\sinh\!\left({{\langle\beta,\phi\rangle} \over 
{2}}\right)\right]^{2 - N} \, \prod_{j=1}^N \, \left[2
\sinh\!\left({{\langle\beta,\phi\rangle} \over {2
a_j}}\right)\right]\,,\quad\quad \alpha \hbox{ regular}\,.}}

To simplify \SFTWYVI\ further, we make an elementary observation
regarding discrete symmetries.  By definition, the measure $[d\phi]$ in
the contour integral is invariant under the Weyl group $\FW$ of $G$.
Moreover, the integrand of \SFTWYVI\ can generally be decomposed as a sum of
terms, each of which transforms in a one-dimensional representation of
$\FW$.  Then since $[d\phi]$ is Weyl invariant, only the Weyl
invariant piece of the integrand actually contributes to the integral
over $\phi$.

Now, of the various factors in that integrand, the quadratic function
$\Tr(\phi^2)$ appearing in the argument of the exponential in
\SFTWYVI\ is obviously Weyl invariant.  Since $\FW$ is
generated by reflections in the root lattice of $G$, the
expression in the last line of \SFTWYVI\ is also Weyl invariant, as it
arises from a product over all positive roots ${\beta > 0}$ of the even
function 
\eqn\EVNFCN{\eqalign{
F_\beta(\phi) \,&=\, \left[2
\sinh\!\left({{\langle\beta,\phi\rangle} \over 
{2}}\right)\right]^{2 - N} \, \prod_{j=1}^N \, \left[2
\sinh\!\left({{\langle\beta,\phi\rangle} \over {2
a_j}}\right)\right]\,,\cr
F_\beta(\phi) \,&=\, F_\beta(-\phi)\,=\,F_{-\beta}(\phi)\,.}}
So in the integrand of \SFTWYVI, we are left to consider the factor
\eqn\BIGS{
S_\alpha(\phi)\,=\,{{\e{\!-\langle\alpha+\rho,\,\phi\rangle}}
\over{\A_\rho(\phi)}}\,.} 

By construction, the Weyl denominator $\A_\rho(\phi)$ is alternating
under $\FW$.  Therefore, only the alternating piece of the numerator
$\exp{\![-\langle\alpha+\rho,\,\phi\rangle]}$ in $S_\alpha(\phi)$
actually contributes to the contour integral over $\phi$ in 
\SFTWYVI.  We immediately recognize that alternating piece to be 
\eqn\AEXP{\eqalign{
\A\!\left[\e{\!-\langle\alpha + \rho,\phi\rangle}\right]
&\equiv\, {1 \over {|\FW|}} \, \sum_{w \in \FW} \, (-1)^w \,
\e{\!\langle w\cdot(\alpha + \rho),\,-\phi\rangle}\,,\cr
&=\, {1 \over {|\FW|}} \A_{\alpha + \rho}(-\phi)\,,\cr
&=\, (-1)^{(\Delta_G - \Delta_{T})/2} \cdot {1 \over {|\FW|}} \,
\A_{\alpha + \rho}(\phi)\,.}}
Here $\A\!\left[\,\cdot\,\right]$ denotes anti-symmetrization under
$\FW$, and we have been careful to divide by the order of $\FW$ to ensure
that the anti-symmetrization in \AEXP\ is properly normalized.
Also, the overall sign $(-1)^{(\Delta_G - \Delta_T)/2}$ in
the last line of \AEXP\ accounts for reflecting $-\phi$ to $\phi$ 
in the argument of $\A_{\alpha+\rho}$, just as in \BIGAA.\foot{The
same sign under reflection can be seen directly, for instance, in the Weyl
denominator formula \ARHO.}

Without loss, we replace $S_\alpha(\phi)$ in the integrand of
\SFTWYVI\ with the Weyl-invariant function 
\eqn\BIGSII{ S_\alpha(\phi)
\buildrel{\,\,\left[\,\cdot\,\right]^\FW}\over{\longmapsto} 
{{(-1)^{(\Delta_G - \Delta_T)/2}}\over{|\FW|}}
\cdot {{\A_{\alpha+\rho}(\phi)}\over{\A_\rho(\phi)}}\,.}
Via the character formula in \WCF, we finally obtain the following 
elegant result for the contribution of ${\{\varrho_{\rm ab}\} \cong
\CO_\alpha/G}$ to the Seifert loop path integral,
\eqn\SFTWYVII{\eqalign{ 
&Z\big(\epsilon;C,R\big)\Big|_{\CO_\alpha/G} \,=\,
\exp{\!\left[-{{i\pi}\over 2} \eta_0(0)\right]} \, {1 \over {|\FW|}}
\, {{(-1)^{(\Delta_G - \Delta_T)/2}} \over{\Vol(T)}} \left({1 \over
{i\sqrt{\RP}}}\right)^{\Delta_T}\times\cr
&\qquad\times\,\int_{\Ft \times \CC^{(0)}}\!\left[d\phi\right] \,
\ch_R\!\left(\e{\!\phi}\right)\,\exp{\!\left[ -{i \over
{2\epsilon_{\rm r}}} \left({d \over \RP}\right) \Tr(\phi^2) 
\right]}\,\times\cr
&\qquad\qquad\times\,\prod_{\beta > 0} \left[2
\sinh\!\left({{\langle\beta,\phi\rangle} \over 
{2}}\right)\right]^{2 - N} \, \prod_{j=1}^N \, \left[2
\sinh\!\left({{\langle\beta,\phi\rangle} \over {2
a_j}}\right)\right]\,,\quad\quad\alpha \hbox{ regular}\,.}}
As claimed, all dependence on the weight $\alpha$ has been subsumed
into the character $\ch_R$, which represents the Seifert loop operator
under localization on $\CO_\alpha/G$.

\bigskip\noindent{\it A Remark on Equivariant Pushdown}\smallskip

Before we proceed further, let us make one theoretical remark about
the interpretation of the Seifert loop operator as the character $\ch_R$.
Though this remark is not strictly necessary, it nicely
foreshadows certain aspects of the computation that we will perform in
Section $7.3$.

As in \FBMCAHS, we consider the distinguished abelian connection
${\{\varrho_{\rm ab}\} \cong \CO_\alpha/G}$ to fiber equivariantly over the
trivial connection ${\{0\} = pt/G}$,
\eqn\FBMCAHSII{\matrix{
&\CO_\alpha/G\cr
&\mskip15 mu\Big\downarrow\lower 0.5ex\hbox{$^\Rq$}\cr
&{pt}/G\cr}\,.}
When $\alpha$ is regular, localization on ${\CO_\alpha = G/T}$
naturally produces an element of the equivariant cohomology ring 
\eqn\EQRNGS{ H^*_G(\CO_\alpha) \,=\, H^*_G(G/T) \,=\, H^*_T({pt})\,,}
which we recognize to be the ring of functions on the Cartan subalgebra 
${\Ft\subset\Fg}$.  Indeed, the particular function on $\Ft$ which we obtain via
localization is precisely the integrand of \SFTWYVI.  Given the equivariant
fibration in \FBMCAHSII, we can then push the result of localization
on ${\{\varrho_{\rm ab}\}\in\SM(C,\alpha)}$ down to the point
${\{0\}\in\SM}$.  But what does this equivariant pushdown mean?

Just as in \EQRNGS, localization at the point $\{0\}$ naturally 
produces an element of $H^*_G({pt})$, which is the ring of invariant
functions on the Lie algebra $\Fg$.  As well-known, the ring of
invariant functions on $\Fg$ is isomorphic to the ring of
Weyl-invariant functions on $\Ft$, so that 
\eqn\DQRNGS{ H^*_G({pt}) \,\cong\, H^*_T({pt})^\FW\,.}
The pushdown ${\Rq_*\!:H^*_G(\CO_\alpha) \rightarrow H^*_G({pt})}$ is then
simply symmetrization under $\FW$, exactly as in \BIGSII.  
So from the theoretical perspective, the character $\ch_R$ that
represents the Seifert loop operator in \SFTWYVII\ should not really
be interpreted as the result of localization at the point
${\{\varrho_{\rm ab}\}\in\SM(C,\alpha)}$, but rather as the result of
localization at $\{\varrho_{\rm ab}\}$ followed by equivariant 
pushdown to ${\{0\}\in\SM}$.

\bigskip\noindent{\it Extension to Irregular Weights}\smallskip

So far we have assumed the weight ${\alpha > 0}$ to be regular.  For
instance, if ${G=SU(2)}$, then all non-zero weights are regular, and
\SFTWYVII\ reproduces the empirical result in \ZLWRZII, at least 
up to an overall phase which we discuss at the end of this section.

However, for groups other than $SU(2)$, some weights are inevitably
irregular.  As an elementary example, the weight associated to the
fundamental $(r+1)$-dimensional representation of $SU(r+1)$ is not
regular when ${r > 1}$.  We certainly want to consider Wilson loops
such as those associated to the fundamental representation of
$SU(r+1)$, so let us quickly consider what happens to the preceding
analysis when $\alpha$ is an irregular weight.

To start, we decompose the roots $\beta$ of $G$ into two sets,
consisting of roots $\beta_{+}$ for which ${(\beta_{+},\alpha) \neq
0}$ and roots $\beta_\perp$ for which ${(\beta_\perp,\alpha) =
0}$, just as in \WEYLINTII. The set of roots $\beta_\perp$ is
empty when $\alpha$ is regular, and the set of roots $\beta_\perp$
runs over all roots when $\alpha$ vanishes.  The Weyl
denominator $\A_\rho$ in \ARHO\ then factorizes as a product over each
set,
\eqn\ARHOII{ \A_\rho(\phi) \,=\, \left[\prod_{(\beta_{+},\alpha) > 0} \, 2
\sinh\!\left({{\langle\beta_{+},\phi\rangle}\over 2}\right)\right] \cdot
\left[ \prod_{\beta_\perp > 0} \, 2
\sinh\!\left({{\langle\beta_\perp,\phi\rangle}\over
2}\right)\right].}
Using \ARHOII, we rewrite the general contour integral in \SFTWYV\ as 
\eqn\SFTWYVIII{\eqalign{ 
&Z\big(\epsilon;C,R\big)\Big|_{\CO_\alpha/G} \,=\,
\exp{\!\left[-{{i\pi}\over 2} \eta_0(0)\right]} \, {1 \over
{|\FW_\alpha|}} \, {{(-1)^{(\Delta_{G_\alpha} - \Delta_T)/2}}
\over{\Vol(T)}} \left({1 \over {i
\sqrt{\RP}}}\right)^{\Delta_T}\times\cr 
&\qquad\qquad\times\,\int_{\Ft \times \CC^{(0)}}\!\left[d\phi\right] \,
{1 \over {\A_\rho(\phi)}} \left[{\prod_{\beta_\perp > 0} \, 2 \, 
\sinh\!\left({{\langle\beta_\perp,\phi\rangle}\over
2}\right)}\right] \exp{\!\left[-\big\langle\alpha +
\rho^{[\alpha]}, \phi\big\rangle\right]}\,\times\cr
&\qquad\qquad\times\,\exp{\!\left[-{i \over {2\epsilon_{\rm r}}} \left({d
\over \RP}\right) \Tr(\phi^2)\right]}\,\prod_{\beta > 0} \left[2
\sinh\!\left({{\langle\beta,\phi\rangle} \over 
{2}}\right)\right]^{2 - N} \, \prod_{j=1}^N \, \left[2
\sinh\!\left({{\langle\beta,\phi\rangle} \over {2
a_j}}\right)\right]\,.}}

Once again, we wish to tease the character $\ch_R$ out of the
integrand in \SFTWYVIII.  To do so, let us introduce the following
function of $\phi$,
\eqn\BIGBB{ \B_\alpha(\phi) \,=\, \e{\!\langle\alpha +
\rho^{[\alpha]},\,\phi\rangle} \cdot \prod_{\beta_\perp > 0} \left[ 2
\sinh\!\left({{\langle\beta_\perp,\phi\rangle}\over 2}\right)\right],}
\countdef\BigB=187\BigB=\pageno
in terms of which we write the factor in the second line of
\SFTWYVIII\ as 
\eqn\BIGSB{\eqalign{
S_\alpha(\phi) \,&=\,{{\e{\!-\langle\alpha+\rho^{[\alpha]},\,\phi\rangle}}
\over{\A_\rho(\phi)}} \cdot \left[{\prod_{\beta_\perp > 0} \, 2 \, 
\sinh\!\left({{\langle\beta_\perp,\phi\rangle}\over
2}\right)}\right],\cr
&=\,  (-1)^{(\Delta_{G_\alpha} - \Delta_T)/2} \cdot 
{{\B_\alpha(-\phi)}\over{\A_\rho(\phi)}}\,.}}
By the same symmetry argument as before, only the Weyl-invariant component of
$S_\alpha(\phi)$, or equivalently the alternating component of
$\B_\alpha(\phi)$, contributes to the contour integral over $\phi$.

The function $\B_\alpha(\phi)$ turns out to have some nice properties,
which make evaluating its alternating component under $\FW$
particularly easy.  For instance, in the extreme case that $\alpha$
vanishes, $\rho^{[\alpha]}$ vanishes as well, and the product over
$\beta_\perp$ in \BIGBB\ runs over all roots of $G$.  Via the
denominator formula in \ARHO, $\B_\alpha(\phi)$ then reduces to
$\A_\rho(\phi)$, as required for \SFTWYVIII\ to reproduce the
contribution from the trivial connection on $M$ to Chern-Simons
partition function.

More generally, $\B_\alpha(\phi)$ satisfies an identity which extends
the denominator formula in \ARHO.  According to this extended
denominator formula, $\B_\alpha(\phi)$ can be rewritten as an alternating
sum over elements $w'$ of the Weyl group $\FW_\alpha$ of the
stabilizer $G_\alpha$, so that 
\eqn\BIGBII{ \B_\alpha(\phi) \,=\, \sum_{w'\in\FW_\alpha} \, (-1)^{w'} \,
\e{\!\langle w'\cdot(\alpha+\rho),\,\phi\rangle}\,.}
When $\alpha$ is regular, the Weyl group $\FW_\alpha$ is trivial, and
the statement in \BIGBII\ is immediate.  Otherwise, $\FW_\alpha$ is
non-trivial for irregular weights, in which case the identity in
\BIGBII\ has non-trivial content.  Though we spare the reader the
details, \BIGBII\ can be proven in exactly the same way as the Weyl
denominator formula, to which \BIGBII\ reduces when ${\alpha = 0}$.
For sake of completeness, we sketch such a proof in Appendix D.

Given the identity in \BIGBII, the alternating component of
$\B_\alpha(\phi)$ is easy to evaluate.  Clearly,
\eqn\BIGBIV{\eqalign{
\A\Big[\B_\alpha(\phi)\Big]\,&=\, {1\over{|\FW|}} \, \sum_{w
\in \FW} \, (-1)^w \, \B_\alpha(w\cdot\phi)\,,\cr
&=\, {1\over{|\FW|}} \, \sum_{w \in \FW} \, \sum_{w' \in \FW_\alpha}
\, (-1)^{(w \, w')} \, \e{\!\langle (w \, w') \cdot
(\alpha+\rho),\,\phi\rangle}\,,\cr
&=\, {1\over{|\FW|}} \, \sum_{w' \in\FW_\alpha} \A_{\alpha+\rho}(\phi)
\,=\, {{|\FW_\alpha|}\over{|\FW|}} \, \A_{\alpha+\rho}(\phi)\,.}}

In complete analogy to \BIGSII, we apply the identity in \BIGBIV\ to
symmetrize $S_\alpha(\phi)$ under $\FW$,
\eqn\BIGBIII{ S_\alpha(\phi)
\,\buildrel{\,\,\left[\,\cdot\,\right]^\FW}\over{\longmapsto}\, 
(-1)^{(\Delta_G - \Delta_{G_\alpha})/2} \cdot
{{|\FW_\alpha|}\over{|\FW|}} \cdot
{{\A_{\alpha+\rho}(\phi)}\over{\A_\rho(\phi)}}\,.}
The sign on the right in \BIGBIII\ again arises after a reflection
from $-\phi$ to $\phi$ in the argument of $\A_{\alpha+\rho}$. 

Via the character formula \WCF, the contour integral in \SFTWYVIII\
then becomes 
\eqn\SFTWYIX{\eqalign{ 
&Z\big(\epsilon;C,R\big)\Big|_{\CO_\alpha/G} \,=\,
\exp{\!\left[-{{i\pi}\over 2} \eta_0(0)\right]} \, {1 \over
{|\FW|}} \, {{(-1)^{(\Delta_G - \Delta_T)/2}}
\over{\Vol(T)}} \left({1 \over {i
\sqrt{\RP}}}\right)^{\Delta_T}\times\cr 
&\qquad\qquad\times\,\int_{\Ft \times \CC^{(0)}}\!\left[d\phi\right] \,
\ch_R\!\left(\e{\!\phi}\right)\,\exp{\!\left[-{i \over {2\epsilon_{\rm
r}}} \left({d \over \RP}\right) \Tr(\phi^2)\right]}\,\times\,\cr
&\qquad\qquad\qquad\times\,\prod_{\beta > 0} \left[2
\sinh\!\left({{\langle\beta,\phi\rangle} \over 
{2}}\right)\right]^{2 - N} \, \prod_{j=1}^N \, \left[2
\sinh\!\left({{\langle\beta,\phi\rangle} \over {2
a_j}}\right)\right],}}
exactly as in the regular case \SFTWYVII.  So regardless of whether
$\alpha$ is regular or irregular, the Seifert loop operator reduces to
the character $\ch_R$ under localization on ${\{\varrho_{\rm ab}\} \cong
\CO_\alpha/G}$ and subsequent pushdown to the trivial connection
$\{0\}$ on $M$.

\bigskip\noindent{\it On Framings and the Phase of
$Z\big(\epsilon;C,R\big)$}\smallskip

We finally wish to discuss the phase of our result \SFTWYIX\ for the
local contribution from ${\{\varrho_{\rm ab}\} \cong \CO_\alpha/G}$ to
$Z(\epsilon;C,R)$.  This discussion extends related remarks at the
very end of \S $5.2$ in \BeasleyVF\ concerning the phase of the
partition function $Z(\epsilon)$.

With no essential loss, we specialize to the case
${G = SU(2)}$, for which we can make a precise comparison between
\SFTWYIX\ and the result of Lawrence and Rozansky in \ZLWRZII.
After parametrizing ${\phi\in\Ft\cong\BR}$ in terms of a coordinate
$z$ via ${\phi = (i/2) \diag(z,-z)}$, we easily see that the
localization result in \SFTWYIX\ agrees exactly with the corresponding
empirical result in \ZLWRZII, at least up to a phase.  However, the
phases of the expressions in \SFTWYIX\ and \ZLWRZII\ do differ.

To be precise, in the case ${G=SU(2)}$ with ${R \cong \CMj}$, the ratio
$\exp{\!\left(i\,\delta\Psi\right)}$ between the canonical phase in
\ZLWRZII\ and the phase in \SFTWYIX\ is given by 
\eqn\CANPHI{\eqalign{
\exp{\!(i\,\delta\Psi)} \,&=\, \exp{\!\left[{{i\pi}\over
4}\!\left({k\over{k+2}}\right) \theta_0\right]} \cdot
\exp{\!\left[{{-i\,\pi}\over{2(k+2)}}\left(j^2-1\right)
\RP\right]},\cr
\theta_0 \,&=\, 3 - {d \over \RP} + 12 \, \sum_{j=1}^N \, s(b_j,
a_j)\,,\cr
\RP \,&=\, \prod_{j=1}^N \, a_j \quad\hbox{if } N \ge 1\,,\qquad \RP
= 1 \quad\hbox{otherwise}\,.}}
In computing $\delta\Psi$, we have used the expression \ETAZII\ for
$\eta_0(0)$ in the case that $M$ is a Seifert homology sphere, or a
cyclic $\BZ_d$ quotient thereof.  Most significantly, we write
$\exp{\!\left(i\,\delta\Psi\right)}$ in \CANPHI\ as a product of two
factors, the first proportional to $\theta_0$ and the second
proportional to $\RP$.  As we now explain, these factors are
associated to the respective choices of framing for the Seifert
manifold $M$ and the embedded curve ${C\subset M}$.

In general, under a shift in the framing of $M$ by $s$ units, the
phase of $Z(\epsilon;C,R)$ transforms as 
\eqn\FRMMI{\eqalign{
Z(\epsilon;C,R) \,&\buildrel s\over\longmapsto\, \exp{\!\left[{{i \pi
c}\over{12}}\,s\right]} \cdot Z(\epsilon;C,R)\,,\qquad\qquad s \in
\BZ\,,\cr
c\,&=\, {{k\,\Delta_G}\over{k \,+\, \check{c}_\Fg}}\,,\qquad\qquad
\Delta_G = \dim G\,.}}
Here the constant $c$ is the central charge of the current algebra for $G$
at level $k$.  Similarly, under a shift in the framing of $C$ by $t$
units, the phase of $Z(\epsilon;C,R)$ transforms as 
\eqn\FRMMII{\eqalign{
Z(\epsilon;C,R)\,&\buildrel t\over\longmapsto\, \exp{\!\left[2\pi i \,
h_R \, t\right]}\cdot Z(\epsilon;C,R)\,,\qquad\qquad t \in \BZ\,,\cr
h_R \,&=\, {{(\alpha, \alpha + 2 \rho)}\over{2(k+\check{c}_\Fg)}}\,.}}
Here $h_R$ is the conformal weight of the current algebra primary
associated to the representation $R$.\foot{As throughout, $\alpha$ in
\FRMMII\ is the highest weight of $R$, $\rho$ is the Weyl
vector for $G$, $\check{c}_\Fg$ is the dual Coxeter number, and
${(\,\cdot\,,\,\cdot\,) = -\Tr(\,\cdot\,)}$ is the invariant 
metric on ${\Fg\cong\Fg^*}$.} 

Because of the intrinsic ambiguity in the phase of $Z(\epsilon;C,R)$ 
under shifts in the framing of the pair $\left(M,C\right)$, we need
only check that the relative phase
$\exp{\!\left(i\,\delta\Psi\right)}$ between the formulae in \ZLWRZII\
and \SFTWYIX\ can be absorbed with suitable choices of $s$ and $t$ in
\FRMMI\ and \FRMMII.  For the case at hand, with ${G=SU(2)}$ and ${R
\cong \CMj}$, the central charge $c$ and the weight $h_R$ become 
\eqn\WTHJ{ c \,=\, {{3 k}\over{k+2}}\,,\qquad\qquad\qquad h_R \,=\, {{j^2 -
1}\over{4(k+2)}}\,.}
Comparing to \CANPHI, we see immediately that the relative phase 
takes the requisite form 
\eqn\CANPHII{ \exp{\!\left(i\,\delta\Psi\right)} \,=\, \exp{\!\left[{{i \pi
c}\over{12}}\,s\right]} \cdot \exp{\!\left[2\pi i \,
h_R \, t\right]}\,,}
provided we set 
\eqn\FRMMIII{ s\,=\,\theta_0\,,\qquad\qquad t\,=\,-\RP\,.}

Though apparently resolving the phase discrepancy in \CANPHI, this
observation begs two immediate questions.  First, at no stage of 
the localization computation did we actually specify a framing for
$\left(M,C\right)$.  So how did we manage to produce a definite phase
for $Z(\epsilon;C,R)$ at all?  

Second and perhaps more technically, although the parameter
${\theta_0 = 3 - d}$ is integral when the Riemann surface $\Sigma$ at
the base of $M$ is smooth, $\theta_0$ is typically only
rational when $\Sigma$ carries a non-trivial orbifold structure.  For
instance, in the case of the Seifert fibration of $S^3$ over ${\Sigma
= \BW\BC\BP^1_{{\bf p},{\bf q}}}$, we computed $\theta_0$ in \QRZEKII\
to be ${\theta_0 = ({\bf p}/{\bf q}) + ({\bf q}/{\bf
p})}$, which is certainly not an integer in general.  (When ${d =
{\bf p} = {\bf q} = 1}$, corresponding to the smooth, degree one
fibration of $S^3$ over $\BC\BP^1$, we note the trivial equality
${\theta_0 = 3 - d = {{\bf p}\over{\bf q}} \,+\, {{\bf q}\over{\bf
p}} = 2}$.)  So how are we to make sense of the shift \FRMMIII\ in
framing by ${s = \theta_0}$ units when $\Sigma$ is an orbifold?

We address these questions in turn.

We largely answered the first question at the end of \S $5.2$ in
\BeasleyVF, for which the crucial observation is the following.  If $M$ is a
Seifert manifold equipped with a locally-free $U(1)_\RR$ action, then
$M$ possesses a pair of distinguished `two-framings'.  By definition,
a two-framing of $M$ is a trivialization of the direct sum
${TM^{\oplus 2} = TM\oplus TM}$ of two copies of the tangent bundle
$TM$.  By way of comparison, a framing of $M$ is just a
trivialization of $TM$ itself.  Like an honest framing, a two-framing
of $M$ suffices to determine a definite phase for the Chern-Simons
partition function $Z(\epsilon)$.

The first of the distinguished two-framings on $M$ was introduced by
Atiyah \AtiyahFM\ and is very well-known.  The Atiyah
two-framing $\tilde\alpha$ is defined for any compact, oriented
three-manifold $M$, not necessarily Seifert, and is characterized
up to homotopy by the following property:  if $W$ is any oriented
four-manifold with boundary $M$, then the signature $\sigma(W)$ of 
$W$ is given by ${\sigma(W) = {1\over 6} \, p_1\!\left(TW^{\oplus
2},\tilde\alpha\right)}$, where $p_1\!\left(TW^{\oplus
2},\tilde\alpha\right)$ is the relative Pontrjagin class computed using
the boundary trivialization by $\tilde\alpha$.  When $M$ is endowed
with the Atiyah two-framing, the Chern-Simons partition function
$Z(\epsilon)$ can always be presented with a canonical phase, as
appears for instance in the empirical formulae for $Z(\epsilon)$ in
\LawrenceRZ\ and \MarinoFK.\countdef\Atiyahtwo=188\Atiyahtwo=\pageno

The second two-framing, introduced in \BeasleyVF, exists only when $M$
admits a Seifert fibration ${\pi:M\rightarrow\Sigma}$.  Briefly,
the Seifert two-framing $\tilde\beta$ is defined as follows.  By
assumption, $M$ is equipped with a nowhere-vanishing vector field 
$\RR$ which generates the given action by $U(1)_\RR$.  Therefore $\RR$
determines a rank-one subbundle $\big[\RR\big]$ of $TM$.  The Seifert
manifold $M$ is also equipped with an invariant contact form
$\kappa$, which satisfies ${\langle\kappa,\RR\rangle=1}$.  Thus the
kernel of $\kappa$ is complementary to $\RR$ in each fiber of $TM$,
and $TM$ decomposes as the direct sum  ${TM \cong \big[\RR\big] \oplus
\ker(\kappa)}$.  Equivalently, 
\eqn\BETFRM{ TM^{\oplus 2} \,\cong\, \big[\RR\big] \oplus
\big[\RR\big] \oplus \ker(\kappa) \oplus \ker(\kappa)\,.}
\countdef\Seiferttwo=189\Seiferttwo=\pageno

Via the isomorphism in \BETFRM, we obtain a trivialization of
$TM^{\oplus 2}$ from any trivialization of the rank-four bundle 
${\ker(\kappa)\oplus\ker(\kappa)}$.  The essential step in the
construction of the Seifert two-framing $\tilde\beta$ is then 
to show that a natural trivialization, unique up to homotopy, exists
for ${\ker(\kappa)\oplus\ker(\kappa)}$.  In a nutshell,
${\ker(\kappa)\oplus\ker(\kappa)}$ is the 
pullback of a corresponding bundle on $\Sigma$, and the bundle on
$\Sigma$ admits a natural trivialization as a $Spin(4)$-bundle for
dimensional reasons --- specifically, the vanishing of the homotopy groups 
$\pi_i(Spin(4))$ for ${i < 3}$.  

For further details on the construction of $\tilde\beta$, we refer the
interested reader to \S $5.2$ of \BeasleyVF.  Later on, we will say a
bit more about the Seifert two-framing in the special case of the
fibration ${\pi:S^3\rightarrow \BW\BC\BP^1_{{\bf p}, {\bf q}}}$.

On a general Seifert manifold $M$, the homotopy types of the respective
Atiyah and Seifert two-framings are distinct,
${\tilde\alpha\neq\tilde\beta}$, leading to two different but equally 
natural ways to define a phase for the partition function
$Z(\epsilon)$.  As we mentioned, empirical results 
such as \ZLWRZII\ are presented with the phase defined by the Atiyah
two-framing $\tilde\alpha$.  On the other hand, when $Z(\epsilon)$
is computed using localization, one might expect the result to 
appear with the phase determined by the Seifert two-framing
$\tilde\beta$.  At least when the base $\Sigma$ of $M$ is smooth, with no
orbifold points, we directly checked this expectation in \BeasleyVF,
where we verified that ${s = \theta_0 = 3 - d}$ in \FRMMIII\ correctly
accounts for the shift in framing from $\tilde\alpha$ to $\tilde\beta$.  

At the end of this section, we reconsider the situation when $\Sigma$ is
an orbifold.  However, before we indulge in orbifold technicalities,
let us discuss the framing of the knot ${C \subset M}$, which is the
essentially new ingredient at present.

\bigskip\noindent{\it The Seifert Framing on $C$}\smallskip

By definition, a framing of $C$ is given by the choice of a
non-vanishing normal vector field along ${C \subset M}$.  Such a
normal vector field determines a new curve $C'$ which is a small
displacement of $C$ along the vector field.  The pair $(C, C')$,
along with the two-framing of $M$, then determines a definite 
phase for the Wilson loop path integral $Z(\epsilon;C,R)$.

Most important here, if $C$ is an arbitrary curve embedded in an
integral homology sphere $M$, then $C$ admits a canonical framing.  This
framing is fixed up to homotopy by the condition that $C$ and its
displacement $C'$ have zero linking number inside $M$,
\eqn\LKFRM{  \lk(C,C') \,=\, 0\,,}
where the linking number in a general homology sphere is
defined by the homology class of $C'$ in the complement of $C$ (or
vice versa),
\eqn\LNKCC{ \lk(C, C') \,=\, [C'] \,\in\, H_1(M^o; \BZ)
\,\cong\,\BZ\,,\qquad\qquad M^o = M - C\,.}\countdef\LkCC=191\LkCC=\pageno
If $M$ is an integral homology sphere, $H_1(M^o;\BZ)$ is
freely-generated by the meridian $\Rm$ of $C$, from which we obtain
the essential isomorphism ${H_1(M^o; \BZ)\,\cong\,\BZ}$ in \LNKCC.
In particular, when $C$ carries the canonical framing \LKFRM, the
class of $C'$ is trivial in $H_1(M^o;\BZ)$.

The canonical framing \LKFRM\ of $C$ is roughly analogous to
the Atiyah two-framing $\tilde\alpha$ of $M$, insofar as both are defined
whether or not the pair $(M, C)$ is Seifert.  On the other hand, when
$M$ is Seifert and ${C\subset M}$ is a generic Seifert fiber, we have
the option to introduce a homotopically distinct Seifert framing on
$C$ analogous to the Seifert two-framing $\tilde\beta$ on $M$.
Up to homotopy, the Seifert framing on $C$ is uniquely characterized
by the condition that it be invariant under the Seifert $U(1)_\RR$
action.  Such an invariant framing on $C$ clearly exists, but for sake
of completeness we spell out its construction explicitly.

To construct the Seifert framing, we first pick a basepoint
on $C$ and a non-vanishing normal vector ${\nu_0 \neq 0}$ in the fiber
of the normal bundle $N_{C/M}$ at that point.  To be
concrete, let us parametrize $C$ with a periodic coordinate ${\tau
\sim \tau + 1}$, so that ${\tau = 0}$ corresponds to the basepoint at
which $\nu_0$ is defined.

By assumption, $C$ is a generic orbit of $U(1)_\RR$, which
therefore acts freely on a neighborhood of $C$ in $M$.  In particular,
each point of $C$ is associated to a unique element of $U(1)_\RR$, and
we are welcome to use the periodic coordinate $\tau$ on $C$ to
parametrize the group $U(1)_\RR$ itself.  In this way we obtain a
family of maps ${f^\tau\!:M\rightarrow M}$, parametrized by
${\tau\in[0,1)}$, which describe the action of $U(1)_\RR$ on $M$.
Each map $f^\tau$ automatically induces a pushforward 
${f^\tau_*\!:TM \rightarrow TM}$ on the tangent bundle $TM$.
Since each $f^\tau$ also preserves the orbit $C$, we obtain by
restriction a pushforward ${f^\tau_*\big|_C\!:TC \rightarrow TC}$
acting on tangent vectors to $C$ itself.  Hence $f^\tau$ also induces a
pushforward ${f^\tau_*\big|_C\!: N_{C/M} \rightarrow N_{C/M}}$ on the
normal bundle $N_{C/M}$ associated to the embedding ${C \subset M}$.

We now use these pushforward maps on $N_{C/M}$ to extend the normal
vector $\nu_0$ defined at the basepoint ${\tau = 0}$ to an invariant,
non-vanishing normal vector field $\nu_\RR$ defined everywhere on $C$.
To define $\nu_\RR$ for all $\tau$, we simply set ${\nu_\RR(\tau) \,=\,
f^\tau_*\big|_C\big(\nu_0\big)}$.  By construction, $\nu_\RR$ is invariant
under the action of $U(1)_\RR$ on $C$.  Since each 
$f^\tau_*$ is a fiberwise isomorphism in a neighborhood of ${C\subset M}$, the
normal vector $\nu_\RR(\tau)$ is also non-vanishing for all $\tau$.
Finally, the only choices in our construction of $\nu_\RR$ were the
choices of $\nu_0$ and the parameter $\tau$, so the Seifert
framing of $C$ is unique up to homotopy.  

The Seifert framing on $C$ is defined for any Seifert manifold $M$,
whether or not $M$ is a homology sphere.  Consequently, using the Seifert
two-framing on $M$ and the Seifert framing on $C$, we can always
present the Seifert loop path integral $Z(\epsilon;C,R)$ with a
canonical phase, which we naturally expect to appear in
computations of $Z(\epsilon;C,R)$ by non-abelian localization.

As a check, let us compare the Seifert framing of $C$ embedded in an
integral homology sphere $M$ to the canonical framing in \LKFRM, for
which $C$ and its displacement $C'$ have zero linking number.  In the
case of the Seifert framing, both $C$ and the normal vector field
$\nu_\RR$ are invariant under $U(1)_\RR$, so the displacement $C'_\RR$
of $C$ along $\nu_\RR$ is naturally invariant as well.  Thus like $C$
itself, $C'_\RR$ is a Seifert fiber of $M$.

In  general, the linking number of two Seifert fibers in $M$ is non-zero.
Hence the canonical framing in \LKFRM\ is related to the Seifert framing
by a shift of ${t = \lk\!\big(C, C'_\RR\big)}$ units.  To make a
quantitative comparison between the canonical and the Seifert
framings, we are left to compute the linking number $\lk\!\big(C,
C'_\RR\big)$ for two Seifert fibers of $M$.

According to the definition in \LNKCC, the linking number
$\lk\!\big(C,C'_\RR\big)$ is given by the homology class of $C'_\RR$ in the
complement ${M^o = M - C}$, where the class of $C'_\RR$ is to be
measured as an appropriate multiple of the class of the meridian
$\Rm$ of $C$.  Of course, $H_1(M^o;\BZ)$ is the abelianization of
$\pi_1(M^o)$, and in \PIMIII\ we have already provided an explicit
description of $\pi_1(M^o)$ via generators $\{\Rc_j, \Rf, \Rm\}$,
${j=1,\ldots,N}$, and relations 
\eqn\PIMIIILK{\eqalign{
 \left[\Rm, \Rf\right] \,&=\, \left[\Rc_j, \Rf\right] \,=\, 1\,,\qquad
 j=1,\ldots,N,\cr
\Rc_j^{a_j} \, \Rf^{b_j} \,&=\, 1\,,\cr
\prod_{j=1}^N \, \Rc_j \,&=\, \Rm \, \Rf^n\,.\cr}}
In interpreting \PIMIIILK, we recall that $\Rf$ corresponds to the
Seifert fiber of $M$, so we identify ${C'_\RR=\Rf}$.  

Abelianizing the relations in \PIMIIILK, we directly obtain 
\eqn\LKSFRT{ \Rf \,=\, \Rm^{-\RP}\,,\qquad\qquad \RP \,=\,
\prod_{j=1}^N {a_j} \,= \left[n \,+\, \sum_{j=1}^N
{{b_j}\over{a_j}}\right]^{-1}.}
In the latter formula for $\RP$, we use the arithmetic condition
\SFRTHOM\ satisfied by the Seifert invariants of the integral homology
sphere $M$.  As a relation between homology classes ${[\Rf] = [C'_\RR]}$ and
$[\Rm]$, the identity in \LKSFRT\ is perhaps better written additively as 
\eqn\LKSFRT{  [\Rf] \,=\, \left[C'_\RR\right]  \,=\, -\RP \cdot [\Rm]
\,\in\, H_1(M^o;\BZ)\,.}
Consequently,
\eqn\FRMMIV{ t \,=\, \lk\!\big(C, C'_\RR\big) \,=\, -\RP\,,}
exactly as we found previously in \FRMMIII. 

\bigskip\noindent{\it Further Remarks on the Seifert Two-Framing}\smallskip

To complete our analysis of phases and framings, we finally consider
the potentially fractional assignment ${s = \theta_0}$ in \FRMMIII,
relevant when the base $\Sigma$ of $M$ is an orbifold.  As a first
step, let us elaborate slightly upon the paradox posed by this result.

We recall that a framing of $M$ is a trivialization of $TM$.
Concretely, such a trivialization is given by three smooth,
non-vanishing vector fields which provide an oriented basis for the
tangent space at each point of $M$.  Any two such framings are related
by a smooth map ${h\!:M\rightarrow SO(3)}$, which takes one
trivialization pointwise to the other.  Up to continuous deformations,
the difference between the given framings is then measured by the homotopy
class of the map $h$.

We actually want to consider not framings but two-framings, which are
given by trivializations of the sum ${TM^{\oplus 2} = TM\oplus TM}$.
The bundle $TM^{\oplus 2}$ carries a natural spin structure, induced
from the diagonal embedding ${SO(3) \subset SO(3)\times SO(3) \subset
SO(6)}$, and any two trivializations of $TM^{\oplus 2}$ as a
$Spin(6)$-bundle are related by a smooth map ${h\!:M\rightarrow
Spin(6)}$.  Again, up to continuous deformations, the difference
between the given two-framings as trivializations of $TM^{\oplus 2}$
is measured by the homotopy class of the map $h$.

By standard arguments, the homotopy class of $h$ takes values
in ${\pi_3(Spin(6))\cong\BZ}$.  We naturally want to identify the
parameter  ${s = \theta_0}$ in \FRMMIII\ with the class
${[h]\in\pi_3(Spin(6))}$ which measures the difference between the
Atiyah two-framing $\tilde\alpha$ and the Seifert two-framing $\tilde\beta$
on $M$.  If the base $\Sigma$ of $M$ is a smooth Riemann surface, then
${\theta_0 = 3 - d}$ is indeed an integer, and we have already checked
in \S $5.2$ of \BeasleyVF\ that the identification ${\theta_0 = [h]}$
is correct.  But if $\theta_0$ is fractional, as can happen when
$\Sigma$ has orbifold points, then ${s=\theta_0}$ cannot
correspond to the homotopy class of any smooth map 
${h\!:M\rightarrow Spin(6)}$.

This paradox has a natural resolution: the Seifert two-framing
$\tilde\beta$ is not necessarily smooth when $\Sigma$ is an orbifold.  By
this statement, we mean that the vector fields on $M$ which describe
the trivialization of $TM^{\oplus 2}$ associated to $\tilde\beta$ are not
themselves smooth but have singularities along the exceptional fibers of
$M$ which sit over orbifold points of $\Sigma$.  As a result, the map
${h\!:M\rightarrow Spin(6)}$ relating the smooth, Atiyah two-framing
$\tilde\alpha$ to the Seifert two-framing $\tilde\beta$ also has
singularities, and the class $[h]$ need not be integral.  Apparently,
despite the orbifold singularities that may occur along exceptional
fibers in $M$, the Seifert two-framing $\tilde\beta$ still suffices to
fix the phase of $Z(\epsilon;C,R)$, as we will see in the example below.

According to our earlier sketch, the construction of $\tilde\beta$ relies
upon the existence of a natural trivialization for a certain
$Spin(4)$-bundle over $\Sigma$.  When $\Sigma$ has orbifold points,
the presence of orbifold singularities in the trivialization over
$\Sigma$ is perhaps not terribly surprising.  Nonetheless, to
illustrate precisely the relation between orbifold points of $\Sigma$
and singularities of $\tilde\beta$, let us analyze the Seifert
two-framing for the very simple Seifert fibration
${\pi:S^3\rightarrow\BW\BC\BP^1_{{\bf p},{\bf q}}}$.

We recall that the parameter ${s=\theta_0}$ for the Seifert 
fibration of $S^3$ over ${\Sigma = \BW\BC\BP^1_{{\bf p},{\bf q}}}$ is
given by  
\eqn\FRMMIV{ \theta_0 \,=\, {{\bf p}\over{\bf q}} \,+\, {{\bf
q}\over{\bf p}}\,,\qquad\qquad \gcd({\bf p},{\bf q})\,=\,1\,.}
So $\theta_0$ is fractional in all but the special case ${{\bf p} =
{\bf q} = 1}$, for which ${\Sigma = \BC\BP^1}$ is smooth.  Beyond merely
establishing that $\tilde\beta$ is singular whenever $\theta_0$ is
fractional, we would also like to understand why $\theta_0$ takes the
particular value in \FRMMIV.

One might worry that the quantitative, technical analysis of singular
two-framings on even a three-manifold as simple as $S^3$ could become
rather involved.  To sidestep many potential complications arising
from singularities in $\tilde\beta$, we follow a slightly indirect
approach, adapted to the presentation of $S^3$ via surgery on ${S^2
\times S^1}$.  Besides its concreteness, the analysis of $\tilde\beta$
via surgery also makes clear the relation between the fractions in
\FRMMIV\ and the orbifold points of orders ${\bf p}$ and ${\bf q}$ in
$\BW\BC\BP^1_{{\bf p},{\bf q}}$.

We first recall that ${S^3 = T' \cup_\delta T}$ can be presented
topologically as a union of solid tori $T'$ and $T$ which are glued
together along their boundaries by a non-trivial element $\delta$ in
the mapping class group $SL(2,\BZ)$ of the two-torus ${S^1 \times S^1}$.
As standard, $\delta$ acts on the lattice ${H_1(S^1\times S^1;\BZ) \cong
\BZ \oplus \BZ}$; to describe $\delta$ as a matrix, we pick a basis
$\big\{\Rm,\Rl\big\}$ for this lattice as follows.  We identify ${S^1\times
S^1}$ with the boundary of ${T = D^2 \times S^1}$, where $D^2$ is the
unit disk in $\BR^2$.  The meridian $\Rm$ is then given by the
boundary of $D^2$, and the longitude $\Rl$ is given by the $S^1$ fiber
over any point on that boundary.  Equivalently, if we parametrize $D^2$
with polar coordinates $(r,\varphi)$ and $S^1$ with a periodic coordinate
${\tau\sim\tau + 1}$, the meridian $\Rm$ is parametrized by $\varphi$,
and the longitude $\Rl$ is parametrized by $\tau$.

If $\delta$ were trivial in $SL(2,\BZ)$, then ${T' \cup_\delta
T}$ would simply be the aforementioned product ${S^2 \times S^1}$,
where $S^2$ arises by gluing together the respective $D^2$ factors in
$T$ and $T'$.  To obtain $S^3$ instead of ${S^2\times S^1}$, we take
$\delta$ to have the form 
\eqn\GLUED{ \delta \,=\, {\bf T}^b \cdot {\bf S} \cdot {\bf
T}^a\,,\qquad\qquad a\,,b\,\in\,\BZ\,.}
Here ${\bf S}$ and ${\bf T}$ are the standard generators of $SL(2,\BZ)$,
\eqn\MODST{ {\bf S} \,=\, \pmatrix{0&-1\cr 1&0}\,,\qquad\qquad
{\bf T} \,=\, \pmatrix{1&1\cr 0&1}\,,}
and $(a, b)$ are arbitrary integers (possibly zero).
\countdef\SLtwoZ=190\SLtwoZ=\pageno

As will be of use momentarily, let us review how the form of $\delta$
in \GLUED\ arises.  We regard $S^3$ as the unit sphere embedded in
$\BC^2$, parametrized by complex coordinates $(X,Y)$ satisfying
${|X|^2 + |Y|^2 = 1}$.  In these coordinates, the solid tori $T$ and
$T'$ in the decomposition ${S^3 = T' \cup_\delta T}$ are given without
loss by the respective subsets of $S^3$ below,\foot{The choice of the
symmetric bound for $|X|^2$ and $|Y|^2$ by $1/2$, as opposed to any
other parameters $x$ and ${1-x}$ in the unit interval, is inessential.}
\eqn\TORIIS{ T \,=\, \Big\{(X,Y) \in S^3 \,\big|\, |X|^2 \le \ha
\Big\}\,,\qquad\qquad T' \,=\, \Big\{(X,Y) \in S^3 \,\big|\, |Y|^2 \le
\ha \Big\}\,.}
In terms of the topological identification ${T = D^2 \times S^1}$, the
disk $D^2$ is parametrized up to an overall rescaling by the complex
variable $X$, and the $S^1$ fiber is parametrized by the phase of the
non-vanishing complex variable $Y$.  Thus on the boundary of $T$,
where ${|X|^2 = |Y|^2 = 1/2}$, the phase of $X$ is identified with
$\varphi$, and the phase of $Y$ is identified with $\tau$.  In the
other solid torus $T'$, the roles of $X$ and $Y$ are reversed.  The
exchange ${X \leftrightarrow Y}$ in passing from $T$ to 
$T'$ is then accomplished by the generator ${\bf S}$ in \GLUED.

Otherwise, the choice of the pair $(a,b)$ is related to the
possibility to perform Dehn twists by ${{\bf T}\in SL(2,\BZ)}$ on the
boundaries of $T$ and $T'$.  Our description of $\delta$ as a matrix
relies upon the choice of the particular basis $\big\{\Rm,\Rl\big\}$
for $H_1(\partial T;\BZ)$.  The meridian $\Rm$ is canonically
determined, at least up to sign, as the generator of the kernel of the
natural map from ${H_1(\partial T;\BZ) \cong \BZ\oplus\BZ}$ to
${H_1(T;\BZ)\cong\BZ}$.  But any ${\Rl' = {\bf T}^a(\Rl) = \Rl +
a\cdot\Rm}$ is equally good as a complementary basis element for
$H_1(\partial T;\BZ)$.  As a result, the possible twists by ${\bf
T}^a$ and ${\bf T}^b$ acting on the respective boundaries of $T$ and
$T'$ do not change the topology of the union ${S^3 = T' \cup_\delta T}$.

However, the parameters $a$ and $b$ are far from irrelevant, since
these data determine a definite two-framing on $S^3$.  As carefully
explained for instance in \FreedRG, the solid tori $T$ and $T'$ can
each be endowed with a standard two-framing, which restricts on
the boundary to the product two-framing determined by the vector
fields $\partial/\partial\varphi$ and $\partial/\partial\tau$.  
These standard two-framings on $T$ and $T'$ then determine a
corresponding two-framing on ${S^3 = T' \cup_\delta T}$ after gluing
by $\delta$.

We now wish to relate the two-framing on $S^3$ determined by $\delta$
to both the Atiyah two-framing $\tilde\alpha$ and the Seifert two-framing
$\tilde\beta$.

The relation to the Atiyah two-framing $\tilde\alpha$ is well-known.
Briefly, with the conventions in \GLUED, the Atiyah two-framing
$\tilde\alpha$ is described by any pair $(a,b)$ such that  ${a+b=0}$.
Else, each Dehn twist by ${\bf T}$ shifts the standard two-framing on the
solid torus by one unit, and the gluing under $\delta$ generally 
produces the two-framing on $S^3$ which is shifted by a total 
${s = a + b}$ units relative to $\tilde\alpha$.  Once again, ${s = [h]}$
is the homotopy class of the map ${h:M\rightarrow Spin(6)}$ which measures
the difference between $\tilde\alpha$ and the two-framing induced by
$\delta$.  That said, throughout  the present analysis, we will really
only keep track of the magnitude of $s$, not the overall sign.
Finally, as a small check that the overall shift by ${s=a+b}$ units is
correct, we consider the case ${b = -a}$.  Then ${\delta = {\bf
T}^{-a}\cdot{\bf S}\cdot{\bf T}^a}$ is conjugate to ${\bf S}$ in
$SL(2,\BZ)$, and the two-framing determined by $\delta$ necessarily
agrees with the two-framing determined by ${\bf S}$ alone.

We are left to describe the Seifert two-framing $\tilde\beta$ similarly
in terms of surgery under $\delta$.  As before, $\tilde\beta$ is
constructed as a trivialization of $TM^{\oplus 2}$ using the evident
isomorphism  
\eqn\BETFRMII{ TM^{\oplus 2} \,\cong\, \big[\RR\big] \oplus
\big[\RR\big] \oplus \ker(\kappa) \oplus \ker(\kappa)\,,}
along with a certain natural trivialization of the rank-four bundle
${\ker(\kappa) \oplus \ker(\kappa)}$.  At first glance, one might
worry that a detailed knowledge of the vector fields on $S^3$
responsible for trivializing ${\ker(\kappa) \oplus \ker(\kappa)}$
would be necessary to describe $\tilde\beta$ via surgery. However, all we
really need to know is that the trivialization of $TM^{\oplus 2}$
associated to $\tilde\beta$ involves the non-vanishing, Seifert vector
field $\RR$.  In the current example, if $S^3$ is parametrized by complex
coordinates $(X,Y)$ satisfying ${|X|^2 + |Y|^2 = 1}$, then $U(1)_\RR$
acts on $X$ and $Y$ with respective charges ${\bf p}$ and ${\bf q}$.  The
generating vector field $\RR$ is then given explicitly by 
\eqn\STHREER{ \RR \,=\, {i\over 2} \left[ {\bf p} \, X
{\partial\over{\partial X}} \,+\,{\bf q} \,Y {\partial\over{\partial
Y}} \,-\, {\bf p} \,\bar X {\partial\over{\partial \bar X}} \,-\,{\bf
q} \, \bar Y {\partial\over{\partial \bar Y}}\right].}

Upon restriction, $\tilde\beta$ determines a two-framing on each of the
solid tori $T$ and $T'$ specified as subsets of $S^3$ in \TORIIS.  So to
characterize $\tilde\beta$ under surgery, we need only compare
these induced two-framings on $T$ and $T'$ to the standard
two-framing.  We first perform the comparison on $T$; the comparison
on $T'$ is entirely similar.

As we mentioned earlier, the standard two-framing on ${T = D^2 \times
S^1}$ restricts on the boundary ${\partial T = S^1 \times S^1}$ to the product
two-framing by the non-vanishing vector fields
$\big\{\partial/\partial\varphi,\, \partial/\partial\tau\big\}$ which  
generate rotations in each $S^1$ factor.  Equivalently, 
the standard two-framing identifies twice the tangent bundle of
${S^1\times S^1}$ with the sum
\eqn\STDTWFR{ T (S^1 \times S^1)^{\oplus 2}
\,\buildrel{\rm std}\over\cong\,
\left[{\partial\over{\partial\varphi}}\right] \oplus   
\left[{\partial\over{\partial\varphi}}\right] \oplus 
\left[{\partial\over{\partial\tau}}\right] \oplus 
\left[{\partial\over{\partial\tau}}\right]\,.}
Just as in \BETFRMII, $[\,\cdot\,]$ indicates the trivial line bundle
generated by the given vector field.

In contrast, if we restrict the Seifert two-framing $\tilde\beta$ to the
boundary of $T$, we obtain a different trivialization, now involving the
Seifert vector field $\RR$.  Up to homotopy, the Seifert two-framing
$\tilde\beta$ identifies twice the tangent bundle of ${S^1\times S^1}$
with the sum 
\eqn\STDTWFRII{ T (S^1 \times S^1)^{\oplus 2}
\,\buildrel{\tilde\beta}\over\cong\,\Big[\RR^\perp\Big] \oplus
\Big[\RR^\perp\Big] \oplus \Big[\RR\Big] \oplus \Big[\RR\Big]\,.}
Here $\RR^\perp$ is a non-vanishing vector field on ${S^1\times
S^1}$ which generates the orthocomplement to $\RR$ as determined by 
any convenient metric.  (The homotopy class of the trivialization does
not depend on the continuous choice of the metric on ${S^1 \times S^1}$.)

Again up to homotopy, any two-framing on the solid torus $T$ is
related to the standard two-framing by the repeated action of the Dehn
twist ${\bf T}$ on the boundary.  To relate the respective trivializations in
\STDTWFR\ and \STDTWFRII, we recall that $\varphi$ is identified on 
${\partial T}$ with the phase of the coordinate $X$, and $\tau$ is
identified with the phase of the coordinate $Y$.  Hence $\RR$ in
\STHREER\ is given on the boundary of $T$ by 
\eqn\STHREERII{ \RR \,=\, {\bf p} \, {\partial\over{\partial\varphi}}
\,+\, {\bf q} \, {\partial\over{\partial\tau}}\,.}
Further, with a suitable choice of metric on ${\partial T}$, we can
always arrange that ${\RR^\perp = \partial/\partial\varphi}$.  The
trivialization \STDTWFRII\ induced by $\tilde\beta$ on $\partial T$
then becomes 
\eqn\STDTWFRIII{ T (S^1 \times S^1)^{\oplus 2}
\,\buildrel{\tilde\beta}\over\cong\,\left[{\partial\over{\partial\varphi}}
\right] \oplus \left[{\partial\over{\partial\varphi}}
\right] \oplus \left[ {\partial\over{\partial\tau}} \,+\, \left({{\bf
p}\over{\bf q}}\right) {\partial\over{\partial\varphi}}\right] \oplus
\left[ {\partial\over{\partial\tau}} \,+\, \left({{\bf
p}\over{\bf q}}\right) {\partial\over{\partial\varphi}}\right]\,.} 
To facilitate the comparison between \STDTWFR\ and \STDTWFRIII, we
have rescaled $\RR$ in \STHREERII\ by an overall factor of ${1/{\bf
q}}$.  Of course, such a scaling has no effect on the line bundle
generated by $\RR$.

Under a general twist by ${\bf T}^a$, the line bundles in the
standard product trivialization \STDTWFR\ transform as 
\eqn\TWIST{ \left[{\partial\over{\partial\varphi}}\right] \,\buildrel{{\bf
T}^a}\over\longmapsto\,
\left[{\partial\over{\partial\varphi}}\right]\,,\qquad\qquad
\left[{\partial\over{\partial\tau}}\right] \,\buildrel{{\bf
T}^a}\over\longmapsto\, \left[{\partial\over{\partial\tau}}\,+\,
a\,{\partial\over{\partial\varphi}}\right]\,.}
From \TWIST, we see immediately that ${\bf T}^a$ for ${a = {\bf p}/{\bf q}}$
formally takes the standard two-framing in \STDTWFR\ to the induced
Seifert two-framing in \STDTWFRIII.

Now, if ${{\bf p} = {\bf q} = 1}$, then ${{\bf T}^a = {\bf T}}$ is the
usual Dehn twist, which takes one smooth two-framing  
on $T$ to another, shifted by one unit.  Otherwise, for
relatively-prime pairs ${{\bf p} > {\bf q} > 1}$, the Seifert
two-framing $\tilde\beta$ on $T$ is related to the standard
two-framing via the fractional twist ${\bf T}^{{\bf p}/{\bf q}}$.  Such
a fractional Dehn twist does not really make sense on a 
smooth three-manifold, but it does make sense on a three-dimensional
orbifold.  Indeed, the exceptional fiber given by ${X=0}$ in $T$ sits
precisely over the orbifold point of order ${\bf q}$ in the base of
the fibration ${\pi:S^3\rightarrow \BW\BC\BP^1_{{\bf p},{\bf q}}}$,
from which the denominator in the exponent ${a = {\bf p}/{\bf q}}$
arises.\foot{Let us briefly check that ${X=0}$ sits over the orbifold
point of order ${\bf q}$, as opposed to ${\bf p}$, in
$\BW\BC\BP^1_{{\bf p},{\bf q}}$.  Since ${|Y|\neq 0}$ in $T$, we can
partially fix a slice for the action of $U(1)_\RR$ by imposing the
condition that ${Y = |Y|}$ be real and positive.  But since $Y$
transforms with charge ${{\bf q}>1}$ under $U(1)_\RR$, the condition
${Y = |Y|}$ still leaves unfixed a cyclic subgroup ${\BZ_{\bf q}
\subset U(1)_\RR}$, whose generator acts on $X$ as ${X \mapsto
\zeta^{\bf p} \cdot X}$, ${\zeta = \exp{\!(2\pi i/{\bf q})}}$.  So
under the quotient by $\BZ_{\bf q}$, the point corresponding to
${X=0}$ in $\BW\BC\BP^1_{{\bf p},{\bf q}}$ is an orbifold point of
order ${\bf q}$.}  Clearly as promised, the fractional twist by 
${\bf T}^{{\bf p}/{\bf q}}$ on the standard two-framing creates
orbifold singularities in $\tilde\beta$ along the exceptional fiber.

On the complementary torus $T'$, the same analysis applies after an 
exchange of the coordinates $X$ and $Y$.  Equivalently, because $X$ and
$Y$ are distinguished only by their charges under $U(1)_\RR$, we
simply exchange ${\bf p}$ and ${\bf q}$.  Thus a twist by
${\bf T}^b$ for ${b = {\bf q}/{\bf p}}$ takes the standard two-framing
on $T'$ to the induced Seifert two-framing $\tilde\beta$.
In total, we deduce that the difference between the Seifert
two-framing $\tilde\beta$ and the Atiyah two-framing $\tilde\alpha$ on
the union ${S^3 = T' \cup_\delta T}$ is given by the sum 
\eqn\FRMMV{ s \,=\, a + b \,=\, {{\bf p}\over{\bf q}} \,+\,
{{\bf q}\over{\bf p}}\,,}
exactly as in \FRMMIV.

\subsec{The Seifert Loop Operator as a Chern Character}

To conclude our analysis of Chern-Simons gauge theory on a Seifert
three-manifold $M$, we now extend our work in Section $7.2$ to
describe the cohomology class of the Seifert loop operator $W_R(C)$ on
any smooth component $\SM_0$ in the moduli space of flat connections
on $M$.  Though the discussion here will be self-contained, the
results in this section also build upon those in \S $5.3$ of \BeasleyVF.

Mostly as a means to streamline the exposition, let us immediately 
make a few simplifying assumptions.  First, we assume throughout that
$M$ is the total space of a circle bundle of degree ${n \ge 1}$ over a 
smooth (non-orbifold) Riemann surface $\Sigma$ of genus ${h \ge 1}$, 
\eqn\SFIBRM{\matrix{
&S^1 \buildrel n\over\longrightarrow \,M\cr
&\mskip 70mu \big\downarrow\lower 0.5ex\hbox{$^\pi$}\cr
&\mskip 60mu \Sigma\cr}\,,\qquad\qquad\qquad
\hbox{genus}(\Sigma)\,=\,h\ge 1\,.}
So unlike the examples in Section $7.2$, for which orbifold points in
genus zero were crucial, $\Sigma$ now has no orbifold points.

Second, because our current focus is on topology rather than group
theory, we restrict attention to the familiar case that the
Chern-Simons gauge group is ${G = SU(r+1)}$.  Similarly, for the
irreducible representation $R$, we assume that the highest weight
${\alpha > 0}$ of $R$ is regular. Hence the stabilizer of $\alpha$
under the adjoint action is ${G_\alpha = T = U(1)^r}$, and
the coadjoint orbit ${\CO_\alpha \cong G/T}$ is the complete flag manifold.
Explicitly, again using the identification ${\Ft \cong \Ft^*}$, we
write 
\eqn\COMPAL{ \alpha \,=\, i\,\diag(\alpha_1,\ldots,\alpha_{r+1})\,,
\qquad\qquad \alpha_1 + \cdots + \alpha_{r+1} \,=\, 0\,,}
in terms of which the regularity condition ${\alpha > 0}$ becomes  
\eqn\REGAL{ \alpha_1 > \alpha_2 > \cdots > \alpha_r > \alpha_{r+1}
\,.}
Later, we will want to characterize the diagonal entries \COMPAL\ of
$\alpha$ in a slightly more invariant fashion.  As in \SUWTII, we
therefore introduce the basic generating weights $\big\{\hat\omega_1, \ldots,
\hat\omega_{r+1}\big\}$ of $SU(r+1)$, in terms of which $\alpha$ can
be expanded as 
\eqn\DYNKINL{ \alpha \,=\, \alpha_1 \, \hat\omega_1 \,+\, \cdots \,
\alpha_{r+1} \, \hat\omega_{r+1}\,.}

Finally, we assume without loss that the representation $R$ is
integrable in the current algebra for $G$ at level $k$.  This
assumption implies that the highest weight ${\alpha>0}$ of $R$ is
bounded in terms of the highest root $\vartheta$ of $G$ by
${(\vartheta,\alpha) =  \alpha_1 - \alpha_{r+1} \le k}$.  Actually,
for technical reasons that will become clear, we need to assume that
$\alpha$ satisfies the slightly stronger, strict bound
${(\vartheta,\alpha) < k}$.  As we indicated in Section $7.1$, the
strict bound on $\alpha$ is conceptually the `bare' version of the
integrability condition expressed in the form
${(\vartheta,\alpha+\rho) < k + \check{c}_\Fg}$.

In actuality, neither the restriction on $G$ nor $R$ is particularly
essential, as exemplified by the preceding computation for general
$\left(G,R\right)$ in Section $7.2$.  Moreover, the result we 
eventually obtain for the cohomology class which describes the Seifert
loop operator $W_R(C)$ on $\SM_0$ will have a natural meaning for
arbitrary pairs $\left(G,R\right)$.  Nonetheless, we prefer to phrase the
discussion here in the language of holomorphic vector bundles, as
opposed to bundles with arbitrary structure group, and we prefer to
avoid the small complications, such as in the discussion surrounding
\SFTWYVIII, which arise when $\alpha$ is not regular.

The computation of the Seifert loop class on $\SM_0$, through
straightforward, is slightly involved.  Rather than plunge into that
computation straightaway, let us first present our eventual result,
which is perhaps more illluminating than the computation itself.
Indeed, the class in $H^*(\SM_0)$ which describes the Seifert loop
operator $W_R(C)$ under localization is not hard to guess, provided we
note the following clues.

To start, the Seifert loop operator is defined solely by the choice of
$R$ and the point ${p\in\Sigma}$ over which $C$ sits as a
Seifert fiber of $M$, so the Seifert loop class can depend {\it a priori}
only upon these data.  Moreover, since the isotopy class of the
embedding ${C \subset M}$ is invariant under continuous deformations
of $p$, the Seifert loop class can really only depend upon the
discrete choice of $R$.

Let us now ask --- what are the natural cohomology classes to consider 
on $\SM_0$?  

The answer to this question is particularly clear if we
recall from Section $7.1$ the identification \IDMZEROII\ of moduli spaces 
\eqn\SMTHIDMNO{ \SM_0 \,\cong\, \wt\SN(P)\,,}
where $\wt\SN(P)$ is a finite, unramified cover of the moduli space
$\SN(P)$ parametrizing flat connections on a principal ${G_{\rm
ad}\big(\!= G/\CZ(G)\big)}$-bundle $P$ over $\Sigma$.  Here the
topology of $P$ is determined by the central fiberwise holonomy
${\varrho(\Rf) \in \CZ(G)}$, constant for all points in $\SM_0$, of
the corresponding irreducible flat $G$-connection on $M$.  For
technical simplicity, we assume throughout that ${\zeta =
\varrho(\Rf)^n}$ in \LEUIII\ generates the center ${\CZ(G) =
\BZ_{r+1}}$, in which case $\SN(P)$ is smooth, and the cover
${\wt\SN(P) \rightarrow \SN(P)}$ has degree ${|\CZ(G)|^{2 h} =
(r+1)^{2 h}}$ for ${G = SU(r+1)}$.

As in Section $5.2$, when $\SN(P)$ is smooth, a universal bundle $\SV$ 
exists as a holomorphic vector bundle of rank ${r+1}$ over the product
${\Sigma \times \SN(P)}$,
\eqn\UNIVBRPT{\matrix{
&\BC^{r+1}\,\longrightarrow\,\SV\cr
&\mskip 85mu\big\downarrow\cr
&\mskip 115mu {\Sigma \times \SN(P)}\cr}\,.}
In this situation, the characteristic classes of $\SV$ define natural
cohomology classes on ${\Sigma \times \SN(P)}$, and those
characteristic classes can be evaluated on homology classes of
$\Sigma$ to obtain associated classes on $\SN(P)$ alone.  From the
perspective of gauge theory, these Atiyah-Bott classes (introduced in
\S $9$ of \AtiyahYM) are the natural, God-given cohomology classes on
$\SN(P)$, and as demonstrated in Theorem $9.11$ of \AtiyahYM, they
suffice to generate multiplicatively the integral cohomology ring
$H^*\!\big(\SN(P);\BZ\big)$.

With this philosophy and the identification ${\SM_0 \cong \wt\SN(P)}$,
we see that the natural cohomology classes to consider on
$\SM_0$ are the pullbacks under the covering map ${\wt\SN(P)
\rightarrow \SN(P)}$ of the Atiyah-Bott classes on $\SN(P)$. To avoid
cluttering the notation any further, we will not attempt to
distinguish between cohomology classes on $\SN(P)$ and their pullbacks
to ${\SM_0 \cong \wt\SN(P)}$.

As we mentioned above, each Atiyah-Bott class is defined by evaluating
a given characteristic class of $\SV$ on a given homology class of
$\Sigma$.  For the requisite homology class, we have already noted
that the Seifert loop operator involves the choice of a 
point ${p \in \Sigma}$.  The Atiyah-Bott classes on $\SN(P)$
associated to the homology class of $p$ are then simply the
characteristic classes of the bundle $\SV$ restricted to ${\{p\}
\times \SN(P)}$, 
\eqn\DEFSVPII{ \SV_p \,\equiv\, \SV|_{\{p\}\times \SN(P)}\,,}
as appeared already in \DEFSVP. 

We are finally left to guess which characteristic class of $\SV_p$
actually describes the Seifert loop operator $W_R(C)$ under
localization on $\SM_0$.  This characteristic class must somehow
incorporate the choice of the representation $R$, for which there is
a more-or-less evident way to proceed.  

Via the Chern-Weil homomorphism, the characteristic classes of a
bundle with structure group $G$ are determined by invariant functions
on the Lie algebra $\Fg$, or equivalently, by Weyl-invariant functions
on the Cartan subalgebra $\Ft$.  When $M$ is a Seifert homology
sphere, we have already demonstrated in Section $7.2$ that the Seifert
loop operator $W_R(C)$ reduces at the trivial connection ${\{0\} =
pt/G}$ to the character $\ch_R$,
\eqn\WRCTRV{ W_R(C)\big|_{\{0\}} \,=\,
\ch_R\big(\e{\phi}\big)\,\in\,H^*_T(pt)^\FW\,,\qquad\qquad
\phi\,\in\,\Ft\,,}
interpreted here as a Weyl-invariant function of ${\phi\in\Ft}$.  So for
localization on ${\SM_0 \cong \wt\SN(P)}$, the most natural -- and 
certainly most elegant -- possibility is that the Seifert loop
operator $W_R(C)$ reduces to the characteristic class of $\SV_p$
determined by the same character $\ch_R$,
\eqn\CHRVRP{ W_R(C)\big|_{\SM_0} \,=\, \ch_R\big(\SV_p\big)\,\in\,
H^*\!\big(\SM_0\big).}

In the remainder of this paper, our goal will be to demonstrate the
identification in \CHRVRP\ directly, by a localization computation on
$\SM_0$ very similar to the computation in Section $7.2$.  However,
before we even begin to compute, let us first provide a more concrete
description of the characteristic class $\ch_R(\SV_p)$, as we will
encounter it in practice.

In general, any characteristic class of holomorphic vector bundle can
be expressed as a symmetric function of its Chern roots.  With a
certain malice aforethought, we were careful to provide a complete 
description of the Chern roots of $\SV_p$ at the end of Section
$5.2$.  As we reviewed there, the extended moduli space
$\SN(P;\lambda)$ for regular ${\lambda > 0}$ serves as a canonical
splitting manifold for $\SV_p$, such that under the fibration 
\eqn\PSDWNI{\matrix{ 
&2\pi\CO_{-\lambda}\,\longrightarrow\,\SN(P;\lambda)\cr
&\mskip 115mu\Big\downarrow\lower 0.5ex\hbox{$^{\Rq}$}\cr
&\mskip 100mu\SN(P)\cr}\,,\qquad\qquad \langle\vartheta,\lambda\rangle
< 1\,,}
the pullback $\Rq^*\SV_p$ decomposes smoothly into a direct sum
of line bundles over $\SN(P;\lambda)$,
\eqn\CHRPSD{ \Rq^*\SV_p \,\cong\, \bigoplus_{j=1}^{r+1} \,
\SL_j\,.}
Hence the total Chern class of $\SV_p$ satisfies 
\eqn\CHRNRTSPSD{ \Rq^*c(\SV_p) \,=\, \prod_{j=1}^{r+1} \left(1 \,+\,
\Ru_j\right),\qquad\qquad  \Ru_j \,=\, c_1(\SL_j) \in
H^2\!\big(\SN(P;\lambda);\BZ\big),}
and $\big\{\Ru_1,\ldots,\Ru_{r+1}\big\}$ realize the Chern roots of
$\SV_p$.  

To encode the Chern roots in a slightly more invariant fashion, we
collect them into a single degree-two class $\Ru$ on $\SN(P;\lambda)$
which is valued in the Cartan subalgebra of $SU(r+1)$,
\eqn\INVCHRNRT{ \Ru \,=\, i \diag\!\big(\Ru_1,\cdots,
\Ru_{r+1}\big) \,\in\, H^2\!\big(\SN(P;\lambda);\BZ\big) \otimes
\Ft\,.}
The individual Chern roots can then be extracted from $\Ru$ using the
basic weights $\big\{\hat\omega_1,\ldots,\hat\omega_{r+1}\big\}$ in
\SUWTII,
\eqn\INVCHRNRTII{ \Ru_j \,=\,
\big\langle\hat\omega_j, \Ru\big\rangle\,,\qquad\qquad
j\,=\,1,\ldots,r+1\,.}
With the conventions in \CHRNRTSII, each $\Ru_j$ restricts fiberwise on
$\CO_\lambda$ to the invariant two-form 
\eqn\CHRNRTSUJ{ \Ru_j\big|_{\CO_\lambda} \,=\,
-{\nu_{{\hat\omega}_j}\over{2\pi}}\,,\qquad\qquad j\,=\,1,\ldots,r+1\,.}

Under the Chern-Weil homomorphism, the degree-two class $\Ru$ in
\INVCHRNRT\ plays exactly the role of the variable ${\phi\in\Ft}$
appearing in \WRCTRV.  As a result, the characteristic class
$\ch_R\big(\SV_p\big)$ is given explicitly in terms of $\Ru$ by 
\eqn\CHRVRPII{ \ch_R\big(\SV_p\big) \,=\, \ch_R\big(\e{\Ru}\big) \,=\,
{{\A_{\alpha + \rho}(\Ru)}\over{\A_\rho(\Ru)}}\,.}
Just as in \BIGA, $\A_\alpha(\Ru)$ is defined for each 
weight $\alpha$ via an alternating sum over the Weyl group $\FW$,
\eqn\BIGARU{ \A_\alpha(\Ru) \,=\, \sum_{w \in \FW} \, (-1)^w \,
\e{\!\langle w\cdot\alpha,\,\Ru\rangle}\,,}
and in passing to the latter expression for $\ch_R\big(\SV_p\big)$ in
\CHRVRPII, we apply the Weyl character formula.  Finally, we
abuse notation slightly in \CHRVRPII.  Most literally,
$\ch_R\big(\SV_p\big)$ is a class downstairs on $\SN(P)$, whereas
$\ch_R\big(\e{\Ru}\big)$ is a class upstairs on the
splitting manifold $\SN(P;\lambda)$.  So more correctly, we should
write ${\Rq^*\ch_R\big(\SV_p\big) = \ch_R\big(\e{\Ru}\big)}$,
recalling from \CHRNRTSPSD\ that any symmetric function of the Chern roots
$\big\{\Ru_1,\ldots,\Ru_{r+1}\big\}$ is the pullback from a corresponding
characteristic class on $\SN(P)$.   We will henceforth avoid such
pedantries, and we will not distinguish between characteristic classes
of $\SV_p$ on $\SN(P)$ and their pullbacks written in terms of $\Ru$
on $\SN(P;\lambda)$.

The characteristic class $\ch_R\big(\SV_p\big)$ admits yet another
description as the Chern character of a universal bundle $\SV^R_p$
associated to the representation $R$.  Here $\SV^R_p$ is the
restriction to ${\{p\} \times \SN(P)}$ of a universal 
bundle $\SV^R$ with fiber $R$ over ${\Sigma \times \SN(P)}$,
\eqn\UNIVBRPTR{\matrix{
&R\,\longrightarrow\,\SV^R\cr
&\mskip 50mu\big\downarrow\cr
&\mskip 80mu {\Sigma \times \SN(P)}\cr}\,.}\countdef\SVR=192\SVR=\pageno
If $R$ is the fundamental, $(r+1)$-dimensional representation of
$SU(r+1)$, then $\SV^R$ is the standard universal bundle $\SV$ appearing in
\UNIVBRPT.  Otherwise, at least for the familiar example ${G =
SU(r+1)}$, the associated bundle $\SV^R$ can be constructed algebraically
from appropriate symmetric and anti-symmetric tensor powers of $\SV$.

By definition, if $E$ is a complex vector bundle with Chern roots
$e_j$ for ${j=1,\ldots,\rk E}$, the Chern character $\ch(E)$ is given
by the sum of exponentials 
\eqn\CHCHRII{ \ch(E) \,=\, \sum_{j=1}^{\rk E} \,
\exp{\!(e_j)}\,,\qquad\qquad \exp{\!(e_j)} \,=\, 1 \,+\, e_j \,+\, \ha
e_j^2 \,+\, \cdots\,.}\countdef\ChernCh=192\ChernCh=\pageno
As throughout, the exponential in \CHCHRII\ is to be understood by the
indicated Taylor series, which eventually terminates at some finite
order due to the nilpotency of the $e_j$.  On the other hand, the Chern
roots of the associated bundle $\SV^R_p$ are  by
construction the eigenvalues of $\Ru$ acting as an element of $\Ft$ on
the representation $R$.  So tautologically, the Chern character of
$\SV^R_p$ is given by 
\eqn\CHRVRPIII{\eqalign{
\ch\big(\SV^R_p\big) \,&=\, \Tr_R\big(\e{\Ru}\big) \,\equiv\,
\ch_R\big(\e{\Ru}\big)\,,\cr
&=\, \ch_R\big(\SV_p\big)\,.}}
From the K-theory perspective adopted in \TelemanCW, the alternative
description of the Seifert loop class as the Chern character
$\ch\big(\SV^R_p\big)$ is possibly more natural.

\bigskip\noindent{\it Warm-Up: Index Theory and the Seifert Loop Operator on 
${S^1 \times \Sigma}$}\smallskip

In order to analyze the Seifert loop path integral via non-abelian
localization, we must assume that the degree $n$ of the bundle ${S^1
\buildrel n\over\longrightarrow M \buildrel\pi\over\longrightarrow
\Sigma}$ is non-zero.  Otherwise, if ${n=0}$ and ${M = S^1 \times
\Sigma}$, the construction of the invariant contact form $\kappa$ in
Section $3.2$ fails --- see especially \CONII\ --- and our analysis of
Chern-Simons theory via localization does not apply.

Regardless, localization or no, Chern-Simons gauge theory simplifies
dramatically on a three-manifold which is a product ${S^1 \times
\Sigma}$.  In this situation, as we quickly review below, the Seifert
loop path integral $Z(\epsilon;C,R)$ merely computes the dimension of
a certain Hilbert space, which is constructed geometrically as the
space of holomorphic sections of a prequantum  line bundle over
$\SN(P;\lambda)$.  Since our identification \CHRVRP\ of the
Seifert loop class as the character $\ch_R\big(\SV_p\big)$ is 
independent of the degree $n$, one might wonder 
whether that identification can be understood more easily in the 
special case ${n=0}$.

Somewhat remarkably, this question has already been answered by
Jeffrey in work \JeffreyPB\ that substantially predates the current
paper.  (See also \Hamilton\ for a related follow-up.)  Indeed,
many of the most important ideas required to analyze the Seifert loop
path integral for general degree ${n\ge 1}$ already appear for the somewhat
degenerate case ${n=0}$.  Therefore, as a very useful prelude to our
localization computations, let us first discuss the Seifert loop path
integral on ${M = S^1 \times \Sigma}$, closely following the relevant
portions of \JeffreyPB.

By way of comparison and contrast, we consider simultaneously both the
partition function $Z(\epsilon)$ and the Seifert loop path integral 
$Z(\epsilon;C,R)$ for Chern-Simons theory at level $k$ on ${M = S^1 \times
\Sigma}$.  The curve $C$ associated to the Seifert loop operator is
then given by ${C = S^1 \times \{p\}}$, where $p$ is a point on
$\Sigma$.

When ${M = S^1 \times \Sigma}$ factorizes, Chern-Simons theory on $M$
can be immediately analyzed via canonical quantization in the
Hamiltonian formalism, just as we applied to the Wilson loop operator
itself in Section $4.1$.  In the Hamitonian formalism, we
interpret the $S^1$ factor of $M$ as a periodic, Euclidean ``time''.
According to the standard axioms of quantum field theory, the
Chern-Simons partition function $Z(\epsilon)$ on $M$ is then generally
given by a quantum mechanical trace of the form  
\eqn\IDXI{ Z(\epsilon) \,=\,  \Tr_{\SH(k)}\, P\exp{\!\left(-i
\oint_{S^1}\!{\bf H}\right)}\,,\qquad\qquad M \,=\, S^1 \times \Sigma\,.}
Here $\SH(k)$ is the finite-dimensional Hilbert space obtained by
quantizing Chern-Simons theory at level $k$ on $\Sigma$, and ${\bf H}$
is the Hamiltonian which acts on $\SH(k)$ to generate infinitesimal
translations around $S^1$.  Because the Chern-Simons action is purely
topological, the Hamiltonian vanishes, 
\eqn\HAMV{ {\bf H} \,=\, 0\,.}
Hence $Z(\epsilon)$ reduces to the trace of the identity operator on
$\SH(k)$,
\eqn\IDXII{ Z(\epsilon) \,=\, \Tr_{\SH(k)} \, {\bf 1} \,=\,
\dim_\BC \SH(k)\,,}
and the Chern-Simons path integral on ${S^1\times\Sigma}$
simply computes the dimension of the Hilbert space $\SH(k)$.

For the Seifert loop path integral, the axiomatic reasoning which leads
to \IDXII\ remains valid.  Again, $Z(\epsilon;C,R)$ is given by
the dimension of a Hilbert space,
\eqn\IDXIIR{ Z(\epsilon;C,R) \,=\, \Tr_{\SH(k;\alpha)} \, {\bf 1}
\,=\, \dim_\BC \SH(k;\alpha)\,,}
where $\SH(k;\alpha)$ is now obtained by quantizing Chern-Simons theory
at level $k$ in the presence of the Seifert loop operator $W_R(C)$
which punctures $\Sigma$ at the point $p$.

As nicely reviewed in \AtiyahKN, each of the Hilbert spaces $\SH(k)$ and
$\SH(k;\alpha)$ can be constructed geometrically as the space of
holomorphic sections of a certain prequantum line bundle over the
appropriate classical phase space of Chern-Simons theory.  In the case
of $\SH(k)$, the relevant phase space is just the moduli space $\SN(P)$ of
flat connections on $\Sigma$, so that\foot{The following caveat is in
order.  If the Chern-Simons gauge group $G$ is simply-connected, then
the $G$-bundle $P$ on $\Sigma$ is topologically trivial, and $\SN(P)$ has
singularities at points corresponding to reducible flat connections on
$\Sigma$.  At present, we proceed somewhat formally and ignore
singularities in $\SN(P)$.  Lest the reader worry, when we return to
this discussion in the context of localization, $P$ will be
non-trivial and $\SN(P)$ will be smooth.}
\eqn\HLBSCS{ \SH(k) \,\cong\, H^0_{\bar\partial}\!\left(\SN(P),
\SL_0^k\right).}\countdef\CurlyHk=194\CurlyHk=\pageno
Here the prequantum line bundle $\SL_0^k$ is the $k$-th power of the basic line
bundle $\SL_0$ which generates the Picard group of $\SN(P)$ and whose
first Chern class is represented on $\SN(P)$ by 
\eqn\SLZERO{ c_1(\SL_0) \,=\, \Omega_0 \,=\, {1 \over
{4\pi^2}}\,\Omega\,.}
\countdef\CurlyLnought=196\CurlyLnought=\pageno
\countdef\BigOmnought=197\BigOmnought=\pageno
We recall that $\Omega$ is the symplectic form appearing in \OMYM.
With our normalization conventions, the periods of $\Omega$ are
given by integral multiples of $4\pi^2$, so the normalization factor
in \SLZERO\ ensures that $\Omega_0$ is an integral generator of 
${H^2\!\big(\SN(P);\BZ\big) \cong \BZ}$.

The geometric description of $\SH(k;\alpha)$ may be only marginally
less familiar.  As we have seen in Section $5.1$, classical solutions
of Chern-Simons theory in the presence of the Seifert loop operator
$W_R(C)$ are described on $\Sigma$ by the extended moduli space
$\SN(P;\lambda)$, where the parameter $\lambda$ is fixed as in
\LAMALK\ by the ratio 
\eqn\LAMALKII{ \lambda \,=\, {\alpha \over k}\,.}
Hence the Hilbert space $\SH(k;\alpha)$ which describes the Seifert
loop operator is constructed by quantizing a new phase space, namely
$\SN(P;\lambda)$, 
\eqn\HLBSCSR{ \SH(k;\alpha) \,\cong\,
H^0_{\bar\partial}\!\left(\SN(P;\lambda),\, \SL^{(k)}_\alpha\right),}
where $\SL_\alpha^{(k)}$ is the requisite prequantum line bundle over
$\SN(P;\lambda)$.\countdef\CurlyHkalpha=195\CurlyHkalpha=\pageno

Unlike the prequantum line bundle $\SL_0^k$ which defines the Hilbert space
$\SH(k)$, the line bundle $\SL_\alpha^{(k)}$ is not generally the $k$-th 
power of any more basic line bundle on $\SN(P;\lambda)$.  However,
again up to factors of $2\pi$, the first Chern class of $\SL_\alpha^{(k)}$ 
is still represented on $\SN(P;\lambda)$ by the $k$-th multiple of
the symplectic form $\Omega_\lambda$ in \OMLAM,
\eqn\CHRNSLA{ c_1\!\left(\SL_\alpha^{(k)}\right) \,=\,
{k\over{4\pi^2}} \, \Omega_\lambda \,=\, k\,\Rq^*\Omega_0 \,+\,
\left\langle\alpha,\Ru\right\rangle.}
Here, along with the elementary identification in \LAMALKII, we use the
description \OMLAMII\ of $\Omega_\lambda$ in terms of the Chern roots
$\Ru$ of $\SV_p$.  This description of $\Omega_\lambda$ is valid so
long as the symplectic fibration
${\Rq:\SN(P;\lambda)\rightarrow\SN(P)}$ in \PSDWNI\ is  
itself valid, for which we require the strict integrability bound 
${(\vartheta,\alpha)<k}$ on the weight $\alpha$.  

According to Theorem $9.12$ in \AtiyahYM, $\SN(P)$ is
simply-connected.  Via the symplectic fibration in \PSDWNI,
$\SN(P;\lambda)$ is simply-connected as well.  Hence our expression
for the Chern class of $\SL_\alpha^{(k)}$ in \CHRNSLA\ suffices to
determine the line bundle itself.  To be precise, since $\Omega_0$
represents the Chern class of the generating line bundle $\SL_0$ on
$\SN(P)$, and since each Chern root $\Ru_j$ represents the 
Chern class of the splitting line bundle $\SL_j$ in  \CHRPSD, the sum
in \CHRNSLA\ implies that $\SL_\alpha^{(k)}$ is given by the tensor
product 
\eqn\PREQLA{ \SL_\alpha^{(k)} \,\cong\, \Rq^*\SL_0^k \,\otimes\,
\bigotimes_{j=1}^{r+1} \,
\SL_j^{\alpha_j}\,,}\countdef\CurlyLalpha=198\CurlyLalpha=\pageno
where $\alpha_j$ for ${j=1,\ldots,r+1}$ are the components of $\alpha$
in \DYNKINL.  The quantization of $\alpha$ as a weight of
$SU(r+1)$ is necessary to ensure that the tensor product of line bundles
$\SL_j^{\alpha_j}$ in \PREQLA\ is well-defined.  To check that the
tensor product is sensible, we observe that the product
${\otimes_{j=1}^{r+1}\,\SL_j^{\alpha_j}}$ can be rewritten as 
${\otimes_{j=1}^r \, \SL_j^{\alpha_j - \alpha_{r+1}}}$, where we
eliminate the redundant bundle $\SL_{r+1}$ via the relation
${\SL_{r+1} \cong \otimes_{j=1}^r \, \SL_j^{-1}}$, which follows from
the elementary identity ${\hat\omega_1 \,+\, \cdots \,+\,
\hat\omega_{r+1} = 0}$.  In our conventions, with ${\alpha > 0}$ a
positive weight of $SU(r+1)$, the differences ${\alpha_j -
\alpha_{r+1} \in \BZ_{>0}}$ for ${j=1,\ldots,r}$ are then strictly
positive integers, and no fractional powers of line bundles appear in
the tensor product ${\otimes_{j=1}^r \, \SL_j^{\alpha_j - \alpha_{r+1}}}$.

To summarize, for Chern-Simons theory on ${M=S^1 \times \Sigma}$, both
$Z(\epsilon)$ and $Z(\epsilon;C,R)$ count holomorphic sections of
appropriate prequantum line bundles on $\SN(P)$ and $\SN(P;\lambda)$,
\eqn\IDXIII{\eqalign{
Z(\epsilon) \,&=\, \dim_\BC H^0_{\bar\partial}\Big(\SN(P),
\SL_0^k\Big),\cr
Z(\epsilon;C,R) \,&=\, \dim_\BC
H^0_{\bar\partial}\Big(\SN(P;\lambda),\,\Rq^*\SL_0^k \,\otimes\,
\bigotimes_{j=1}^{r+1} \, \SL_j^{\alpha_j}\Big).}}

Crucially, for both prequantum line bundles appearing in \IDXIII, the
higher Dolbeault cohomology vanishes.  As noted for instance in \S
$11$ of \JeffreyKW, the  vanishing of $H^q_{\bar\partial}\big(\SN(P),
\SL_0^k\big)$ for ${q > 0}$ follows from the Kodaira vanishing
theorem, in combination with the identification of the canonical
bundle of $\SN(P)$ as $\SL_0^{-2}$.  The vanishing of
$H^q_{\bar\partial}\big(\SN(P;\lambda), \SL_\alpha^{(k)}\big)$ for
${q > 0}$ is apparently a bit more subtle, but it has been proven by
Teleman in $(9.6)$ of \TelemanQC.  As a result, both $Z(\epsilon)$ and
$Z(\epsilon;C,R)$ are alternatively given by the holomorphic Euler
characters
\eqn\IDXIV{\eqalign{
Z(\epsilon) \,&=\, \chi\Big(\SN(P),
\SL_0^k\Big),\cr
Z(\epsilon;C,R) \,&=\, \chi\Big(\SN(P;\lambda),\,\Rq^*\SL_0^k \,\otimes\,
\bigotimes_{j=1}^{r+1} \, \SL_j^{\alpha_j}\Big).}}

Let us recall one basic fact about the holomorphic Euler character.
In general, if $E$ is a holomorphic vector bundle over a compact
complex manifold $X$, the holomorphic Euler character $\chi(X,E)$ is
defined by the alternating sum 
\eqn\HOLEULER{ \chi(X,E) \,=\, \sum_{q} \,\, (-1)^q \, \dim_\BC
H^q_{\bar\partial}(X, E)\,.}\countdef\HolEuler=199\HolEuler=\pageno
The Atiyah-Singer index theorem, in its Hirzebruch-Riemann-Roch
form, then provides an entirely cohomological description for
the Euler character,
\eqn\INDXXE{ \chi(X, E) \,=\, \int_X \ch(E) \cdot 
\Td(X)\,.}
We have already encountered the Chern character $\ch(E)$ in \CHCHRII.
Otherwise, $\Td(X)$ is the Todd class of $X$.  Explicitly, if
$x_j$ for ${j=1,\ldots,\dim_\BC X}$ are the Chern roots of the
complex tangent bundle of $X$, such that 
\eqn\CHRNRTSX{  c(X) \,=\,
\prod_{j=1}^{\dim_\BC X} \left(1 \,+\, x_j\right)\,,}
then $\Td(X)$ is given by the product 
\eqn\TDX{ \Td(X) \,=\, \prod_{j=1}^{\dim_\BC X} \, {{x_j}\over{1
\,-\, \exp{\!\left(-x_j\right)}}}.}
\countdef\ToddCl=200\ToddCl=\pageno

Applying the Hirzebruch-Riemann-Roch formula in \INDXXE\ to \IDXIV,
we thereby express $Z(\epsilon)$ and $Z(\epsilon;C,R)$ as integrals
over the respective moduli spaces $\SN(P)$ and $\SN(P;\lambda)$,
\eqn\INDXV{\eqalign{
Z\!\big(\epsilon\big) \,&=\, \int_{\SN(P)} \exp{\!\big[ k \, \Omega_0
\big]} \cdot \Td\!\big(\SN(P)\big)\,,\cr
Z\!\big(\epsilon;C,R\big) \,&=\, \int_{\SN(P;\lambda)} \exp{\!\Big[
k \, \Rq^* \Omega_0 \,+\, \langle\alpha,\Ru\rangle \Big]} \cdot
\Td\!\big(\SN(P;\lambda)\big)\,.}}
In obtaining \INDXV, we trivially note that ${\ch(E) =
\exp{\!\big[c_1(E)\big]}}$ when $E$ is a line bundle, and we
substitute for the respective Chern classes using \SLZERO\ and
\CHRNSLA.

Although \INDXV\ provides the sought-after cohomological description
for $Z(\epsilon;C,R)$, we still need to tease out a description of the
Seifert loop class itself,
\eqn\SLOOPCL{ W_R(C)\big|_{\SN(P)} \,\in\, H^*\!\big(\SN(P)\big)\,.}
Intuitively, the Seifert loop class $W_R(C)\big|_{\SN(P)}$ should play
the same role in the presentation of $Z(\epsilon;C,R)$ as an integral
over the finite-dimensional moduli space $\SN(P)$ that the
Seifert loop operator $W_R(C)$ itself plays in the full Seifert loop
path integral over the infinite-dimensional affine space $\CA$ of
connections,
\eqn\PZCSWLTRV{\eqalign{
Z\!\big(\epsilon; C, R\big) \,&=\, {1 \over {\Vol(\CG)}} \, \left({1
\over {2 \pi \epsilon}}\right)^{\Delta_{\CG}} \, \int_\CA \! \CD \! A
\;\; W_R(C) \cdot \exp{\left[{i \over {2 \epsilon}} \int_M \! \Tr\!\left( A \^
d A + {2 \over 3} A \^ A \^ A \right)\right]}\,,\cr
\epsilon &= {{2 \pi} \over k}\,,\qquad \Delta_{\CG} = \dim \CG\,.}}
Here the Seifert loop path integral \PZCSWLTRV\ is obtained by
multiplying the exponential of the Chern-Simons action by the 
operator $W_R(C)$, interpreted as a classical functional of the
connection $A$.  

Similarly, the Seifert loop class will be
characterized as the element of $H^*\!\big(\SN(P)\big)$ by which we
multiply the index theory integrand of $Z(\epsilon)$ in \INDXV\ to
obtain a corresponding cohomological formula for $Z(\epsilon;C,R)$,
\eqn\INDXVI{ Z\!\big(\epsilon;C,R\big) \,=\, \int_{\SN(P)} \,
W_R(C)\big|_{\SN(P)} \cdot \exp{\!\big[ k \, \Omega_0
\big]} \cdot \Td\!\big(\SN(P)\big).}
Crucially, $Z(\epsilon;C,R)$ is expressed in \INDXVI\ as an integral
over the basic moduli space $\SN(P)$, as opposed to the extended
moduli space $\SN(P;\lambda)$.  To compute the Seifert loop class, we
must therefore push the index density on $\SN(P;\lambda)$ in  \INDXV\
down to $\SN(P)$ by integrating over the fiber $2\pi\CO_{-\lambda}$ of the
map ${\Rq:\SN(P;\lambda)\rightarrow\SN(P)}$,
\eqn\INDXXVI{ Z\!\big(\epsilon;C,R\big) \,=\, \int_{\SN(P)}
\Rq_*\!\left[\exp{\!\Big( k \, \Rq^* \Omega_0 \,+\,
\langle\alpha,\Ru\rangle \Big)} \cdot
\Td\!\big(\SN(P;\lambda)\big)\right].}
After simplifying the expression in \INDXXVI\ and comparing the result
to \INDXVI, we ultimately determine $W_R(C)\big|_{\SN(P)}$.  As will be
clear, this calculation nicely parallels the more elementary
manipulations leading to \SFTWYVII\ in Section $7.2$.

To begin to simplify \INDXXVI, let us express the Todd class of
$\SN(P;\lambda)$ in a more revealing form.  In general, for any
holomorphic fibration of smooth complex manifolds 
\eqn\FIBRFXB{\matrix{
&F \longrightarrow X\cr
&\mskip 60mu\big\downarrow\lower 0.5ex\hbox{$^{\pi}$}\cr
&\mskip 48mu B}\,,}
the Todd class of $X$ factorizes as 
\eqn\TDXII{ \Td(X) \,=\, \Td(F) \cdot \pi^*\Td(B)\,.}
As standard, the factorization in \TDXII\ follows from the
corresponding exact sequence of bundles 
\eqn\TDXIII{0 \,\longrightarrow\, TF \,\longrightarrow\, TX
\,\buildrel\pi_*\over\longrightarrow\, TB \,\longrightarrow\, 0\,,}
along with the manifestly multiplicative definition \TDX\ of
$\Td(X)$ as a product over the individual Chern roots of the
holomorphic tangent bundle $TX$.  For the fibration
${\Rq:\SN(P;\lambda)\rightarrow\SN(P)}$ with the natural complex
structures on all spaces involved, we obtain 
\eqn\TDCLSNPL{ \Td\!\big(\SN(P;\lambda)\big) \,=\,
\Td\!\big(\CO_{-\lambda}\big) \cdot
\Rq^*\Td\!\big(\SN(P)\big)\,.}

Let us make a comment, not inessential, about the appearance of
$\CO_{-\lambda}$ as opposed to $\CO_\lambda$ in \TDCLSNPL.  According
to \PSDWNI, the fiber of the map $\Rq$ is symplectically the orbit
$2\pi\CO_{-\lambda}$.  For the purpose of computing the fiberwise 
Todd class, the overall normalization of the orbit by $2\pi$ is
irrelevant, so we have dropped that normalization in \TDCLSNPL.
However, if we are going to be careful about signs and orientations,
the distinction between $\lambda$ and $-\lambda$ in \TDCLSNPL\ is
important.  

According to \HOLTN, when ${\lambda > 0}$ is positive, the
complex structure on $\CO_\lambda$ compatible with the symplectic form
$\nu_\lambda$ is such that the holomorphic tangent space to
$\CO_\lambda$ at the identity is given by ${\Fg^{(1,0)}\!=\Fg_{+}}$, the
positive rootspace of $\Fg$.  But because the sign of the symplectic form on
$\CO_{-\lambda}$ is reversed relative to $\CO_\lambda$, the compatible
complex structure on $\CO_{-\lambda}$ is also reversed.  That is, the
holomorphic tangent space to $\CO_{-\lambda}$ at the identity is given
by the negative rootspace $\Fg_{-}$ (for a fixed splitting ${\Fg_\BC\!
\ominus \Ft_\BC = \Fg_{+}\!\oplus \Fg_{-}}$).  So even though
$\CO_\lambda$ and $\CO_{-\lambda}$ agree at the level of topology,
they carry opposite complex structures.  Eventually, this distinction
will be manifested through various signs in
$\Td\!\big(\CO_{-\lambda}\big)$ relative to
$\Td\!\big(\CO_\lambda\big)$, associated to the exchange ${\Fg_{+}
\!\leftrightarrow \Fg_{-}}$.

Via \TDCLSNPL, the index density on $\SN(P;\lambda)$ itself
factorizes, implying 
\eqn\INDXVII{ Z(\epsilon;C,R) \,=\, \int_{\SN(P;\lambda)} \,
\exp{\!\big[\langle\alpha,\Ru\rangle\big]} \cdot
\Td\!\big(\CO_{-\lambda}\big) \cdot \Rq^*\!\Big(\!\exp{\!\big[k \,
\Omega_0\big]} \Td\!\big(\SN(P)\big)\Big)\,.}
In particular, comparing to \INDXV, we see the pullback of the index 
density on $\SN(P)$ appears as the final factor in the integrand of \INDXVII. 
Trivially, terms which pull back from the base $\SN(P)$ do not contribute
to the integral over the fiber of $\SN(P;\lambda)$.  Consequently, the
Seifert loop class in \INDXVI\ is given by the pushdown of the
first two factors in the integrand of \INDXVII,
\eqn\SLOOPCLII{ W_R(C)\big|_{\SN(P)} \,=\, \Rq_* S_\alpha(\Ru) \,\in\,
H^*\big(\SN(P)\big)\,,}
where we set 
\eqn\BIGSA{ S_\alpha(\Ru) \,=\,
\exp{\!\big[\langle\alpha,\Ru\rangle\big]} \cdot
\Td\!\big(\CO_{-\lambda}\big).}
Evaluating ${\Rq_* S_\alpha(\Ru)}$ will be easy as soon as we
determine the Todd class of the orbit $\CO_{-\lambda}$.  We accomplish
this task directly, by computing the Chern roots of $\CO_{-\lambda}$
and then applying the definition in \TDX.  

Let us first make a simple observation about signs.  If $x_j$ for
${j=1,\ldots,\dim_\BC \CO_\lambda}$ are the Chern roots of the
coadjoint orbit $\CO_\lambda$ with the standard complex structure in
\HOLTN, 
\eqn\CHRRTSCO{ c\big(\CO_\lambda\big) \,=\, \prod_{j=1}^{\dim_\BC
\CO_\lambda} \!\!\left(1 \,+\, x_j\right),}
then the Chern roots of the conjugate orbit $\CO_{-\lambda}$, which carries the
opposite complex structure, are given by ${-x_j}$ for
${j=1,\ldots,\dim_\BC \CO_\lambda}$,
\eqn\CHRRTSCOII{ c\big(\CO_{-\lambda}\big) \,=\, \prod_{j=1}^{\dim_\BC
\CO_\lambda} \!\!\left(1 \,-\, x_j\right).}
Because of this relation, we have decided to phrase our initial
discussion in terms of the more familiar Chern roots $x_j$ of
$\CO_\lambda$.  We then reverse signs to compute the Chern roots of
$\CO_{-\lambda}$.

The computation of the Chern roots of the complete flag manifold ${\CO_\lambda
\cong G/T}$ is a classic result, so we will be brief.  See \S $14.2$ of
\Hirzebruch\ for a textbook discussion.  Because $\CO_\lambda$ is a
homogeneous space for $G$, the Chern roots of $\CO_\lambda$ are
determined by any smooth splitting of the holomorphic tangent space
${\Fg^{(1,0)} \!= \Fg_+}$ at the identity.  The obvious splitting at
hand is the eigenspace decomposition of $\Fg_+$ under the action of
the maximal torus $T$, 
\eqn\POSROOTSII{ \Fg_+ \,\cong\, \bigoplus_{\beta>0} \,
\Fe_{\beta}\,,\qquad\qquad \beta \,\in\,\FR\,,}
where $\Fe_\beta$ is the one-dimensional rootspace associated to
the root $\beta$ of $G$.  Each rootspace $\Fe_\beta$ then varies
globally over ${\CO_\lambda \cong G/T}$ as the fiber of the
corresponding homogeneous line bundle $\FL(\beta)$, as in
\BGFLA.  So via the splitting in \POSROOTSII, the Chern classes of the
line bundles $\FL(\beta)$ for all ${\beta > 0}$ realize the Chern
roots of $\CO_\lambda$,
\eqn\CHRRTSCOIII{ c\big(\CO_\lambda\big) \,=\, \prod_{\beta > 0}
\Big[1 \,+\, c_1\!\big(\FL(\beta)\big)\Big],\qquad\qquad
c_1\!\big(\FL(\beta)\big) \,=\, {{\nu_\beta}\over{2\pi}}.}
Equivalently, after reversing signs to account for the reversal of
complex structure on $\CO_{-\lambda}$,
\eqn\CHRRTSCOIV{ c\big(\CO_{-\lambda}\big) \,=\, \prod_{\beta > 0}
\Big[1 \,-\, c_1\!\big(\FL(\beta)\big)\Big].}
 
As a check, if ${G = SU(2)}$, then ${\CO_\lambda = \BC\BP^1}$.
Previously in the example in Section $4.1$, we identified the line
bundle of degree $m$ on $\BC\BP^1$ with the homogeneous line bundle
$\FL\big(m\,\hat\omega\big)$, where ${\hat\omega>0}$ is the highest weight of
the fundamental, two-dimensional representation of $SU(2)$.
Observing that ${\beta = 2 \, \hat\omega}$ in \CHRRTSCOIII, we
obtain the standard result that the first Chern class of $\BC\BP^1$
has degree two.

To compute $\Td(\CO_{-\lambda})$ as a class on $\SN(P;\lambda)$, we
now express the Chern class of each line bundle $\FL(\beta)$ in terms of
the fiberwise components of the fundamental Chern roots $\Ru$ of $\SV_p$.
According to \BIGB\ and \CHRNRTSII, 
\eqn\CHCLB{ c_1\!\big(\FL(\beta)\big) \,=\, {{\nu_\beta}\over{2\pi}}
\,=\, -\langle\beta,\Ru\rangle\Big|_{\CO_\lambda}\,.}
Substituting into \CHRRTSCOIV, we then obtain the simple result 
\eqn\CHRRTSCOV{ c\big(\CO_{-\lambda}\big) \,=\, \prod_{\beta > 0}
\Big(1 \,+\, \langle\beta,\Ru\rangle\Big),}
where the signs in \CHRRTSCOIV\ and \CHCLB\ nicely cancel.

Consequently,
\eqn\TDCOALPH{\eqalign{
\Td(\CO_{-\lambda}) \,&=\, \prod_{\beta > 0} \, 
{{\langle\beta,\Ru\rangle}\over{1 \,-\,
\exp{\!\left[-\langle\beta,\Ru\rangle\right]}}}\,,\cr
&=\, \prod_{\beta > 0} \, {{\e{\!\langle\beta,\Ru\rangle/2}}\over{2
\sinh\!\big({{\langle\beta,\Ru\rangle/2}}\big)}} \cdot
\langle\beta,\Ru\rangle\,,\cr
&=\,{{\e{\!\langle\rho,\Ru\rangle}}\over{\A_\rho(\Ru)}} \cdot 
\prod_{\beta > 0} \, \langle\beta,\Ru\rangle\,,\qquad\qquad \rho \,=\,
\ha\,\sum_{\beta>0}\,\beta\,.}}
The second equality in \TDCOALPH\ is an elementary algebraic identity,
and in passing to the last line of \TDCOALPH, we apply the Weyl
denominator formula \ARHO\ for $\A_\rho$.

With this description of $\Td(\CO_{-\lambda})$, the class
$S_\alpha(\Ru)$ in \BIGSA\ becomes 
\eqn\BIGSAII{ S_\alpha(\Ru) \,=\,
{{\e{\!\langle\alpha + \rho,\Ru\rangle}}\over{\A_\rho(\Ru)}} \cdot 
\prod_{\beta > 0} \, \langle\beta,\Ru\rangle\,.}
To finish our computation of the Seifert loop class
$W_R(C)\big|_{\SN(P)}$, we are left to push $S_\alpha(\Ru)$ down to
$\SN(P)$ under the map ${\Rq:\SN(P;\lambda)\rightarrow\SN(P)}$.

To orient the reader, let us quickly recall how pushdown works.  See
\S $6$ of \BottT\ for a more thorough discussion in the special case of
vector bundles.  As in \FIBRFXB, we suppose that we have a smooth
fibration of compact manifolds ${F \,\rightarrow\, X
\,\buildrel\pi\over\rightarrow B}$.\foot{We phrase the discussion here in
the context of differential topology, and we do not necessarily 
assume that $F$, $X$, and $B$ are complex manifolds.}  Locally, with
respect to a given trivialization over a patch of $B$, any differential
form $\psi$ on $X$ can be written as a finite sum of terms 
\eqn\RANDPSI{ \psi \,=\, \sum_i \, \pi^*\xi_{(i)} \^\, \eta_{(i)}\,,} 
where each $\xi_{(i)}$ is a differential form on $B$, and each
$\eta_{(i)}$ is a vertical differential form on $X$.  That is,
$\eta_{(i)}$ has components only along the fiber $F$, or equivalently,
the contraction of $\eta_{(i)}$ with any horizontal tangent vector
along $B$ vanishes.  

As a map from differential forms on $X$ to differential forms on $B$,
the pushdown $\pi_*$ is given by the naive operation of
integrating each vertical component $\eta_{(i)}$ of $\psi$ fiberwise
over $F$,
\eqn\PUSHDWN{ \pi_* \psi \,=\, \sum_i \, \xi_{(i)} \cdot \int_F
\eta_{(i)}\,.}
By definition, if $\eta_{(i)}$ has no component of top degree on $F$,
the integral over $F$ is set to zero.  As in \S $6$ of \BottT, one
can check that the operation in \PUSHDWN\ does not depend upon the
particular choice of local trivialization used to obtain the fiberwise
decomposition of $\psi$ in \RANDPSI, so long as the trivialization
respects the relative orientations of $X$ and $F$.  Hence $\pi_*$ is
well-defined globally on $X$.  Moreover, $\pi_*$ commutes with the exterior
derivative $d$ and thus induces a map on de Rham cohomology ${\pi_*\!: H^*(X)
\rightarrow H^{*-\dim F}(B)}$, where we observe that the integration in
\PUSHDWN\ reduces the degree of a form on $X$ by the dimension of $F$.

We now return to the specific problem of computing the pushdown 
$\Rq_* S_\alpha(\Ru)$, for which the essential step is simply to
write $S_\alpha(\Ru)$ in the form \RANDPSI.  Let us therefore consider
the Weyl action by permutations on the entries of  ${\Ru =
i \diag(\Ru_1,\ldots,\Ru_{r+1})}$.  Because
$(\Ru_1,\ldots,\Ru_{r+1})$ are the Chern roots of the bundle $\SV_p$ 
over $\SN(P)$, any symmetric function of $(\Ru_1,\cdots,\Ru_{r+1})$ is 
tautologically the pullback from $\SN(P)$ of a corresponding function
of the Chern classes of $\SV_p$.  

On the other hand, as we have already seen in \CHRRTSCOV, the
fiberwise components of the classes $\langle\beta,\Ru\rangle$ for all
${\beta > 0}$ represent the Chern roots of $\CO_{-\lambda}$.  Let us
focus attention on the product over $\langle\beta,\Ru\rangle$
occurring as the final factor in \BIGSAII, for which we introduce the
shorthand 
\eqn\TOPOA{ \eta \,=\, \prod_{\beta > 0}\,\langle\beta,\Ru\rangle\,.}
This differential form $\eta$ on $\SN(P;\lambda)$ has two important features.  

First, since the Weyl group $\FW$ is generated by reflections in the
simple roots, $\eta$ is manifestly alternating under the action of $\FW$.  

Second, $\eta$ restricts to a non-vanishing top-form on each fiber of 
$\SN(P;\lambda)$.  For this claim, we note that each Chern root
$\Ru_j$ restricts via \CHRNRTSII\ to an invariant two-form on
$\CO_{-\lambda}$, so $\eta$ is also invariant.  As a result, $\eta$
will be non-vanishing on $\CO_{-\lambda}$ so long as the integral of
$\eta$ over $\CO_{-\lambda}$ is itself non-vanishing.  But the
product of Chern roots in \TOPOA\ represents the top Chern class of
${\CO_{-\lambda}}$, so the fiberwise integral of $\eta$ computes the
topological Euler character $\chi_{\rm top}$ of ${\CO_{-\lambda}
\cong G/T}$,
\eqn\ELROA{\eqalign{
\int_{\CO_{-\lambda}} \!\eta \,&=\, \chi_{\rm top}(G/T)\,,\cr
&= |\FW|\,.}}\countdef\TopEuler=201\TopEuler=\pageno
As we review at the conclusion of this discussion, for any compact
simple Lie group $G$, the Euler character of $G/T$ is given by the
order of the Weyl group $\FW$, which is certainly non-zero.  In the
case ${G = SU(r+1)}$ of present relevance, $\FW$ is the symmetric
group acting by permutations on $\{\Ru_1,\ldots,\Ru_{r+1}\}$, so
${|\FW| = (r+1)!}$.

As apparent from these observations, in any fiberwise decomposition of
$S_\alpha(\Ru)$ as in \RANDPSI, each top-degree vertical form
appearing in the decomposition must be proportional to $\eta$ and is thus
alternating under $\FW$.  Otherwise, each horizontal form which pulls
back from $\SN(P)$ under $\Rq^*$ is automatically invariant under
$\FW$.  Consequently, in the fiberwise decomposition of
$S_\alpha(\Ru)$, only those summands which are themselves 
alternating under $\FW$ can contribute under the pushdown $\Rq_*$.

To pick out the alternating piece of  
\eqn\BIGSAIII{ S_\alpha(\Ru) \,=\, {{\e{\!\langle\alpha +
\rho,\Ru\rangle}}\over{\A_\rho(\Ru)}} \cdot \eta\,,}
we note that both $\eta$ and $\A_\rho(\Ru)$ are already
alternating.  So the alternating piece of $S_\alpha(\Ru)$ is obtained
by anti-symmetrizing the remaining exponential factor exactly as in \AEXP,
\eqn\AEXPII{\eqalign{
\A\!\left[\e{\!\langle\alpha + \rho,\Ru\rangle}\right]
&=\, {1 \over {|\FW|}} \, \sum_{w \in \FW} \, (-1)^w \,
\e{\!\langle w\cdot(\alpha + \rho),\,\Ru\rangle}\,,\cr
&=\, {1 \over {|\FW|}} \, \A_{\alpha + \rho}(\Ru)\,.}}
Thus we write $S_\alpha(\Ru)$ as 
\eqn\BIGSAIV{\eqalign{
S_\alpha(\Ru) \,&=\, {1 \over {|\FW|}} \,\,
{{\A_{\alpha+\rho}(\Ru)}\over{\A_\rho(\Ru)}} \cdot \eta \,+\,
\cdots\,,\cr
&=\, {1 \over {|\FW|}} \,\, \Rq^*\ch_R\big(\SV_p\big) \cdot \eta \,+\,
\cdots\,,}}
where the `$\cdots$' indicate terms which are annihilated by $\Rq_*$,
and in the second line of \BIGSAIV\ we apply the character formula as
in \CHRVRPII.

Finally, integrating $\eta$ over the fiber of $\SN(P;\lambda)$ via
\ELROA, we obtain the promised result \CHRVRP\ for the Seifert loop class on
$\SN(P)$, albeit in the special case ${n=0}$,
\eqn\SLOOPCLIII{\eqalign{
W_R(C)\big|_{\SN(P)} \,&=\, \Rq_* S_\alpha(\Ru)\,,\cr
&=\,  {1 \over {|\FW|}} \,\, \ch_R\big(\SV_p\big) \cdot
\int_{\CO_{-\lambda}} \!\eta \,,\cr
&=\, \ch_R\big(\SV_p\big)\,.}}
In passing to the last line of \SLOOPCLIII, we see that integral of
$\eta$ over $\CO_{-\lambda}$ precisely cancels the factor ${1/|\FW|}$
which appears under the anti-symmetrization in \AEXPII.  

Equivalently via \INDXVI,
\eqn\INDXVIII{ Z(\epsilon;C,R) \,=\, \int_{\SN(P)} \,
\ch_R\big(\SV_p\big) \cdot \exp{\!\big[ k \, \Omega_0
\big]} \cdot \Td\!\big(\SN(P)\big)\,.}

To tie up a loose end, let us quickly sketch a computation of the
Euler character of $G/T$.  One approach, valid for any compact simple
Lie group, is to use the Bruhat decomposition of the complex group 
$G_\BC$.  In a nutshell (see \S $23.4$ of \Fulton\ for details), the
Bruhat decomposition of $G_\BC$ provides a distinguished covering of
$G/T$ by disjoint open affine cells, each cell being labeled by an
element of the Weyl group $\FW$ of $G$.  As a concrete example, if
$G=SU(2)$, ${\FW \cong \BZ_2}$, and ${G/T = \BC\BP^1}$, the Bruhat
decomposition implies the cellular presentation ${\BC\BP^1 = \{pt\}
\cup D^2}$, where $D^2$ is the open disc realizing the complement of
the given point.  Since each Bruhat cell generally has even dimension,
the Euler character simply counts the number of cells, which is given
by $|\FW|$.

For the special case ${G = SU(r+1)}$, the Euler character of $G/T$ can
also be computed by more elementary means.  For each rank ${r\ge 1}$,
we have the standard fiber bundle 
\eqn\FBRI{\matrix{
&SU(r) \,\longrightarrow\, SU(r+1)\cr
&\mskip 120mu \Big\downarrow\cr
&\mskip 120mu S^{2r+1}}\,.}
Dividing $SU(r+1)$ in \FBRI\ by the maximal torus ${T = U(1)^r}$, we obtain 
\eqn\FBRII{\matrix{
&SU(r)/U(1)^{r-1} \,\longrightarrow\, SU(r+1)/U(1)^r\cr
&\mskip 138mu \Big\downarrow\cr
&\mskip 175mu S^{2r+1}\!/U(1) \cong \BC\BP^r}\,.}
Because we are about to compute the Euler character inductively in
$r$, let us set 
\eqn\BIGXR{ X_r \,=\, SU(r)/U(1)^{r-1}\,.} 
The fibration in \FBRII, together with the multiplicativity of the
Euler character, then implies 
\eqn\ELROAR{\eqalign{
\chi_{\rm top}\big(X_{r+1}\big) \,&=\, \chi_{\rm
top}\big(X_r\big) \cdot \chi_{\rm top}\big(\BC\BP^r\big)\,,\cr
&=\, \chi_{\rm top}\big(X_r\big) \cdot (r+1)\,.}}
Trivially, $X_1$ is a point, so that ${\chi_{\rm top}(X_1) = 1}$.  The
relation in \ELROAR\ thus yields by induction 
\eqn\ELROARII{ \chi_{\rm top}\big(X_{r+1}\big) \,=\, (r+1)! \,=\,
|\FW|\,,\qquad\qquad G \,=\, SU(r+1)\,.}

\bigskip\noindent{\it A Symplectic Model for Localization on
$\SM_0\big(C,\alpha\big)$}\smallskip

The identification \SLOOPCLIII\ of the Seifert loop class with the
character ${\ch_R\big(\SV_p\big)}$ is certainly an elegant result, but
one might wonder whether this identification is somehow special to
Chern-Simons theory on the product ${S^1 \times \Sigma}$.  After
all, only in this case does the Seifert loop path integral
$Z(\epsilon;C,R)$ compute a quantity as simple as the dimension of a
Hilbert space, an observation which provided the starting point
\IDXIIR\ for the preceding analysis.  Indeed, given the intricacies of
that analysis, one might question whether the Seifert loop operator
still reduces to the character $\ch_R\big(\SV_p\big)$ even when $M$ is
the total space of a non-trivial $S^1$-bundle over $\Sigma$ of
positive degree ${n \ge 1}$.

To complete our analysis of Chern-Simons gauge theory on Seifert 
manifolds, we now establish the universal description \CHRVRP\ of the
Seifert loop class as $\ch_R\big(\SV_p\big)$ for all degrees ${n \ge 1}$.
In essence,  our strategy is direct --- we apply the general non-abelian
localization formula in \ZEV\ to the Seifert loop path integral on
$M$.  Rather than plunge immediately into details, however, let us quickly 
indicate the structure of that localization computation, which has two
main steps.

According to our discussion of the classical Seifert loop operator in
Section $7.1$, the Seifert loop path integral localizes onto
the extended moduli space $\SM(C,\alpha)$ of flat connections on $M$
with fixed monodromy around the meridian of $C$, as in \MCALPHA.  For the
first step in our computation, we use the localization formula in
\ZEV\ to reduce the Seifert loop path integral over the
infinite-dimensional space ${\bar\CA_\alpha = \bar\CA \times \epsilon
L\CO_\alpha}$ to the integral of an appropriate de Rham cohomology
class $[d\mu]$ on each smooth component $\SM_0(C,\alpha)$ of the
moduli space $\SM(C,\alpha)$.  Schematically,
\eqn\SCHSL{ Z\big(\epsilon;C,R\big)\Big|_{\SM_0(C,\alpha)} \,=\,
\int_{\SM_0(C,\alpha)} [d\mu]\,,\qquad\qquad
[d\mu] \in H^*\!\big(\SM_0(C,\alpha)\big)\,,}
\countdef\Dmu=202\Dmu=\pageno
where the class $[d\mu]$ generally depends upon the discrete parameters
$(n, k, \alpha)$ which specify respectively the degree of the
$S^1$-bundle, the Chern-Simons level, and the highest weight of the
irreducible representation $R$.  As will hopefully be clear, the
computation of $[d\mu]$ closely resembles the localization computation on
$\CO_\alpha$ in Section $7.2$.

According to \CHRVRP, the universal identification of the Seifert loop
class with the character $\ch_R\big(\SV_p\big)$ is to hold on each smooth
component $\SM_0$ of the basic moduli space $\SM$ of flat
connections, as opposed to the extended moduli space
$\SM(C,\alpha)$.  For the second step in our calculation, we therefore 
push the class $[d\mu]$ on $\SM_0(C,\alpha)$ down to $\SM_0$ 
via the map ${\Rq:\SM_0(C,\alpha) \rightarrow \SM_0}$ in the
symplectic fibration \FBMCA\ of moduli spaces.  Not surprisingly,
given the identifications 
\eqn\MODIDS{ \SM_0 \,\cong\, \wt\SN(P)\,,\qquad\qquad \SM_0(C,\alpha)
\,\cong\, \wt\SN(P;\lambda)\,,\qquad\qquad \lambda \,=\, {\alpha\over k}\,,}
the pushdown under $\Rq_*$ proceeds along exactly the same
lines as the previous index theory computation for the Seifert loop
operator on ${M = S^1 \times \Sigma}$.  So for the second step in the
computation, in which we push $[d\mu]$ down to $\SM_0$, we effectively
recycle the results of Jeffrey in \JeffreyPB.

Of the two main steps in our calculation of the Seifert loop class,
only the first step, the localization computation of $[d\mu]$ on 
$\SM_0(C,\alpha)$, requires additional work.  

To make use of the general formula in \ZEV\ for localization on
$\SM_0(C,\alpha)$, we must first specify the symplectic model for a
small neighborhood in ${\bar\CA_\alpha = \bar\CA \times \epsilon
L\CO_\alpha}$ of any pair $\big(A_{\rm cl},\, U_{\rm cl}\big)$ which
satisfies the classical equations of motion in \EOMGGV\ and thereby
represents a point in $\SM_0(C,\alpha)$.  Because we assume
$\SM_0(C,\alpha)$ to be smooth, $A_{\rm cl}$ is an irreducible flat
connection on the knot complement ${M^o = M - C}$, and the 
only gauge transformations which fix the pair $\big(A_{\rm cl},\, U_{\rm
cl}\big)$ are constant gauge transformations taking values in the
center ${\CZ(G) = \BZ_{r+1}}$ of the simply-connected gauge group ${G
= SU(r+1)}$.  Of course, here we note that the center of $G$ always
acts trivially in the adjoint representation.

Without loss, we additionally assume that the pair $\big(A_{\rm cl}, U_{\rm
cl}\big)$ is invariant under the Seifert action by $U(1)_\RR$ on $M$.
Hence the covariant derivative determined by $A_{\rm cl}$ commutes
with the Lie derivative $\lie_\RR$, 
\eqn\ACLINVII{ \big[d_{A_{\rm cl}},\, \lie_\RR\big] \,=\, 0\,,}  
and the fluctuating modes of the fields $\big(A,\, U\big)$ about the
basepoint $\big(A_{\rm cl},\, U_{\rm cl}\big)$ admit a covariant Fourier
decomposition with respect to $U(1)_\RR$.  For the defect field $U$,
which we mostly focus on here, we will make the covariant Fourier
decomposition explicit in a moment.

Following the ansatz in Section $6.2$, we again model an equivariant
neighborhood $N$ of the point ${\big(A_{\rm cl},\, U_{\rm cl}\big) \in
\bar\CA_\alpha}$ on a symplectic fibration of the general form 
\eqn\NLOOPII{\matrix{
&\SF^\alpha \longrightarrow\, N\cr\noalign{\vskip 2 pt}
&\mskip 77mu\Big\downarrow\lower 0.5ex\hbox{$^{\rm pr}$}\cr
&\mskip 65mu \SM_0(C,\alpha)\cr}\,,}
where $\SF^\alpha$ is the total space of a homogeneous vector bundle
over a quotient ${\CH/H_0^\alpha}$,
\eqn\FLOOPII{ \SF^\alpha \,=\, \CH \times_{H_0^{\alpha}}
\!\left(\Fh^\perp \oplus \CE_1^{\alpha}\right), \qquad\qquad
\Fh^\perp \,\equiv\, \Fh \ominus \Fh_0^{\alpha} \ominus
\CE_0^{\alpha}\,.}
Here $H_0^\alpha$ is the stabilizer of the point $\big(A_{\rm cl},\,
U_{\rm cl}\big)$ in the full Hamiltonian group 
\eqn\SLHAMCH{ \CH \,=\, U(1)_\RR \ltimes \wt\CG_0\,,}
and $\big(\CE_0^\alpha, \CE_1^\alpha\big)$ are vector spaces upon
which $H_0^\alpha$ acts.  

As in Section $7.2$, we are left to identify $H_0^\alpha$,
$\CE_0^\alpha$, and $\CE_1^\alpha$ in the canonical local model
\FLOOPII\ for $\SF^\alpha$.

First, since the pair $\big(A_{\rm cl},\, U_{\rm cl}\big)$ is
irreducible and invariant under $U(1)_\RR$, the stabilizer
${H_0^\alpha \subset \CH}$ is given immediately by the product 
\eqn\SMHNOUGHT{ H_0^\alpha \,=\, U(1)_\RR \times \CZ(G) \times
U(1)_\RZ\,\equiv\, H_0\,,}
where we recall that the central ${U(1)_\RZ \subset \wt\CG_0}$ acts
trivially on all of $\bar\CA_\alpha$.  As indicated in \SMHNOUGHT,
because $H_0^\alpha$ does not in fact depend upon $\alpha$ (so long as
${\alpha>0}$ is regular), we abbreviate ${H_0^\alpha \equiv H_0}$
throughout the following analysis.

As for the vector spaces $\big(\CE_0^\alpha,\,\CE_1^\alpha\big)$, we
proceed in close analogy to the discussion surrounding the
similar identifications in \ENOUGHTA\ and \EONEA\ of 
Section $7.2$.  Since we consider a neighborhood of $\big(A_{\rm
cl},\, U_{\rm cl}\big)$ in the product ${\bar\CA_\alpha \,=\, \bar\CA
\times \epsilon L\CO_\alpha}$, both $\CE_0^\alpha$ and $\CE_1^\alpha$
decompose {\it a priori} into pair of subspaces, one subspace
associated to the affine space $\bar\CA$ and one subspace associated
to the loopspace $L\CO_\alpha$.  In the case ${\alpha = 0}$, for
which $L\CO_\alpha$ reduces to a point, we have already determined in
Section $5.1$ of \BeasleyVF\ the summands associated to $\bar\CA$ alone,
\eqn\SLEONE{\eqalign{
\CE_0 \,&=\, \bigoplus_{t \ge 1} \, H^0_{\bar\partial}\Big(\Sigma, \ad(P)
\otimes \big(\CL^t \oplus \CL^{-t}\big)\Big)\,,\cr
\CE_1 \,&=\, \bigoplus_{t \ge 1} \, H^1_{\bar\partial}\Big(\Sigma, \ad(P)
\otimes \big(\CL^t \oplus \CL^{-t}\big)\Big)\,.\cr}}
The pair $\big(\CE_0,\,\CE_1\big)$ in \SLEONE\ should be compared to
the corresponding pair in \EONE, for which we have merely replaced the
Lie algebra $\Fg$ of $G$ with the adjoint bundle $\ad(P)$ over
$\Sigma$.  As throughout, $P$ is the principal $G_{\rm ad}$-bundle
appearing in the diffeomorphism ${\SM_0 \cong \wt\SN(P)}$, and $\CL$
is the positive line bundle over $\Sigma$ whose unit-circle bundle
describes the Seifert manifold $M$.
 
Given \SLEONE, our job here is only to capture the additional
symplectic geometry in $\SF^\alpha$ which derives from the loopspace 
$L\CO_\alpha$.  As in Section $7.2$, we identify the tangent space to
$L\CO_\alpha$ at the point $\big[U_{\rm cl}\big]$ with the space of
maps ${\delta U:C\rightarrow \Fg\ominus\Ft}$, and again we
decompose $\delta U$ into eigenmodes of $\lie_\RR$ as 
\eqn\FDUTWO{ \delta U \,=\, \sum_{t=-\infty}^{+\infty} \, \delta
U_t\,,\qquad\qquad \lie_\RR \, \delta U_t \,=\, -2\pi i\, t \cdot
\delta U_t\,.}
Here, in identifying each eigenmode $\delta U_t$ geometrically with a
map from $C$ to ${\Fg\ominus\Ft}$, we  trivialize both the line bundle
$\CL$ and the adjoint bundle $\ad(P)$ at the basepoint ${p\in\Sigma}$
under the Seifert fiber ${C \subset M}$.  For this reason, and in
contrast to the description of $\big(\CE_0, \CE_1\big)$ in \SLEONE,
neither the adjoint bundle $\ad(P)$ nor the line bundle $\CL$ play any
role in \FDUTWO.

Now, in the description \ENOUGHTA\ of $\CE_0^\alpha$ in Section $7.2$,
we were careful to include as an additional summand a copy of
$\Fg^{(1,0)}$ to account for directions tangent to the orbit  
${\CO_\alpha \subset L\CO_\alpha}$ parametrizing constant loops.  But for
the case at hand, $\CO_\alpha$ already appears (up to normalization) as
a submanifold of the extended moduli space $\SM_0(C,\alpha)$ sitting at
the base of the symplectic fibration in \NLOOPII.  Thus for
localization on $\SM_0(C,\alpha)$, we do not need to account
separately for $\CO_\alpha$ in the vector space $\CE_0^\alpha$.  With
nothing further to include, $\CE_0^\alpha$ is given simply by 
\eqn\BENOUGHTA{ \CE_0^\alpha \,=\, \CE_0\,,\qquad\qquad \alpha > 0
\quad \hbox{regular}\,.}

Otherwise, via the Fourier decomposition in \FDUTWO, the normal
directions to $\CO_\alpha$ inside $L\CO_\alpha$ are given by a 
countable sum of copies of ${\Fg\ominus\Ft}$ graded by the non-zero
integer ${t\neq 0}$.  As in \NLOC, we introduce the holomorphic normal bundle
$\CN_\alpha$ for the embedding ${\CO_\alpha \subset L\CO_\alpha}$,
\eqn\BNLOC{ \CN_\alpha \,=\, \bigoplus_{t \ge 1} \Big[\Fg^{(1,0)}_{t}
\oplus \Fg^{(1,0)}_{-t}\Big]\,.}
Once more, in the process of identifying $\CE_1^\alpha$, we must be
careful about the complex structure on $\CN_\alpha$.  Repeating
verbatim the discussion which surrounds the corresponding presentation of
$\CE_1^\alpha$ in \EONEA, we see that consistency with the convention in
\HOLCONV\ requires that we identify $\CE_1^\alpha$ as the direct
sum of $\CE_1$ in \SLEONE\ with the conjugate normal bundle 
\eqn\CCBNLOC{ \bar\CN_\alpha \,=\, \bigoplus_{t \ge 1} \Big[\Fg^{(0,1)}_{t}
\oplus \Fg^{(0,1)}_{-t}\Big]\,.}
Hence
\eqn\BEONEA{ \CE_1^\alpha \,=\, \CE_1 \oplus \bar\CN_\alpha\,.}

Together, \SMHNOUGHT, \BENOUGHTA, and \BEONEA\ specify the symplectic
model for localization on $\SM_0(C,\alpha)$.

\bigskip\noindent{\it Non-Abelian Localization on $\SM_0(C,\alpha)$}\smallskip 

We now possess all the ingredients required to apply the non-abelian
localization formula in \ZEV\ to compute the cohomology class $[d\mu]$
in \SCHSL.  

Immediately,
\eqn\SCHSLZ{\eqalign{
Z\big(\epsilon;C,R\big)\Big|_{\SM_0(C,\alpha)} &=\,
{{(2\pi\epsilon)} \over {|\CZ(G)|}} \, \int_{\Fh_0 \times
\SM_0(C,\alpha)} \left[{{d p} \over {2\pi}}\right] \left[{{d a} \over
{2\pi}}\right] \,\, {{e_{H_0}\!\Big(\SM_0\big(C,\alpha\big),
\CE_0^\alpha\Big)}\over{e_{H_0}\!\Big(\SM_0\big(C,\alpha\big),
\CE_1^\alpha\Big)}} \, \times\cr
&\qquad\qquad\qquad\qquad\qquad \times\,\exp{\!\Big[\Omega_\lambda
\,+\, i \epsilon n \Theta \,-\, i a \,+\, i \epsilon p a\Big]}\,.}}
Here the prefactor involving $\epsilon$ arises for the same reason as
the corresponding prefactor in \SFTWYI.  Otherwise, the semiclassical 
contribution to $Z\big(\epsilon;C,R\big)$ from $\SM_0(C,\alpha)$
reduces to an integral over the abelian Lie algebra ${\Fh_0 \cong
\BR \oplus \BR}$ of the stabilizer $H_0$, as well as an integral over
$\SM_0(C,\alpha)$ itself.  Following the notation in Section $3$, we
parametrize $\Fh_0$ with coordinates $(p,a)$.

As for the integrand in \SCHSLZ, beyond the ratio of equivariant Euler
classes associated to $\big(\CE_0^\alpha, \CE_1^\alpha\big)$, we
recognize in the argument of the exponential the equivariant
symplectic form $\big[\Omega_\lambda - i a\big]$ on $\SM_0(C,\alpha)$,
as appears more generally in the first line of \HHZM.  Similarly,
$\big[n \Theta \,+\, p a\big]$ is the  equivariant characteristic
class of degree four on $\SM_0(C,\alpha)$ appearing in the second line
of \HHZM.  The same degree-four class enters the localization
computation in Section $5.3$ of \BeasleyVF, and according to the
discussion there, $\Theta$ is given concretely in terms of the Chern
roots $\Ru$ of $\SV_p$ by 
\eqn\THETRU{ \Theta \,=\, -\ha \big(\Ru, \Ru\big)\,\in\,
H^4\big(\SM_0(C,\alpha)\big)\,.}
\countdef\BigThetaMC=203\BigThetaMC=\pageno
Since $\Theta$ is manifestly a symmetric function of the Chern roots,
$\Theta$ is the pullback from $\SM_0$ of a corresponding degree-four
characteristic class of the universal bundle $\SV_p$.  Again, we find
it convenient to abuse notation slightly, and we will not usually attempt to
distinguish $\Theta$ as a characteristic class on $\SM_0$ from its
pullback \THETRU\ to the splitting manifold $\SM_0(C,\alpha)$.

The attentive reader may note that $\Theta$ is multiplied in
\SCHSLZ\ by a factor $+i\,n$.  The $+i$ arises trivially from the
$+i$ that multiplies the Chern-Simons action in the original path
integral.  Otherwise, as we observed previously in $(5.110)$
of \BeasleyVF, the degree $n$ arises from the geometric identity 
\eqn\PULLBKG{ \int_M \kappa\^d\kappa \, \Tr\big(\phi^2\big) \,=\, n
\int_\Sigma \omega \, \Tr\big(\phi^2\big)\,,}
where $\phi$ is any element in the Lie algebra of $\CG_0$ which
pulls back from $\Sigma$, and $\omega$ is a unit-volume symplectic
form on $\Sigma$ satisfying ${d\kappa = n \, \pi^*(\omega)}$.  Under
localization on ${\SM_0 \cong \wt\SN(P)}$, the quadratic function of
$\phi$ on the right of \PULLBKG\ reduces to ${n \, \Theta}$, from
which the factor of $n$ in \SCHSLZ\ arises.

As in Section $7.2$, our main task here is to evaluate the ratio of
equivariant Euler classes associated to the bundles $\big(\CE_0^\alpha,
\CE_1^\alpha\big)$ over $\SM_0(C,\alpha)$.  Using the multiplicative
property of the Euler class and the identification ${\CE_1^\alpha
= \CE_1 \oplus \bar\CN_\alpha}$ in \BEONEA, we immediately factor 
the ratio in \SCHSLZ\ as 
\eqn\RATEL{ {{e_{H_0}\!\Big(\SM_0\big(C,\alpha\big),
\CE_0^\alpha\Big)}\over{e_{H_0}\!\Big(\SM_0\big(C,\alpha\big),
\CE_1^\alpha\Big)}} \,=\, \Rq^*\!\!\left[{{e_{H_0}\!\Big(\SM_0,
\CE_0\Big)}\over{e_{H_0}\!\Big(\SM_0, \CE_1\Big)}}\right] \cdot {1 \over
{e_{H_0}\!\Big(\SM_0\big(C,\alpha\big),\,\bar\CN_\alpha\Big)}}\,.}
In obtaining \RATEL, we observe that  $\CE_0$ and $\CE_1$ are
defined in \SLEONE\ as equivariant bundles on $\SM_0$ which pull back
to $\SM_0(C,\alpha)$, implying that the ratio of Euler classes pulls
back as well.

Evaluating the ratio of equivariant Euler classes associated to
$\CE_0$ and $\CE_1$ on $\SM_0$ turns out to be fairly tricky.  Luckily, in
$(5.168)$ of \BeasleyVF\ we have already computed that ratio.  For
sake of brevity, we merely state the answer,
\eqn\RATELMZERO{\eqalign{
{{e_{H_0}\!\Big(\SM_0,\CE_0\Big)}\over{e_{H_0}\!\Big(\SM_0,
\CE_1\Big)}} \,&=\, \exp{\!\left[-{{i\pi}\over 2} \eta_0(0) \,+\, 
{\pi\over p} \, c_1\big(\SM_0\big) \,+\, {{i n
\check{c}_\Fg}\over{2\pi p^2}} \, \Theta\right]} \cdot
\prod_{j=1}^{\dim_\BC \SM_0} \,
{{\varpi_j}\over{2\sinh(\pi\varpi_j/p)}}\,,\cr
\eta_0(0) \,&=\, -{{n \, \Delta_G}\over 6}\,.}}
Here $\varpi_j$ for ${j = 1,\ldots,\dim_\BC \SM_0}$ are the Chern
roots of the complex tangent bundle of $\SM_0$, so that 
\eqn\CHRNRTMZERO{ c(\SM_0) \,=\, \prod_{j=1}^{\dim_\BC \SM_0} \big(1
\,+\, \varpi_j\big),\qquad\qquad c_1(\SM_0) \,=\, \sum_{j=1}^{\dim_\BC
\SM_0} \varpi_j\,,}
and we recall from \ETAZ\ the expression for the constant $\eta_0(0)$.

Given \RATELMZERO, the only new calculation required is to evaluate the
equivariant Euler class of the bundle $\bar\CN_\alpha$ in \CCBNLOC\ as
it fibers equivariantly over $\SM_0(C,\alpha)$.  According to the
general description in \HEQE,  the equivariant Euler class of
$\bar\CN_\alpha$ is given by the formal product  
\eqn\ELCCN{ e_{H_0}\!\Big(\SM_0(C,\alpha),\,\bar\CN_\alpha\Big) \,=\,
\prod_{t \neq 0} \, \prod_{\beta > 0} \left(-i t p \,+\,
\langle\beta,\Ru\rangle\right).}
Here the dependence on $p$ arises exactly as for the corresponding 
determinant in \DETEI.  Otherwise, as ${\beta > 0}$ ranges over the
positive roots of $G$, the two-forms $\langle\beta,\Ru\rangle$
represent the Chern roots of each summand ${\Fg^{(0,1)}_t \!\subset
\CN_\alpha}$ as it fibers over $\SM_0(C,\alpha)$.  Not coincidentally, we
encountered the same Chern roots in the index theory computation on
${M = S^1 \times \Sigma}$, as in \CHRRTSCOV\ and \TDCOALPH.

To make sense of the infinite product in \ELCCN, we must regularize it
in some way.  See the remarks beginning at $(5.150)$ in \BeasleyVF\ for a
general discussion of the ways in which \ELCCN\ can be reasonably
defined.  For sake of time, we proceed here in a more {\it ad hoc}
fashion and simply use zeta/eta-function regularization to define the
respective norm and phase of the product in \ELCCN.  

Thankfully, the evaluation of the product in \ELCCN\ using
zeta/eta-function regularization is also a calculation which we have, in
effect, already accomplished.  In \DETEI\ of Section $7.2$, we encountered
a product which is formally identical to the product in \ELCCN,
provided we make the natural Chern-Weil substitution 
\eqn\CHRNWEIL{ \phi \,\longmapsto\, 2\pi i\, \Ru\,.}
The computation of the equivariant Euler class of the bundle $\bar\CN_\alpha$
then proceeds along exactly the same lines which led previously to
the formula in \DETEIII.  So either by a direct computation completely
analogous to the computation following \DETEI\ (which we omit), or
just by substituting $2\pi i\,\Ru$ for $\phi$ in our previous result 
\DETEIII, we obtain 
\eqn\ELCCNII{ e_{H_0}\!\Big(\SM_0(C,\alpha),\,\bar\CN_\alpha\Big)
\,=\, \exp{\!\left(-{{2\pi\langle\rho,\Ru\rangle}\over p}\right)}
\cdot \prod_{\beta>0} \, {2 \over {\langle\beta,\Ru\rangle}} \, 
\sinh\!\left({{\pi \langle\beta,\Ru\rangle}\over p}\right),}
where $\rho$ is the usual Weyl vector.

Combining the formulae in \RATELMZERO\ and \ELCCNII, we see that the ratio
of equivariant Euler classes in \RATEL\ becomes 
\eqn\TOTRATEL{\eqalign{
{{e_{H_0}\!\Big(\SM_0\big(C,\alpha\big),
\CE_0^\alpha\Big)}\over{e_{H_0}\!\Big(\SM_0\big(C,\alpha\big),
\CE_1^\alpha\Big)}} \,&=\, \exp{\!\left(-{{i\pi}\over 2} \eta_0(0) \,+\,
{{2\pi\langle\rho,\Ru\rangle}\over p}\right)} \cdot \prod_{\beta>0} \,
{{\langle\beta,\Ru\rangle}\over{2\sinh\!\big({\pi
\langle\beta,\Ru\rangle/p}\big)}}\,\times\cr
&\times\,\Rq^*\!\left[\exp{\!\left({\pi\over p} \, c_1\big(\SM_0\big)
\,+\, {{i n \check{c}_\Fg}\over{2\pi p^2}} \, \Theta\right)} \cdot
\prod_{j=1}^{\dim_\BC \SM_0} {{\varpi_j} \over
{2\sinh(\pi\varpi_j/p)}}\right].}}
Of particular note, the ratio in \TOTRATEL\ depends only on the
coordinate $p$, not $a$, in the Lie algebra ${\Fh_0 \cong \BR \oplus
\BR}$.  We could have predicted this occurence at the outset, since
$a$ parametrizes the Lie algebra of the central $U(1)_\RZ$ which acts
trivially on $\bar\CA_\alpha$.  Just as in Section $7.2$, the integral
over $a$ in \SCHSLZ\ then yields a delta-function $2\pi\,\delta(1 -
\epsilon p)$, and the integral over the remaining coordinate $p$
amounts to setting ${p = 1/\epsilon}$ in \TOTRATEL.

Thus we express $Z(\epsilon;C,R)\big|_{\SM_0(C,\alpha)}$ solely as an
integral over the classical Seifert loop moduli space $\SM_0(C,\alpha)$,
\eqn\SCHSLZII{\eqalign{
&Z\big(\epsilon;C,R\big)\Big|_{\SM_0(C,\alpha)} \,=\,
{1 \over {|\CZ(G)|}} \, \exp{\!\left(-{{i\pi}\over 2} \eta_0(0)\right)}\,\,
\times\,\cr
&\qquad\qquad\times\,\int_{\SM_0(C,\alpha)}
\exp{\!\Big(2\pi\epsilon\,\big\langle\alpha+\rho,\Ru\big\rangle\Big)} \cdot
\prod_{\beta>0} \, {{\langle\beta,\Ru\rangle} \over
{2\sinh\!\big({\pi\epsilon\langle\beta,\Ru\rangle}\big)}}\,\,\times\cr
&\qquad\qquad\times\,\Rq^*\!\left[\exp{\!\left(\Omega \,+\, \pi\epsilon\,
c_1(\SM_0) \,+\, i \epsilon n \Big(1 + {{\epsilon
\check{c}_\Fg}\over{2\pi}}\Big) \Theta\right)} \cdot
\prod_{j=1}^{\dim_\BC \SM_0} {{\varpi_j} \over
{2\sinh\!\big(\pi\epsilon\varpi_j\big)}}\right].}}
Since we eventually want to push the integrand in \SCHSLZII\ down to
$\SM_0$, we have been careful to factor out those classes which manifestly
pull back  from $\SM_0$ under $\Rq^*$.  In particular, when obtaining
\SCHSLZII, we have used the symplectic decomposition 
\eqn\OMLAMAK{ \Omega_\lambda \,=\, \Rq^*\Omega \,+\,
2\pi\epsilon\left\langle\alpha,\Ru\right\rangle,\qquad\qquad \lambda
\,=\, \alpha/k\,,}
as follows directly from \OMLAMII\ upon setting ${\epsilon=2\pi/k}$.

To make the cohomological interpretation of \SCHSLZII\ more
transparent, let us rescale each element in the cohomology ring of
$\SM_0(C,\alpha)$ by a factor $(2\pi\epsilon)^{-q/2}$, where $q$ is the
degree of the given class.  For instance, the Chern roots
$\varpi_j$ and $\Ru$, each of degree two, scale by 
\eqn\SCALCOH{\eqalign{
\varpi_j \,&\longmapsto\, {1\over{2\pi\epsilon}}\,\varpi_j,\cr
\Ru \,&\longmapsto\, {1\over{2\pi\epsilon}}\,\Ru\,.}}
To preserve the value of the integral over $\SM_0(C,\alpha)$, we
simultaneously scale the integral itself by an overall factor
$(2\pi\epsilon)^d$, where ${d = \dim_\BC \SM_0(C,\alpha)}$.  After
this change of variables to clear away extraneous factors of
$\epsilon$, $Z(\epsilon;C,R)\big|_{\SM_0(C,\alpha)}$ becomes 
\eqn\SCHSLZIII{\eqalign{
&Z\big(\epsilon;C,R\big)\Big|_{\SM_0(C,\alpha)} \,=\,
{1 \over {|\CZ(G)|}} \, \exp{\!\left(-{{i\pi}\over 2}
\eta_0(0)\right)}\,\,\times\cr
&\qquad\times\,\int_{\SM_0(C,\alpha)} {{\e{\!\langle\alpha +
\rho,\Ru\rangle}}\over{\A_\rho(\Ru)}} \, \prod_{\beta > 0} \,
\langle\beta,\Ru\rangle \cdot
\Rq^*\!\left[\exp{\!\left({1\over{2\pi\epsilon}}\,\Omega  
\,+\, \ha\, c_1(\SM_0) \,+\, i {n \over {4\pi^2\epsilon_{\rm r}}}
\Theta\right)} \, \widehat{A}\big(\SM_0\big)\right],}}
where we recall the definition of the renormalized coupling 
\eqn\EPSILONRII{ \epsilon_{\rm r} \,=\, {{2\pi}\over{k+\check{c}_\Fg}}\,.}

The expression for $Z(\epsilon;C,R)\big|_{\SM_0(C,\alpha)}$ in
\SCHSLZIII\ deserves a number of comments.  First, we recognize in the integral
over $\SM_0(C,\alpha)$ our friend the Weyl denominator $\A_\rho(\Ru)$,
\eqn\AROOFO{ \A_\rho(\Ru) \,=\, \prod_{\beta > 0}
2\sinh\!\left({{\langle\beta,\Ru\rangle}\over 2}\right).}
Second, we see that the integrand of \SCHSLZIII\ contains the
$\widehat{A}$-genus of $\SM_0$.  In general, if $X$ is a smooth
complex manifold, the $\widehat{A}$-genus of $X$ is given by the following
product over the Chern roots $x_j$ for ${j=1,\ldots,\dim_\BC X}$ of
the complex tangent bundle $TX$,
\eqn\AROOFX{ \widehat{A}(X) \,=\, \prod_{j=1}^{\dim_\BC X} {{x_j}
\over {2\sinh\!\big(x_j/2\big)}}\,.}
\countdef\AroofX=204\AroofX=\pageno
Directly upon specialization,
\eqn\AROOFM{ \widehat{A}(\SM_0) \,=\, \prod_{j=1}^{\dim_\BC \SM_0}
{\varpi_j\over{2\sinh\!\big(\varpi_j/2\big)}}\,.}

The alert reader may also recognize the $\widehat{A}$-genus of
the orbit $\CO_{-\lambda}$ (with ${\lambda = \alpha/k}$) appearing  as
a factor in \SCHSLZIII,
\eqn\AROOFOL{ \widehat{A}\big(\CO_{-\lambda}\big) \,=\,
{1\over\A_\rho(\Ru)} \cdot \prod_{\beta > 0}  \,
\langle\beta,\Ru\rangle\,.}
In comparing the general definition \AROOFX\ of the
$\widehat{A}$-genus to \AROOFOL, we recall that the classes
$\langle\beta,\Ru\rangle$ as ${\beta > 0}$ ranges over the positive
roots of $G$ are precisely the Chern roots of $\CO_{-\lambda}$.  

The appearance of the $\widehat{A}$-genus of the orbit
$\CO_{-\lambda}$ in \SCHSLZIII\ is no accident.  Just as in \PSDWNI,
we have a holomorphic fibration of complex manifolds 
\eqn\PSDWNMCA{\matrix{ 
&2\pi\CO_{-\lambda}\,\longrightarrow\,\SM_0(C,\alpha)\cr
&\mskip 135mu\Big\downarrow\lower 0.5ex\hbox{$^{\Rq}$}\cr
&\mskip 130mu\SM_0\cr}\,,\qquad\qquad \lambda \,=\, {\alpha\over
k}\,.}
The description of the $\widehat{A}$-genus in \AROOFX, like the
description of the Todd class in \TDX, is manifestly multiplicative.  
Hence the fibration of $\SM_0(C,\alpha)$ over $\SM_0$ in \PSDWNMCA\
implies the relation
\eqn\AROOFTOT{ \widehat{A}\big(\SM_0(C,\alpha)\big) \,=\,
\widehat{A}\big(\CO_{-\lambda}\big) \cdot
\widehat{A}\big(\SM_0\big)\,.}
The individual factors in \AROOFM\ and \AROOFOL\ therefore combine in
the integrand of \SCHSLZIII\ to describe the $\widehat{A}$-genus of
the extended moduli space $\SM_0(C,\alpha)$.

Given the secret appearance of $\widehat{A}\big(\SM_0(C,\alpha)\big)$
in the integrand of $Z(\epsilon;C,R)\big|_{\SM_0(C,\alpha)}$, let us
work backwards a bit to express the entire integral over
$\SM_0(C,\alpha)$ in terms of natural classes on the extended moduli 
space.  In addition to the identities in \OMLAMAK\ and \AROOFTOT, we
note that the first Chern class of $\SM_0(C,\alpha)$ is given by the sum 
\eqn\CONESMCA{\eqalign{
 c_1\big(\SM_0(C,\alpha)\big) \,&=\, \Rq^*c_1\big(\SM_0\big) \,+\,
 c_1\big(\CO_{-\lambda}\big)\,,\cr
&=\, \Rq^*c_1\big(\SM_0\big) \,+\, \sum_{\beta>0} \,
\langle\beta,\Ru\rangle \,=\, \Rq^*c_1(\SM_0) \,+\,
2\,\langle\rho,\Ru\rangle\,,}}
as enters the argument of the exponentials in \SCHSLZIII.  Using
\OMLAMAK, \AROOFTOT, and \CONESMCA, we then rewrite
$Z(\epsilon;C,R)\big|_{\SM_0(C,\alpha)}$ in terms of
classes defined intrinsically on $\SM_0(C,\alpha)$,  
\eqn\AROOFZCM{\eqalign{
&Z\big(\epsilon;C,R\big)\Big|_{\SM_0(C,\alpha)} \,=\,
{1 \over {|\CZ(G)|}} \, \exp{\!\left(-{{i\pi}\over 2}
\eta_0(0)\right)}\,\,\times\cr
&\qquad\times\,\int_{\SM_0(C,\alpha)}
\widehat{A}\big(\SM_0(C,\alpha)\big) \cdot
\exp{\!\left[{1\over{2\pi\epsilon}} \, \Omega_\lambda \,+\, \ha\,
c_1\big(\SM_0(C,\alpha)\big) \,+\, i {n \over {4\pi^2\epsilon_{\rm
r}}} \Rq^*\Theta\right]}.}}
We emphasize above that the degree-four characteristic class $\Theta$
pulls back from $\SM_0$ under the map
${\Rq:\SM_0(C,\alpha)\to\SM_0}$.

The integral over $\SM_0(C,\alpha)$ in \AROOFZCM\ should be compared
to the expression for the partition function
$Z(\epsilon)\big|_{\SM_0}$ in $(5.172)$ of \BeasleyVF,
\eqn\SCHZZIII{ Z(\epsilon)\big|_{\SM_0} =\,
{1 \over {|\CZ(G)|}} \, \exp{\!\left(-{{i\pi}\over 2}
\eta_0(0)\right)} \, \int_{\SM_0} \widehat{A}(\SM_0) \cdot 
\exp{\!\left[{1\over{2\pi\epsilon}} \, \Omega \,+\, \ha\,
c_1(\SM_0) \,+\, i {n \over {4\pi^2\epsilon_{\rm r}}} \Theta\right]}.}
Just as for the cohomological formulae in \INDXV, we see that the
respective contributions from the extended moduli space $\SM_0(C,\alpha)$ to
$Z(\epsilon;C,R)$ and from the basic moduli space $\SM_0$ to
$Z(\epsilon)$ are structurally identical, insofar as the same
characteristic classes appear in the respective integrands of each.
In hindsight, the beautiful correlation between \AROOFZCM\ and
\SCHZZIII\ was bound to occur, since in both cases we apply the same
localization formula in the same smooth setting.  The structural
agreement between \AROOFZCM\ and \SCHZZIII\ therefore provides an
additional check on our computations.

Finally, returning our attention to \SCHSLZIII, we note that the
renormalized coupling $\epsilon_{\rm r}$ naturally appears in the
coefficient of $\Theta$ when we rescale the integral in \SCHSLZII.
Indeed, based upon our experience with Chern-Simons theory, we expect
the integrand in \SCHSLZIII\ to depend on the Chern-Simons level $k$
only through the renormalized coupling $\epsilon_{\rm r}$, as opposed
to the bare coupling $\epsilon$.  

To recast our result \SCHSLZIII\ entirely in terms of $\epsilon_{\rm
r}$, we apply a theorem of Drezet and Narasimhan \Drezet\ which
determines $c_1\big(\SM_0\big)$ in the case ${G = SU(r+1)}$ to be 
\eqn\CONEMO{ c_1\big(\SM_0\big) \,=\, 2 (r+1) \, \Omega_0\,,\qquad\qquad
\Omega_0 \,=\, {1\over{4\pi^2}} \Omega\,.}
Since ${\check{c}_\Fg = r+1}$ as well, the local contribution from
$\SM_0(C,\alpha)$ to $Z(\epsilon;C,R)$ becomes 
\eqn\SCHSLZIV{\eqalign{
&Z\big(\epsilon;C,R\big)\Big|_{\SM_0(C,\alpha)} \,=\,
{1 \over {|\CZ(G)|}} \, \exp{\!\left(-{{i\pi}\over 2}
\eta_0(0)\right)}\,\,\times\cr
&\qquad\times\,\int_{\SM_0(C,\alpha)} {{\e{\!\langle\alpha +
\rho,\Ru\rangle}}\over{\A_\rho(\Ru)}} \, \prod_{\beta > 0} \,
\langle\beta,\Ru\rangle \cdot
\Rq^*\!\left[\exp{\!\left({1\over{2\pi\epsilon_{\rm r}}}\left(\Omega
\,+\, i {n \over {2\pi}} \Theta\right)\right)} \cdot 
\widehat{A}\big(\SM_0\big)\right],}}
and all dependence on $k$ has been absorbed into the renormalized
coupling $\epsilon_{\rm r}$.

According to \SCHSL, the integrand in \SCHSLZIV\ is the class ${[d\mu]
\in H^*\big(\SM_0(C,\alpha)\big)}$ which describes the local
contribution from $\SM_0(C,\alpha)$ to the Seifert loop path integral
$Z(\epsilon;C,R)$,
\eqn\DMU{\eqalign{
[d\mu] \,&=\, {1 \over {|\CZ(G)|}} \, \exp{\!\left(-{{i\pi}\over 2}
\eta_0(0)\right)} \,\times\cr
&\qquad\,\times\, {{\e{\!\langle\alpha +
\rho,\Ru\rangle}}\over{\A_\rho(\Ru)}} \cdot \prod_{\beta > 0} \,
\langle\beta,\Ru\rangle \cdot
\Rq^*\!\left[\exp{\!\left({1\over{2\pi\epsilon_{\rm r}}}\left(\Omega
\,+\, i {n \over {2\pi}} \Theta\right)\right)} \cdot 
\widehat{A}\big(\SM_0\big)\right].}}
To obtain a cohomological formula for the Seifert loop class itself,
we are left to push $[d\mu]$ down to $\SM_0$ via the map
${\Rq:\SM_0(C,\alpha)\to\SM_0}$ in \PSDWNMCA.

\bigskip\noindent{\it Pushdown to $\SM_0$ and Relation to Index
Theory}\smallskip

Like the pair of expressions in \AROOFZCM\ and \SCHZZIII, the
integral over $\SM_0(C,\alpha)$ in \SCHSLZIV\ should be 
compared to the corresponding localization result for the partition
function $Z(\epsilon)|_{\SM_0}$ in $(5.174)$ of \BeasleyVF,
\eqn\SCHZZIV{ Z(\epsilon)\Big|_{\SM_0} \,=\, {1 \over {|\CZ(G)|}} \,
\exp{\!\left(-{{i\pi}\over 2} \eta_0(0)\right)} \, \int_{\SM_0}
\exp{\!\left[{1\over{2\pi\epsilon_{\rm r}}}\left(\Omega \,+\, i {n
\over {2\pi}} \Theta\right)\right]} \cdot \widehat{A}\big(\SM_0\big).}
The Seifert loop class $W_R(C)\big|_{\SM_0}$ is then the element of
$H^*(\SM_0)$ such that the pushdown $\Rq_*[d\mu]$ is given by the product of
$W_R(C)\big|_{\SM_0}$ with the integrand of the partition function
$Z(\epsilon)|_{\SM_0}$ in \SCHZZIV, such that 
\eqn\SLOOPCLMCA{\eqalign{
Z\big(\epsilon;C,R\big)\Big|_{\SM_0} \,&=\,
{1 \over {|\CZ(G)|}} \, \exp{\!\left(-{{i\pi}\over 2}
\eta_0(0)\right)}\,\times\cr
&\qquad\,\times\, \int_{\SM_0} W_R(C)\Big|_{\SM_0} \!\!\cdot 
\exp{\!\left[{1\over{2\pi\epsilon_{\rm r}}}\left(\Omega \,+\, i {n
\over {2\pi}} \Theta\right)\right]} \cdot \widehat{A}\big(\SM_0\big).}}
Comparing the Seifert integrand ${[d\mu]\in
H^*\big(\SM_0(C,\alpha)\big)}$ in \DMU\ to the preceding formula
\SLOOPCLMCA\ for $Z(\epsilon;C,R)|_{\SM_0}$, we deduce 
\eqn\SLOOPCLIV{ W_R(C)\Big|_{\SM_0} =\, \Rq_*
S_\alpha(\Ru)\,,\qquad\qquad S_\alpha(\Ru) \,=\,
{{\e{\!\langle\alpha + \rho,\Ru\rangle}}\over{\A_\rho(\Ru)}} \cdot 
\prod_{\beta > 0} \, \langle\beta,\Ru\rangle\,.}

The class $S_\alpha(\Ru)$ in \SLOOPCLIV\ is precisely the same class
that appeared previously in \BIGSAII\ when we considered the Seifert
loop operator in Chern-Simons theory on ${S^1\times\Sigma}$.  Moreover,
under the identification of moduli spaces ${\SM_0(C,\alpha) \cong
\wt\SN(P;\lambda)}$ and ${\SM_0 \cong \wt\SN(P)}$, the calculation of
the pushdown $\Rq_* S_\alpha(\Ru)$ proceeds just as before.
So without further ado, recycling the result in \SLOOPCLIII, we find 
the promised general description for the Seifert loop class,
\eqn\SLOOPCLIV{ W_R(C)\Big|_{\SM_0} =\, \ch_R\big(\SV_p\big).}
Equivalently, 
\eqn\SCHSLZV{ Z\big(\epsilon;C,R\big)\Big|_{\SM_0} =\,
{1 \over {|\CZ(G)|}} \, \exp{\!\left(-{{i\pi}\over 2}
\eta_0(0)\right)}\,\int_{\SM_0} \ch_R\big(\SV_p\big) \cdot
\exp{\!\left[{1\over{2\pi\epsilon_{\rm r}}}\!\left(\Omega \,+\, i {n
\over {2\pi}} \Theta\right)\right]} \cdot \widehat{A}\big(\SM_0\big).}

The expression for $Z(\epsilon;C,R)\big|_{\SM_0}$ in \SCHSLZV\ is
very similar to the cohomological formula \INDXVIII\
derived from the index theorem when ${M = S^1 \times \Sigma}$.  To
make the relation to \INDXVIII\ more transparent, we recall that the
Todd class and the $\widehat{A}$-genus of a complex manifold $X$
generally satisfy the relation 
\eqn\AROOFTD{ \Td(X) \,=\, \exp{\!\left[\ha \, c_1(X)\right]} \cdot
\widehat{A}(X)\,,}
as follows directly by comparison of the formulae in \TDX\ and
\AROOFX.  This identity, applied to $\widehat{A}(\SM_0)$ in
\SCHSLZIII, implies that the 
contribution from $\SM_0$ to $Z(\epsilon;C,R)$ can be alternatively
presented as 
\eqn\SCHSLZVI{ Z\big(\epsilon;C,R\big)\Big|_{\SM_0} =\,
{1 \over {|\CZ(G)|}} \, \exp{\!\left(-{{i\pi}\over 2}
\eta_0(0)\right)}\,\int_{\SM_0} \ch_R\big(\SV_p\big) \cdot
\exp{\!\left[ k \, \Omega_0 \,+\, i {n \over {4\pi^2\epsilon_{\rm r}}} 
\Theta\right]} \cdot \Td\big(\SM_0\big).}
Under the identification ${\SM_0 \cong \wt\SN(P)}$, the integrand of
\SCHSLZVI\ now manifestly reproduces the index density in \INDXVIII\
when ${n=0}$.  Otherwise, if ${n > 0}$, our localization formula 
for $Z(\epsilon;C,R)\big|_{\SM_0}$ amounts to a surprisingly simple
deformation away from the index theory result.

\bigskip\noindent{\it The Yang-Mills Limit of the Seifert Loop
Operator}\smallskip

Given the smooth Seifert fibration ${S^1 \buildrel n\over
\rightarrow M  \buildrel\pi\over\rightarrow \Sigma}$ of degree ${n >
0}$, another interesting regime in which to consider the Seifert loop
path integral $Z(\epsilon;C,R)$ is the limit ${n \rightarrow
\infty}$.  Intuitively, as $n$ becomes large, the non-trivial Fourier
modes of the gauge field $A$ along the circle fiber of $M$ decouple,
and Chern-Simons theory on $M$ effectively reduces to Yang-Mills
theory on $\Sigma$.  In this limit, according to the discussion
surrounding \LOCA, we expect the Seifert loop operator $W_R(C)$ to
reduce to the Yang-Mills monodromy operator $\RV_\lambda(p)$, where
${\lambda = \alpha/k}$.

To check the latter statement, let us quickly apply non-abelian
localization to the monodromy operator path integral
$Z(\epsilon_{\rm ym};p,\lambda)$.  Here to avoid confusion with the
Chern-Simons coupling parameter ${\epsilon = 2\pi/k}$, we introduce
a distinct notation ${\epsilon_{\rm ym} = g_{\rm ym}^2}$ for the
Yang-Mills coupling.\countdef\Epsym=205\Epsym=\pageno

As we demonstrated in \PZYMIII, $Z(\epsilon_{\rm ym};p,\lambda)$ takes
the canonical form determined by the Hamiltonian action of $\CG(P)$ on
the symplectic space ${\CA(P)_\lambda \,=\, \CA(P) \times
2\pi\CO_\lambda}$.  When $\CG(P)$ acts freely near the vanishing locus
of the moment map ${\mu = \CF_A}$ in $\CA(P)_\lambda$, the extended
moduli space $\SN(P;\lambda)$ is smooth, and the very simple version
\ZEVI\ of the non-abelian localization formula applies.  Immediately
by \ZEVI,
\eqn\ZPLAMLOC{ Z\!\big(\epsilon_{\rm
ym};p,\lambda\big)\big|_{\SN(P;\lambda)} \,=\, \int_{\SN(P;\lambda)}
\exp{\!\Big[ \Omega_\lambda \,+\, \epsilon_{\rm ym} \Theta\Big]}.}
Again, $\Omega_\lambda$ is the symplectic form on $\SN(P;\lambda)$ in
\OMLAMII, and $\Theta$ is the degree-four characteristic class of $\SV_p$
in \THETRU.  

If one wishes, the integrand in \ZPLAMLOC\ can be pushed
down to $\SN(P)$ to describe the monodromy operator $\RV_\lambda(p)$
as a class on $\SN(P)$.  For sake of brevity, we will not do so here.
We refer the interested reader to \WittenWE\ and to \BlauHJ\ for a
more detailed analysis of $Z\big(\epsilon_{\rm ym};p,\lambda\big)$ in
the special case ${\epsilon_{\rm ym} = 0}$, for which the integral in
\ZPLAMLOC\ merely computes the symplectic volume of $\SN(P;\lambda)$.

We now compare our localization result for $Z(\epsilon;C,R)$ in \SCHSLZII\ to
the Yang-Mills formula in \ZPLAMLOC.  To obtain a sensible limit for
$Z(\epsilon;C,R)$ as $n$ goes to infinity, we take the
parameters ${k, \alpha}$ to infinity with the ratios ${\epsilon_{\rm
ym} = n/k}$ and ${\lambda = \alpha/k}$ held fixed,
\eqn\YMSCAL{  n,\,\,k,\,\,\alpha \,\to\, \infty\,,\qquad\qquad \epsilon_{\rm ym}
\,=\, {n\over k},\quad \lambda \,=\, {\alpha\over
k}\,\quad\,\hbox{fixed}\,.}
To obtain a finite result in the limit \YMSCAL, we also scale
$Z(\epsilon;C,R)$ by an overall prefactor $(2\pi\epsilon)^d$, where
${d = \dim_\BC \SM_0(C,\alpha)}$,
\eqn\YMSCALII{ Z(\epsilon;C,R) \,\to\, (2\pi\epsilon)^d \,
Z(\epsilon;C,R)\,.}
Under this scaling, the products over the Chern roots
$\langle\beta,\Ru\rangle$ and $\varpi_j$ in \SCHSLZII\ reduce to the
identity, and the respective quantum shifts by $\check{c}_\Fg$ and
$\rho$ both vanish.  With the final identification ${\SM_0(C,\alpha) =
\wt\SN(P;\lambda)}$, the localization formula for Chern-Simons theory
in \SCHSLZII\ then reproduces the Yang-Mills formula in \ZPLAMLOC, at
least up to overall normalization.

\bigbreak\bigskip\bigskip\centerline{{\bf Acknowledgements}}\nobreak

I take great pleasure in thanking Dima Belov, Dan Freed, Amir-Kian
Kashani-Poor, Martin Ro\v cek, Yongbin Ruan, Amit Sever, George Thompson, and 
Jonathan Weitsman for stimulating discussions on the matters herein.  I
also thank the organizers and participants of the Bonn workshop
``Chern-Simons Gauge Theory:~20 Years After'' and the 2007 Simons
Workshop, where  portions of this paper were presented. Last but not
least, I thank Cliff Taubes for very helpful comments on a preliminary
draft, and I especially thank Edward Witten, both for our prior
collaboration on the subject and for posing the question which
sparked this work.

This work was supported in part under DOE grant DE-FG02-92ER40697.  Any
opinions, findings, and conclusions or recommendations expressed in
this material are those of the author and do not necessarily reflect
the views of the Department of Energy.

\vfill\eject

\appendix{A}{Index of Notation}

\halign to 5.5in{\tabskip=3em plus2em
minus.5em#\hfill&#\hfill\tabskip=0pt\cr
$M$&compact oriented three-manifold, \number\M\cr
$A$&gauge field, \number\AA\cr
${k\in\BZ}$&Chern-Simons level, \number\k\cr
$Z(k)$&Chern-Simons partition function, \number\Zk\cr
$\Sigma$&Riemann surface, possibly with orbifold points, \number\Sig\cr
${C \subset M}$&closed oriented curve, \number\C\cr
$G$&compact, connected, simple Lie group,\cr
&\quad often simply-connected, \number\G\cr
$R$&irreducible representation of $G$, \number\R\cr
$W_R(C)$&Wilson loop operator, \number\WRC\cr
$Z(k;C,R)$&Wilson loop path integral, \number\ZkCR\cr
$\Fg$&Lie algebra of $G$, \number\LieG\cr
$\Tr$&negative-definite invariant form on $\Fg$, \number\TR\cr
$P$&principal $G$-bundle over $\Sigma$, \number\P\cr
$\CA(P)$&affine space of connections on $P$, \number\AP\cr
$\CG(P)$&group of gauge transformations on $P$, \number\GP\cr
${d_A = d + A}$&covariant derivative, \number\dA\cr
${F_A = dA + A\^A}$&curvature of $A$, \number\FA\cr
$\omega$&symplectic form on $\Sigma$, \number\SmOm\cr
${\Omega \,=\, -\int_\Sigma\Tr\big(\delta A\^\delta
A\big)}$&symplectic form on $\CA(P)$, \number\Om\cr 
$\langle\,\cdot\,,\,\cdot\,\rangle$&canonical dual pairing on a vector
space, \number\BRC\cr
$\mu$&moment map, \number\Mom\cr
$(\,\cdot\,,\,\cdot\,)$&invariant quadratic form on a vector space,
\number\PR\cr
$\CA$&affine space of connections on $M$, \number\CurlyA\cr
$\CG$&group of gauge transformations acting on $\CA$,
\number\CurlyG\cr
${\epsilon = 2\pi/k}$&(bare) Chern-Simons coupling, \number\Eps\cr
$\CS$&group of shift symmetries acting on $\CA$, \number\CurlyS\cr
${\RC\RS}(\,\cdot\,)$&Chern-Simons functional, \number\CSFun\cr 
$\kappa$&contact form on $M$, \number\Contact\cr
$h$&genus of $\Sigma$, \number\genus\cr
$n$&degree of the bundle ${S^1\buildrel n\over\rightarrow M
\buildrel\pi\over\rightarrow \Sigma}$, \number\degree\cr 
$\big[h;n;(a_1,b_1), \ldots, (a_N,b_N)\big]$&Seifert invariants of
$M$, \number\Seifert\cr
$\CL\,\, \big(\widehat{\CL}\big)$&(orbifold) line bundle over $\Sigma$
associated to the\cr
&\quad Seifert presentation of $M$, \number\CurlyL\cr
${\bar\CA = \CA/\CS}$&quotient of $\CA$ by $\CS$, \number\BarA\cr
${\Omega \,=\, -\int_M \kappa\^\Tr\big(\delta A\^\delta A\big)}$&symplectic
form on $\bar\CA$, \number\OmBarA\cr
$\CG_0$&identity component of $\CG$, \number\CurlyGZero\cr
$c(\,\cdot\,,\,\cdot\,)$&cocycle on the Lie algebra of $\CG$,
\number\LieCoc\cr
${LG = \Map(S^1, G)}$&loop group of $G$, \number\LoopG\cr
${L\Fg = \Map(S^1, \Fg)}$&Lie algebra of $LG$, \number\Loopg\cr
$\wt{{\CG}}_0$&central extension of $\CG_0$ defined by
$c(\,\cdot\,,\,\cdot\,)$, \number\TildG\cr
${U(1)_\RZ \subset \wt{\CG}_0}$&central $U(1)$ subgroup of $\wt\CG_0$,
\number\UoneZ\cr
$U(1)_\RR$&locally-free, Seifert $U(1)$ acting on $M$, \number\UoneR\cr
$\RR$&vector field on $M$ generating $U(1)_\RR$, \number\BoldR\cr
$\lie_\RR$&Lie derivative along $\RR$, \number\PoundR\cr
${\CH = U(1)_\RR \ltimes \wt{\CG}_0}$&Hamiltonian group acting on
$\bar\CA$, \number\CurlyH\cr
$\FR$&set of roots of $G$, \number\Roots\cr
${\FR_\pm \subset \FR}$&positive/negative roots of $G$,
\number\PosNegRoots\cr
${\alpha\in\Ft^*\cong\Ft}$&highest weight of the representation
$R$, \number\Aleph\cr
${T \subset G}$&maximal torus, \number\MaxT\cr
${\Ft \subset \Fg}$&Cartan subalgebra, \number\Maxt\cr
${\lambda \in \Ft}$&element of $\Ft$, \number\Littlelam\cr
${\CO_\lambda \subset \Fg}$&adjoint orbit through $\lambda$,
\number\BigOl\cr
${G_\lambda \subseteq G}$&stabilizer of $\lambda$, \number\BigGl\cr
${\Fg_\lambda\subseteq\Fg}$&Lie algebra of $G_\lambda$, \number\Biggl\cr
${\theta = g^{-1} dg}$&left-invariant Cartan form on $G$,
\number\CartanForm\cr
${\Theta_\lambda = -\Tr(\lambda \, \theta)}$&pre-symplectic one-form on
$G$, \number\PreSym\cr
${\nu_\lambda = d\Theta_\lambda}$&coadjoint symplectic form on
$\CO_\lambda$, \number\Coadj\cr
$\SJ$&invariant complex structure on $\CO_\lambda$, \number\CurlyJ\cr
${\Fg_\BC \!= \Fg \otimes \BC}$&complexification of $\Fg$, \number\LieGc\cr
${\Ft_\BC \!= \Ft \otimes \BC}$&complexification of $\Ft$, \number\Maxtc\cr
${\beta \in \Ft^*}$&root of $G$, \number\Betas\cr
$\RC_+$&positive Weyl chamber, \number\PosWyl\cr
${\Fe_\beta \subset \Fg_\BC\!\ominus\Ft_\BC}$&rootspace associated to
$\beta$, \number\eRtspace\cr 
${\Fg^{(1,0)} \!= \Fg_{+}}$&holomorphic tangent space to 
$\CO_\lambda$, \number\LieGplus\cr
${\Fg^{(0,1)} \!= \Fg_{-}}$&anti-holomorphic tangent space to 
$\CO_\lambda$, \number\LieGplus\cr
${\Gamma_{\rm wt} = \Hom\!\big(T, U(1)\big) \subset \Ft^*}$&weight
lattice of $G$, \number\Gamwt\cr
${\Gamma_{\rm cochar} = \Hom\!\big(U(1), T\big) \subset
\Ft}$&cocharacter lattice of $G$, \number\Gamch\cr
${\Gamma_{\rm cort} \subset \Ft}$&coroot lattice of $G$; identical to
${\Gamma_{\rm cochar}}$\cr
&\quad when $G$ is simply-connected, \number\Gamrt\cr
${U\negthinspace:C\rightarrow \CO_\alpha}$&sigma model field on $C$ with target
$\CO_\alpha$, \number\BigU\cr
${\Rc\Rs}_\alpha(\,\cdot\,)$&topological, Chern-Simons-type action for
$U$, \number\CSalpha\cr
$L\CO_\alpha$&free (unbased) loopspace of $\CO_\alpha$,
\number\LoopOalpha\cr
$\delta_C$&two-form with delta-function support which is\cr
&\quad Poincar\'e dual to ${C\subset M}$, \number\DeltaC\cr
$\SM$&moduli space of flat connections on $M$, \number\Flat\cr
${M^o = M - C}$&complement of ${C \subset M}$, \number\KnotComp\cr
${\Rm \in \pi_1(M^o)}$&meridian of $C$, \number\Meridian\cr
${\FC_{\lambda} = \Cl\!\big[\!\exp{\!(2\pi\lambda)}\big]}$&conjugacy
class in $G$ containing ${\Lambda = \exp(2\pi\lambda)}$, 
\number\ConjCl\cr
$\SM(C,\alpha)$&moduli space of flat connections on $M^o$ with \cr
&\quad holonomy around $\Rm$ in $\FC_{\alpha/k}$, \number\ExtFlat\cr
${\CF_A = F_A \,+\, \epsilon \, (g \alpha g^{-1}) \,
\delta_C}$&generalized curvature in the presence of $W_R(C)$,
\number\CurlyF\cr
${\bar\CA_\alpha = \bar\CA \times \epsilon L\CO_\alpha}$&symplectic space
associated to $Z(\epsilon;C,R)$, \number\BarAalpha\cr
$\Upsilon_\alpha$&coadjoint symplectic form on $L\CO_\alpha$\cr
&\quad induced from $\nu_\alpha$, \number\BigUpsilon\cr
$\Xi_\alpha$&pre-symplectic one-form on $LG$\cr
&\quad such that ${\Upsilon_\alpha = \delta\Xi_\alpha}$, \number\BigXi\cr
${\Omega_\alpha = \Omega + \epsilon \Upsilon_\alpha}$&symplectic form
on $\bar\CA_\alpha$, \number\BigOmalpha\cr
$\RV_\lambda(p)$&monodromy operator inserted at ${p\in\Sigma}$ with\cr
&\quad parameter ${\lambda \in \Ft}$, \number\MonoOp\cr
$\FW$&Weyl group of $G$, \number\WeylGp\cr
${\FW_{\rm aff} = \FW \ltimes \Gamma_{\rm cochar}}$&affine Weyl group
of $G$, \number\AffWeylGp\cr
$Z(\epsilon;p,\lambda)$&monodromy operator path integral, \number\Zepsplam\cr
$\wt G$&simply-connected form of the Yang-Mills\cr
&\quad gauge group $G$, \number\Gtilde\cr
$\CZ(\wt G),\, \CZ(G)$&center of $\wt G$, respectively $G$, \number\Center\cr
$\delta_p$&two-form with delta-function support which is\cr
&\quad Poincar\'e dual to ${p \in \Sigma}$, \number\deltap\cr
${\RD_+ \subset \RC_+}$&fundamental Weyl alcove, \number\FundWeyl\cr
$\vartheta$&highest root of $G$, \number\Highestrt\cr
${\CA(P)_\lambda  = \CA(P) \times \CO_{2\pi\lambda}}$&symplectic space
associated to $Z(\epsilon;p,\lambda)$, \number\CurlyAPlam\cr
${\Omega_\lambda = \Omega \,+\, 2\pi\nu_\lambda}$&symplectic form on
$\CA(P)_\lambda$, \number\Omlam\cr
$\SN(P)$&moduli space of flat connections\cr
&\quad on the $G$-bundle $P$ over $\Sigma$, \number\CurlyNP\cr
${\SN(P;\lambda) \equiv \SN(P;p,\lambda)}$&moduli space of flat
connections on $P$ with\cr 
&\quad monodromy at $p$ in ${\FC_\lambda =
\Cl[\exp{\!(2\pi\lambda)}]}$, \number\CurlyNPlam\cr
${\Sigma^o = \Sigma - \{p\}}$&punctured Riemann surface,
\number\Sigmao\cr
${\varrho^o:\pi_1(\Sigma^o)\rightarrow G}$&homomorphism from the
fundamental group\cr
&\quad of $\Sigma^o$ to $G$, \number\Rhoo\cr
${\Sigma^{oo} = \Sigma - \{p\} - \{q\}}$&doubly-punctured Riemann
surface, \number\Sigmaoo\cr
${\varrho^{oo}:\pi_1(\Sigma^{oo})\rightarrow G}$&homomorphism from the
fundamental group\cr
&\quad of $\Sigma^{oo}$ to $G$, \number\Rhooo\cr
$\wt\SN(P;\lambda)$&unramified, degree $|\wt G\!:\!G|^{2 h}$ cover of
$\SN(P;\lambda)$, \number\tildeCurlyNP\cr
$\Omega,\,\,\Omega_\lambda$&symplectic forms on $\SN(P)$ and
$\SN(P;\lambda)$\cr
&\quad induced from those on $\CA(P)$ and
$\CA(P)_\lambda$, \number\OmCurlyNP\cr
$\Re_\lambda$&closed two-form on $\SN(P;\lambda)$ which restricts\cr
&\quad fiberwise to $\nu_\lambda$ on $\CO_\lambda$, \number\Relam\cr 
${\CZ_\Lambda \subseteq G}$&centralizer of the element $\Lambda$ in
$G$, \number\Centralizerlam\cr
$\SV$&universal bundle over ${\Sigma \times \SN(P)}$,
\number\CurlyV\cr
$\SV_p$&restriction of $\SV$ to ${\{p\}\times \SN(P)}$ for ${p \in
\Sigma}$, \number\CurlyVp\cr
${\SL_j,\,\, j=1,\ldots,r+1}$&splitting line for the pullback of $\SV_p$ to
$\SN(P;\lambda)$, \number\CurlyLj\cr
${\Ru_j \,=\, c_1\!\left(\SL_j\right)}$&Chern root of $\SV_p$,
\number\Ruj\cr
${\hat\omega_1,\ldots,\hat\omega_{r+1}}$&standard generators for
the weight lattice\cr
&\quad of $SU(r+1)$, with ${\hat\omega_1 + \cdots + 
\hat\omega_{r+1} = 0}$, \number\Hatomj\cr
${\Ru = i \diag\!\left(\Ru_1,\cdots,\Ru_{r+1}\right)}$&element of
$H^2\!\left(\SN(P;\lambda);\BZ\right)\otimes\Ft$ encoding\cr
&\quad the Chern roots of $\SV_p$, \number\Rut\cr 
$X$&symplectic manifold, \number\BigX\cr
$\Omega$&symplectic form on $X$, \number\BigOmX\cr
$H$&connected Lie group acting in a\cr
&\quad Hamiltonian fashion on $X$, \number\BigH\cr
$\Delta_H$&(real) dimension of $H$, \number\BigDelH\cr
$\Fh$&Lie algebra of $H$, \number\Gothh\cr
$\Fh^*$&dual of $\Fh$, \number\Gothhdual\cr
${\mu:X\rightarrow \Fh^*}$&moment map for $H$ acting on $X$,
\number\MomentX\cr
${\epsilon\in\BR}$&coupling parameter, \number\EpsilonR\cr
${S=\ha(\mu,\mu)}$&norm-square of moment map, \number\ActionS\cr
$V(\phi)$&vector field on $X$ determined by action of ${\phi \in \Fh}$,
\number\Vphi\cr
${D = d\,+\, i\, \iota_{V(\phi)}}$&BRST operator/Cartan differential,
\number\CartanD\cr
$H^*_H(X)$&$H$-equivariant cohomology ring of $X$ with\cr
&\quad coefficients in $\BR$, \number\HequivX\cr
${{\bf J}:TX\rightarrow TX}$&almost-complex structure on
$X$ compatible with $\Omega$, \number\AcJX\cr
${\CC\subset X}$&connected component in the critical locus\cr
&\quad of ${S = \ha(\mu,\mu)}$ on $X$, \number\CurlyC\cr
${H_0 \subset H}$&stabilizer at an arbitrary basepoint on 
$\CC$, \number\Hnought\cr
$\Fh_0$&Lie algebra of $H_0$, \number\GothHnought\cr
${\SM = \CC/H}$&smooth quotient of $\CC$ by $H$, \number\CurlyM\cr
${\gamma_0 \in \Fh_0^* \cong \Fh_0}$&value of the moment map $\mu$ at the
basepoint on $\CC$, \number\Littlegamnought\cr
${F = H \times_{H_0} \!\big(\Fh^\perp \oplus E_1\big)}$&symplectic
fiber over $\SM$ in a neighborhood of ${\CC \subset X}$, \number\SympF\cr
${E_0 \subset \Fh}$&subspace of $\Fh$, preserved under the adjoint
action of $\Fh_0$,\cr
&\quad on which $\gamma_0$ acts non-degenerately, \number\Enought\cr
${\Fh^\perp = \Fh \ominus \Fh_0 \ominus E_0}$&orthocomplement to
${\Fh_0 \oplus E_0}$ inside $\Fh$, \number\Gothhperp\cr
$E_1$&symplectic vector space on which $H_0$ acts in a\cr
&\quad Hamiltonian fashion and $\gamma_0$ acts non-degenerately,
\number\Eone\cr
$Z(\epsilon)\big|_\SM$&local contribution to $Z(\epsilon)$ from ${\CC
\subset X}$, \number\ZepsSM\cr
$\Theta$&degree-four characteristic class on $\SM$ derived from\cr
&\quad $-\ha(\phi,\phi)$ under the Chern-Weil homomorphism,
\number\BigTheta\cr
${e_{H_0}\!\big(\SM, E_0\big),\,\, e_{H_0}\!\big(\SM,
E_1\big)}$&$H_0$-equivariant Euler classes of the complex vector\cr
&\quad bundles over $\SM$ associated to $E_0$ and $E_1$,
\number\EquivEuler\cr
${\Delta_G = \dim G}$&dimension of $G$, \number\DeltaG\cr
${\Delta_{G_\lambda} = \dim G_\lambda}$&dimension of ${G_\lambda
\subseteq G}$, \number\DeltaGlam\cr
${\Ra_\ell,\, \Rb_\ell,\, \Rc_j \,\in\,\pi_1(M)}$&generators derived
from cycles in the orbifold base $\Sigma$\cr
&\quad of the Seifert manfold $M$, \number\Orbgen\cr
${\Rf \,\in\, \pi_1(M)}$&generator associated to the circle fiber of the\cr
&\quad Seifert manifold $M$, \number\Fibgen\cr
${G_{\rm ad} = G/\CZ(G)}$&adjoint form of the simply-connected group
$G$, \number\Gad\cr
$\SM_0$&smooth component of $\SM$, \number\SMnought\cr
$\SM_0(C,\alpha)$&smooth component of $\SM(C,\alpha)$,
\number\SMalpha\cr
${\Rq:\SN(P;\lambda)\to \SN(P)}$&symplectic fibration of smooth moduli
spaces, \number\Romanq\cr
${\Rq:\SM_0(C,\alpha)\to \SM_0}$&symplectic fibration of smooth moduli
spaces, \number\Romanq\cr
$\check{c}_\Fg$&dual Coxeter number of $\Fg$, \number\CoxG\cr
${\rho \,=\, \ha \sum_{\beta > 0}\,\beta}$&Weyl vector of $G$,
\number\Weylv\cr
${d = |H_1(M)|}$&order of $H_1(M)$ when $M$ is a rational homology\cr
&\quad sphere, \number\HoneM\cr
${\varrho_{\rm ab}:\pi_1(M^o)\to G}$&maximally-reducible, abelian
representation\cr
&\quad of $\pi_1(M^o)$, \number\Rhoab\cr
$\SK_{{\bf p},{\bf q}}$&$({\bf p},{\bf q})$-torus knot in $S^3$,
\number\SKpq\cr
${\epsilon_r = 2\pi/(k + \check{c}_\Fg)}$&renormalized
Chern-Simons coupling, \number\EpsR\cr
${\RP = \prod_{j=1}^N a_j}$&product of orders of orbifold points 
on $\Sigma$, \number\RomanP\cr
$s(b,a)$&Dedekind sum, \number\Dedekind\cr
$\CMj$&irreducible representation of $SU(2)$ with\cr
&\quad dimension $j$, \number\SUrep\cr
${\ch_\CMj(z) = {{\sinh(j z)}/{\sinh(z)}}}$&character for the
$SU(2)$ representation $\CMj$,
\number\SUrep\cr
${\CC^{(0)} = \e{{i\pi}\over 4} \times \BR}$&diagonal contour through
the origin, \number\Czero\cr
${\CC^{(l)} = \CC^{(0)} - 4\pi i \,l\, (\RP/d)}$&shifted contour,
\number\Cell\cr
$\Res\!\big(f(z)\big)\big|_{z = z_0}$&residue of the function $f(z)$
at ${z=z_0}$, \number\Resf\cr
$Z(\epsilon;\SK_{{\bf p},{\bf q}}, \CMj)$&$SU(2)$ Wilson loop path integral
for the torus knot\cr
&\quad ${\SK_{{\bf p},{\bf q}}\subset S^3}$ decorated with the
representation $\CMj$, \number\Ztorus\cr
$Z(\epsilon;\SK_{{\bf p},{\bf q}}, \CMj)\big|_{\{0\}}$&contribution to
$Z(\epsilon;\SK_{{\bf p},{\bf q}}, \CMj)$ from the trivial\cr
&\quad connection on $S^3$, \number\Ztriv\cr
$Z(\epsilon;\SK_{{\bf p},{\bf q}}, \CMj)_{\rm res}$&residue contribution to
$Z(\epsilon;\SK_{{\bf p},{\bf q}}, \CMj)$, \number\Zres\cr
$V_\SK(\Rt)$&Jones polynomial of the knot $\SK$, \number\Jones\cr
${\Rt = \exp{\!\left[- 2 \pi i/(k+2)\right]}}$&formal variable in
$V_\SK$, \number\tVariable\cr
${\SF^\alpha = \CH \times_{H_0^{\alpha}} \!\left(\Fh^\perp \oplus
\CE_1^{\alpha}\right)}$&symplectic fiber describing a neighborhood 
of\cr
&\quad ${\CO_\alpha \subset \bar\CA_\alpha}$, \number\SFalpha\cr
${H_0^\alpha = U(1)_\RR \times G_\alpha \times U(1)_\RZ}$&stabilizer
in $\CH$ of a point in $\CO_\alpha$, \number\BigHalpha\cr
${\Fh_0^\alpha \cong \BR \oplus \Fg_\alpha \oplus \BR}$&Lie algebra of
$H_0^\alpha$, \number\Littlehalpha\cr
$\big(\CE_0^\alpha,\, \CE_1^\alpha\big)$&infinite-dimensional complex
vector spaces\cr
&\quad which determine $\SF^\alpha$, \number\CurlyEalpha\cr
${\Fh^\perp = \Fh \ominus \Fh_0^\alpha \ominus
\CE_0^\alpha}$&orthocomplement to ${\Fh_0^\alpha \oplus \CE_0^\alpha}$
inside $\Fh$, \number\hyperpal\cr
${\big(\CE_0,\, \CE_1\big)}$&infinite-dimensional complex vector 
spaces\cr
&\quad which describe a neighorhood of ${\{0\} \in \bar\CA}$, 
\number\CurlyEs\cr
${\CN_\alpha = \bigoplus_{t \ge 1} \Big[\Fg^{(1,0)}_{t}
\oplus \Fg^{(1,0)}_{-t}\Big]}$&graded normal bundle to $\CO_\alpha$
inside $L\CO_\alpha$, \number\CurlyN\cr
${\bar\CN_\alpha = \bigoplus_{t \ge 1} \Big[\Fg^{(0,1)}_{t}
\oplus \Fg^{(0,1)}_{-t}\Big]}$&conjugate of $\CN_\alpha$, \number\CurlyNbar\cr
$Z\big(\epsilon;C,R\big)\big|_{\CO_\alpha/G}$&local contribution to the
Seifert loop path integral from\cr
&\quad the point ${\{\varrho_{\rm ab}\}\cong \CO_\alpha/G}$,
\number\ZOalphaG\cr
${e(\bar\CA),\, e(L\CO_\alpha)}$&equivariant Euler classes associated 
to the normal\cr
&\quad directions to $\CO_\alpha$ inside ${\bar\CA \times \epsilon L\CO_\alpha}$,
\number\eqEulers\cr
$\eta_0(0)$&adiabatic eta-invariant on the Seifert manifold $M$,
\number\Etainv\cr
${\rho^{[\alpha]} \,=\, \ha\!\lower
1.0ex\hbox{$\sum\atop{(\beta,\alpha)>0}$}\!\beta}$
&generalized Weyl vector, \number\Rhoalpha\cr
${\A_\alpha = \sum_{w \in \FW}\,(-1)^w\,\e{w\cdot\alpha}}$&alternating
sum of exponentials, \number\BigA\cr
$\ch_R$&character of the representation $R$ of $G$, \number\ChR\cr
$\B_\alpha$&generalized Weyl denominator, \number\BigB\cr
$\tilde\alpha$&Atiyah two-framing on $M$, \number\Atiyahtwo\cr
$\tilde\beta$&Seifert two-framing on $M$, \number\Seiferttwo\cr
$\lk(C,C')$&linking number of curves $C$ and $C'$\cr
&\quad in the integral homology sphere $M$, \number\LkCC\cr
${{\bf S},\,{\bf T}}$&standard generators of $SL(2,\BZ)$,
\number\SLtwoZ\cr
$\SV^R$&associated universal bundle with fiber $R$\cr
&\quad over ${\Sigma \times \SN(P)}$, \number\SVR\cr
$\ch(E)$&Chern character of a complex vector bundle $E$,
\number\ChernCh\cr
$\SH(k)$&Hilbert space of Chern-Simons theory at level $k$\cr
&\quad on $\Sigma$, \number\CurlyHk\cr
$\SL_0$&line bundle generating the Picard group of $\SN(P)$,
\number\CurlyLnought\cr
${\Omega_0 = \Omega/4\pi^2}$&first Chern class of $\SL_0$,
\number\BigOmnought\cr
$\SH(k;\alpha)$&Hilbert space of Chern-Simons theory at level $k$\cr
&\quad with a Wilson line decorated by $\alpha$ puncturing
$\Sigma$, \number\CurlyHkalpha\cr
$\SL^{(k)}_\alpha$&prequantum line bundle over $\SN(P;\lambda)$ for 
${\lambda=\alpha/k}$, \number\CurlyLalpha\cr
$\chi(X,E)$&Euler character of a holomorphic vector bundle $E$\cr
&\quad over a complex manifold $X$, \number\HolEuler\cr
$\Td(X)$&Todd class of $X$, \number\ToddCl\cr
$\chi_{\rm top}$&topological Euler character, \number\TopEuler\cr
${[d\mu]\,\in\,H^*\!\big(\SM_0(C,\alpha)\big)}$&de Rham class on
$\SM_0(C,\alpha)$ describing $Z(\epsilon;C,R)\big|_{\SM_0(C,\alpha)}$\cr
&\quad under localization, \number\Dmu\cr
${\Theta = -\ha\left(\Ru,\Ru\right)}$&degree-four characteristic
class of $\SV_p$, \number\BigThetaMC\cr
$\widehat{A}(X)$&$\widehat{A}$-genus of a complex manifold $X$,
\number\AroofX\cr
${\epsilon_{\rm ym} = g_{\rm ym}^2}$&two-dimensional Yang-Mills
coupling parameter, \number\Epsym\cr
}

\appendix{B}{Vanishing Lemma for a Gaussian Sum}

In this appendix, we consider the Gaussian sum which appears in
$Z\big(\epsilon;\SK_{{\bf p},{\bf q}},\CMj\big)_{\rm res}$ in \JPSUM,
\eqn\BIGI{\eqalign{
I \,&=\, \sum_{t=1}^{2{\bf p}{\bf q} -
1} \, (-1)^{t (j + 1)}\,\sin{\!\left({\pi t}\over{\bf
p}\right)}\,\sin{\!\left({\pi t}\over{\bf
q}\right)}\,\exp{\!\left({{-i \pi k_r}\over{2 {\bf p} {\bf q}}} \, 
t^2\right)}\,,\cr
k_r \,&\equiv\, k+2\,.\cr}}
We recall that ${\bf p}$ and ${\bf q}$ are positive, relatively-prime
integers which label the torus knot $\SK_{{\bf p},{\bf q}}$, and $j$
is the dimension of the irreducible $SU(2)$ representation $\CMj$.

Our goal is now to show that ${I = 0}$.  The proof is elementary, but
as often the case with arithmetic arguments of this sort, the proof does 
require a tiny bit of work.  For instance, $I$ need not vanish if
${\gcd({\bf p},{\bf q}) > 1}$, as one can easily verify in examples.

To start, since ${\bf p}$ and ${\bf q}$ are relatively-prime, there exist 
unique integers ${\bf r}$ and ${\bf s}$ such that 
\eqn\RS{ {\bf p}{\bf s} \,-\, {\bf q}{\bf r} \,=\, 1\,,\qquad 
0 < {\bf s} < {\bf q}\,,\qquad 0 < {\bf r} < {\bf p}\,,}
or 
\eqn\RSII{ {{\bf s}\over{\bf q}} \,-\, {{\bf r}\over{\bf p}} \,=\,
{1\over{{\bf p}{\bf q}}}\,.}
Hence the sum in \BIGI\ can be immediately rewritten as 
\eqn\BIGII{
I \,=\, \sum_{t=0}^{2{\bf p}{\bf q} -
1} \, (-1)^{t (j + 1)}\,\sin{\!\left({\pi t}\over{\bf
p}\right)}\,\sin{\!\left({\pi t}\over{\bf
q}\right)}\,\exp{\!\left({{-i \pi k_r {\bf s}}\over{2 {\bf q}}} \, 
t^2\right)}\,\exp{\!\left({{i \pi k_r {\bf r}}\over{2 {\bf p}}} \, 
t^2\right)}\,.}
We include the summand for ${t=0}$ without loss (since that term trivially
vanishes), and we note that the summand of \BIGII\ is otherwise manifestly
periodic in $t$ with period $2 {\bf p} {\bf q}$.  

Let us expand each sine in \BIGII\ as a sum of exponentials, so that  
\eqn\BIGIII{ I \,=\, I^{(+ +)} \,+\, I^{(- -)} \,-\, I^{(+ -)} \,-\,
I^{(- +)}\,,}
where
\eqn\BIGIPM{
I^{(\pm \pm)} \,=\,\sum_{t\,\in\,\BZ_{2 {\bf p}{\bf q}}} \,
-{1\over 4}\,(-1)^{t (j+1)}\,\exp{\!\left({{-i\pi k_r {\bf s}}\over{2
{\bf q}}} t^2 \,\pm\, {{i\pi}\over{\bf q}} t\right)} \,
\exp{\!\left({{i\pi k_r {\bf r}}\over{2 {\bf p}}} t^2 
\,\pm\, {{i\pi}\over{\bf p}} t\right)}\,.}
Because the summand in $I^{(\pm \pm)}$ depends only on the value of $t$
modulo $2{\bf p}{\bf q}$, we regard the sum in \BIGIPM\ as running
over elements of the cyclic group $\BZ_{2{\bf p}{\bf q}}$.  We note
that trivially 
\eqn\BIGIPME{ I^{(+ +)} \,=\, I^{(- -)}\,,\quad\quad I^{(+ -)} \,=\,
I^{(- +)}\,,}
where these identifications follow by sending ${t \mapsto -t}$ in \BIGIPM.
To prove that $I$ vanishes, we will similarly demonstrate that 
\eqn\BIGIPMEI{ I^{(+ +)} \,=\, I^{(+ -)}\,.}

Our argument depends on the value of $k_r$ mod\foot{The roles of ${\bf
p}$ and ${\bf q}$ are entirely symmetric, but we will focus on ${\bf
p}$ for concreteness.} ${\bf p}$.  Without loss, we assume that ${{\bf
p} > 1}$, since $I$ vanishes trivially if ${{\bf p}=1}$.  We now
consider two cases, depending upon whether or not $k_r$ is relatively-prime
to ${\bf p}$.  

Let us first assume that 
\eqn\GCDKP{ \gcd(k_r, {\bf p}) \,=\, d \,>\, 1\,,}
so that ${\bf p}$ and $k_r$ are {\sl not} relatively-prime.  We
therefore introduce the reduced integer ${\bar{\bf p} \,=\, {{\bf p} / d}}$.
On one hand, a overall shift ${t \mapsto t \,+\, 2 \, \bar {\bf p} \, {\bf
q}}$ in the sum over elements of $\BZ_{2{\bf p}{\bf q}}$ trivially
leaves ${I^{(\pm \pm)}}$ invariant.  On the other hand, under this
shift the summand of \BIGIPM\ transforms with the phase ${\zeta \,=\,
\exp{\!(\pm 2 \pi i \, {\bf q} / d)}}$, where the sign in $\zeta$ is
determined by the sign in the term $\exp{\!(\pm i \pi t/ {\bf p})}$ in
\BIGIPM.  Thus 
\eqn\GCDKPII{ I^{(\pm \pm)} \,=\, \zeta \cdot I^{(\pm \pm)}\,.}
Finally, since $d$ divides ${\bf p}$ and $\gcd({\bf p},{\bf q}) = 1$,
the integers $d$ and ${\bf q}$ must themselves be relatively-prime.  Thus
${\zeta \neq 1}$, and \GCDKPII\ implies that each of the four summands
$I^{(\pm \pm)}$ separately vanishes when \GCDKP\ holds.

We are left to consider the case that ${\bf p}$ and $k_r$ are
relatively-prime, 
\eqn\GCDKPII{ \gcd(k_r, {\bf p}) \,=\, 1\,.}
In this case, there again exist positive integers $a$ and $b$ such
that 
\eqn\AB{ a \, k_r \,-\, b \, {\bf p} \,=\, 1\,,\qquad 0 < a < {\bf
p}\,,\qquad 0 < b < k_r\,.}
That is, $a$ defines an inverse for $k_r$ mod ${\bf p}$, so that
${k_r^{-1} \equiv a \, \mod {\bf p}}$.  Similarly, according to \RS, ${{\bf
q}^{-1} \equiv -{\bf r} \, \mod {\bf p}}$.  

We now consider shifting the summation variable $t$ in $I^{(+ +)}$ by ${t
\mapsto t \,+\, 2\,{\bf q}\,a}$.  Under this shift, the factors
$(-1)^{t(j+1)}$ and $\exp{\!\left[\left(-i \pi k_r {\bf s} \, t^2
\,+\, 2 \pi i \, t\right)\!/2 {\bf q}\right]}$ appearing in the summand of
\BIGIPM\ are invariant, but 
\eqn\SHFEXP{ \exp{\!\left({{i\pi k_r {\bf r}}\over{2 {\bf p}}} t^2 
\,+\, {{i\pi}\over{\bf p}} t\right)} \,\longmapsto\, \zeta(t) 
\cdot \exp{\!\left({{i\pi k_r {\bf r}}\over{2 {\bf p}}} t^2 
\,+\, {{i\pi}\over{\bf p}} t\right)}\,,}
where 
\eqn\NEWZET{\eqalign{
\zeta(t) \,&=\, \exp{\!\left[{{2 \pi i}\over{\bf p}}\!\left(k_r \, {\bf
r} \, {\bf q} \, a \, t \,+\,  k_r \, {\bf r} \, {\bf q}^2 \, a^2
\,+\, {\bf q} \, a\right)\right]}\,,\cr
&= \exp{\!\left[-{{2 \pi i}\over{\bf p}}\,t\right]}\,.\cr}}
In passing to the second line of \NEWZET, we recall that
${k_r \, a \equiv 1 \, \mod {\bf p}}$ and ${{\bf q} \, {\bf r} \equiv -1 \,
\mod {\bf p}}$.  As a result,
\eqn\BIGIPP{\eqalign{
I^{(+ +)} \,&=\, \sum_{t\,\in\,\BZ_{2 {\bf p}{\bf q}}} \,
-{1\over 4}\,(-1)^{t (j+1)}\,\zeta(t)\,\exp{\!\left({{-i\pi k_r {\bf s}}\over{2
{\bf q}}} t^2 \,+\, {{i\pi}\over{\bf q}} t\right)} \, 
\exp{\!\left({{i\pi k_r {\bf r}}\over{2 {\bf p}}} t^2 
\,+\, {{i\pi}\over{\bf p}} t\right)}\,,\cr
&=\,\sum_{t\,\in\,\BZ_{2 {\bf p}{\bf q}}} \,
-{1\over 4}\,(-1)^{t (j+1)}\,\exp{\!\left({{-i\pi k_r {\bf s}}\over{2
{\bf q}}} t^2 \,+\, {{i\pi}\over{\bf q}} t\right)} \, 
\exp{\!\left({{i\pi k_r {\bf r}}\over{2 {\bf p}}} t^2 
\,-\, {{i\pi}\over{\bf p}} t\right)}\,\cr
&=\, I^{(+ -)}\,.\cr}}
So via \BIGIII\ and \BIGIPME, the Gaussian sum $I$ vanishes when ${\bf
p}$ and ${\bf q}$ are relatively-prime.

\appendix{C}{The Adiabatic Eta-Invariant on a Seifert Manifold}

In this appendix we explain how the adiabatic eta-invariant
$\eta_0(0)$ is defined and computed on a general Seifert
three-manifold $M$.  Although the discussion here will be reasonably
self-contained, it will also be only a sketch, and we refer the reader
to \S $5.2$ of \BeasleyVF\ along with \Nicolaescu\ for further details.

As usual for computations of functional determinants, we introduce an
eta-function \APS\ to define the phase of the formal expression in
\DETSPH,
\eqn\BIGEA{
e\big(\bar\CA\big) \,=\, \det\!\left({{\psi} \over {2
\pi}}\Big|_{\CE_0}\right) \det\!\left({{\psi} \over {2
\pi}}\Big|_{\CE_1}\right)^{-1}\,,\qquad \psi \,\equiv\, (p,\phi,a) 
\,\in\, \BR\oplus\Fg_\alpha\oplus\BR\,.}
Concretely, the generator $\psi$ acts on elements of both $\CE_0$ and
$\CE_1$ as the first-order differential operator\foot{We note that the
sign of $\phi$ in $\CD_{(p,\phi)}$ is opposite to the corresponding sign in
\BeasleyVF, a difference which can be traced to our convention in \DGA.}
\eqn\OPPSIC{ \CD_{(p,\phi)} \,=\, p \, \lie_\RR \,+\, [\phi,\,\cdot\,]\,.}
If $\CL$ is the line bundle over $\Sigma$ associated to the Seifert
manifold $M$, then $\lie_\RR$ acts on sections of $\CL^t$ with
eigenvalue $-2\pi i \, t$.  Hence $e(\bar\CA)$ can be written concretely
as a product over the non-zero eigenvalues of $\lie_\RR$ as 
\eqn\DAIII{ e(\bar\CA) \,=\, \prod_{t \neq 0} \, \det\left[\left(-i t p +
{{\left[ \phi, \,\cdot\, \right]} \over {2
\pi}}\right)\Bigg|_{\Fg}\right]^{\chi(\CL^t)}.}
In \DAIII\ we correctly account for the ratio of determinants associated to
$\CE_0$ and $\CE_1$ by introducing the Euler character
\eqn\ELLT{ \chi(\CL^t) \,=\, \dim_\BC H^0_{\bar\partial}(\Sigma,\CL^t) - 
\dim_\BC H^1_{\bar\partial}(\Sigma,\CL^t)\,.}

Without loss, we take $\phi$ to lie in a Cartan subalgebra ${\Ft
\subset \Fg}$.  As standard, we diagonalize the adjoint action of
$\phi$ on $\Fg$ by introducing the roots ${\beta\in\Ft^*}$.  By
definition, if $x_\beta$ is an element of the associated 
rootspace ${\Fe_\beta\subset\Fg_\BC}$, then ${[\phi, x_\beta] \,=\, i\,
\langle\beta,\phi\rangle\,x_\beta}$.  Hence in terms of the roots
$\beta$,
\eqn\DETG{\eqalign{
\det{\left( -i t p + {{\left[ \phi, \,\cdot\, \right]} \over {2
\pi}}\right)}\Bigg|_{\Fg} \,&=\, \left(-i t p\right)^{\Delta_G} \,
\prod_{\beta} \left(1 - {{\langle\beta, \phi\rangle} \over {2 \pi t
p}}\right)\,,\qquad\qquad \Delta_G \,=\, \dim G\,,\cr
&=\, \left(-i t p\right)^{\Delta_G} \, \prod_{\beta > 0} \left(1 -
\left( {{\langle\beta, \phi\rangle} \over {2 \pi t
p}}\right)^2\right)\,,\cr}}
where in the second line of \DETG\ we rewrite the product over all
roots as a product over only the positive roots ${\beta>0}$.  

Using \DETG, we express $e(\bar\CA)$ as 
\eqn\DAIV{ e(\bar\CA) \,=\, \exp{\!\left(-{{i \pi} \over 2}
\eta\right)} \cdot \prod_{t \ge 1} \, \left|\left(t p\right)^{\Delta_G} \,
\prod_{\beta > 0} \left(1 - \left({{\langle\beta, \phi\rangle} \over
{2 \pi t p}}\right)^2\right)\right|^{\chi(\CL^t) + \chi(\CL^{-t})}\,.}
In this appendix, we are most interested in the phase $\exp{\!(-{{i
\pi}\over 2} \eta)}$, which involves an infinite product of factors
$\pm i$.  The norm of $e(\bar\CA)$ is relatively straightforward to
evaluate, as done in \S $5.2$ of \BeasleyVF\ using zeta-function
regularization.

At least formally, the phase $\eta$ in \DAIV\ is given by the 
difference 
\eqn\BIGOPI{ \eta \, \approx \, \sum_{\lambda_{(0)} \neq
0} \, \sgn\!\big(\lambda_{(0)}\big) - \sum_{\lambda_{(1)} \neq 0} \,
\sgn\!\big(\lambda_{(1)}\big)\,,}
where $\lambda_{(0)}$ and $\lambda_{(1)}$ range over the eigenvalues
of the operator ${1 \over {2\pi i}} \CD_{(p,\phi)}$ acting respectively 
on $\CE_0$ and $\CE_1$.  These eigenvalues are real, but they do not
generally have a definite sign.  

We are careful not to write \BIGOPI\ with an equality, because the
sums appearing on the right in \BIGOPI\ are ill-defined without a
regulator.  To regulate these sums, we introduce the eta-function 
\eqn\BIGETA{ \eta_{(p, \phi)}(s) \,=\,
\sum_{\lambda_{(0)} \neq 0} \, \sgn\big(\lambda_{(0)}\big) \,
|\lambda_{(0)}|^{-s} - \sum_{\lambda_{(1)} \neq 0} \,
\sgn\big(\lambda_{(1)}\big) \, |\lambda_{(1)}|^{-s}\,.} Here $s$ is a
complex variable.  When the real part of $s$ is sufficiently
large, the sums in \BIGETA\ are absolutely convergent, so that
$\eta_{(p, \phi)}(s)$ is defined in that case.  Otherwise,
$\eta_{(p, \phi)}(s)$ is defined by analytic continuation in the
$s$-plane. Assuming that the limit ${s \to 0}$ exists, we then set
\eqn\DETA{ \eta = \eta_{(p, \phi)}(0)\,.}

As shown in \S $5.2$ of \BeasleyVF, $\eta_{(p,\phi)}(s)$ retains a
non-trivial dependence on $p$ and $\phi$ even in the limit ${s \to
0}$, due to divergences in the naive sums in \BIGOPI.  Specifically,
\eqn\BIGETAII{ \eta_{(p,\phi)}(0) \,=\, \eta_0(0) \,-\,
{{\check{c}_\Fg} \over {2 (\pi p)^2}} \left({d \over P}\right)
\Tr(\phi^2)\,,}
where $\check{c}_\Fg$ is the dual Coxeter number of the simply-laced
Lie algebra $\Fg$.  The constant $\eta_0(0)$ in \BIGETAII\ is
given by the value at ${s=0}$ of the $(p,\phi)$-independent
eta-function 
\eqn\ETAZ{\eqalign{
\eta_0(s) \,&=\, \sum_{t \neq 0} \sum_\beta \, \chi(\CL^t)
\, \sgn(t) \, |t|^{-s}\,,\cr
&=\, \sum_{t \ge 1} \sum_{\beta} \, {{\chi(\CL^t) - \chi(\CL^{-t})} \over
{t^s}}\,.\cr}}
To remedy the small omission in \BeasleyVF, we now provide a general
computation of $\eta_0(0)$, following the proof of Proposition
$1.4$ in \Nicolaescu.

\bigskip\noindent{\it Evaluating $\eta_0(0)$}\smallskip

By assumption, $\CL$ is the orbifold line bundle which describes the
Seifert manifold $M$ with Seifert invariants 
\eqn\SFRTIIC{\Big[h;n;(a_1,b_1), \ldots, (a_N,b_N)\Big]\,,\qquad\qquad
\gcd(a_j,b_j) = 1\,.}
Here $h$ is the genus of the orbifold $\Sigma$, and ${n = \deg(\CL)}$
is the degree of $\CL$.  Just as for line bundles on a smooth Riemann
surface, the Riemann-Roch theorem for orbifolds \Kawasaki\ states that
the Euler character $\chi(\CL)$ is given in terms of the degree and
the genus by 
\eqn\RRORB{ \chi(\CL) \,=\, n \,+\, 1 \,-\, h\,,}
and similarly
\eqn\RRORBII{ \chi(\CL^t) \,=\, \deg(\CL^t) \,+\, 1 \,-\ h\,.}

From \RRORBII, we see that the difference of Euler characters
appearing in \ETAZ\ is independent of $h$ and given by 
\eqn\DIFFCH{ \delta(t) \,=\, \chi(\CL^t) - \chi(\CL^{-t}) \,=\,
\deg(\CL^t) \,-\, \deg(\CL^{-t}) \,.}
If $\CL$ is a line bundle of degree $d$ over a smooth Riemann surface,
then ${\deg(\CL^t) \,=\, t d}$, and ${\delta(t) \,=\, 2 t d}$.
However, on an orbifold the degree is not generally multiplicative, 
\eqn\ORBDEG{ \deg(\CL^t) \,\neq\, t \, \deg(\CL)\,,}  
and $\delta(t)$ is a more complicated arithmetic function which we
must compute.

If $\CL$ is characterized by isotropy invariants $b_j$ on $\Sigma$ as
in \SFRTIIC, then $\CL^t$ is characterized by isotropy invariants
$b_j^{(t)}$ determined by the conditions 
\eqn\BJT{ b_j^{(t)} \,\equiv\, t \, b_j \,\mod\, a_j\,,\qquad\qquad 0
\,\le\, b_j^{(t)} \,<\, a_j\,.}
In particular,
\eqn\BJTII{ {{b_j^{(t)}}\over {a_j}} \,=\, \left\{{t \, b_j}\over
a_j\right\}\,,}
where $\{\,\cdot\,\}$ denotes the fractional\foot{For ${x\in \BR}$,
$\{x\}$ denotes the element in the interval $[0,1)$ such that ${x -
\{x\} \in \BZ}$.} part of the given argument.  

Now, unlike the degree, the first Chern class is multiplicative even
for orbifold line bundles,
\eqn\CILM{ c_1(\CL^t) \,=\, t \, c_1(\CL)\,.}  
We also recall that the first Chern class of $\CL^t$ is given by 
\eqn\CILCII{ c_1(\CL^t) \,=\, \deg(\CL^t) \,+\, \sum_{j=1}^N \,
{{b_j^{(t)}}\over {a_j}}\,.}
From \BJTII\ and \CILM\ we therefore obtain 
\eqn\DEGLT{ \deg(\CL^t) \,=\, t \, c_1(\CL) \,-\, \sum_{j=1}^N \, 
\left\{{t \, b_j}\over a_j\right\}\,,}
implying 
\eqn\DIFFCHII{ \delta(t) \,=\, 2 t \, c_1(\CL) \,-\, \sum_{j=1}^N \, 
\left[ \left\{{t \, b_j}\over a_j\right\} \,-\,  \left\{-{{t \,
b_j}\over a_j}\right\} \right]\,.}

Let us introduce another function $(\!(\,\cdot\,)\!)$ related
to $\{\,\cdot\,\}$ by 
\eqn\DBLPR{ (\!(x)\!) \,=\, \Biggr\{ \eqalign{&\{x\} - \ha \qquad \hbox{
if } x\in \BR - \BZ\cr &0 \qquad\qquad\quad\hbox{ if } x\in \BZ}\,\,.} 
Clearly ${(\!(x+1)\!) = (\!(x)\!)}$ and ${(\!(-x)\!) = - (\!(x)\!)}$.
In terms of $(\!(\,\cdot\,)\!)$, the sum in \DIFFCHII\ can be rewritten as 
\eqn\DIFFCHIII{ \delta(t) \,=\, 2 t \, c_1(\CL) \,-\, 2 \sum_{j=1}^N
\, \left(\!\left({{t \, b_j}\over {a_j}}\right)\!\right)\,.}
Thus $\eta_0(s)$ is given explicitly by 
\eqn\ETAZCII{
\eta_0(s) \,=\, 2 \Delta_G \, \sum_{t = 1}^\infty \left[
{{c_1(\CL)} \over {t^{s-1}}} \,-\, \sum_{j=1}^N
\, \left(\!\left({{t \, b_j}\over {a_j}}\right)\!\right) {1 \over
{t^s}}\right]\,.}
In passing to \ETAZCII\ from \ETAZ, we simply replace the sum
over all roots $\beta$ (including those trivial roots associated to
the Cartan subalgebra itself) by the prefactor ${\Delta_G = \dim G}$.

To evaluate $\eta_0(s)$ at ${s=0}$, we use the Hurwitz zeta-function
$\zeta(s,a)$, defined by the sum 
\eqn\HURZ{ \zeta(s, a) \,=\, \sum_{m=0}^\infty \, {1 \over {(m + a)^s}}\,.}
Here $s$ is a complex variable, and $a$ is a real parameter in the
interval ${0 < a \le 1}$.  As before, when the real part of $s$ is
sufficiently large, the sum in \HURZ\ is absolutely convergent.  Otherwise,
$\zeta(s,a)$ is defined by analytic continuation in the $s$-plane.
Clearly if ${a=1}$, then ${\zeta(s,1) \equiv \zeta(s)}$ reduces to the 
standard Riemann zeta-function.

The Hurwitz zeta-function plays a role in evaluating \ETAZCII\ for the
following elementary reason.  If $f$ is any periodic function with integral
period ${L \in \BZ}$, then 
\eqn\HURZID{\eqalign{
\sum_{t=1}^\infty \, {{f(t)}\over{t^s}} \,&=\, \sum_{l=1}^L \,
\sum_{m=0}^\infty \, {{f(l \,+\, m \, L)} \over{(l \,+\, m \, L)^s}}\,,\cr
&=\, \sum_{l=1}^L \, {{f(l)}\over{L^s}} \, \zeta(s, {l \over L})\,.}}
In passing to the second line in \HURZID, we use the assumption 
${f(l \,+\, m \, L) = f(l)}$ and the definition of $\zeta(s,a)$ in \HURZ.

Since the function ${f_j(t) = \left(\!\left({{t\,b_j} \over
{a_j}}\right)\!\right)}$ is invariant under the shift ${t \mapsto t +
a_j}$, we apply the identity in \HURZID\ to obtain 
\eqn\ETAZIII{ \eta_0(s) \,=\, 2 \Delta_G \!\left[ c_1(\CL) \, \zeta(s-1)
\,-\, \sum_{j=1}^N \, \sum_{l=1}^{{a_j-1}} \, \left(\!\left({{l\,
b_j}\over{a_j}}\right)\!\right) {1 \over {a_j^s}} \, \zeta\big(s, {l\over
{a_j}}\big)\right]\,.}
Naively, the sum in \ETAZIII\ runs over ${l=1,\ldots,a_j}$, but the
summand vanishes identically when ${l=a_j}$, so we omit that term.

To evaluate $\eta_0(s)$ at ${s=0}$, we recall that ${\zeta(-1) =
-1/12}$ and ${\zeta(0,a) = 1/2 - a}$.  See for instance \S $12$ of
\ApostolII\ for a proof of the latter identity.  Consequently, again
using the function $(\!(\,\cdot\,)\!)$ in \DBLPR,
\eqn\ETAZIV{ \eta_0(0) \,=\, {\Delta_G \over 6} \!\left[ -c_1(\CL) 
\,+\, 12 \sum_{j=1}^N \, \sum_{l=1}^{a_j-1} \left(\!\left({l \over
{a_j}}\right)\!\right)
\left(\!\left({{l\,b_j}\over{a_j}}\right)\!\right)\right]\,.}
Finally, as explained for instance in \S $3$ of \Apostol, the Dedekind
sum $s(b,a)$ can be described not only trigonometrically as in \DEDEK\
but also arithmetically in terms of $(\!(\,\cdot\,)\!)$ as 
\eqn\DEDEKII{ s(b, a) \,=\, \sum_{l=1}^{a-1} \left(\!\left({l \over
a}\right)\!\right) \left(\!\left({{l \, b}\over a}\right)\!\right)\,.}
Hence 
\eqn\ETAZV{ \eta_0(0) \,=\, {\Delta_G \over 6} \!\left[ -c_1(\CL) 
\,+\, 12 \sum_{j=1}^N \, s(b_j, a_j)\right]\,.}
For the Seifert homology spheres considered in Section $7.2$,
${c_1(\CL) = d/\RP}$, so we obtain the formula for $\eta_0(0)$ in
\ETAZII.

\appendix{D}{A Note on the Extended Denominator Formula}

In this appendix, we establish the slight extension of the Weyl
denominator formula needed to identify the two expressions for
$\B_\alpha(\phi)$ in \BIGBB\ and \BIGBII.  

Briefly, the extended denominator formula states 
\eqn\ARHOALII{ \sum_{w\in\FW_\alpha} \, (-1)^w \, \e{\!\langle
w\cdot(\alpha+\rho),\,\phi\rangle} \,=\, \e{\!\langle\alpha +
\rho^{[\alpha]},\,\phi\rangle} \cdot \prod_{\beta_\perp > 0} \left[ 2
\sinh\!\left({{\langle\beta_\perp,\phi\rangle}\over
2}\right)\right],\qquad \phi \in \Ft\,.}
Here ${\alpha>0}$ is a positive weight of $G$, $\rho$ is the standard Weyl
vector given by half the sum of the positive roots of $G$, and the
alternating sum on the left side of \ARHOALII\ runs over elements of
the Weyl group $\FW_\alpha$ of the stabilizer ${G_\alpha \subseteq G}$
of $\alpha$ under the adjoint action of $G$.  If $\alpha$ is regular
so that ${G_\alpha = T}$ is a maximal torus, then $\FW_\alpha$ is
trivial by convention.

On the right side of \ARHOALII, $\rho^{[\alpha]}$ is defined as in
\RHOAL\ by
\eqn\RHOALAD{
\rho^{[\alpha]} \,=\, \ha \, \sum_{(\beta_{+},\alpha)>0} \beta_{+}\,,}
where the sum runs over all roots $\beta_{+}$ of $G$ such that
${(\beta_{+},\alpha)>0}$.  If $\alpha$ is regular, $\rho^{[\alpha]}$ is
therefore the standard Weyl vector $\rho$.  At the other extreme, if
${\alpha = 0}$, then ${\rho^{[\alpha]} = 0}$ as well.  Finally, the
product on the right side of \ARHOALII\ runs over all positive roots
${\beta_\perp > 0}$ of $G$ such that ${(\beta_\perp,\alpha) = 0}$.
For future reference, we note that the roots $\beta_\perp$ are
precisely the roots of $G_\alpha$.  When $\alpha$ is regular, the
product over ${\beta_\perp>0}$ is taken to be $1$.  With this
convention, the identity in \ARHOALII\ holds trivially for regular
weights.

At the opposite extreme, for ${\alpha = \rho^{[\alpha]} = 0}$, the
identity in \ARHOALII\ reduces to the standard product formula for the
Weyl denominator $\A_\rho$,
\eqn\ARHOAD{ \A_\rho(\phi) \,=\,
\sum_{w\in\FW}\,(-1)^w\,\e{\!\langle w\cdot\rho,\,\phi\rangle} \,=\, 
\prod_{\beta > 0} \left[2 \sinh\!\left({{\langle\beta,\phi\rangle}\over
2}\right)\right],\qquad \phi \in \Ft\,.}
where the sum in \ARHOAD\ runs over the full Weyl group $\FW$ of $G$
and the product runs over all positive roots ${\beta > 0}$.

Our proof of the extended denominator formula \ARHOALII\ is an
immediate generalization of the proof of \ARHOAD\ offered in Lemma $24.3$
of \Fulton.  Nonetheless, for sake of completeness we reproduce that
argument here.

Let us first introduce some temporary notation for the left and right
sides of \ARHOALII,
\eqn\ARHOALI{\eqalign{
\B_\alpha \,&=\, \sum_{w\in\FW_\alpha} \, (-1)^w \,
\e{\!w\cdot(\alpha+\rho)}\,,\cr  
\B \,&=\, \e{\!\left(\alpha + \rho^{[\alpha]}\right)} \cdot \prod_{\beta_\perp
> 0} \left[ 2 \sinh\!\left({{\beta_\perp}\over 2}\right)\right],}}
where for sake of brevity we omit the variable ${\phi \in \Ft}$
henceforth.  

Clearly $\B_\alpha$ is alternating under the action of $\FW_\alpha$.
We also observe that $\B$ is alternating under $\FW_\alpha$.  By
definition, $\alpha$ and $\rho^{[\alpha]}$ are invariant under
$\FW_\alpha$, and the product ${\prod_{\beta_\perp > 0} \left[ 2
\sinh\!\left(\beta_\perp/2\right)\right]}$ otherwise appearing in $\B$
is alternating.  Here we recall that $\FW_\alpha$ is generated by
reflections in the simple roots of $G_\alpha$.  Hence the quotient
$\B_\alpha/\B$ is invariant under $\FW_\alpha$.

Next, we observe trivially that 
\eqn\PRODB{\eqalign{
\B \,&=\, \e{\!\left(\alpha + \rho^{[\alpha]}\right)} \cdot \left[
\exp{\!\left(\ha\sum_{\beta_\perp>0} \beta_\perp\right)} \cdot
\prod_{\beta_\perp > 0} \left(1 \,-\, \e{\!-\beta_\perp}\right)\right]\,,\cr
&=\, \e{\!(\alpha + \rho)} \cdot \prod_{\beta_\perp>0} \left(1 \,-\,
\e{\!-\beta_\perp}\right),}}
where we use the identity 
\eqn\IDRHOA{ \rho \,=\, \rho^{[\alpha]} \,+\, \ha
\sum_{\beta_\perp > 0} \beta_\perp\,.}
Inverting the latter expression for $\B$ in \PRODB, we find 
\eqn\PRODBII{ \B^{-1} \,=\, \e{\!-(\alpha + \rho)} \cdot
\prod_{\beta_\perp > 0} \left[\sum_{m=0}^\infty \, \e{\!-m
\beta_\perp}\right],}
where we use the standard series presentation for ${1/\big(1 -
\e{-\beta_\perp}\big)}$.  So as a formal series, the quotient
$\B_\alpha/\B$ becomes 
\eqn\QUOTB{ {{\B_\alpha}\over\B} \,=\, \sum_{w \in \FW_\alpha} 
(-1)^w \, \e{\!(w\cdot\rho \,-\, \rho)} \cdot \prod_{\beta_\perp>0}
\left[\sum_{m=0}^\infty \, \e{\!-m\beta_\perp}\right]\,.}
In arriving at \QUOTB, we use that ${w\cdot(\alpha+\rho) \,=\, \alpha
+ w\cdot\rho}$, as $\alpha$ is invariant under $\FW_\alpha$.

We now observe that ${(w\cdot\rho - \rho) < 0}$ is strictly negative
for all non-trivial elements ${w \in \FW_\alpha}$.  Thus the
expression on the right in \QUOTB\ is a sum of exponentials, all of
whose arguments, if non-trivial, are strictly negative weights of
$G_\alpha$.\foot{Since $\alpha$ is fixed by $\FW_\alpha$, the inner-product 
${(\alpha,\,w\cdot\rho - \rho)}$ vanishes for ${w\in\FW_\alpha}$, so
the difference $\left(w\cdot \rho - \rho\right)$ is
indeed a weight of the subgroup ${G_\alpha \subseteq G}$.}  On the
other hand, the sum of exponentials in \QUOTB\ is by construction
invariant under $\FW_\alpha$.  Since $\FW_\alpha$ acts simply
transitively on the Weyl chambers of $G_\alpha$, these two
observations are consistent only if ${\B_\alpha/\B}$ is constant as a
function on $\Ft$.

Finally, comparing the leading terms of the sum and product in
\ARHOALI, we easily verify that ${\B_\alpha = \B}$.

\listrefs

\end